\documentclass[12pt]{article}

\usepackage[utf8]{inputenc}
\usepackage[T1]{fontenc}
\usepackage{babel}

\usepackage{cite}
\usepackage{float} 
\usepackage{graphicx}
\usepackage{amsfonts}
\usepackage{amsmath}
\usepackage{longtable}
\usepackage{mathtools}
\usepackage{physics}
\usepackage{ccaption}
\usepackage{subcaption}
\usepackage{placeins}
\usepackage{csquotes}
\usepackage{comment}
\input{epsf}

\usepackage[colorlinks,linkcolor=blue,citecolor=red,hyperindex]{hyperref}

\newcommand{\half}{\frac{1}{2}}

\numberwithin{equation}{section}

\def\br{\begin{eqnarray}}
\def\er{\end{eqnarray}}
\def\be{\begin{equation}}
\def\ee{\end{equation}}

\def\({\left(}
\def\){\right)}
\newcommand{\keywords}[1]{%
  \par\medskip\noindent\textbf{Keywords:} #1\par}

\relax

\def\pa{\partial}

\def\tp0{\Theta_{+}^{(0)}}
\def\tm0{\Theta_{-}^{(0)}}

\def\u2{\mid u\mid^2}

\def\vp{\varphi}

\def\rlx{\relax\leavevmode}
\def\inbar{\vrule height1.5ex width.4pt depth0pt}
\def\IZ{\rlx\hbox{\sf Z\kern-.4em Z}}
\def\IR{\rlx\hbox{\rm I\kern-.18em R}}
\def\IC{\rlx\hbox{\,$\inbar\kern-.3em{\rm C}$}}
\def\IN{\rlx\hbox{\rm I\kern-.18em N}}
\def\IO{\rlx\hbox{\,$\inbar\kern-.3em{\rm O}$}}
\def\IP{\rlx\hbox{\rm I\kern-.18em P}}
\def\IQ{\rlx\hbox{\,$\inbar\kern-.3em{\rm Q}$}}
\def\IF{\rlx\hbox{\rm I\kern-.18em F}}
\def\IG{\rlx\hbox{\,$\inbar\kern-.3em{\rm G}$}}
\def\IH{\rlx\hbox{\rm I\kern-.18em H}}
\def\II{\rlx\hbox{\rm I\kern-.18em I}}
\def\IK{\rlx\hbox{\rm I\kern-.18em K}}
\def\IL{\rlx\hbox{\rm I\kern-.18em L}}

\def\one{\hbox{{1}\kern-.25em\hbox{l}}}
\def\0#1{\relax\ifmmode\mathaccent''7017{#1}%
B        \else\accent23#1\relax\fi}

\begin{document}

\begin{titlepage}
        \vspace*{-1cm}

        \vskip 1.5cm

        \vspace{.2in}
        \begin{center}
                {\large\bf Platonic self-dual solitons in a (1+1)-dimensional $\IC\IP^1$ model}
        \end{center}

        \vspace{.5cm}

        \begin{center}
                L. A. Ferreira$^{\dagger ,}$\footnote{laf@ifsc.usp.br}, and G. Marcelino$^{\dagger\; \ddag ,}$\footnote{giumarcelino@fisica.ufmt.br}

                \vspace{.2 in}
                \small

                \par \vskip .2in \noindent
                $^{\dagger}$Instituto de F\'\i sica de S\~ao Carlos; IFSC/USP;\\
                Universidade de S\~ao Paulo, USP  \\
                CEP 13566-590, S\~ao Carlos-SP, Brasil\\
				\par \vskip .2in \noindent
                $^{\ddag}$Instituto de F\'\i sica; IF/UFMT;\\
                Universidade Federal de Mato Grosso, UFMT  \\
                CEP 78060-900, Cuiabá-MT, Brasil\\

                \vskip 2cm

\begin{abstract}
We construct a family of self-dual $(1+1)$-dimensional $\IC\IP^1$
sigma models whose potentials $V$ are determined from  pre-potentials $U$ by Platonic geometry.
Using the round metric on $\mathbb{S}^2\simeq\IC\IP^1$ and the
generalized self-duality formalism, we build symmetry-invariant
pre-potentials $U$ from the unit vertices of each Platonic solid intersecting 
the unit circumsphere.
The regular-lattice sites associated with vertex, face-center, and
edge-midpoint directions are critical points of $U$ and degenerate
vacua of the scalar potential $V$.
Dual solids share the same regular lattice
but generate different pre-potentials, charges, potentials, and
metric-gra\-dient flows. Numerical integration reveals distinct self-dual
trajectories and energy-density profiles, including multi-peak
structures for paths passing near saddle points, while solutions with
the same endpoint difference $\Delta U$ have the same total energy.
The positive maxima of $V$ are also organized into symmetry orbits.
This construction provides a geometric prescription for controlling
the vacuum structure and self-dual sectors of $\IC\IP^1$ models with a
potential.

\medskip
\end{abstract}

                \normalsize
        \end{center}
\end{titlepage}

\tableofcontents

\section{Introduction}
\label{sec:introduction}
\setcounter{equation}{0}

Nonlinear field theories support localized configurations whose
properties cannot, in general, be inferred from small fluctuations
around a single vacuum. The Fermi--Pasta--Ulam recurrence and the
solitary-wave solutions of the Korteweg--de Vries equation were early
examples of the structured behavior produced by nonlinear dynamics
\cite{FPU,KdV,zabusky1965}. In field theory, solitons provide an
important class of such configurations. When their existence is tied
to nontrivial boundary conditions or topology, they appear as
topological solitons, including kinks, vortices, monopoles, Skyrmions,
and instantons in different spacetime dimensions
\cite{shnirTop,TopSol,lafBPS}. Perturbative expansions around a vacuum
remain useful for many questions, but they do not in general describe
finite-amplitude configurations interpolating between distinct vacua.

A particularly tractable class of topological configurations is formed
by self-dual, or Bogomolny, solutions \cite{bogbound}. Solutions of the
first-order self-duality equations saturate a lower
bound for the static energy and satisfy the corresponding
second-order Euler--Lagrange equations. The
generalized self-duality framework of \cite{lafBPS} makes it possible
to construct theories with self-dual sectors starting from a
topological charge. In $(1+1)$ dimensions this procedure applies to
models with several real scalar fields and extends the familiar
single-field construction \cite{laf_1+1}. The first-order formulation
is especially useful when the resulting nonlinear equations do not
admit known analytic solutions.

An essential ingredient of this construction is the choice of the
pre-potential $U$, since its critical points determine the vacua and
its differences between asymptotic configurations determine the
self-dual energies. Previous constructions have used algebraic data to
organize such critical points \cite{laf_1+1}. Here we instead use the
geometry of the target space itself. We consider
$\mathbb{S}^2\simeq\IC\IP^1$ with its round metric and use the
rotational symmetry of the Platonic solids to define finite regular
lattices on the sphere. Their sites are the intersections of the
symmetry axes with $\mathbb{S}^2$, equivalently the directions of the
vertices, face centers, and edge midpoints. Dual Platonic solids share
the same set of symmetry axes and therefore the same regular lattice
\cite{coxeter_regular}.

For each solid, we construct a smooth pre-potential from its unit
vertex vectors and then express it in two independent fields by
stereographic projection. The induced metric is the standard
$\IC\IP^1$ metric, and the scalar potential is fixed by generalized
self-duality. The rotational invariance of $U$ makes it constant on
symmetry orbits and places the regular-lattice sites among its critical
points. In the individual models studied here, the vertices are
minima, the dual vertices are maxima, and the remaining edge-midpoint
sites are saddle points. All of these points are degenerate vacua of the
physical potential $V$. The self-dual charge is
$Q=U(+\infty)-U(-\infty)$; in the present models it need not be an
integer and does not uniquely label a pair of asymptotic vacua.
Consequently, equal energy should be understood in terms of the
endpoint values of $U$ and the actual self-dual connections, rather
than as evidence by itself for the multiplicity of solutions.

We construct the five Platonic models and solve their self-duality
equations nu\-me\-ri\-cally. The solutions show that different trajectories
and local energy-density profiles can connect vacua with the same
pre-potential difference and hence have the same total energy. Paths
that approach saddle points can display separated peaks, while direct
connections may produce a single localized peak. The two nontrivial
dual pairs, octahedron--cube and icosahedron--dodecahedron, share their
vacuum positions but have different pre-potentials, so the associated
metric-gradient flows, charges, and energy-density profiles are not
the same. We also identify positive maxima of $V$ forming symmetry
orbits whose multiplicities reflect the corresponding rotation groups.
These results exhibit how Platonic geometry can be used to organize the
vacuum structure and self-dual sectors of $(1+1)$-dimensional
$\IC\IP^1$ models with a potential.

The paper is organized as follows. In Section \ref{sec:self-dual} we use the Platonic solids to construct regular lattices on the two-sphere, and define the pre-potentials $U$ with critical points on that lattice, and the corresponding potentials $V$. In Section \ref{sec:models} we construct the models associated to each Platonic solid and in Section  \ref{sec:numerical} we calculate examples of the self-dual solutions numerically. Our conclusions are given in Section \ref{sec:conclusion}. The coordinates of the maxima of the potential are tabulated in Appendix~\ref{app:maxV}.

\section{Self-dual sectors in $\mathbb{S}^2$}
\label{sec:self-dual}

We consider a $(1+1)$-dimensional $\IC\IP^1$ sigma model with a potential,
described by the action
\begin{equation}
    S = \int \dd t\,\dd x\,
    \qty[\half\eta_{ab}(\vp)\pa_\mu\vp_a\pa^\mu\vp_b - V(\vp_c)],
    \label{eq:action1}
\end{equation}
where $a,b=1,2$, repeated indices are summed, and the spacetime metric
has signature $(+,-)$. The target-space metric is
\begin{equation}
    \eta_{ab} = \dfrac{4}{\qty(1+r^2)^2}\delta_{ab},
    \qquad r^2 = \vp_1^2 + \vp_2^2.
    \label{eq:metric_chi}
\end{equation}
The fields $\vp_a$ are stereographic coordinates on the unit sphere
$\mathbb{S}^2\simeq\IC\IP^1$, related to a three-component unit vector by
\begin{equation}
    \vb*{\phi}
    = \dfrac{1}{1+r^2}\qty(2\vp_1,\,2\vp_2,\,r^2-1),
    \qquad r^2 = \vp_1^2 + \vp_2^2.
    \label{eq:change_coordinates}
\end{equation}
Thus, $\vb*{\phi}\cdot\vb*{\phi}=1$, and
\eqref{eq:metric_chi} is the round metric induced on the sphere in these
coordinates. Introducing the complex coordinate $u=\vp_1+i\vp_2$, the
action \eqref{eq:action1} takes the $\IC\IP^1$ form
\begin{equation}
    S = \int \dd t\,\dd x\,
    \qty[\dfrac{2\pa_\mu\bar{u}\pa^\mu u}{\qty(1+\abs{u}^2)^2}
    -V(u,\bar{u})],
    \label{eq:action_metric_cp1}
\end{equation}
where $V(u,\bar{u})$ denotes the same potential expressed in complex
coordinates. The coefficient of the kinetic term follows from the
normalization adopted in \eqref{eq:action1} and \eqref{eq:metric_chi}.

For static configurations, the energy is
\begin{equation}
    E = \int_{-\infty}^{\infty}\dd x\,
    \qty[\half\eta_{ab}\dv{\vp_a}{x}\dv{\vp_b}{x}+V(\vp_c)].
    \label{eq:func_E}
\end{equation}
Equivalently, in complex coordinates,
\begin{equation}
    E = \int_{-\infty}^{\infty}\dd x\,
    \qty[\dfrac{2\pa_x\bar{u}\pa_x u}{\qty(1+\abs{u}^2)^2}
    +V(u,\bar{u})].
    \label{eq:energy_cp1}
\end{equation}
We use the real-coordinate expressions in terms of $\eta_{ab}$ to
construct the self-dual sector below.
In one spatial dimension, the gradient and potential contributions to
\eqref{eq:func_E} scale oppositely under spatial dilations. A nonzero
potential is therefore necessary to balance these contributions and allow
nontrivial localized static solutions, in accordance with Derrick's
scaling argument.

To construct a self-dual sector of this model, we follow the generalized
self-duality framework of \cite{lafBPS,laf_1+1}. The self-duality equations are first-order differential equations whose solutions satisfy the second-order static Euler--Lagrange differential equations. What allows for one integration less is the existence of a topological charge which admits an integral representation. In the cases under consideration the relevant topological (homotopic) charge is given by 
\begin{equation}
    Q = \int_{-\infty}^{\infty}\dd x\,
    \dv{\vp_a}{x}\pdv{U}{\vp_a}
    = U\qty(x\rightarrow+\infty)-U\qty(x\rightarrow-\infty),
    \label{eq:Q_1+1}
\end{equation}
where $U$ is a smooth real function
$U:\mathbb{S}^2\rightarrow\IR$, the so-called pre-potential. The approach works for those cases where the potential $V$ takes the form 
\begin{equation}
    V = \half\eta^{-1}_{ab}\pdv{U}{\vp_a}\pdv{U}{\vp_b}.
    \label{eq:potential}
\end{equation}
The self-duality equations are given by
\begin{equation}
    \dv{\vp_a}{x}
    = \pm\eta^{-1}_{ab}\pdv{U}{\vp_b},
    \qquad a=1,2.
    \label{eq:bps1d}
\end{equation}

The static Euler--Lagrange equations following from \eqref{eq:action1}, or equivalently \eqref{eq:func_E}, are
\begin{equation}
    \dv{x}\qty(\eta_{ac}\dv{\vp_a}{x})
    -\eta^{-1}_{ab}\pdv{U}{\vp_a}{\vp_c}\pdv{U}{\vp_b}
    =\half\pdv{\eta_{ab}}{\vp_c}\dv{\vp_a}{x}\dv{\vp_b}{x}
    -\half\eta^{-1}_{ad}\pdv{\eta_{de}}{\vp_c}\eta^{-1}_{eb}
    \pdv{U}{\vp_a}\pdv{U}{\vp_b}.
    \label{eq:elequation}
\end{equation}
Using the self-duality equations \eqref{eq:bps1d}, one gets that
\begin{equation}
    \dv{x}\qty(\eta_{ac}\dv{\vp_a}{x})
    -\eta^{-1}_{ab}\pdv{U}{\vp_a}{\vp_c}\pdv{U}{\vp_b}=0
\end{equation}
and
\begin{equation}
    \pdv{\eta_{ab}}{\vp_c}\dv{\vp_a}{x}\dv{\vp_b}{x}
    -\eta^{-1}_{ad}\pdv{\eta_{de}}{\vp_c}\eta^{-1}_{eb}
    \pdv{U}{\vp_a}\pdv{U}{\vp_b}=0.
\end{equation}
So, indeed, the self-duality equations imply the static Euler--Lagrange equations.

In order to have finite-energy solutions we need the fields to go to a constant at spatial infinity, i.e. $\dv{\vp_a}{x}\rightarrow 0$, for $x\rightarrow \pm \infty$. Therefore, from the self-duality equations we have that at spatial infinity the fields should go to configurations $\vp_b=\vp_b^{(vac)}$, such that
\begin{equation}
    \eval{\pdv{U}{\vp_a}}_{\vp_b=\vp_b^{(vac)}}=0,
    \qquad a,b=1,2.
    \label{eq:dU_zero}
\end{equation}
Since the target-space metric is positive definite, the condition \eqref{eq:dU_zero} implies that the critical points of $U$ are zeros of $V$, 
i.e.
\begin{equation}
    \eval{V}_{\vp_b=\vp_b^{(vac)}}=0,
    \qquad b=1,2.
    \label{eq:V_zero}
\end{equation}

Using \eqref{eq:potential}, we can write  \eqref{eq:func_E} as 
\begin{equation}
        E = \half\int_{-\infty}^{\infty}\dd x\,
        \qty[\dv{\vp_a}{x}\mp\eta^{-1}_{ac}\pdv{U}{\vp_c}]\,\eta_{ab}\,
        \qty[\dv{\vp_b}{x}\mp\eta^{-1}_{bd}\pdv{U}{\vp_d}]
        \pm Q.
    \label{eq:igualdade_carga_energia}
\end{equation}
The self-duality equations \eqref{eq:bps1d} imply that
\begin{equation}
    \dv{\vp_a}{x}\pdv{U}{\vp_a}
    = \pm\eta^{-1}_{ab}\pdv{U}{\vp_a}\pdv{U}{\vp_b}.
\end{equation}
The left-hand side is the integrand in \eqref{eq:Q_1+1}. Since $\eta_{ab}$,
and hence $\eta^{-1}_{ab}$, is positive definite, this integrand is
non-negative for the choice of the plus sign and non-positive otherwise.
Therefore, $Q\geq 0$ for the plus sign and $Q\leq 0$ for the minus sign, so in
both cases the last term in \eqref{eq:igualdade_carga_energia} equals $\abs{Q}$.
Since the integral in \eqref{eq:igualdade_carga_energia} is non-negative, the
energy has lower bound
\begin{equation}
    E\geq\abs{Q},
\end{equation}
which is saturated by the self-dual solutions.

Because $Q$ is fixed by the boundary conditions, the self-dual configurations
minimize the static energy and their energy 
$E=\abs{Q}$ is determined entirely by the values of the pre-potential at the
endpoints.

The model therefore realizes the $(1+1)$-dimensional self-duality
construction of \cite{lafBPS,laf_1+1} for two real fields, with the metric
$\eta_{ab}$ fixed by the geometry of $\IC\IP^1$. This allows us to use
the first-order equations \eqref{eq:bps1d} to obtain static solutions and
the topological bound \eqref{eq:igualdade_carga_energia} to determine
their energies. In the construction developed below, the geometry of the
Platonic solids enters through the choice of $U$, while
\eqref{eq:potential} determines $V$ from this pre-potential and the
target-space metric. The resulting theories are therefore $\IC\IP^1$
models with potentials constructed from the Platonic geometry through
generalized self-duality. Their vacuum structure and possible asymptotic
sectors are specified by the chosen pre-potential. We now turn to this
geometric construction, based on regular lattices on $\mathbb{S}^2$.

\subsection{Regular lattices in $\mathbb{S}^2$}
\label{sec:regular_lattices}

Our aim is to construct a pre-potential whose critical points form a
regular lattice on $\mathbb{S}^2$. Through \eqref{eq:potential}, these
points become degenerate vacua of $V$, providing possible endpoints for
self-dual solitons. This follows the motivation of the Lie-algebra-based
construction in \cite{laf_1+1}, where the symmetry properties of the
pre-potential organize its critical points. Here, the geometry of the
target space motivates the use of the Platonic solids to construct the
regular lattice.

In this work, a regular lattice on $\mathbb{S}^2$ is the finite set of
points at which the rotational symmetry axes of a Platonic solid intersect
its unit circumsphere. Its sites lie in the directions of the vertices,
face centers, and edge midpoints. The rotational symmetry group maps
the lattice into itself and organizes its sites into symmetry orbits;
different geometric types may belong to different orbits. Dual Platonic
solids share the same symmetry axes and therefore define the same regular
lattice, with their vertex and face-center sites interchanged
\cite{coxeter_regular}.

The critical-point condition is intrinsic to the sphere: it requires the
gradient tangent to $\mathbb{S}^2$ to vanish. The stereographic fields
$\vp_a$ provide independent local coor\-di\-na\-tes in which this condition
takes the form \eqref{eq:dU_zero}. We seek pre-potentials invariant under
the rotational group of the chosen solid. This invariance maps critical
points into critical points and makes $U$ constant on each symmetry
orbit. The values of $U$ on these orbits consequently constrain the
charges and energies of self-dual solutions connecting the corresponding
vacua. In the following, we introduce the relevant Platonic-solid
geometry and construct a pre-potential for which the regular-lattice
sites are critical points.

\subsection{Platonic solids}
\label{sec:platonic_solids}

\begin{figure}[!htbp]
	\begin{center}
		\includegraphics[width=\textwidth]{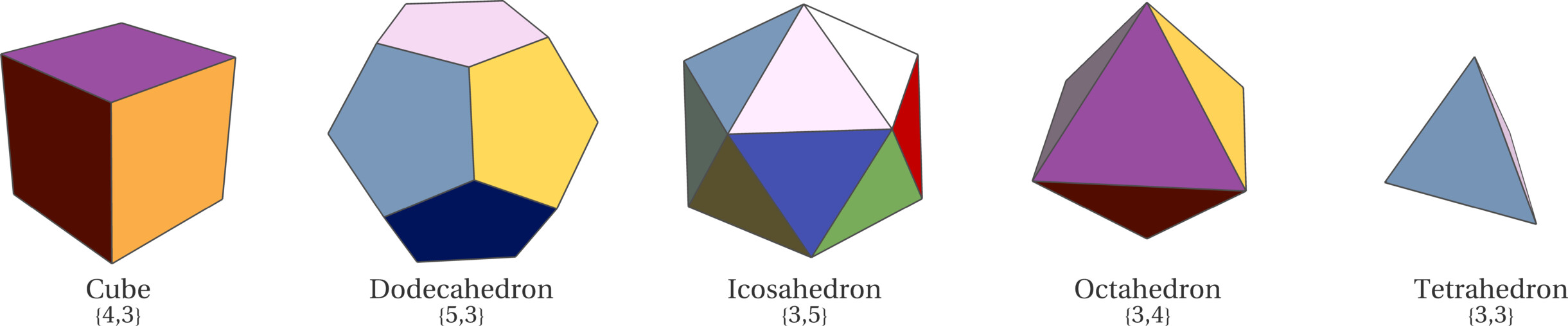}
	\end{center}
	\caption{Platonic solids.}
	\label{fig:Platonic_solids}
	
\end{figure}

The five Platonic solids, shown in Figure~\ref{fig:Platonic_solids},
are convex regular polyhedra with congruent regular polygonal faces
and identical vertex configurations \cite{coxeter_regular}.
In our construction, each solid is centered at
the origin and its vertices are normalized to lie on the unit sphere
$\mathbb{S}^2$.

Polyhedral duality exchanges vertices and faces. The vertices of the dual
solid are obtained by radially projecting the face centers of the original
solid onto the same unit sphere. The tetrahedron is self-dual in this
geometric sense, while the remaining solids form two dual pairs.
We consider the proper rotation group $G$ of each solid, whose elements
permute its vertices. Three such groups occur, as summarized in
Table~\ref{tab:platonic_groups} \cite{coxeter_regular}.

\begin{table}[htbp]
    \centering
    \begin{tabular}{lcc}
        \hline
        Dual pair & Rotation group $G$ & Order $|G|$ \\
        \hline
        Tetrahedron--tetrahedron & Tetrahedral $T$ & $12$ \\
        Cube--octahedron & Octahedral $O$ & $24$ \\
        Dodecahedron--icosahedron & Icosahedral $I$ & $60$ \\
        \hline
    \end{tabular}
    \caption{Dual pairs of Platonic solids and their proper rotation groups.}
    \label{tab:platonic_groups}
\end{table}

Members of each dual pair share the same rotation group and symmetry
axes, hence the same regular lattice introduced in
Section~\ref{sec:regular_lattices}. Their distinct vertex sets provide
different choices for the construction of the pre-potential. We now use
these sets to define $U$ and establish the symmetry and critical-point
properties relevant to the self-dual models.

\subsection{Pre-potential}
\label{sec:U_platonic}
Let $\{v_a\}_{a=1}^{n}$ be the unit vertex vectors of a Platonic solid.
Inspired by Klein polynomials \cite{klein2003lectures}, we choose the
pre-potential
\begin{equation}
    U(\vb*{\phi})=C\prod_{a=1}^{n}\qty(1-v_a\cdot\vb*{\phi}),
    \qquad \vb*{\phi}\cdot\vb*{\phi}=1.
    \label{eq:U_plato}
\end{equation}
We orient the solid so that no vertex lies at the north pole
$\vb*{e}_3=(0,0,1)$ and fix the normalization by
\begin{equation}
    C=\prod_{a=1}^{n}\frac{1}{1-(v_a)_3}.
    \label{eq:constant_C}
\end{equation}
Thus, $C>0$ and $U(\vb*{e}_3)=1$. The product defines a smooth,
nonnegative function on $\mathbb{S}^2$, whose zeros are precisely the
vertices of the chosen solid. These points are therefore global minima
of $U$.

\begin{figure}[!htbp]
    \centering
    \includegraphics[width=0.4\linewidth]{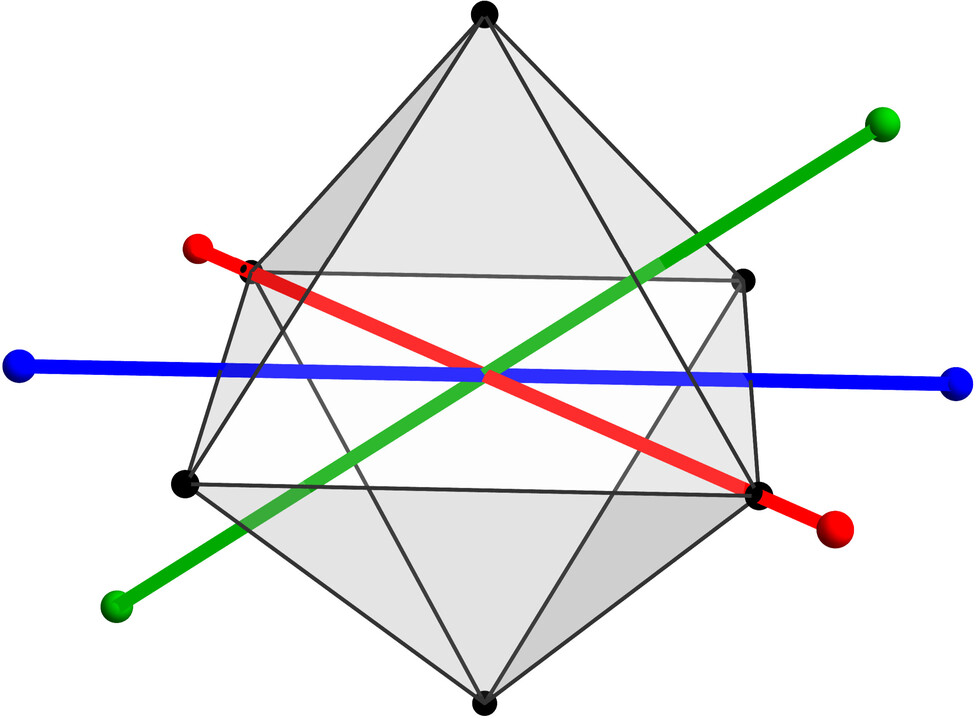}
    \caption{Representative rotational symmetry axes of the octahedron.
    The intersections of all its rotational symmetry axes with the unit
    circumsphere define the regular lattice.}
    \label{fig:axis_octa}
    
\end{figure}

Every rotation $R\in G$ permutes the vertex set. Since $R^{-1}=R^T$,
the pre-potential inherits the rotational symmetry of the solid:
\begin{equation}
    U(R\vb*{\phi})
    =C\prod_{a=1}^{n}\qty(1-(R^Tv_a)\cdot\vb*{\phi})
    =U(\vb*{\phi}).
    \label{eq:inv_U}
\end{equation}
Let $\nabla_{\mathbb{S}^2}U$ denote the gradient tangent to the unit
sphere. Differentiating \eqref{eq:inv_U} and projecting onto the tangent
plane gives
\begin{equation}
    \nabla_{\mathbb{S}^2}U(R\vb*{\phi})
    =R\nabla_{\mathbb{S}^2}U(\vb*{\phi}).
    \label{eq:gradient_equivariance}
\end{equation}
Representative symmetry axes of the octahedron are shown in
Figure~\ref{fig:axis_octa}. At a regular-lattice site $\vb*{u}$, a nonidentity rotation about the
corresponding symmetry axis fixes $\vb*{u}$. The gradient must then be
both tangent to the sphere and invariant under this rotation. The only
such tangent vector is zero, so
\begin{equation}
    \nabla_{\mathbb{S}^2}U(\vb*{u})=0,
    \qquad \vb*{u}\in\mathcal{U},
    \label{eq:cond_crit}
\end{equation}
where $\mathcal{U}$ is the regular lattice defined in
Section~\ref{sec:regular_lattices}. Hence every lattice site is a
critical point of $U$ and, by \eqref{eq:potential}, a vacuum of the
associated potential $V$.

The regular lattice contains the vertex, face-center, and edge-midpoint
directions. For the individual pre-potentials considered below, our
numerical analysis identifies the face-center sites, which are the
vertices of the dual solid, as maxima, and the edge-midpoint sites as
saddle points. Together with the vertex minima established above, these
give the critical-point types used in the subsequent analysis. Rotations
map critical points into critical points of the same type, and
\eqref{eq:inv_U} makes $U$ constant on each symmetry orbit. The values on
these orbits determine the charges and energies of self-dual solutions
through \eqref{eq:Q_1+1} and \eqref{eq:igualdade_carga_energia}.

Dual solids share the same regular lattice and proper rotation group,
although their vertex sets define different pre-potentials. If
$\hat{U}$ and $\tilde{U}$ correspond to a dual pair in this common
orientation, their linear combination
\begin{equation}
    U=\alpha\qty(\cos\theta\,\hat{U}+\sin\theta\,\tilde{U}),
    \qquad \alpha\in\mathbb{R}\setminus\{0\},
    \quad 0\leq\theta<2\pi,
    \label{eq:combined_U}
\end{equation}
is also invariant under $G$, and every site of $\mathcal{U}$ remains
critical. These sites therefore remain vacua of $V$. This argument does
not exclude additional critical points, and the critical-point types
and possible degeneracies can depend on the mixing parameters.

The three dual pairs yield the families
\begin{equation}
    \begin{aligned}
        U_1&=\alpha\qty(\cos\theta\,U_T+\sin\theta\,U_{-T}),\\
        U_2&=\alpha\qty(\cos\theta\,U_O+\sin\theta\,U_C),\\
        U_3&=\alpha\qty(\cos\theta\,U_I+\sin\theta\,U_D).
    \end{aligned}
    \label{eq:U_possibilidades}
\end{equation}
Here the subscripts denote the tetrahedron and its dual, the octahedron
and cube, and the icosahedron and dodecahedron, respectively; $-T$ has
the vertices of $T$ reflected through the origin. The parameter $\alpha$
sets the overall scale and sign of $U$, while $\theta$ controls the
relative contributions of the dual pair. Up to the factor $\alpha$,
the individual pre-potentials are recovered at $\theta=0$ and
$\theta=\pi/2$.

Expressing these pre-potentials in the stereographic coordinates
\eqref{eq:change_coordinates} and using \eqref{eq:potential} yields the
$\IC\IP^1$ models with potentials determined by the Platonic geometry
through self-duality. We now examine the individual models, their
vacuum structure, and their self-dual solutions.

\section{Models}
\label{sec:models}

We study the five individual Platonic-solid models contained in
\eqref{eq:U_possibilidades}, taking $\alpha=1$, with $\theta=0$ for the
tetrahedron and $\theta=0,\pi/2$ for the octahedron--cube and
icosahedron--dodecahedron pairs. Each model is specified by its vertex
set and the construction \eqref{eq:U_plato}--\eqref{eq:constant_C}.
We describe the resulting regular lattices and the values of $U$ at
their sites, which determine the possible topological charges and
self-dual energies through \eqref{eq:Q_1+1} and
\eqref{eq:igualdade_carga_energia}.

We retain the tetrahedron in the orientation given below. For the other
four solids, we apply the same rotation through $\pi/6$ about
$\vb*{n}=(1,1,0)/\sqrt{2}$, with matrix \eqref{eq:rot_mat}.
This places their regular-lattice sites at finite stereographic
coordinates and preserves the common orientation of each dual pair.
The normalization \eqref{eq:constant_C} is evaluated using the chosen
vertex orientation. A rigid rotation preserves the spherical symmetry
and the critical-point types; imposing the same normalization at the
north pole after a rotation can additionally rescale $U$ and the
self-dual energies.

In the figures, black filled circles, white circles, and black squares
denote minima, maxima, and saddle points of $U$, respectively; all are
vacua of $V$. The arrows in the planar plots represent the metric
gradient $\eta^{-1}_{ab}\pdv{U}{\vp_b}$, whose integral curves describe
self-dual trajectories \cite{laf_beauty}. Interactive versions of the
three-dimensional graphs are available in the accompanying
repository.\footnote{\url{https://github.com/giuseppe-braz/Dissertation-Repository.git}.
The interactive files can be viewed with Wolfram Player.}
\subsection{Tetrahedron}
The vertex set is
\begin{equation}
    V_T=\frac{1}{\sqrt{3}}
    \qty{(1,1,1),(1,-1,-1),(-1,1,-1),(-1,-1,1)}.
    \label{eq:vertices_tetra_sphere}
\end{equation}
With $C=9/4$, the pre-potential takes the form
\begin{equation}
    \begin{aligned}
        U_T(\vb*{\phi})
        &=\frac{9}{4}\prod_{a=1}^{4}\qty(1-v_a\cdot\vb*{\phi})\\
        &=1-\phi_1^2\phi_2^2-\phi_2^2\phi_3^2-\phi_3^2\phi_1^2
        -2\sqrt{3}\phi_1\phi_2\phi_3,
    \end{aligned}
    \label{eq:U_tetra}
\end{equation}
where the second equality uses $\vb*{\phi}\cdot\vb*{\phi}=1$.
Its expression in the stereographic fields follows directly from
\eqref{eq:change_coordinates}, and the associated potential is given by
\eqref{eq:potential}.

The tetrahedral regular lattice consists of four minima at $V_T$, four
maxima at the dual vertices $V_{dual(T)}=-V_T$, and six saddle points at
\begin{equation}
    P_{saddle}=\qty{(\pm1,0,0),(0,\pm1,0),(0,0,\pm1)}.
\end{equation}
These fourteen sites are vacua of $V$, and the corresponding
pre-potential values are
\begin{equation}
    U_{min}=0,\qquad U_{saddle}=1,\qquad U_{max}=\frac{4}{3}.
    \label{eq:valores_U_tetra}
\end{equation}

\begin{figure}[!htbp]
    \centering
    \begin{subfigure}{0.45\textwidth}
        \centering
        \includegraphics[width=\textwidth]{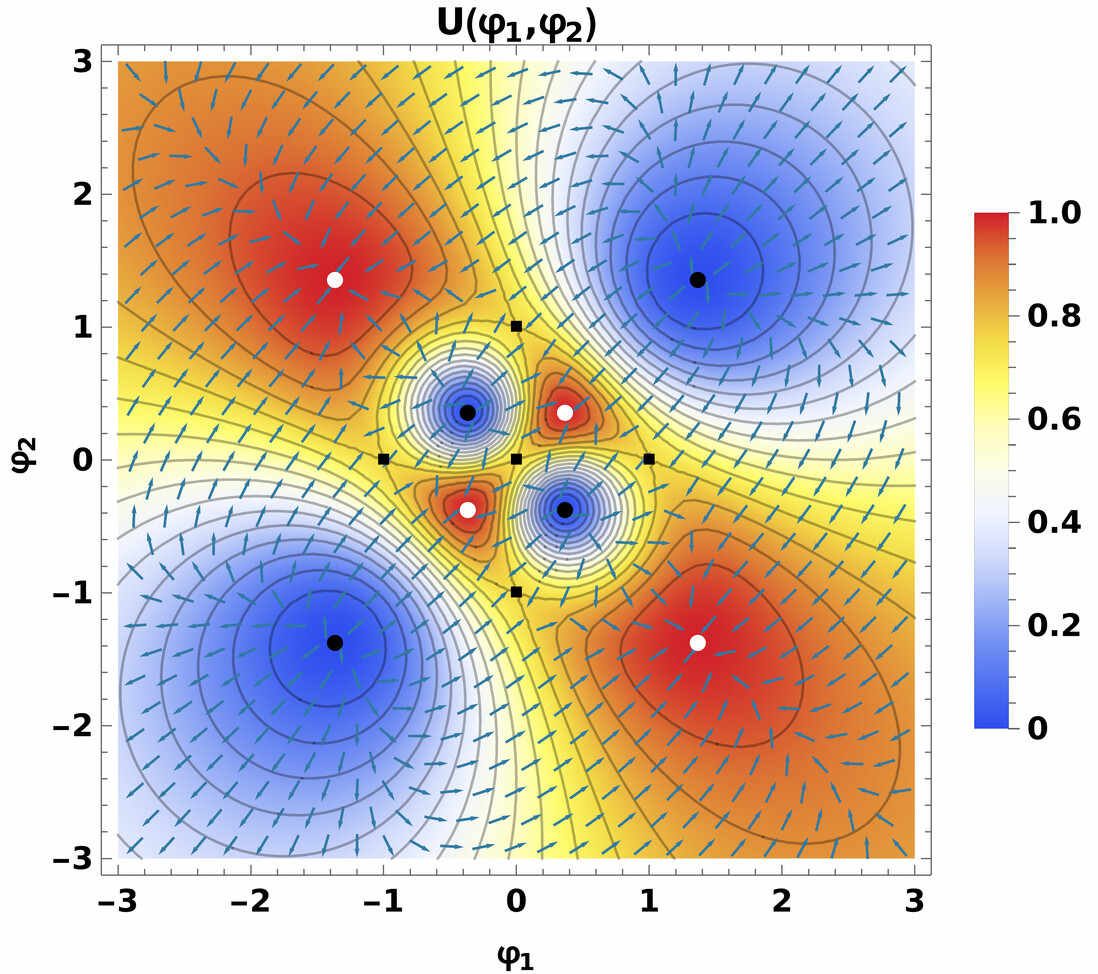}
        \caption{}
        \label{U_2d_tetra}
    \end{subfigure}
    \hfill
    \begin{subfigure}{0.45\textwidth}
        \centering
        \includegraphics[width=\textwidth]{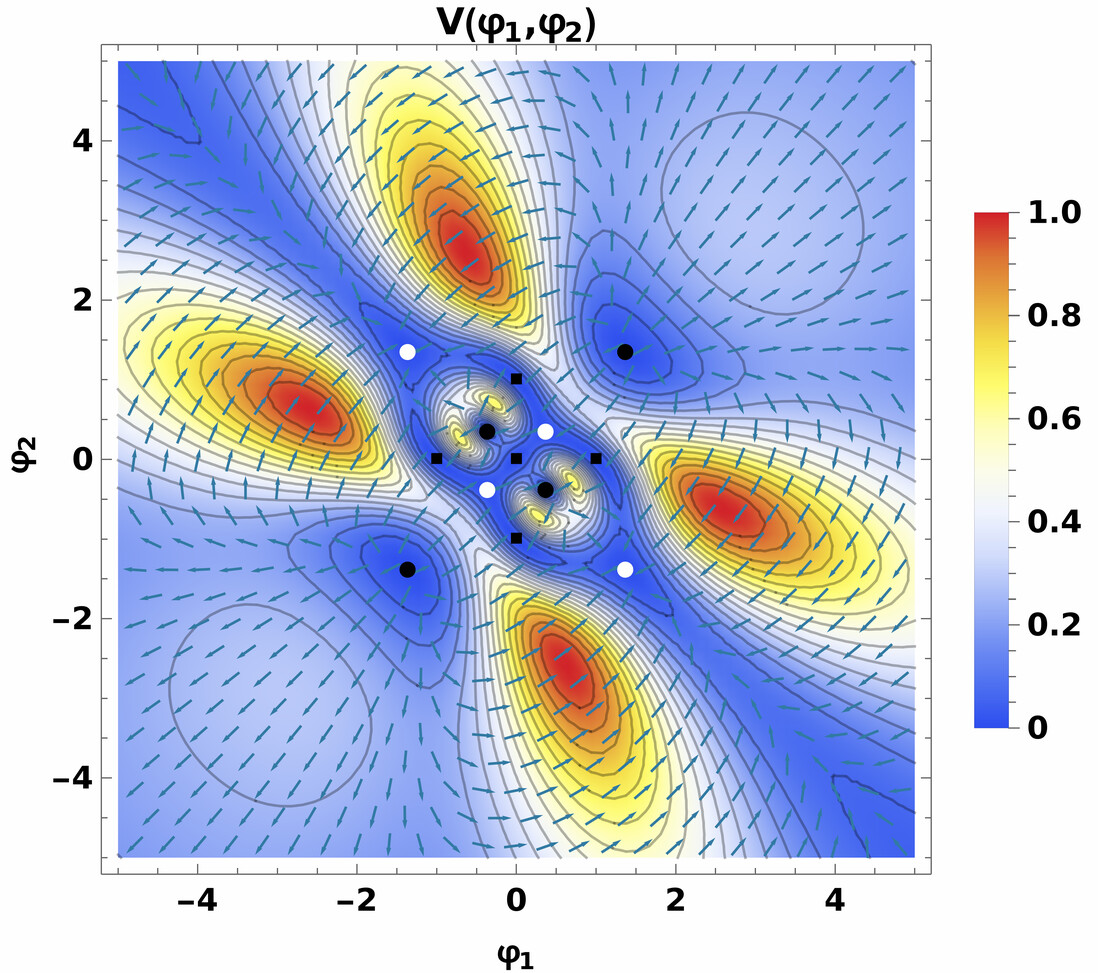}
        \caption{}
        \label{V_2d_tetra}
    \end{subfigure}
    \par\vspace{0.15cm}
    \begin{subfigure}{0.45\textwidth}
        \centering
        \includegraphics[width=\textwidth]{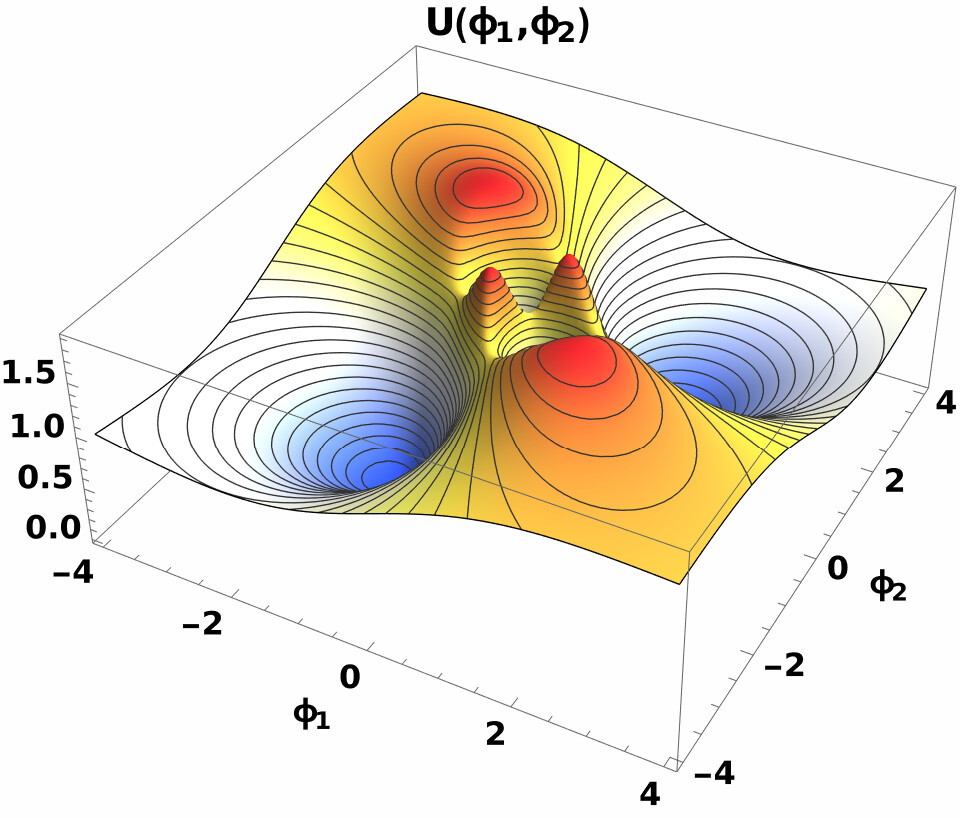}
        \caption{}
    \end{subfigure}
    \hfill
    \begin{subfigure}{0.45\textwidth}
        \centering
        \includegraphics[width=\textwidth]{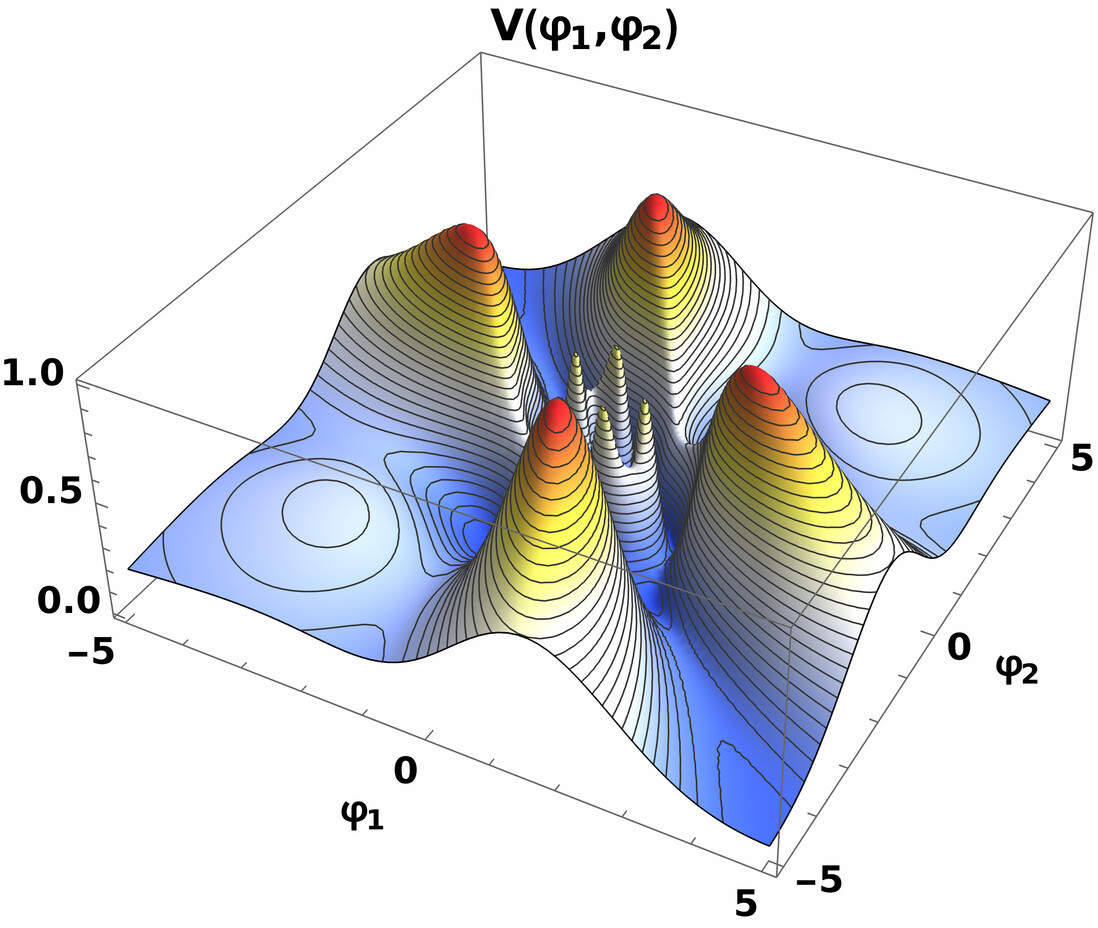}
        \caption{}
        \label{V_3d_tetra}
    \end{subfigure}
    \caption{Tetrahedral pre-potential $U_T$ in (a,c) and potential $V$
    in (b,d), shown as planar contour plots and three-dimensional
    surfaces. In (a,b), markers identify the vacua and arrows represent
    the metric gradient of $U_T$.}
    \label{fig:U_V_tetra}
    
\end{figure}

Figure~\ref{fig:U_V_tetra} shows the pre-potential, potential, and metric
gradient in stereographic coordinates. The saddle point at the north
pole $(0,0,1)$ is mapped to infinity and therefore does not appear among
the thirteen finite vacuum points. Trajectories approaching this saddle
reach unbounded stereographic coordinates.

\begin{figure}[!htbp]
    \centering
    \begin{subfigure}{0.45\textwidth}
        \centering
        \includegraphics[width=\textwidth]{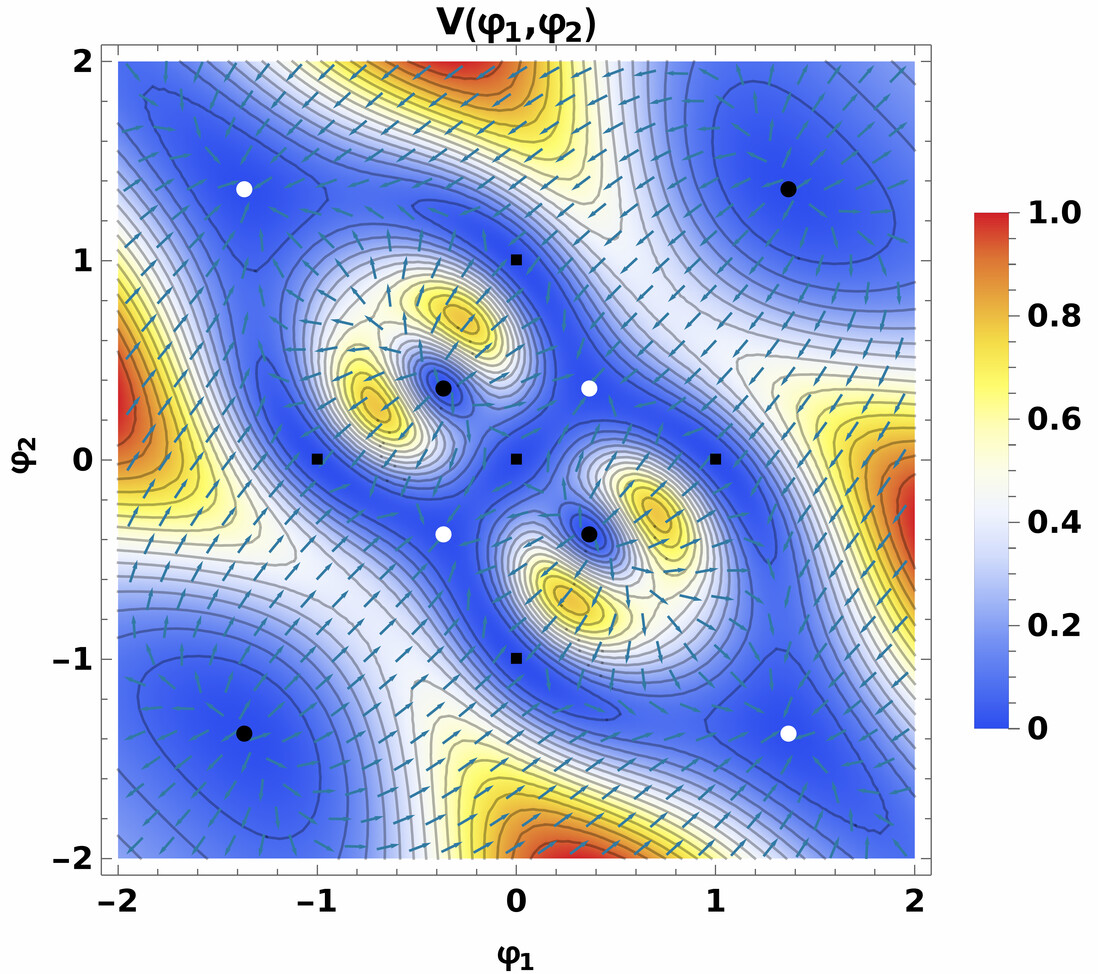}
        \caption{}
    \end{subfigure}
    \hfill
    \begin{subfigure}{0.45\textwidth}
        \centering
        \includegraphics[width=\textwidth]{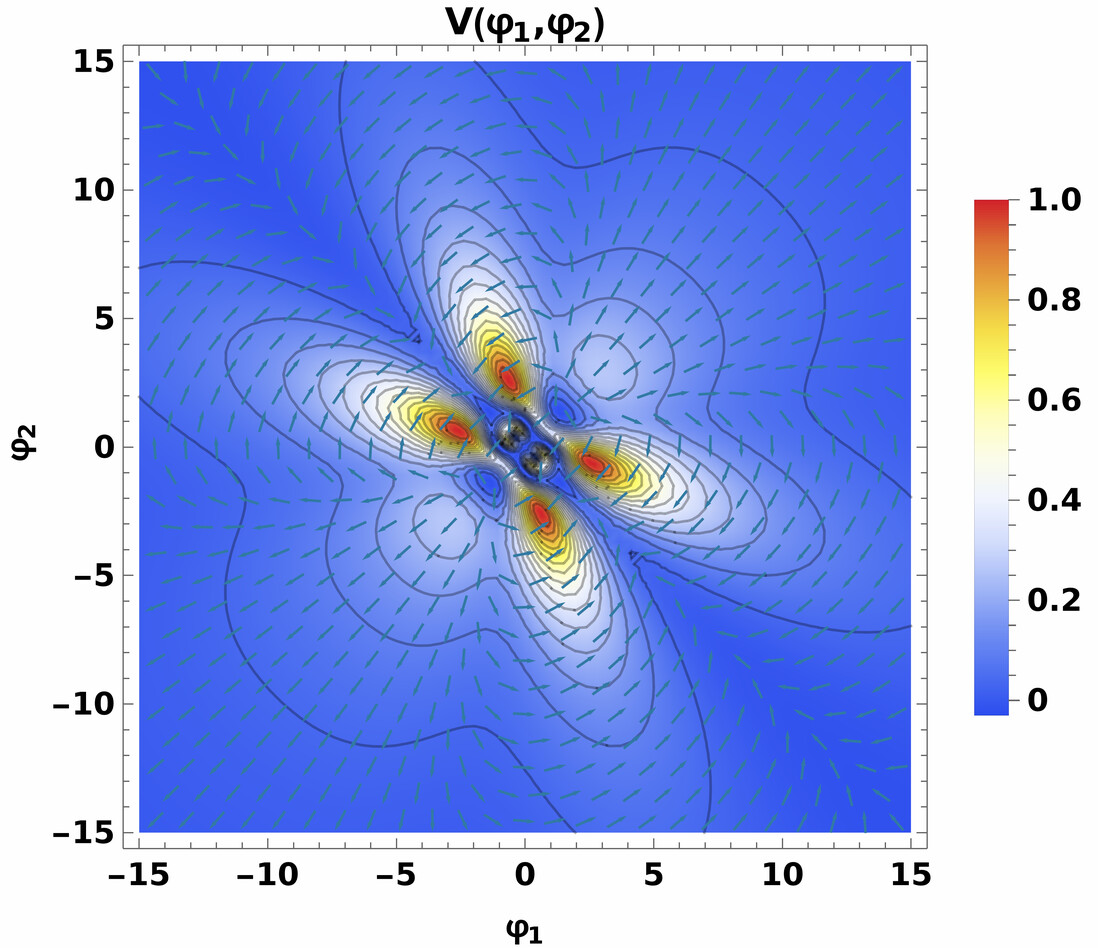}
        \caption{}
    \end{subfigure}
    \par\vspace{0.15cm}
    \begin{subfigure}{0.45\textwidth}
        \centering
        \includegraphics[width=\textwidth]{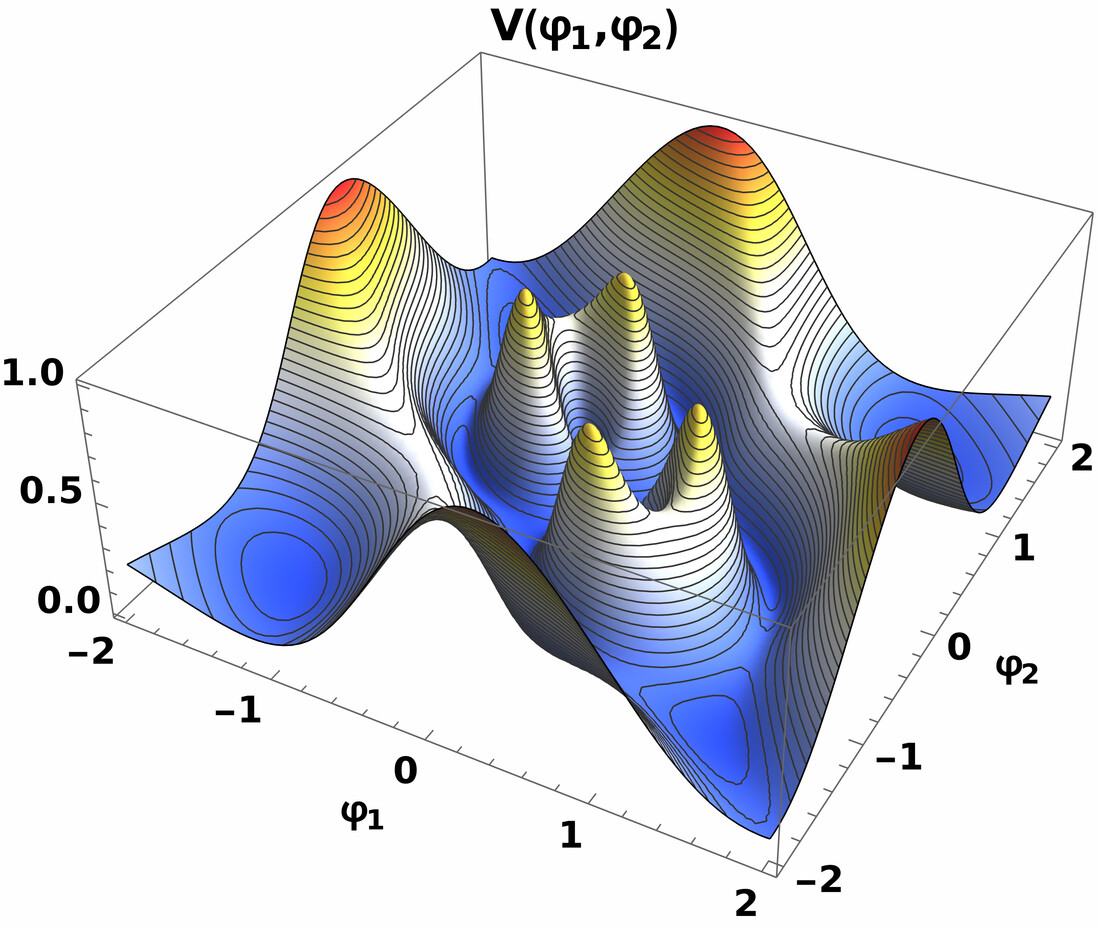}
        \caption{}
    \end{subfigure}
    \hfill
    \begin{subfigure}{0.45\textwidth}
        \centering
        \includegraphics[width=\textwidth]{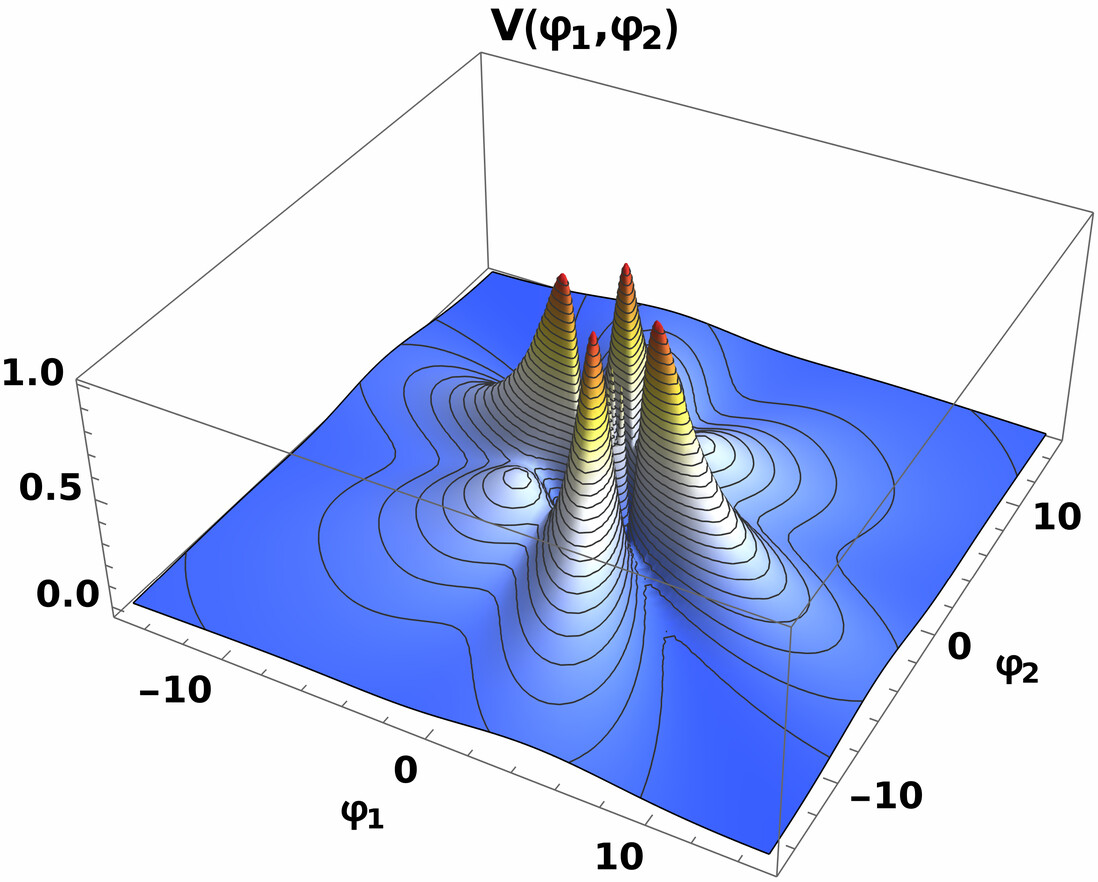}
        \caption{}
    \end{subfigure}
    \caption{Tetrahedral potential $V$ at two scales: (a,c) a close
    view near the origin and (b,d) a wider view of the stereographic
    plane. The upper panels show contours and the metric gradient of
    $U_T$; the lower panels show the corresponding surfaces.
    Markers follow the convention of Figure~\ref{fig:U_V_tetra}.}
    \label{fig:V_tetra2}
    
\end{figure}

Figure~\ref{fig:V_tetra2} resolves the potential near the origin and
over a larger range of field values. The close views distinguish the
nearby vacua and the potential barriers between them, while the wider
views show the decay towards $V=0$ as the stereographic radius tends to
infinity, corresponding to the north-pole vacuum.

\begin{figure}[!htbp]
    \centering
    \begin{subfigure}{0.27\textwidth}
        \centering
        \includegraphics[width=\textwidth]{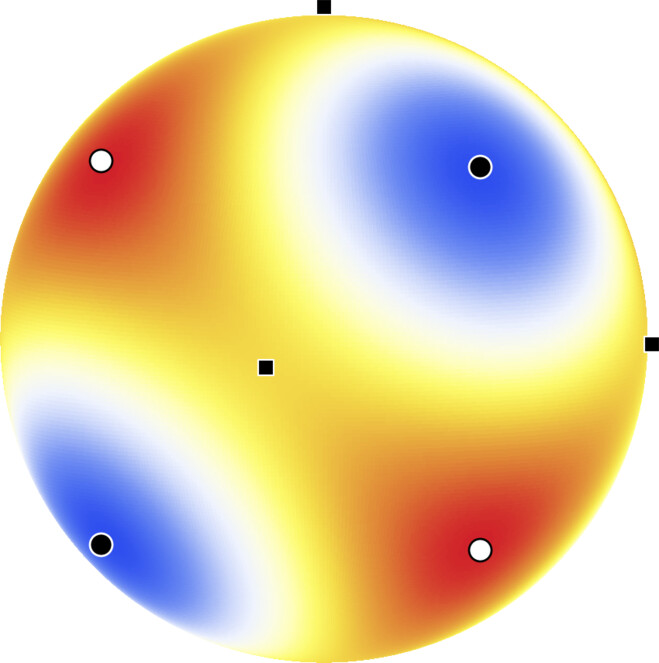}
        \caption{}
        \label{U_tetra_sphere}
    \end{subfigure}
    \hspace{0.03\textwidth}
    \begin{subfigure}{0.27\textwidth}
        \centering
        \includegraphics[width=\textwidth]{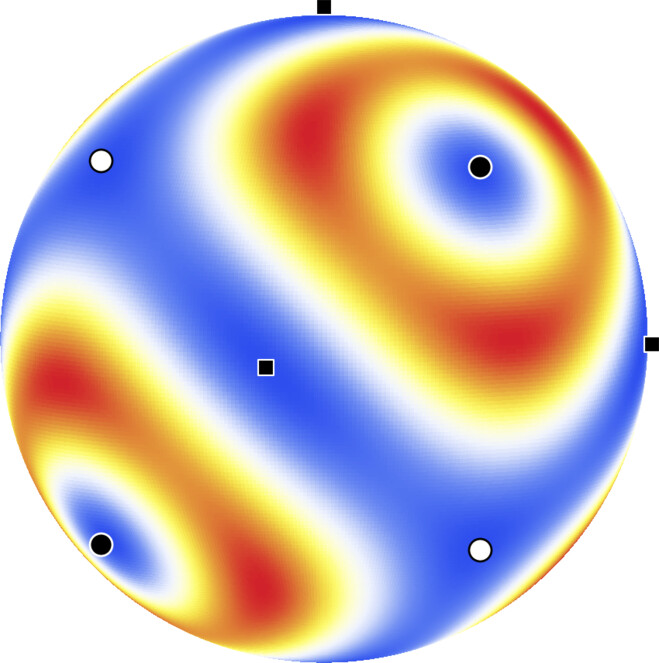}
        \caption{}
        \label{V_tetra_sphere}
    \end{subfigure}
    \caption{Tetrahedral (a) pre-potential $U_T$ and (b) potential $V$
    represented by color on $\mathbb{S}^2$. Black filled circles, white
    circles, and black squares mark minima, maxima, and saddle points
    of $U_T$, respectively. All marked sites are vacua of $V$.}
    \label{fig:U_V_tetra_sphere}
\end{figure}

Figure~\ref{fig:U_V_tetra_sphere} displays both functions on the unit
sphere, making their tetrahedral symmetry and the geometry of the
regular lattice directly apparent. The north-pole saddle is included
in this representation, and all three types of critical points of
$U_T$ correspond to degenerate minima of $V$.
\FloatBarrier

\subsection{Octahedron}
\begin{figure}[!htbp]
    \centering
    \begin{subfigure}{0.45\textwidth}
        \centering
        \includegraphics[width=\textwidth]{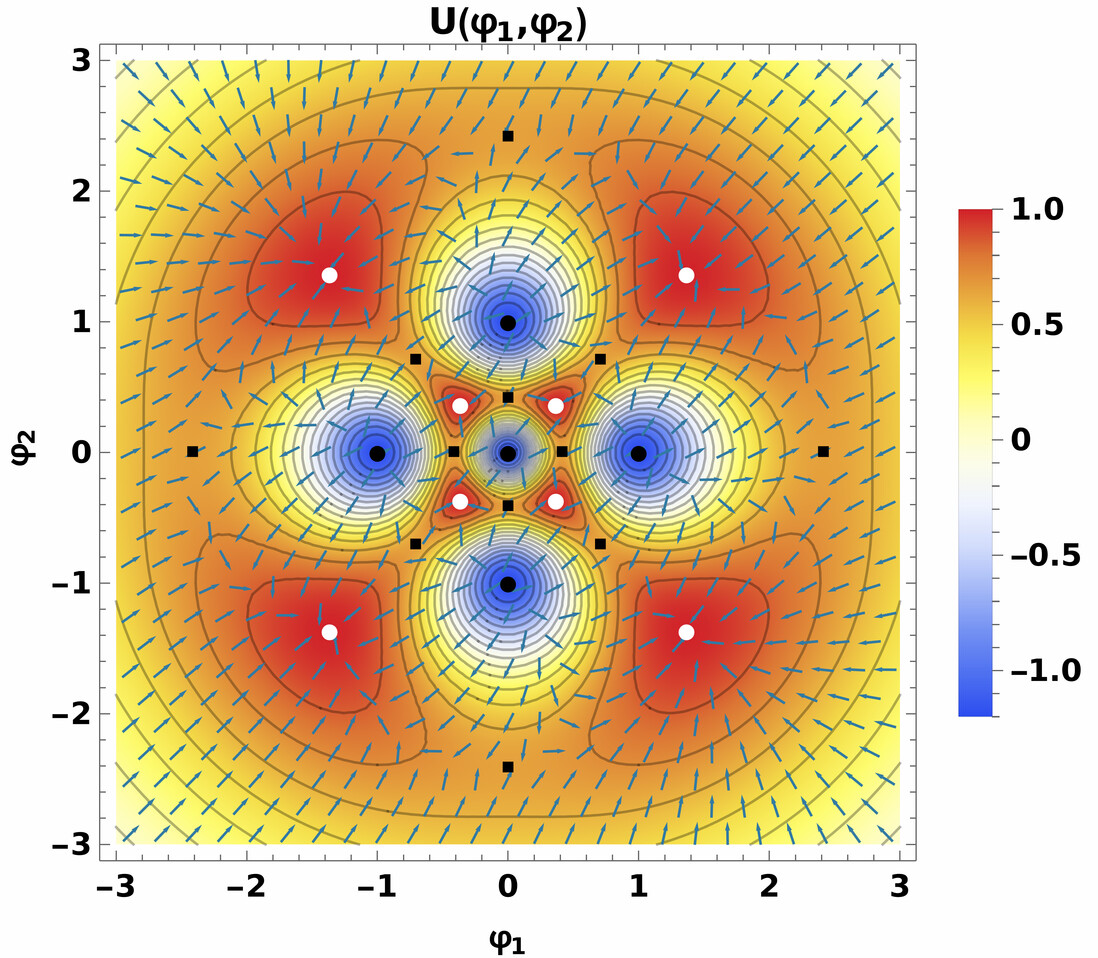}
        \caption{}
        \label{U_2d_octa_antes}
    \end{subfigure}
    \hfill
    \begin{subfigure}{0.45\textwidth}
        \centering
        \includegraphics[width=\textwidth]{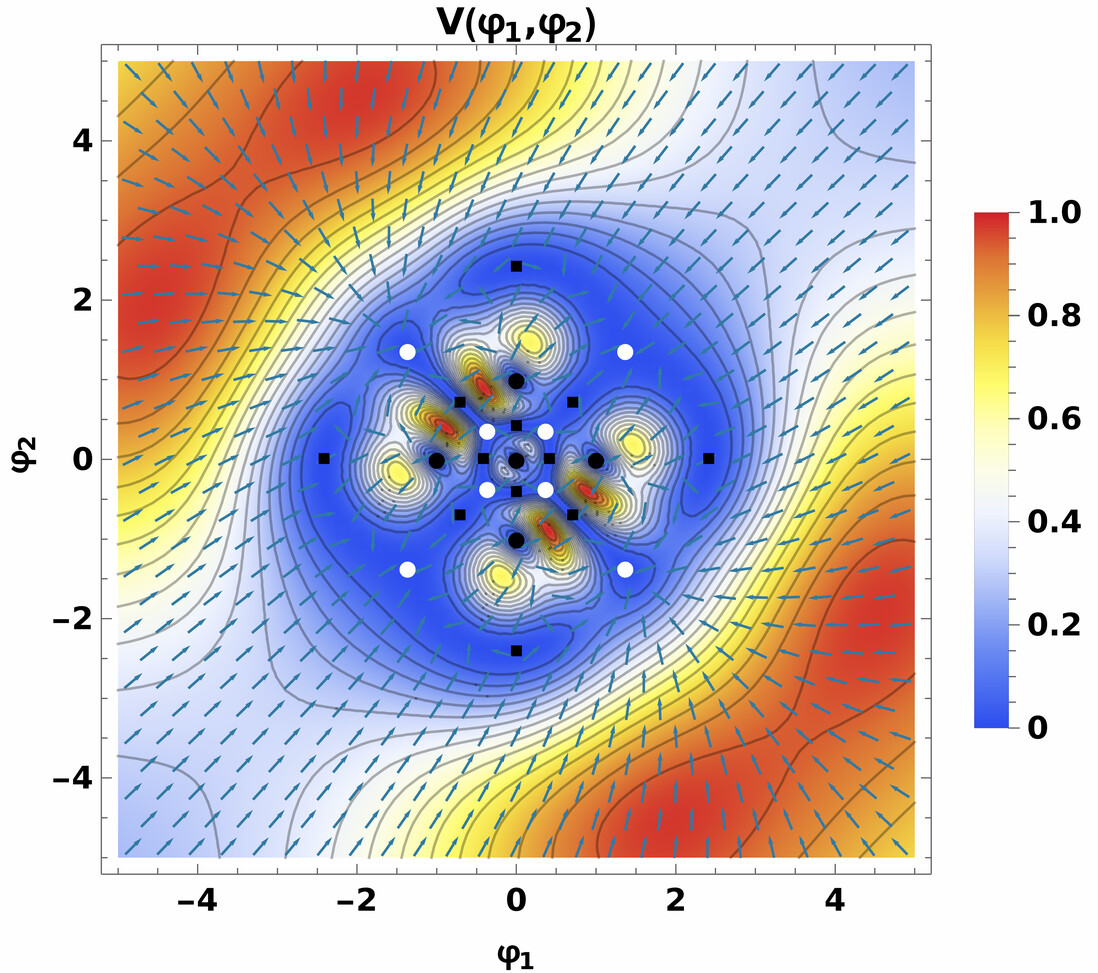}
        \caption{}
        \label{V_2d_octa_antes}
    \end{subfigure}
    \par\vspace{0.15cm}
    \begin{subfigure}{0.45\textwidth}
        \centering
        \includegraphics[width=\textwidth]{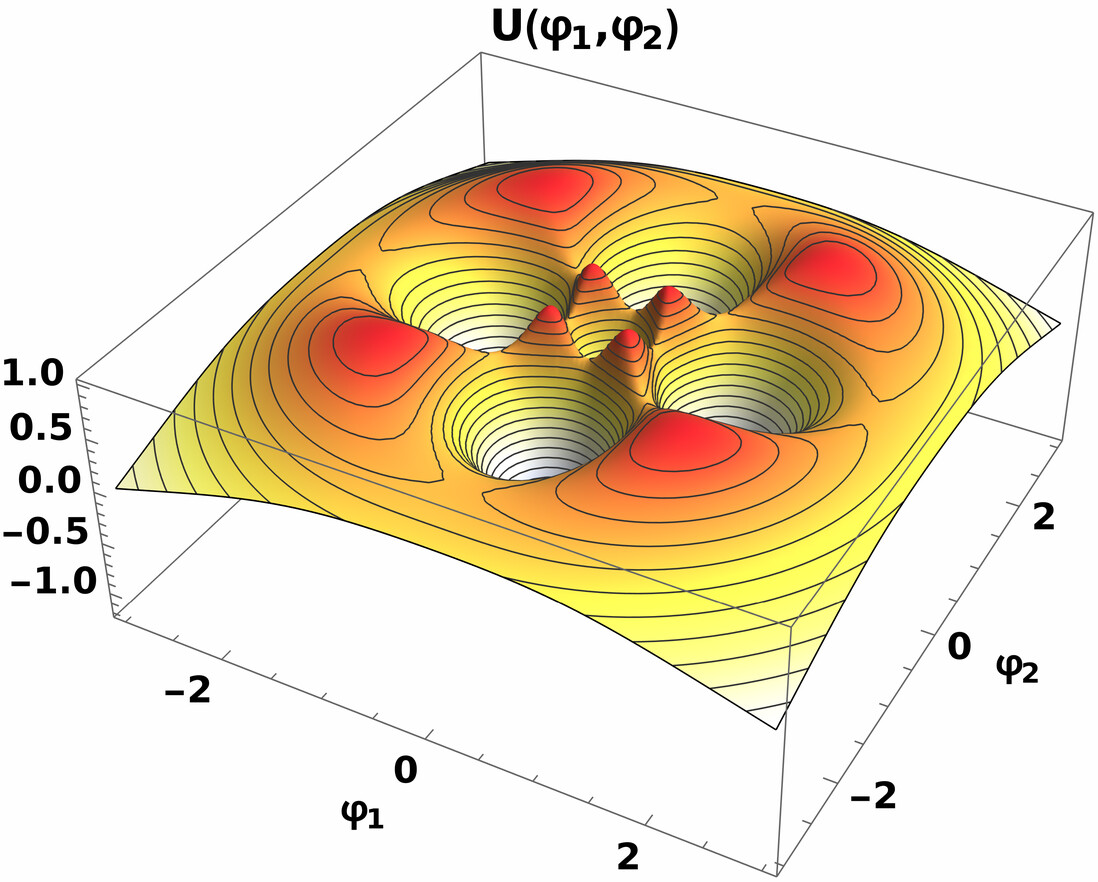}
        \caption{}
    \end{subfigure}
    \hfill
    \begin{subfigure}{0.45\textwidth}
        \centering
        \includegraphics[width=\textwidth]{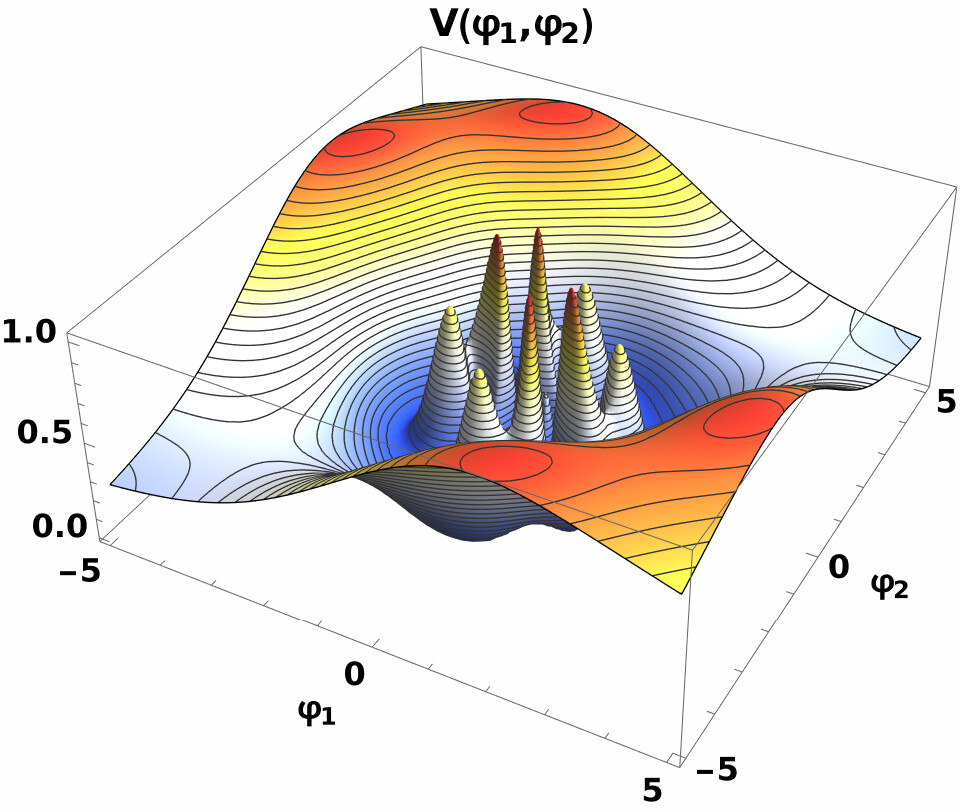}
        \caption{}
    \end{subfigure}
    \caption{Reference plots in the unrotated octahedral orientation:
    (a,c) pre-potential and (b,d) potential. The upper panels show
    contours, critical-point markers, and metric-gradient arrows;
    the lower panels show the corresponding surfaces.}
    \label{fig:U_V_octa_antes}
    
\end{figure}

The six vertices in the reference orientation are
\begin{equation}
    V_O=\qty{(\pm1,0,0),(0,\pm1,0),(0,0,\pm1)}.
\end{equation}
Figure~\ref{fig:U_V_octa_antes} illustrates this orientation before the
rotation. The north pole is a vertex, so it lies outside the finite
stereographic chart and the normalization \eqref{eq:constant_C} is
undefined in this orientation. An unrotated reference pre-potential
can instead be defined with a finite positive normalization constant;
its critical-point positions are independent of that constant.

For the model analyzed below, we use the rotation specified in the
opening of this section,
\begin{equation}
    R=R_{\vb*{n}}\qty(\frac{\pi}{6})=
    \begin{pmatrix}
        \dfrac{2+\sqrt{3}}{4} & \dfrac{2-\sqrt{3}}{4} & \dfrac{1}{2\sqrt{2}}\\[3pt]
        \dfrac{2-\sqrt{3}}{4} & \dfrac{2+\sqrt{3}}{4} & -\dfrac{1}{2\sqrt{2}}\\[3pt]
        -\dfrac{1}{2\sqrt{2}} & \dfrac{1}{2\sqrt{2}} & \dfrac{\sqrt{3}}{2}
    \end{pmatrix},
    \qquad \vb*{n}=\frac{1}{\sqrt{2}}(1,1,0),
    \label{eq:rot_mat}
\end{equation}
so that the vertex set becomes
\begin{equation}
    V'_O=R V_O.
    \label{eq:vertices_octa_sphere}
\end{equation}
This rotation is not an element of the original octahedral group $O$.
It moves the vertices and the regular lattice, placing all its sites
at finite stereographic coordinates. Applying
\eqref{eq:U_plato}--\eqref{eq:constant_C} to $V'_O$ gives
\begin{equation}
    U_O(\vb*{\phi})=\frac{256}{49}
    \prod_{a=1}^{6}\qty(1-v_a\cdot\vb*{\phi}),
    \qquad v_a\in V'_O.
    \label{eq:U_octa}
\end{equation}
The planar expression follows from \eqref{eq:change_coordinates}.

To distinguish rotation from normalization, let $U_0$ and $U_R$ be
constructed from an initial vertex set and its rotated image, with
finite positive constants $C_0$ and $C_R$, respectively. Orthogonality
of $R$ gives
\begin{equation}
    U_R(\vb*{\phi})=\lambda U_0(R^{-1}\vb*{\phi}),
    \qquad \lambda=\frac{C_R}{C_0}>0.
    \label{eq:rotated_prepotential}
\end{equation}
Thus a critical point $\vb*{\phi}_*$ moves to
$\vb*{\phi}'_*=R\vb*{\phi}_*$, retaining its critical-point type, and
\begin{equation}
    U_R(\vb*{\phi}'_*)=\lambda U_0(\vb*{\phi}_*).
    \label{eq:igualdade_U_rot}
\end{equation}
The new pre-potential has the conjugate rotation group:
\begin{equation}
    G_R=R O R^{-1},\qquad
    U_R\qty((R g R^{-1})\vb*{\phi})=U_R(\vb*{\phi}),
    \quad g\in O.
    \label{eq:octa_conjugate_group}
\end{equation}
The original critical-point set is therefore rotated, while its
polyhedral symmetry is realized about the rotated axes. At fixed
normalization, $\lambda=1$, the critical values and self-dual energies
are unchanged. If the normalization is adjusted, both scale by
$\lambda$, as follows from \eqref{eq:Q_1+1}.

\begin{figure}[!htbp]
    \centering
    \begin{subfigure}{0.45\textwidth}
        \centering
        \includegraphics[width=\textwidth]{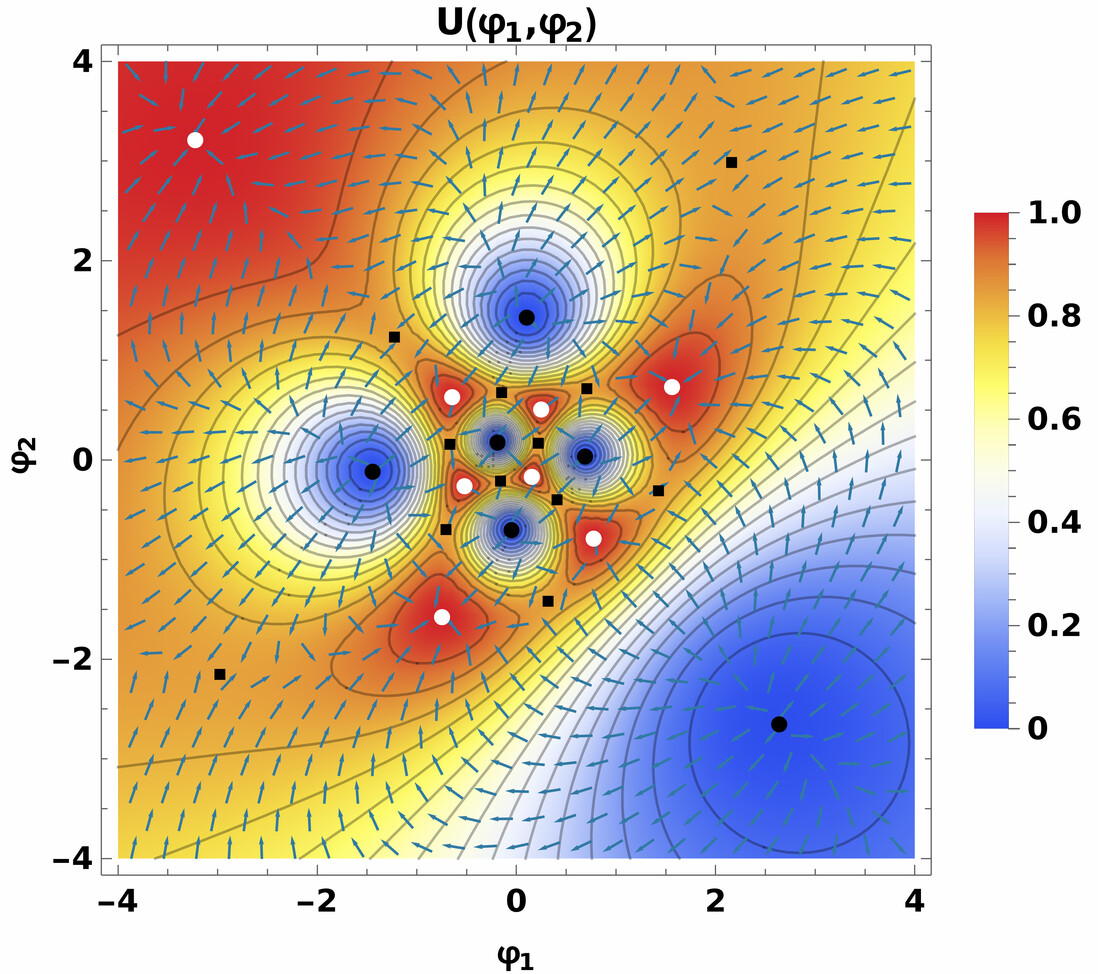}
        \caption{}
        \label{U_2d_octa}
    \end{subfigure}
    \hfill
    \begin{subfigure}{0.45\textwidth}
        \centering
        \includegraphics[width=\textwidth]{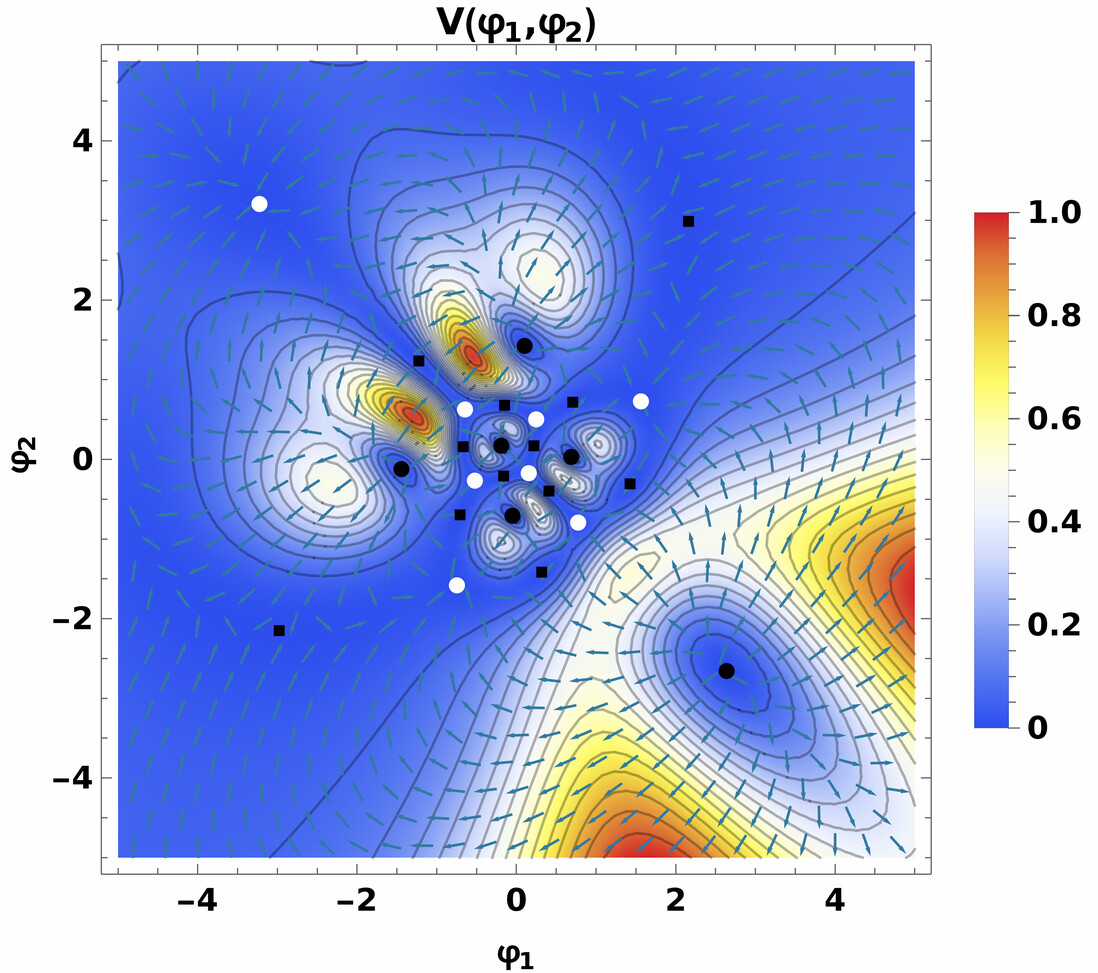}
        \caption{}
        \label{V_2d_octa}
    \end{subfigure}
    \par\vspace{0.15cm}
    \begin{subfigure}{0.45\textwidth}
        \centering
        \includegraphics[width=\textwidth]{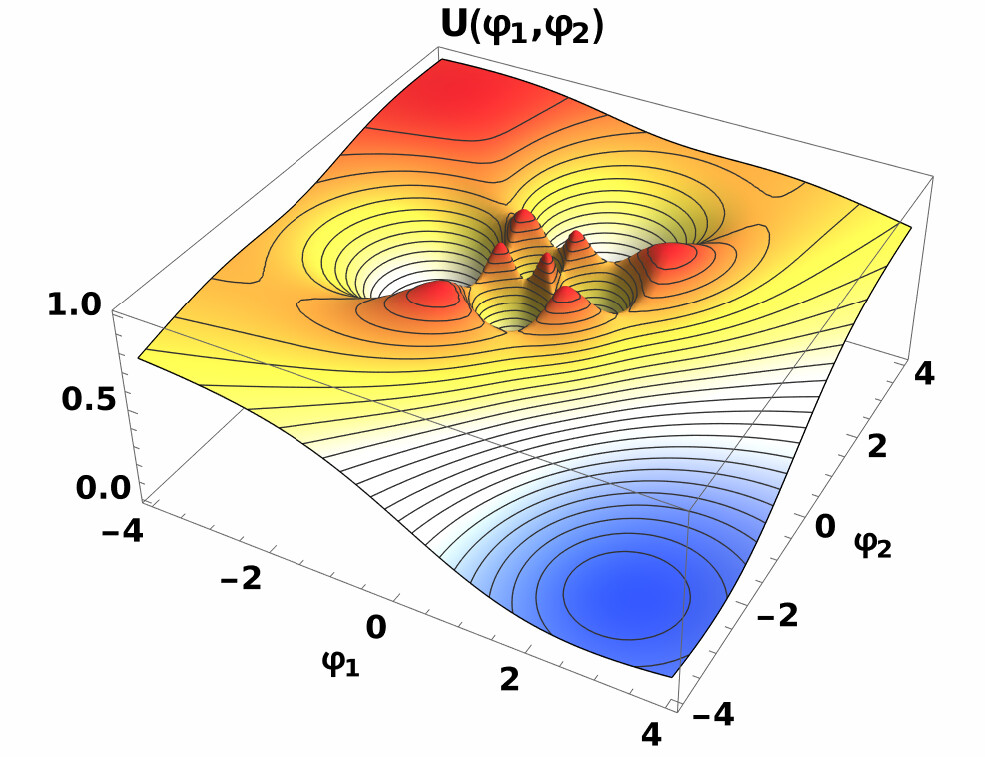}
        \caption{}
    \end{subfigure}
    \hfill
    \begin{subfigure}{0.45\textwidth}
        \centering
        \includegraphics[width=\textwidth]{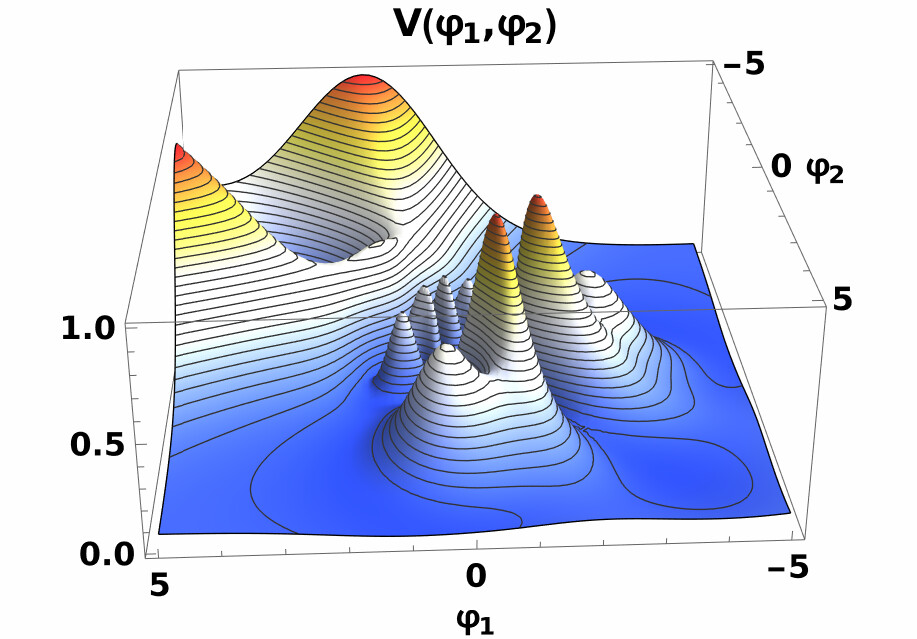}
        \caption{}
        \label{V_3d_octa}
    \end{subfigure}
    \caption{Octahedral pre-potential $U_O$ in (a,c) and potential $V$
    in (b,d), after the rotation \eqref{eq:rot_mat}. The upper panels
    show contours, vacuum markers, and the metric gradient of $U_O$;
    the lower panels show the corresponding surfaces.}
    \label{fig:U_V_octa}
    
\end{figure}

The octahedral regular lattice contains six minima at $V'_O$, eight
maxima at the vertices of the correspondingly rotated cube, and twelve
saddle points in the rotated edge-midpoint directions. Their values are
\begin{equation}
    U_{min}=0,\qquad U_{saddle}\approx1.306,\qquad U_{max}\approx1.548.
    \label{eq:valores_U_octa}
\end{equation}
Figure~\ref{fig:U_V_octa} shows the rotated model. Its planar
representations differ from those in
Figure~\ref{fig:U_V_octa_antes}, although the critical-point
classification is preserved. Both $U_O$ and $V$ are invariant under
the planar reflection
$(\vp_1,\vp_2)\mapsto(-\vp_2,-\vp_1)$.

\begin{figure}[!htbp]
    \centering
    \begin{subfigure}{0.45\textwidth}
        \centering
        \includegraphics[width=\textwidth]{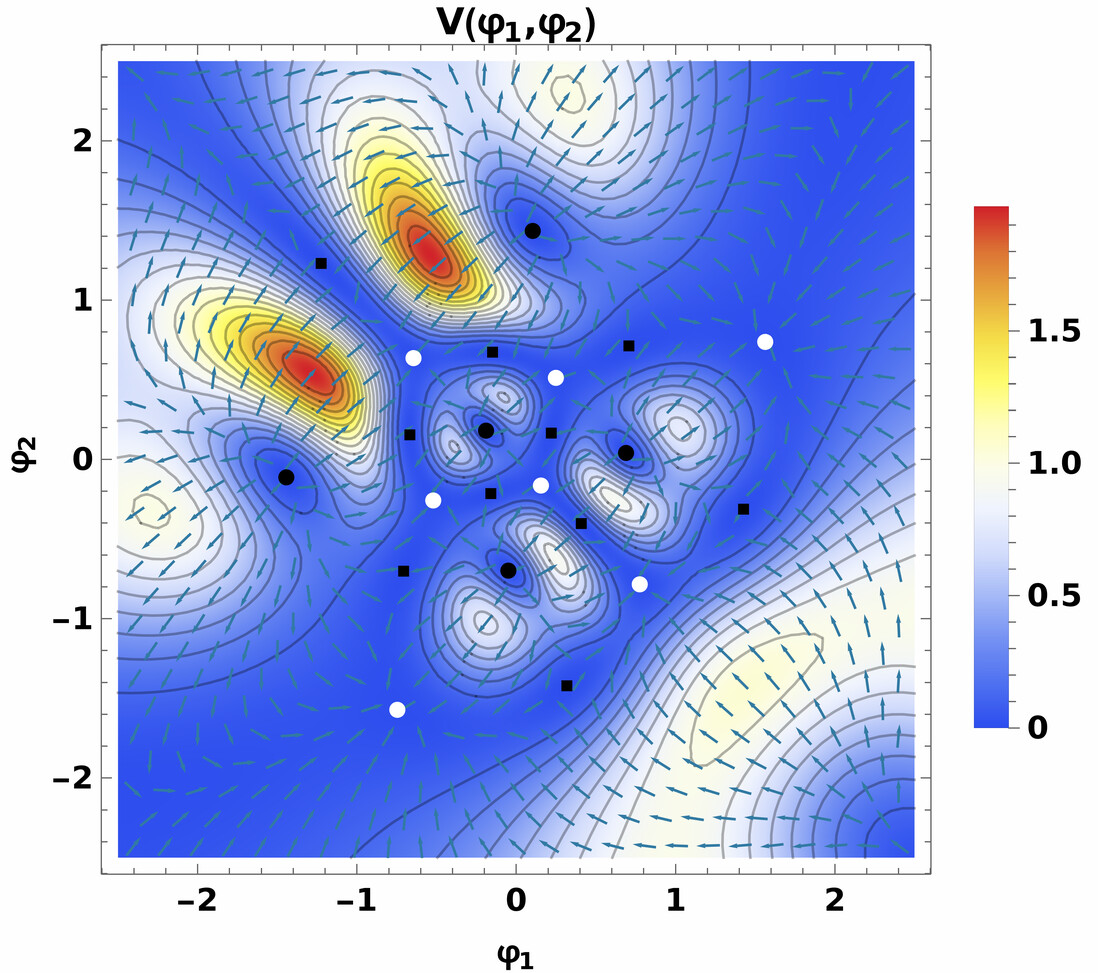}
        \caption{}
    \end{subfigure}
    \hfill
    \begin{subfigure}{0.45\textwidth}
        \centering
        \includegraphics[width=\textwidth]{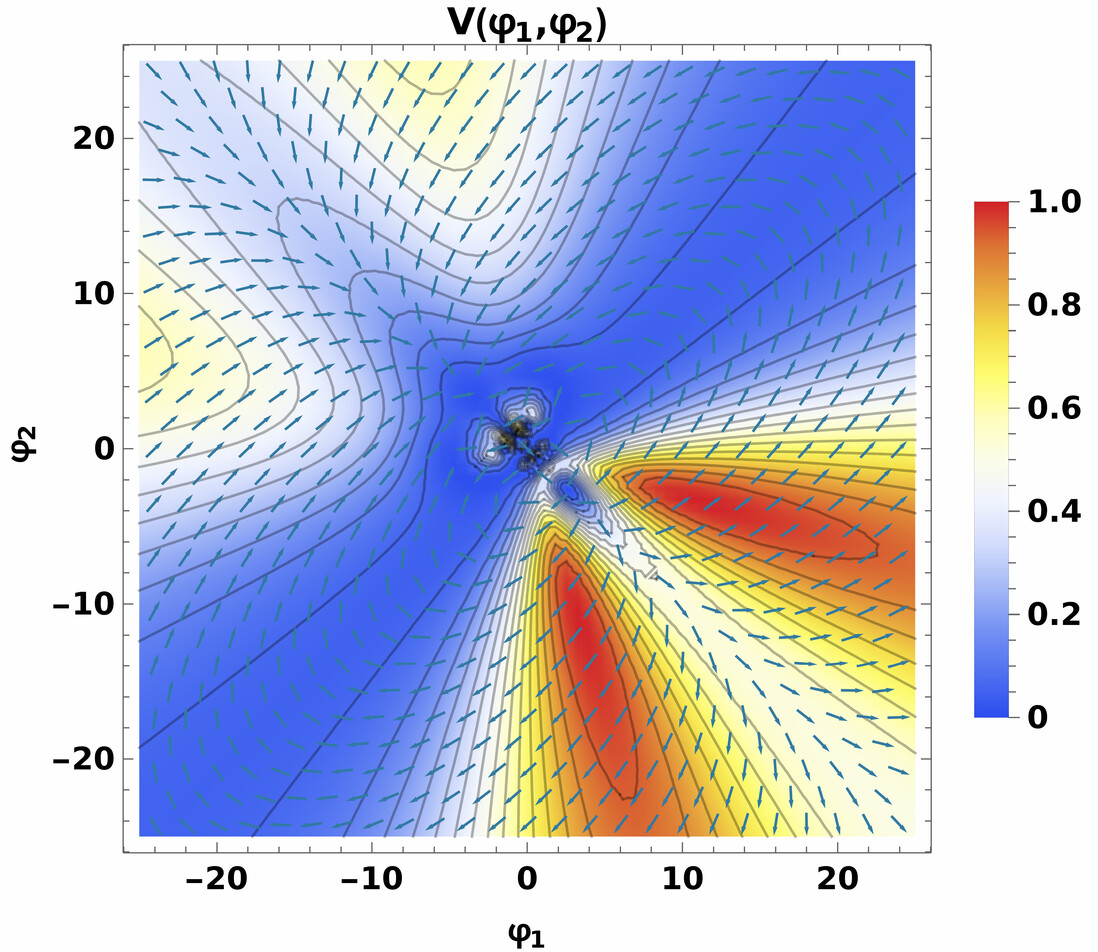}
        \caption{}
    \end{subfigure}
    \par\vspace{0.15cm}
    \begin{subfigure}{0.45\textwidth}
        \centering
        \includegraphics[width=\textwidth]{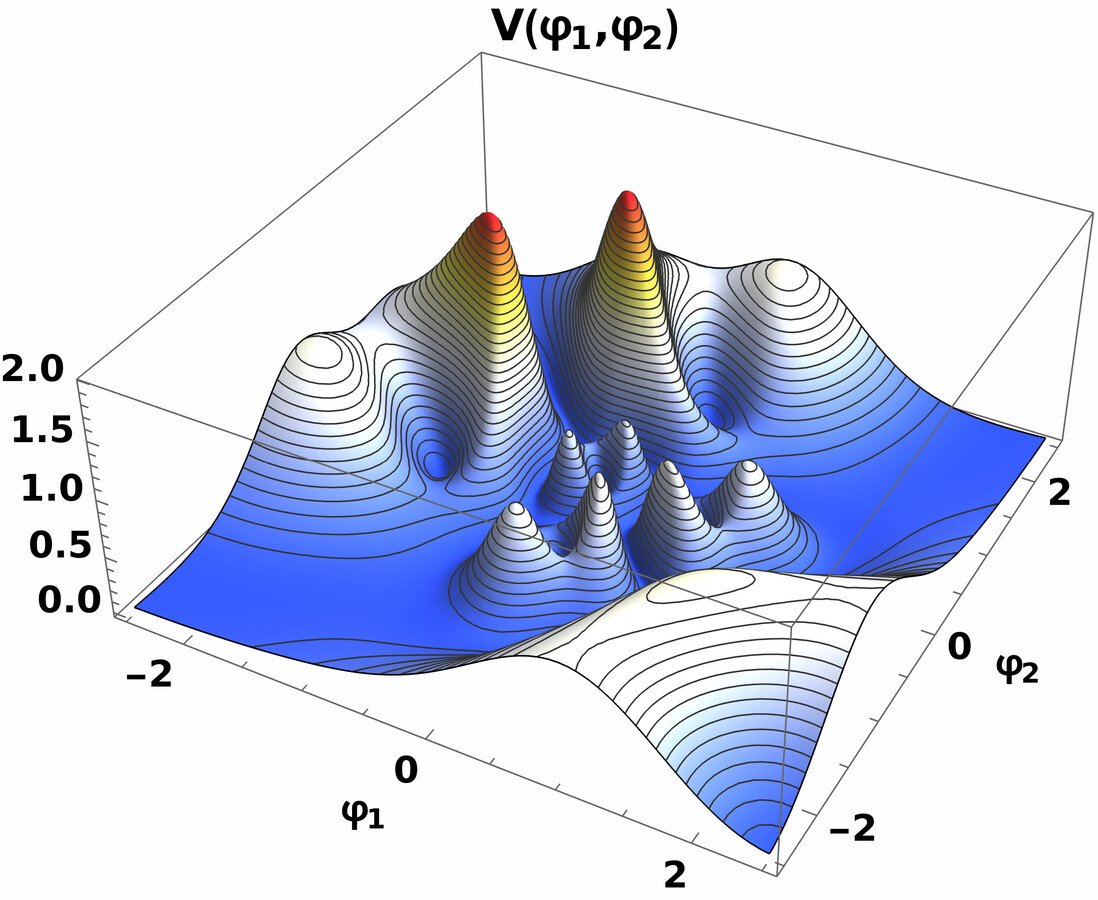}
        \caption{}
    \end{subfigure}
    \hfill
    \begin{subfigure}{0.45\textwidth}
        \centering
        \includegraphics[width=\textwidth]{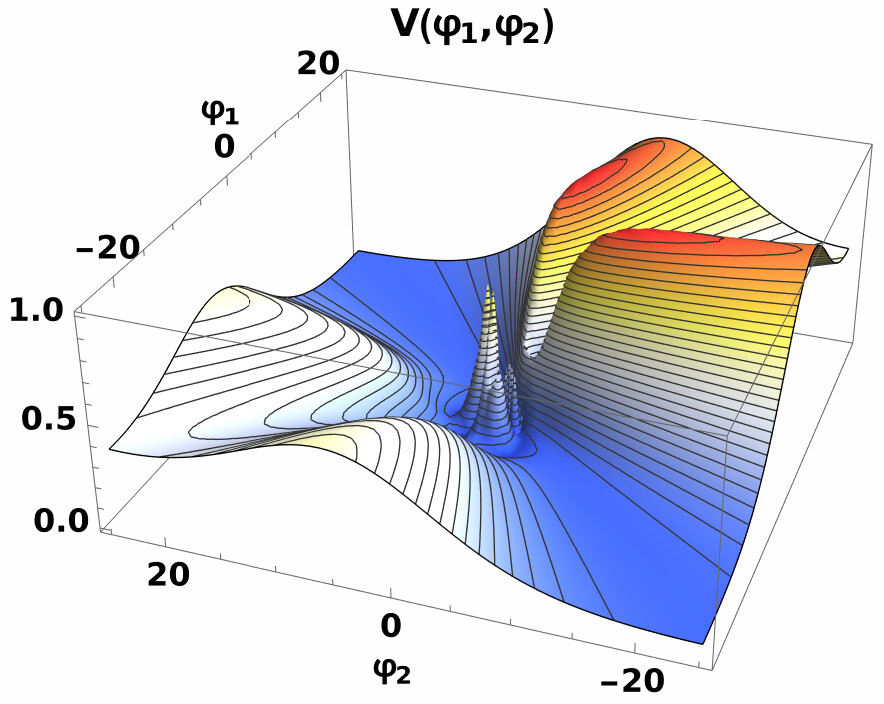}
        \caption{}
        \label{V_3d_octa_longe}
    \end{subfigure}
    \caption{Octahedral potential $V$ at two scales: (a,c) a close view
    near the origin and (b,d) a wider view of the stereographic plane.
    The upper panels include vacuum markers and the metric gradient
    of $U_O$; the lower panels show the corresponding surfaces.}
    \label{fig:V_octa2}
    
\end{figure}

Figure~\ref{fig:V_octa2} resolves the closely spaced vacua and potential
barriers near the origin, and shows the same potential over a wider
range of field values. The metric-gradient arrows indicate the
possible self-dual trajectories through these regions.

\begin{figure}[!htbp]
    \centering
    \begin{subfigure}{0.27\textwidth}
        \centering
        \includegraphics[width=\textwidth]{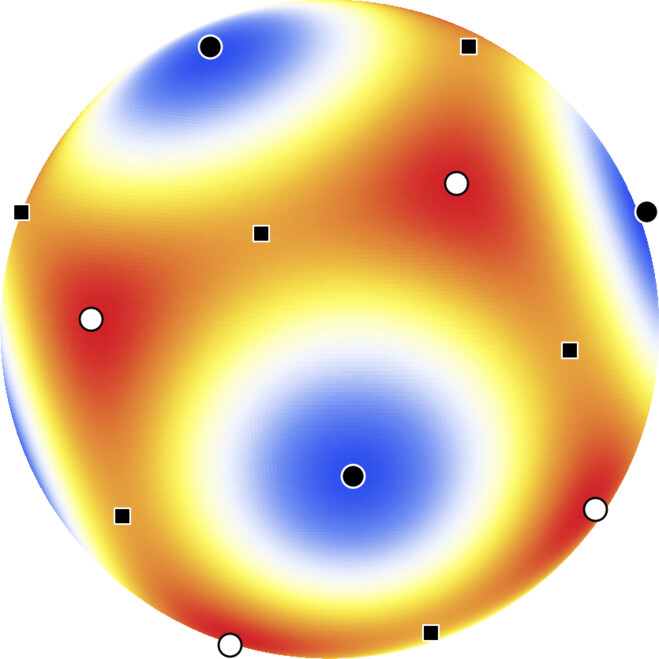}
        \caption{}
        \label{U_octa_sphere}
    \end{subfigure}
    \hspace{0.03\textwidth}
    \begin{subfigure}{0.27\textwidth}
        \centering
        \includegraphics[width=\textwidth]{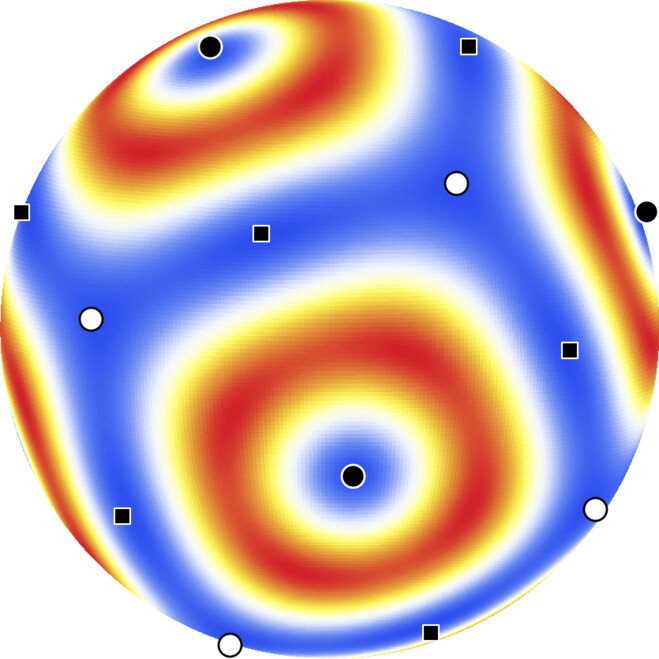}
        \caption{}
        \label{V_octa_sphere}
    \end{subfigure}
    \caption{Octahedral (a) pre-potential $U_O$ and (b) potential $V$
    represented by color on $\mathbb{S}^2$. Black filled circles,
    white circles, and black squares mark minima, maxima, and saddle
    points of $U_O$, respectively; all are vacua of $V$.}
    \label{fig:U_V_octa_sphere}
    
\end{figure}

In Figure~\ref{fig:U_V_octa_sphere}, the spherical representation makes
the rotated octahedral symmetry and the regular-lattice geometry
apparent. All three classes of critical points of $U_O$ appear as
vacua of $V$.
\FloatBarrier
\subsection{Cube}
\begin{figure}[!htbp]
    \centering
    \begin{subfigure}{0.45\textwidth}
        \centering
        \includegraphics[width=\textwidth]{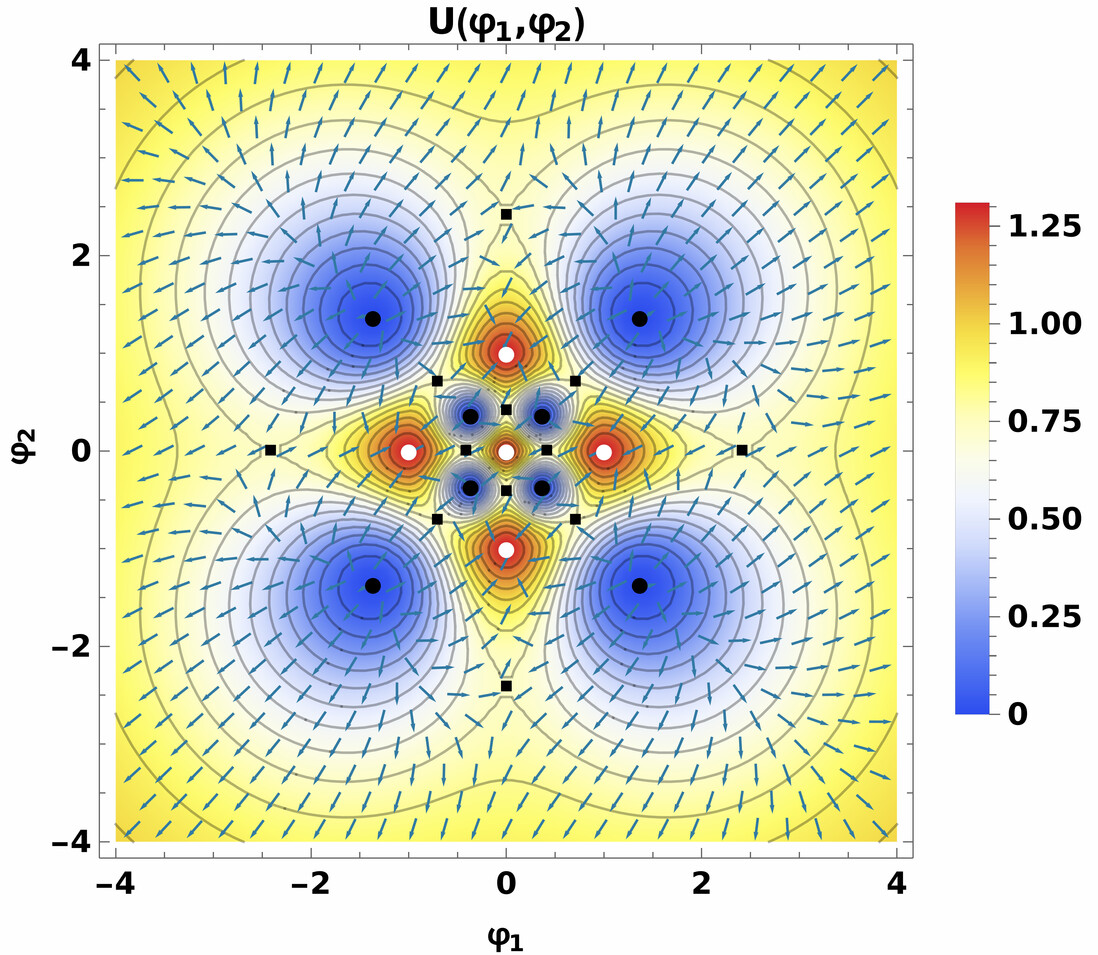}
        \caption{}
        \label{U_2d_cubo_antes}
    \end{subfigure}
    \hfill
    \begin{subfigure}{0.45\textwidth}
        \centering
        \includegraphics[width=\textwidth]{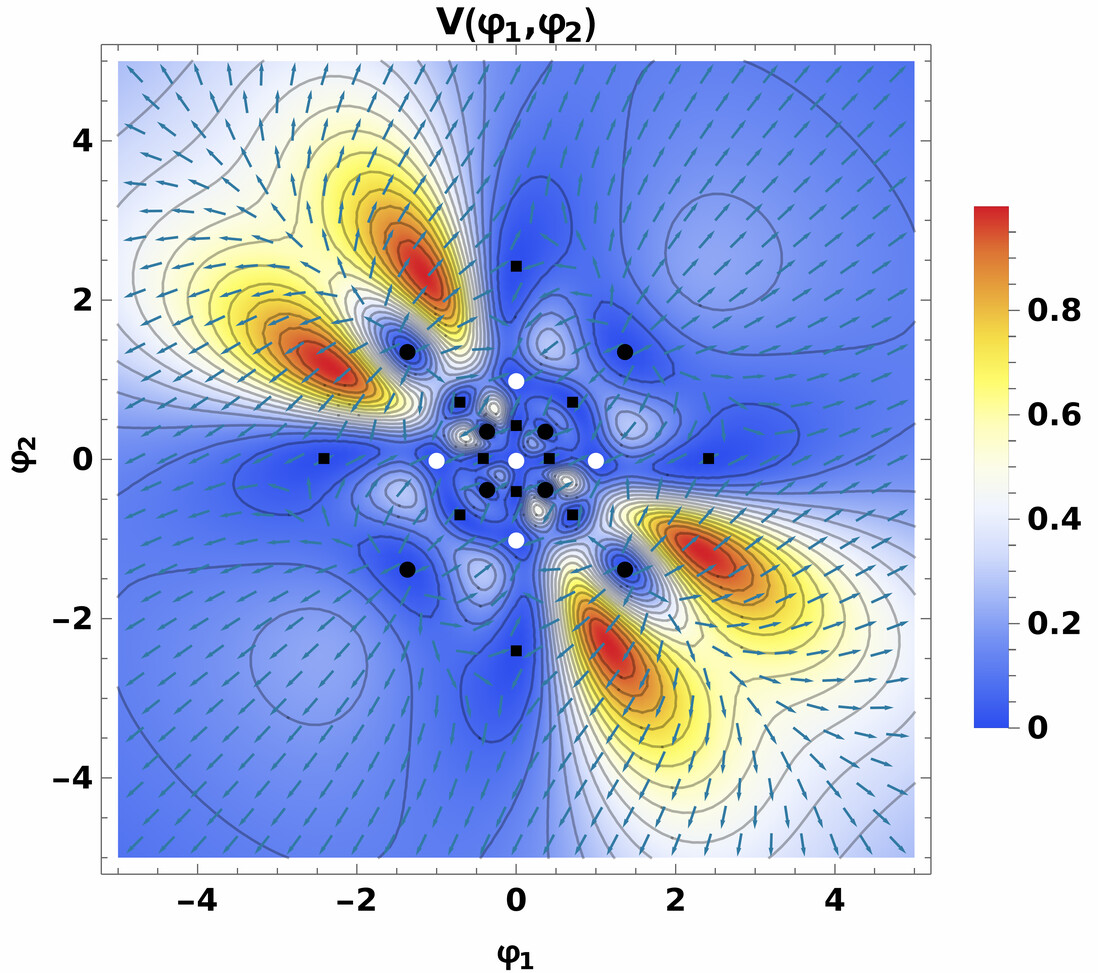}
        \caption{}
        \label{V_2d_cubo_antes}
    \end{subfigure}
    \par\vspace{0.15cm}
    \begin{subfigure}{0.45\textwidth}
        \centering
        \includegraphics[width=\textwidth]{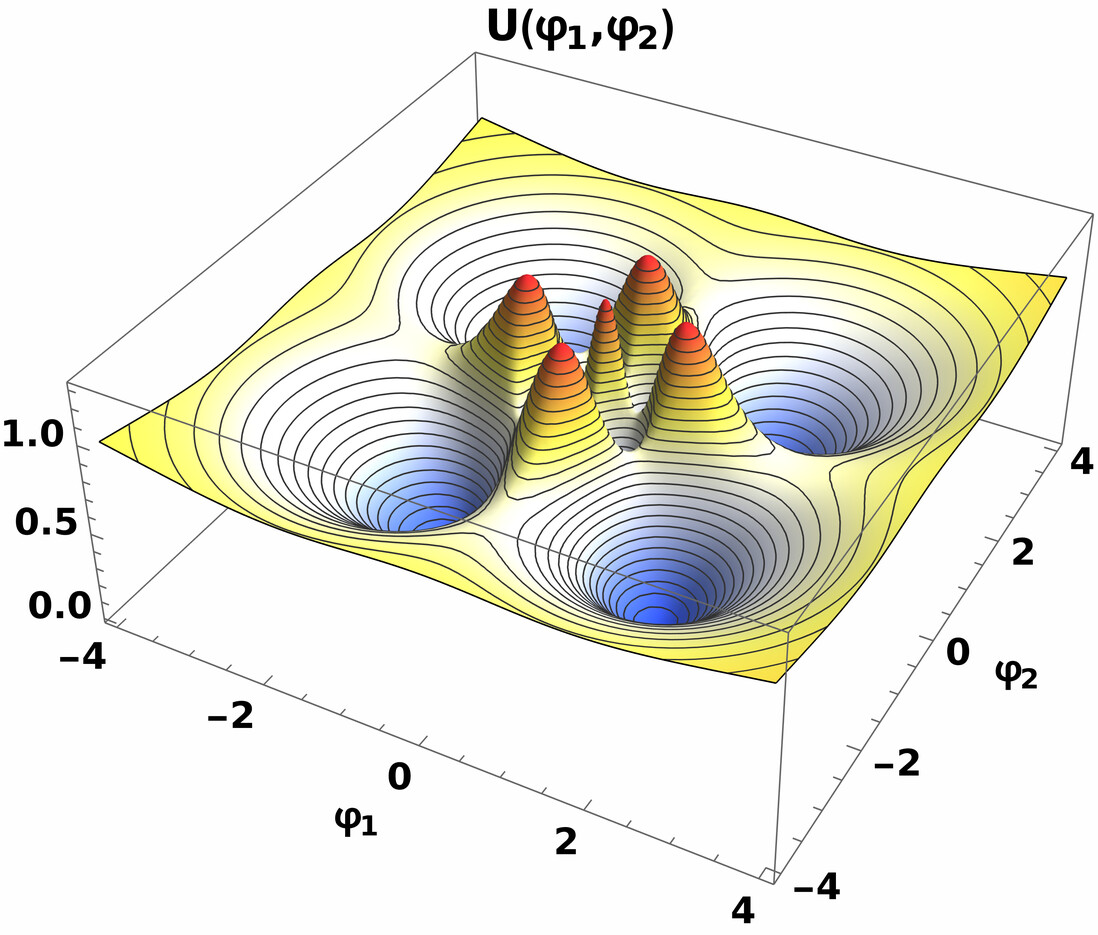}
        \caption{}
    \end{subfigure}
    \hfill
    \begin{subfigure}{0.45\textwidth}
        \centering
        \includegraphics[width=\textwidth]{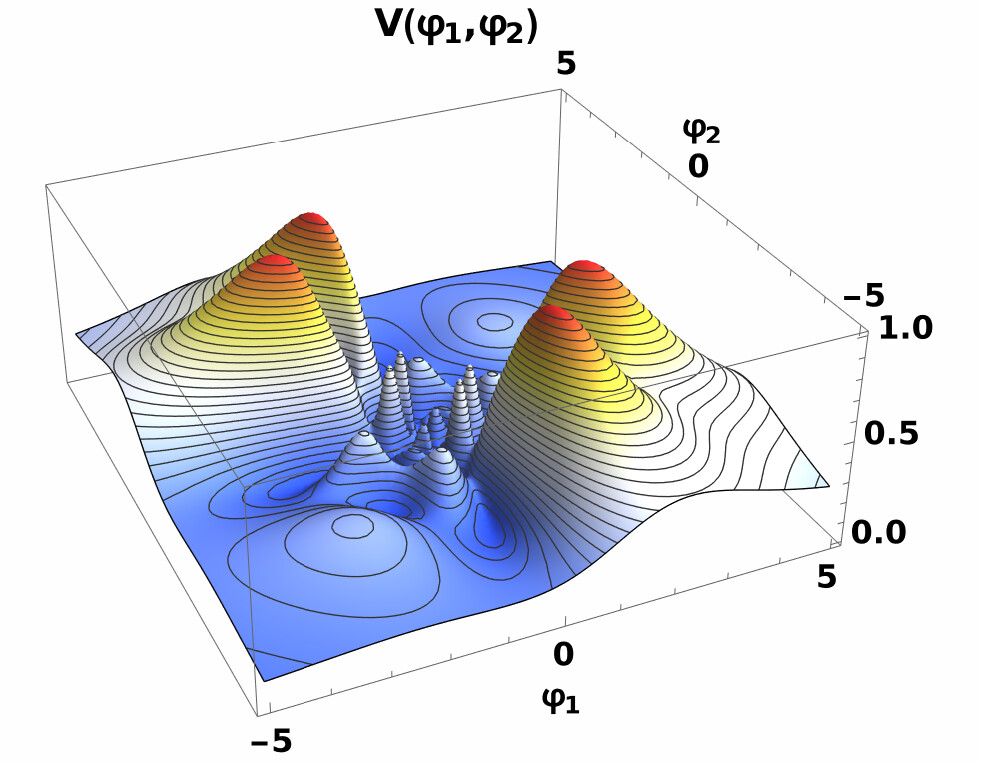}
        \caption{}
    \end{subfigure}
    \caption{Reference plots in the unrotated cubic orientation:
    (a,c) pre-potential and (b,d) potential. The upper panels show
    contours, critical-point markers, and metric-gradient arrows;
    the lower panels show the corresponding surfaces.}
    \label{fig:U_V_cube_antes}
    
\end{figure}

The eight vertices in the reference orientation are
\begin{equation}
    V_C=\frac{1}{\sqrt{3}}\qty{(\pm1,\pm1,\pm1)},
\end{equation}
with all signs chosen independently.
Figure~\ref{fig:U_V_cube_antes} shows the unrotated reference plots.
Here the north pole is a face-center direction and a maximum of the
pre-potential, corresponding to a vacuum at stereographic infinity.
To preserve the common orientation of the dual pair, we apply the same
rotation \eqref{eq:rot_mat}:
\begin{equation}
    V'_C=R V_C.
    \label{eq:vertices_cubo_sphere}
\end{equation}

Using this vertex set, the normalized cubic pre-potential is
\begin{equation}
    U_C(\vb*{\phi})=\frac{256}{25}
    \prod_{a=1}^{8}\qty(1-v_a\cdot\vb*{\phi}),
    \qquad v_a\in V'_C,
    \label{eq:U_cube}
\end{equation}
whose planar form again follows from \eqref{eq:change_coordinates}.
It is invariant under the same conjugate rotation group as $U_O$.
The two pre-potentials share the 26 regular-lattice sites, with minima
and maxima interchanged and the twelve saddle sites in common, as
summarized in Table~\ref{tab:octa_cube_lattice}.
\begin{table}[htbp]
    \centering
    \begin{tabular}{lccc}
        \hline
        Sites in the common orientation & Number & $U_O$ & $U_C$\\
        \hline
        Octahedron vertices $V'_O$ & 6 & Minima & Maxima\\
        Cube vertices $V'_C$ & 8 & Maxima & Minima\\
        Rotated edge-midpoint directions & 12 & Saddles & Saddles\\
        \hline
    \end{tabular}
    \caption{Classification of the shared regular-lattice sites of the
    octahedron--cube pair. All 26 sites are vacua of both associated
    potentials.}
    \label{tab:octa_cube_lattice}
\end{table}

\begin{figure}[!htbp]
    \centering
    \begin{subfigure}{0.45\textwidth}
        \centering
        \includegraphics[width=\textwidth]{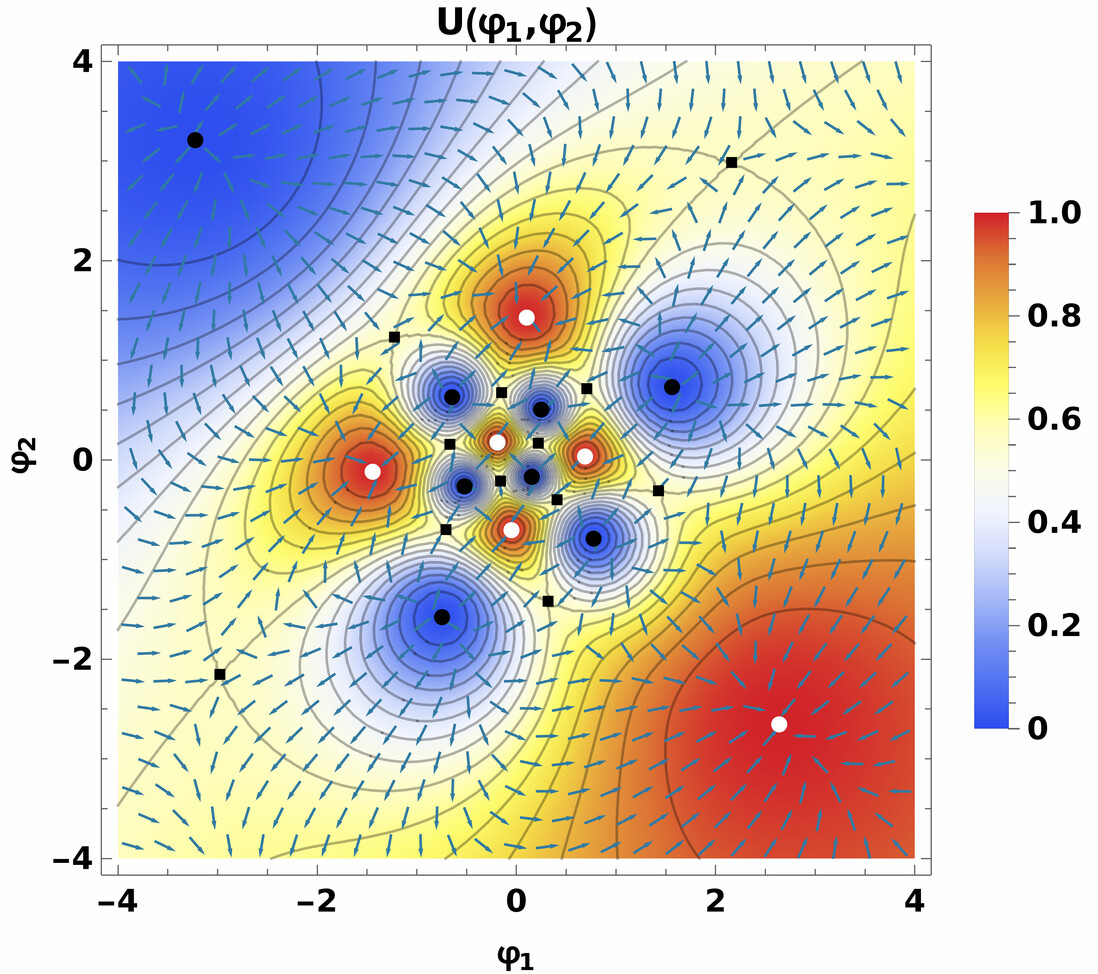}
        \caption{}
        \label{U_2d_cubo}
    \end{subfigure}
    \hfill
    \begin{subfigure}{0.45\textwidth}
        \centering
        \includegraphics[width=\textwidth]{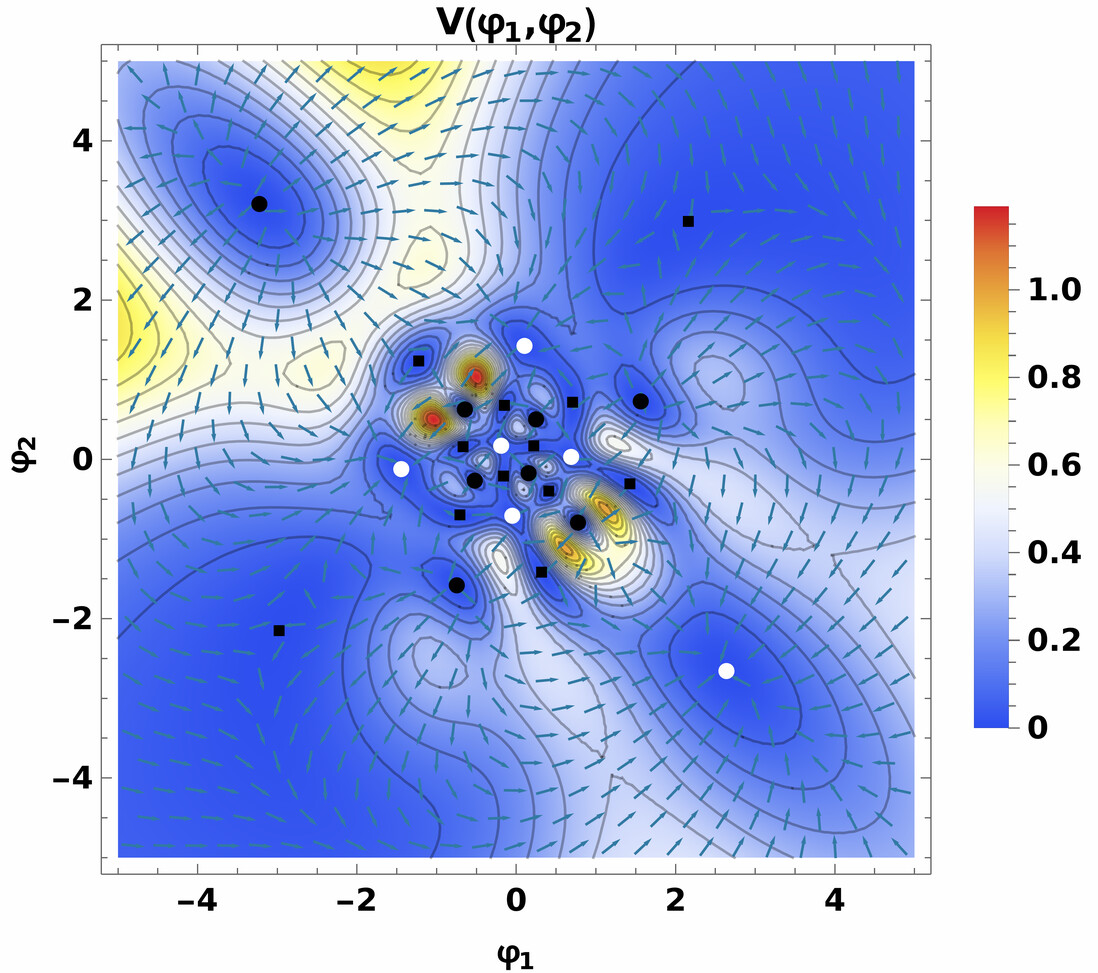}
        \caption{}
        \label{V_2d_cubo}
    \end{subfigure}
    \par\vspace{0.15cm}
    \begin{subfigure}{0.45\textwidth}
        \centering
        \includegraphics[width=\textwidth]{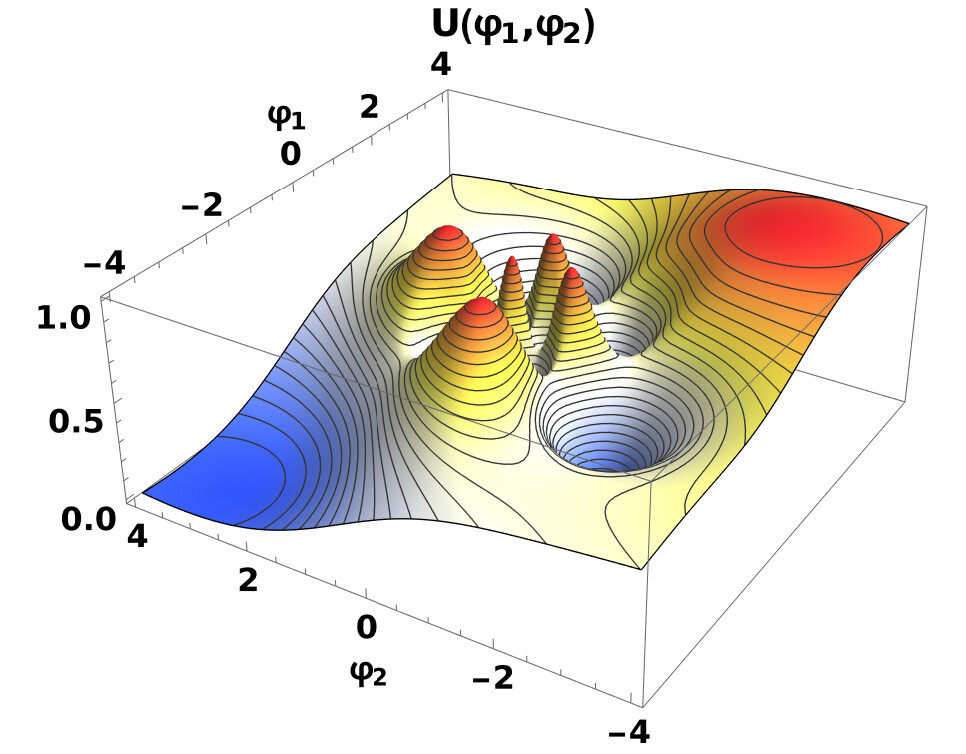}
        \caption{}
    \end{subfigure}
    \hfill
    \begin{subfigure}{0.45\textwidth}
        \centering
        \includegraphics[width=\textwidth]{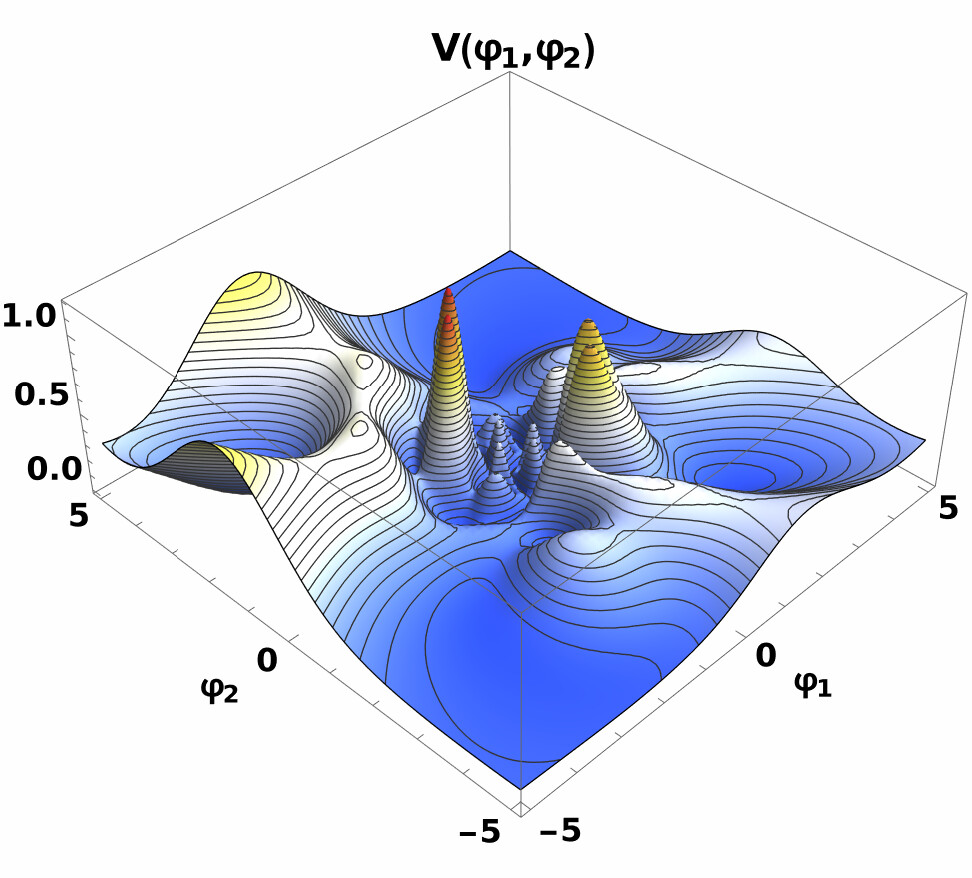}
        \caption{}
        \label{V_3d_cube}
    \end{subfigure}
    \caption{Cubic pre-potential $U_C$ in (a,c) and potential $V$
    in (b,d), after the common rotation \eqref{eq:rot_mat}. The upper
    panels show contours, vacuum markers, and the metric gradient of
    $U_C$; the lower panels show the corresponding surfaces.}
    \label{fig:U_V_cubo}
    
\end{figure}

For the cubic model, the critical values are
\begin{equation}
    U_{min}=0,\qquad U_{saddle}\approx1.138,\qquad U_{max}\approx2.023.
    \label{eq:valores_U_cube}
\end{equation}
Compared with \eqref{eq:valores_U_octa}, these values show that the
common vacuum positions support different pre-potential differences
and hence different possible self-dual energies. Com\-pa\-ring
Figures~\ref{fig:U_V_octa} and \ref{fig:U_V_cubo} also reveals the
interchange of minima and maxima and the resulting differences in the
metric-gradient fields. The reflection
$(\vp_1,\vp_2)\mapsto(-\vp_2,-\vp_1)$ remains a symmetry of both $U_C$
and its associated potential.

\begin{figure}[!htbp]
    \centering
    \begin{subfigure}{0.45\textwidth}
        \centering
        \includegraphics[width=\textwidth]{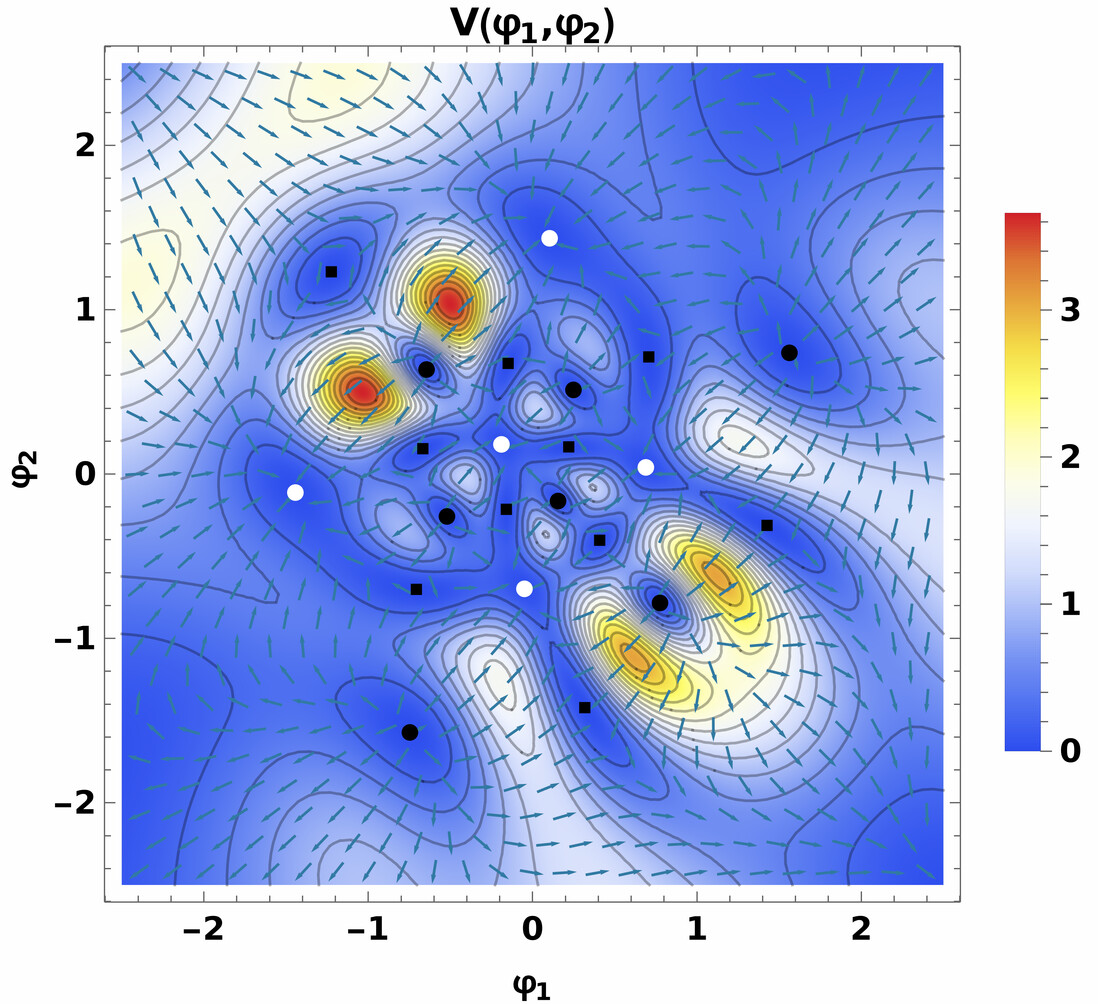}
        \caption{}
    \end{subfigure}
    \hfill
    \begin{subfigure}{0.45\textwidth}
        \centering
        \includegraphics[width=\textwidth]{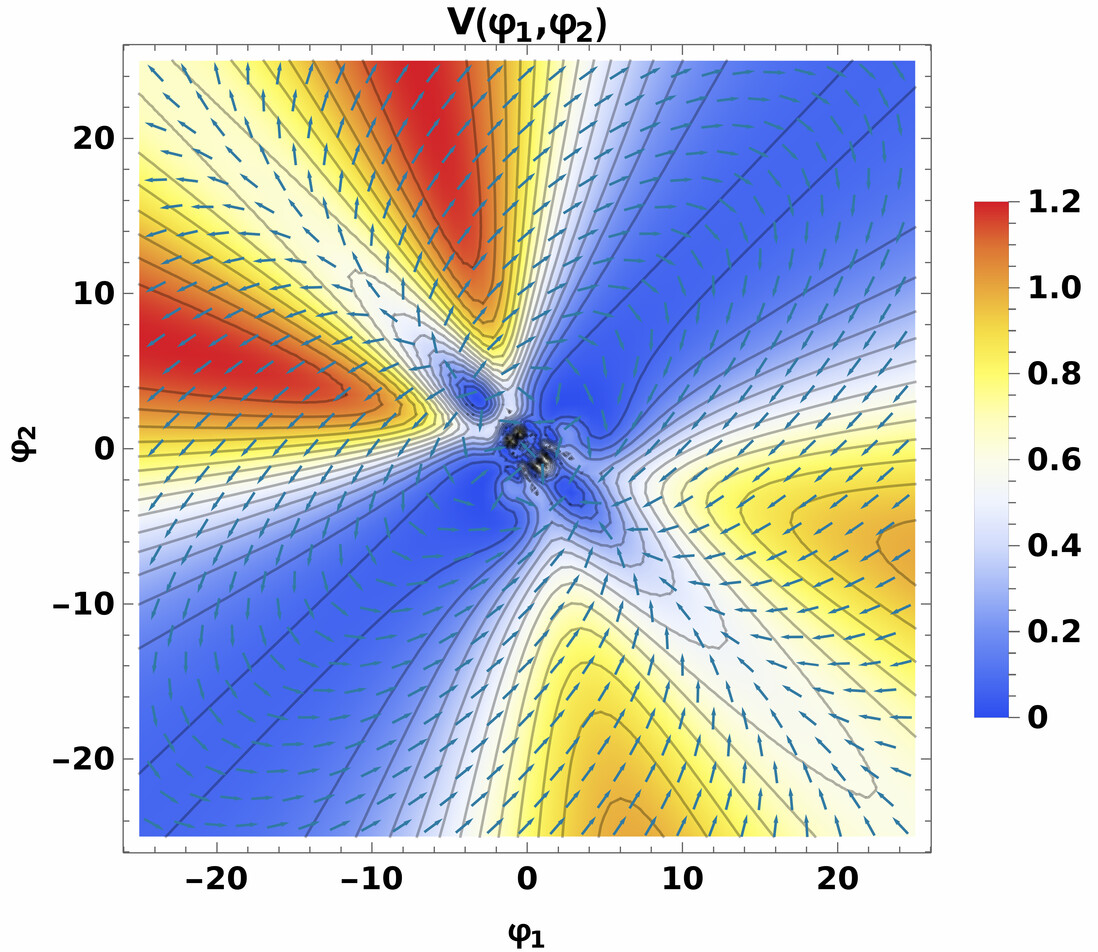}
        \caption{}
    \end{subfigure}
    \par\vspace{0.15cm}
    \begin{subfigure}{0.45\textwidth}
        \centering
        \includegraphics[width=\textwidth]{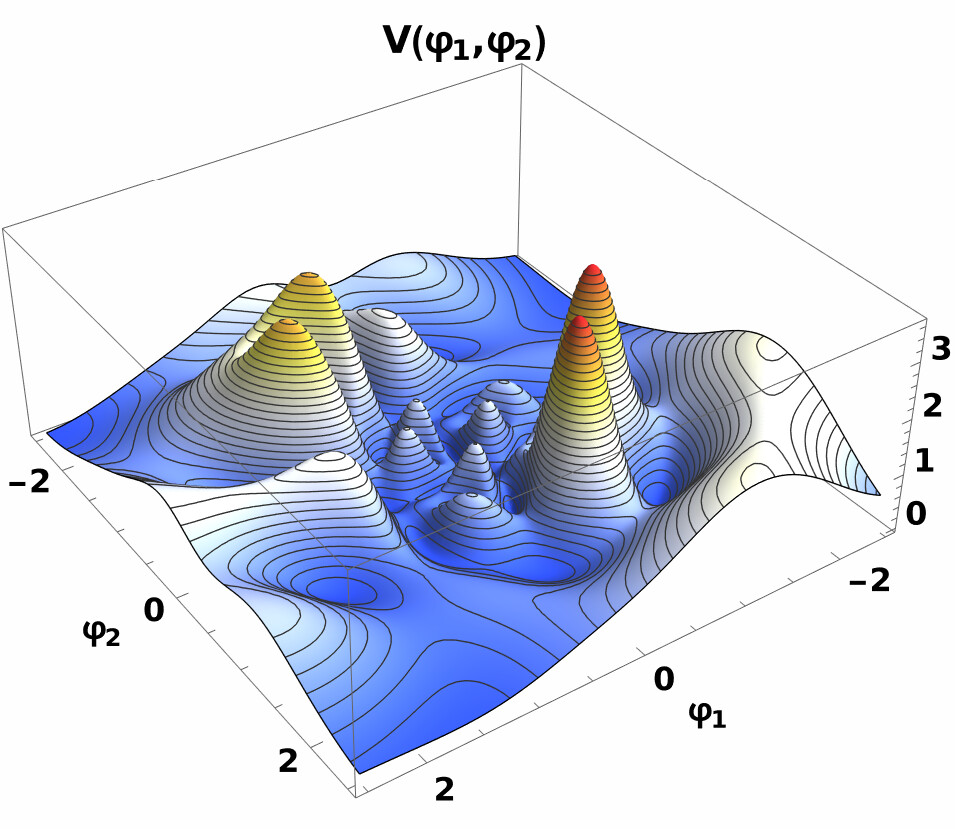}
        \caption{}
    \end{subfigure}
    \hfill
    \begin{subfigure}{0.45\textwidth}
        \centering
        \includegraphics[width=\textwidth]{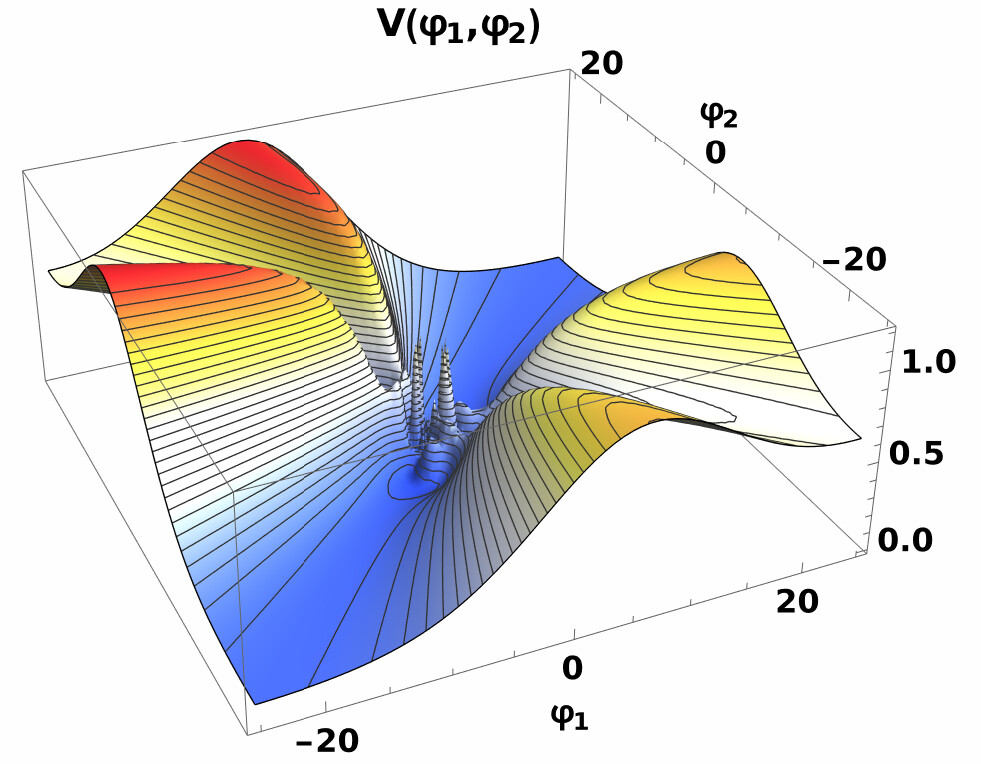}
        \caption{}
    \end{subfigure}
    \caption{Cubic potential $V$ at two scales: (a,c) a close view
    near the origin and (b,d) a wider view of the stereographic plane.
    The upper panels include vacuum markers and the metric gradient
    of $U_C$; the lower panels show the corresponding surfaces.}
    \label{fig:V_cube2}
    
\end{figure}

Figure~\ref{fig:V_cube2} shows the potential near the clustered vacua
and over a wider field range. These views resolve the barriers and
metric-gradient patterns that distinguish the cubic model from the
octahedral model despite their common vacuum positions.

\begin{figure}[!htbp]
    \centering
    \begin{subfigure}{0.27\textwidth}
        \centering
        \includegraphics[width=\textwidth]{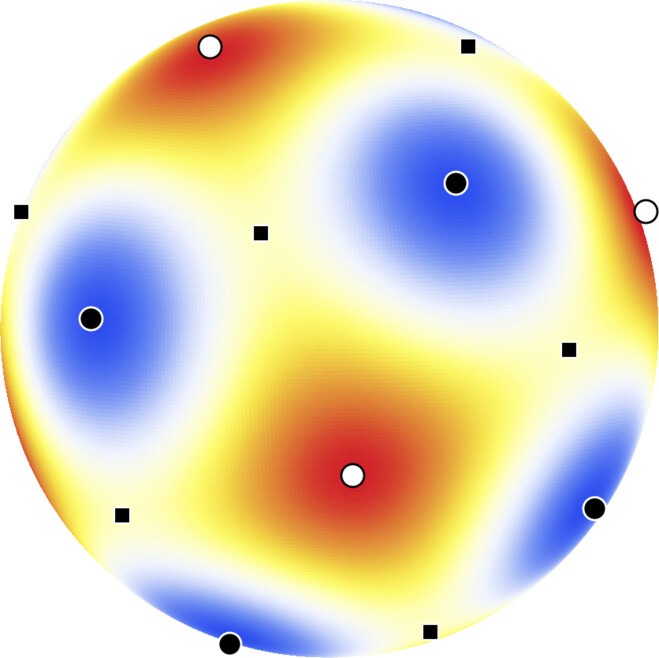}
        \caption{}
        \label{U_cube_sphere}
    \end{subfigure}
    \hspace{0.03\textwidth}
    \begin{subfigure}{0.27\textwidth}
        \centering
        \includegraphics[width=\textwidth]{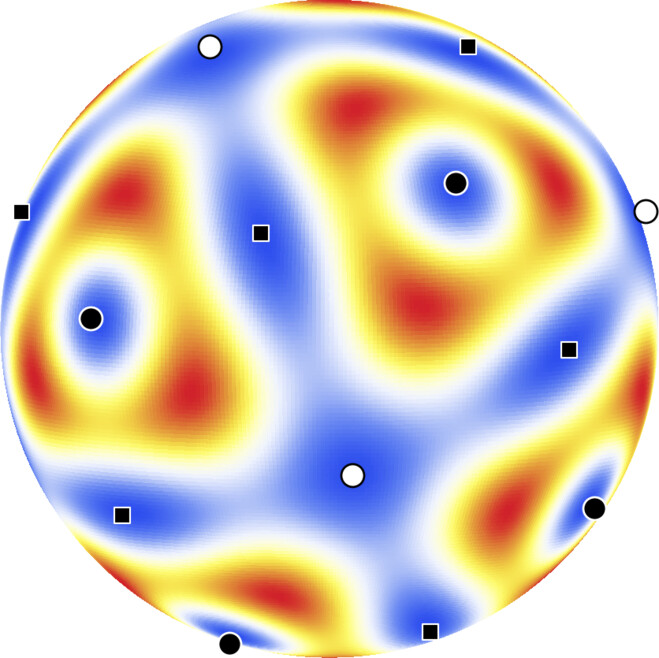}
        \caption{}
        \label{V_cube_sphere}
    \end{subfigure}
    \caption{Cubic (a) pre-potential $U_C$ and (b) potential $V$
    represented by color on $\mathbb{S}^2$. Black filled circles,
    white circles, and black squares mark minima, maxima, and saddle
    points of $U_C$, respectively; all are vacua of $V$.}
    \label{fig:U_V_cube_sphere}
    
\end{figure}

The spherical plots in Figure~\ref{fig:U_V_cube_sphere} make the dual
relation particularly clear when compared with
Figure~\ref{fig:U_V_octa_sphere}: maxima of one pre-potential coincide
with minima of the other. For the associated potentials, the common
lattice instead consists entirely of degenerate minima. Their maxima
need not coincide; the different spatial patterns are visible in
panels (b) of the two spherical figures.
\FloatBarrier

\subsection{Icosahedron}
\begin{figure}[!htbp]
    \centering
    \begin{subfigure}{0.45\textwidth}
        \centering
        \includegraphics[width=\textwidth]{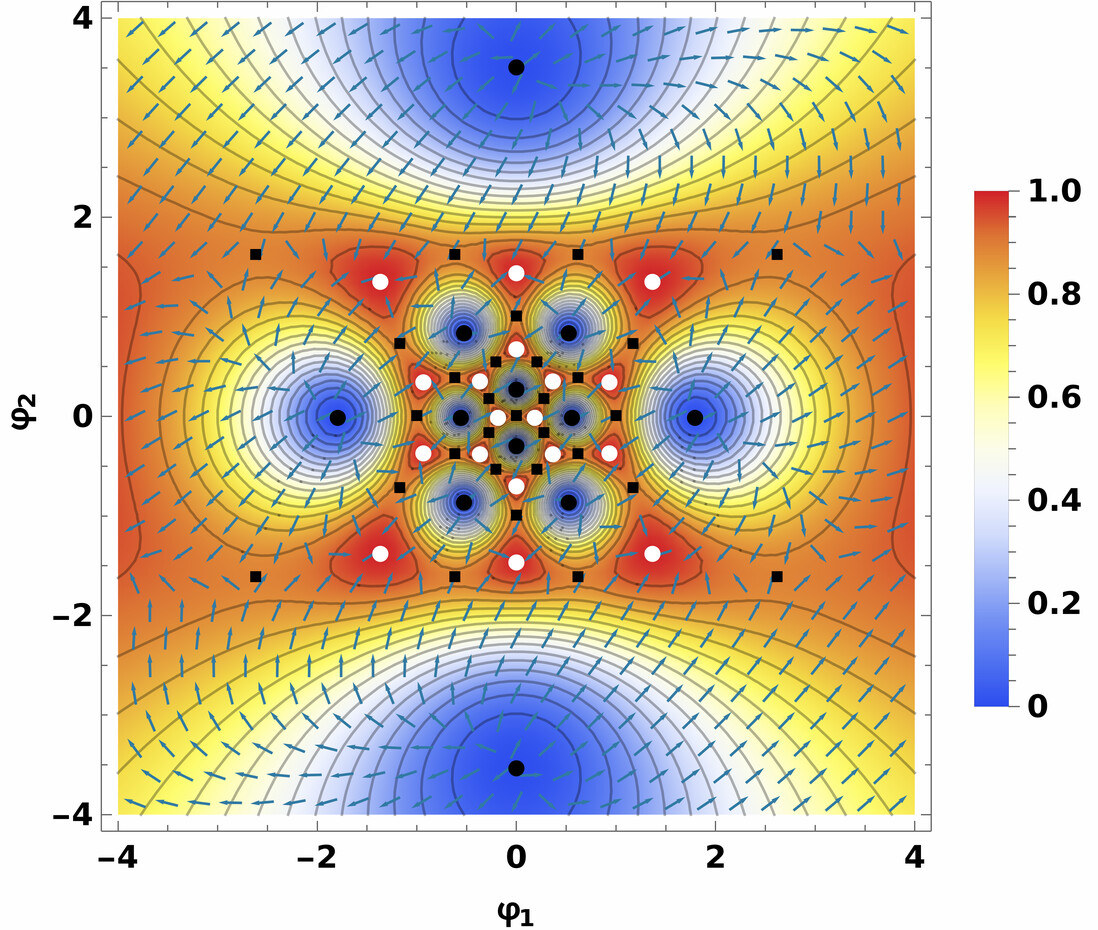}
        \caption{}
        \label{U_2d_icosa_antes}
    \end{subfigure}
    \hfill
    \begin{subfigure}{0.45\textwidth}
        \centering
        \includegraphics[width=\textwidth]{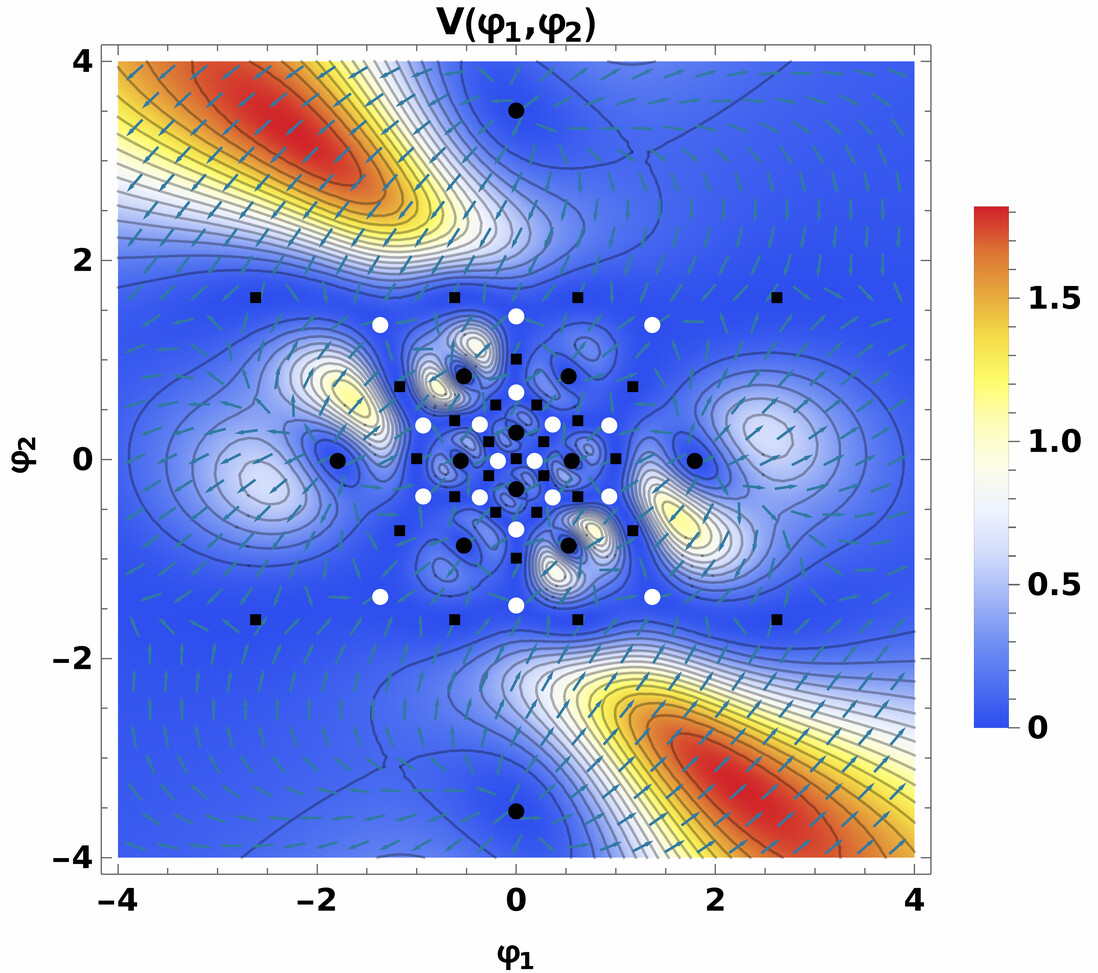}
        \caption{}
        \label{V_2d_icosa_antes}
    \end{subfigure}
    \par\vspace{0.15cm}
    \begin{subfigure}{0.45\textwidth}
        \centering
        \includegraphics[width=\textwidth]{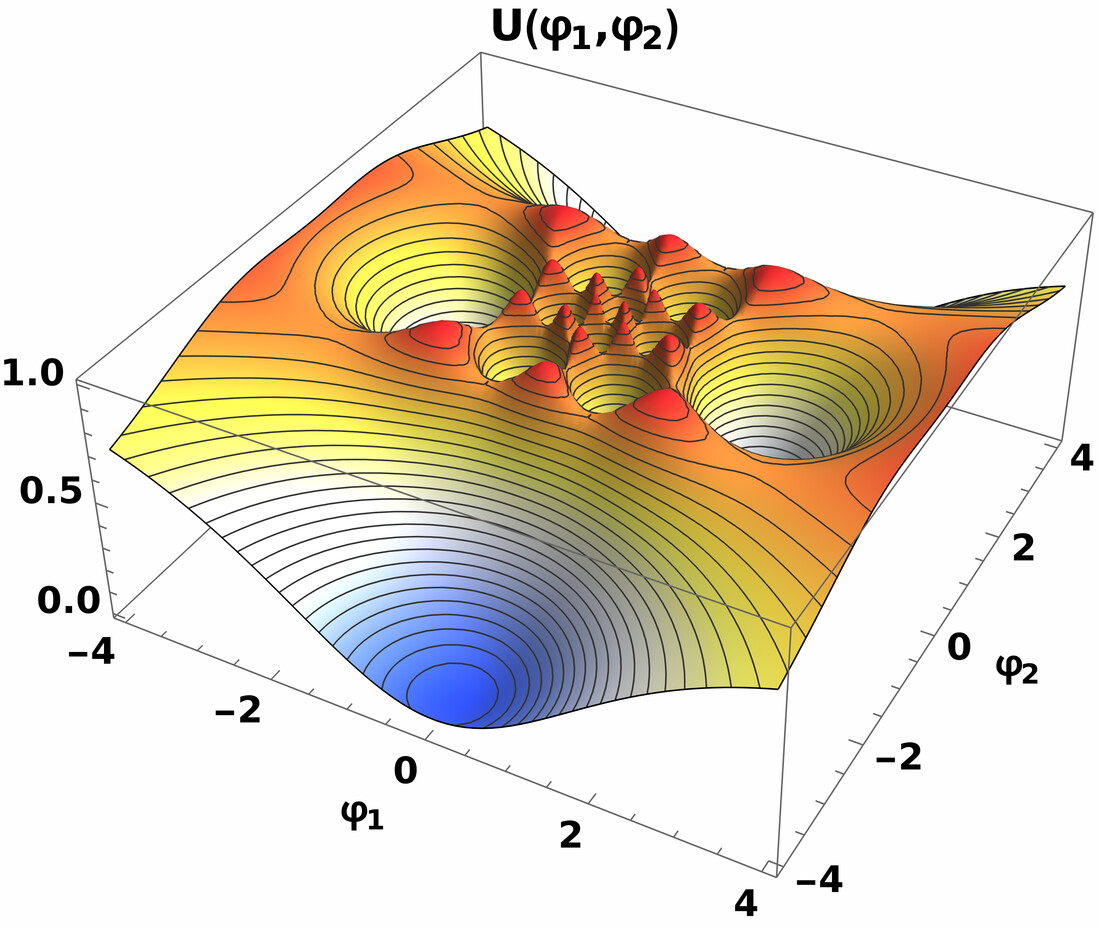}
        \caption{}
    \end{subfigure}
    \hfill
    \begin{subfigure}{0.45\textwidth}
        \centering
        \includegraphics[width=\textwidth]{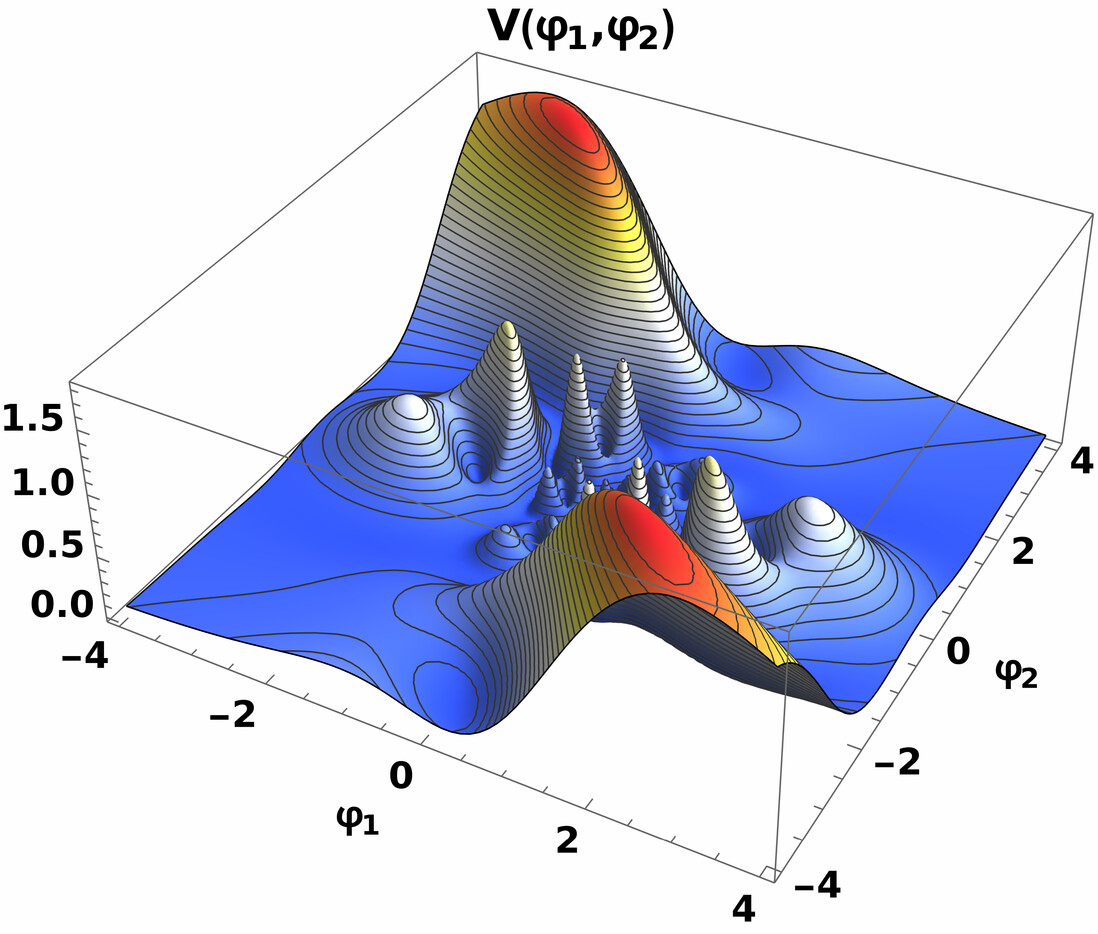}
        \caption{}
    \end{subfigure}
    \caption{Reference plots in the unrotated icosahedral orientation:
    (a,c) pre-potential and (b,d) potential. The upper panels show
    contours, critical-point markers, and metric-gradient arrows;
    the lower panels show the corresponding surfaces.}
    \label{fig:U_V_icosa_antes}
    
\end{figure}

The twelve vertices in the reference orientation are
\begin{equation}
    V_I=\dfrac{1}{\sqrt{1+\tau^2}}
    \qty{(0,\pm1,\pm\tau),(\pm1,\pm\tau,0),(\pm\tau,0,\pm1)},
    \qquad \tau=\dfrac{1+\sqrt{5}}{2},
\end{equation}
with all signs chosen independently and $\tau$ the golden ratio.
Figure~\ref{fig:U_V_icosa_antes} shows the reference plots in this
orientation. The north pole is not a vertex here: it lies on the
two-fold symmetry axis through the midpoint of the edge joining
$(0,\pm1,\tau)/\sqrt{1+\tau^2}$, and is therefore a saddle point of the
pre-potential. The normalization \eqref{eq:constant_C} is thus well
defined in this orientation, but the corresponding regular-lattice site
is mapped to stereographic infinity.

\begin{figure}[!htbp]
    \centering
    \begin{subfigure}{0.45\textwidth}
        \centering
        \includegraphics[width=\textwidth]{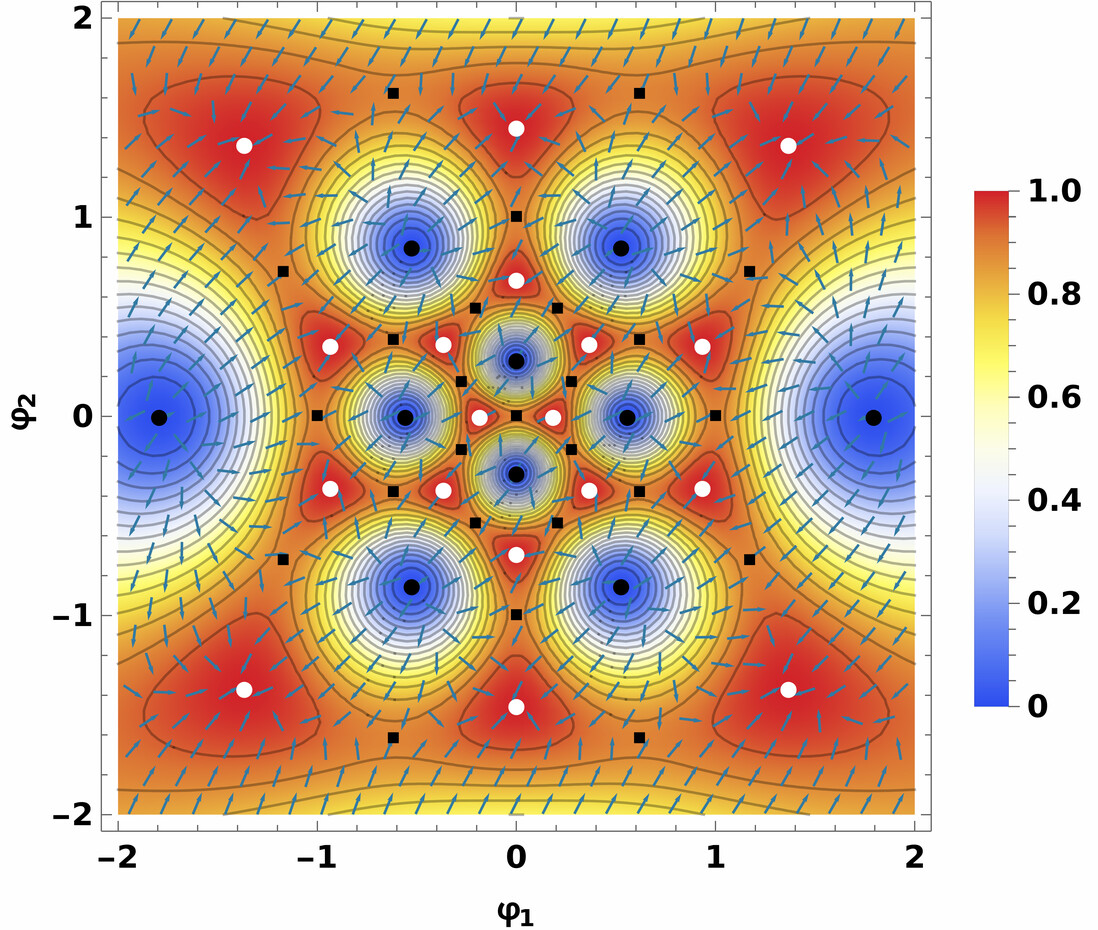}
        \caption{}
        \label{U_2d_icosa_antes_perto}
    \end{subfigure}
    \hfill
    \begin{subfigure}{0.45\textwidth}
        \centering
        \includegraphics[width=\textwidth]{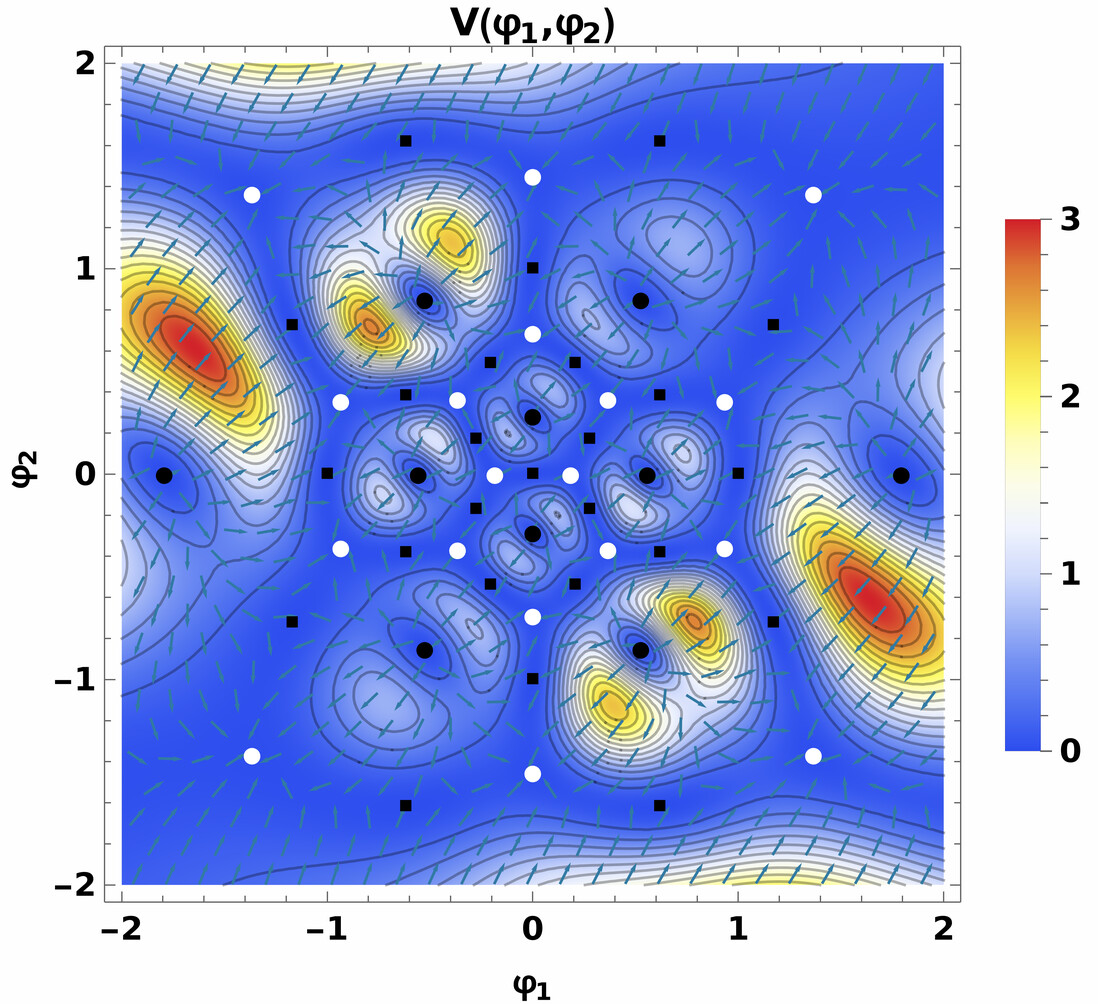}
        \caption{}
        \label{V_2d_icosa_antes_perto}
    \end{subfigure}
    \par\vspace{0.15cm}
    \begin{subfigure}{0.45\textwidth}
        \centering
        \includegraphics[width=\textwidth]{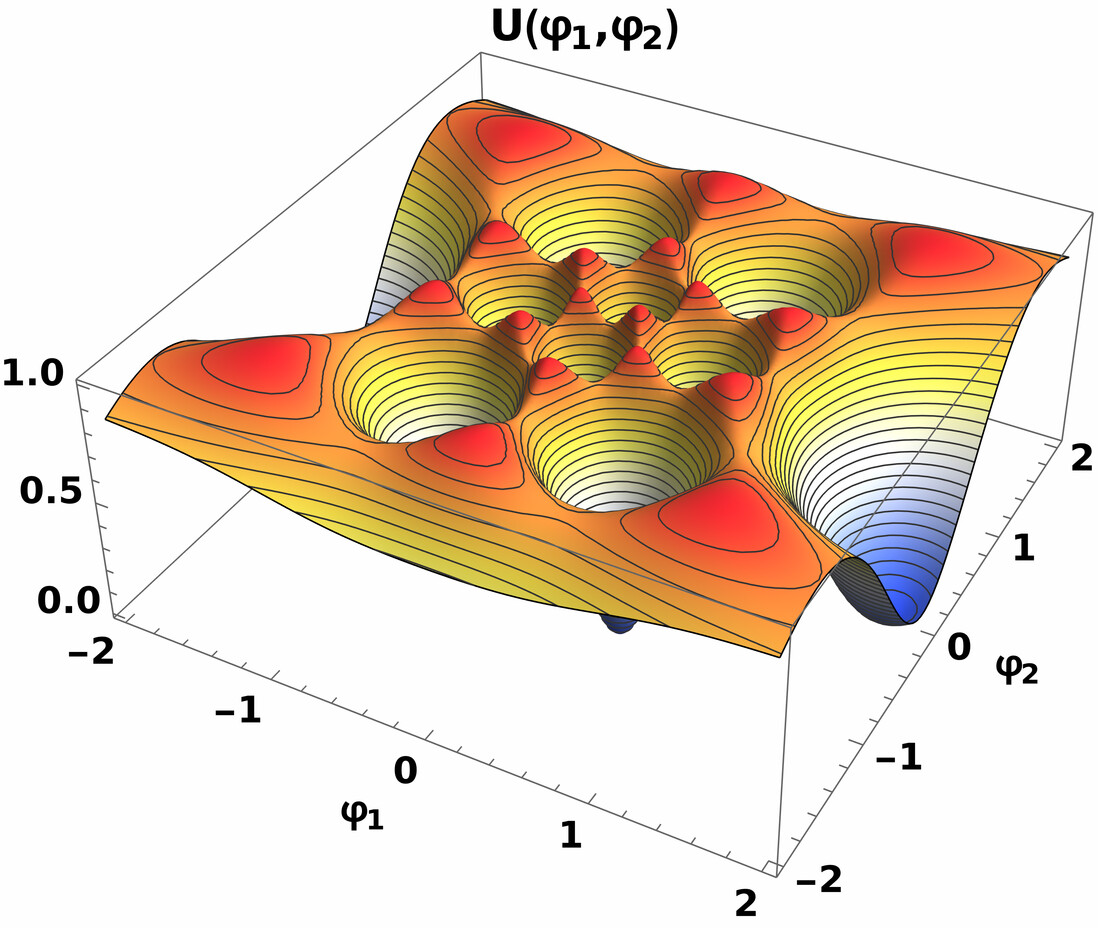}
        \caption{}
    \end{subfigure}
    \hfill
    \begin{subfigure}{0.45\textwidth}
        \centering
        \includegraphics[width=\textwidth]{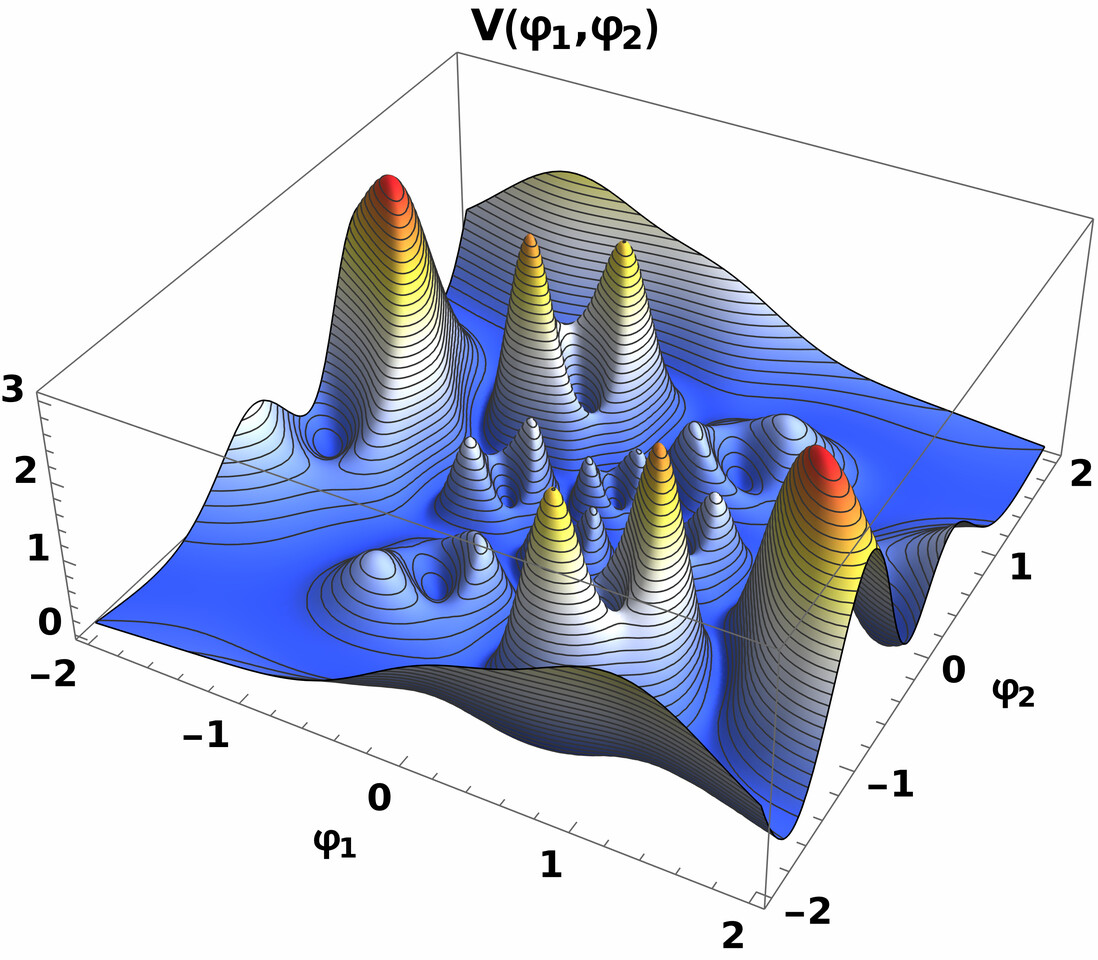}
        \caption{}
    \end{subfigure}
    \caption{Central region of the unrotated icosahedral model:
    (a,c) pre-potential and (b,d) potential.}
    \label{fig:U_V_icosa_antes_perto}
    
\end{figure}

\begin{figure}[!htbp]
    \centering
    \begin{subfigure}{0.45\textwidth}
        \centering
        \includegraphics[width=\textwidth]{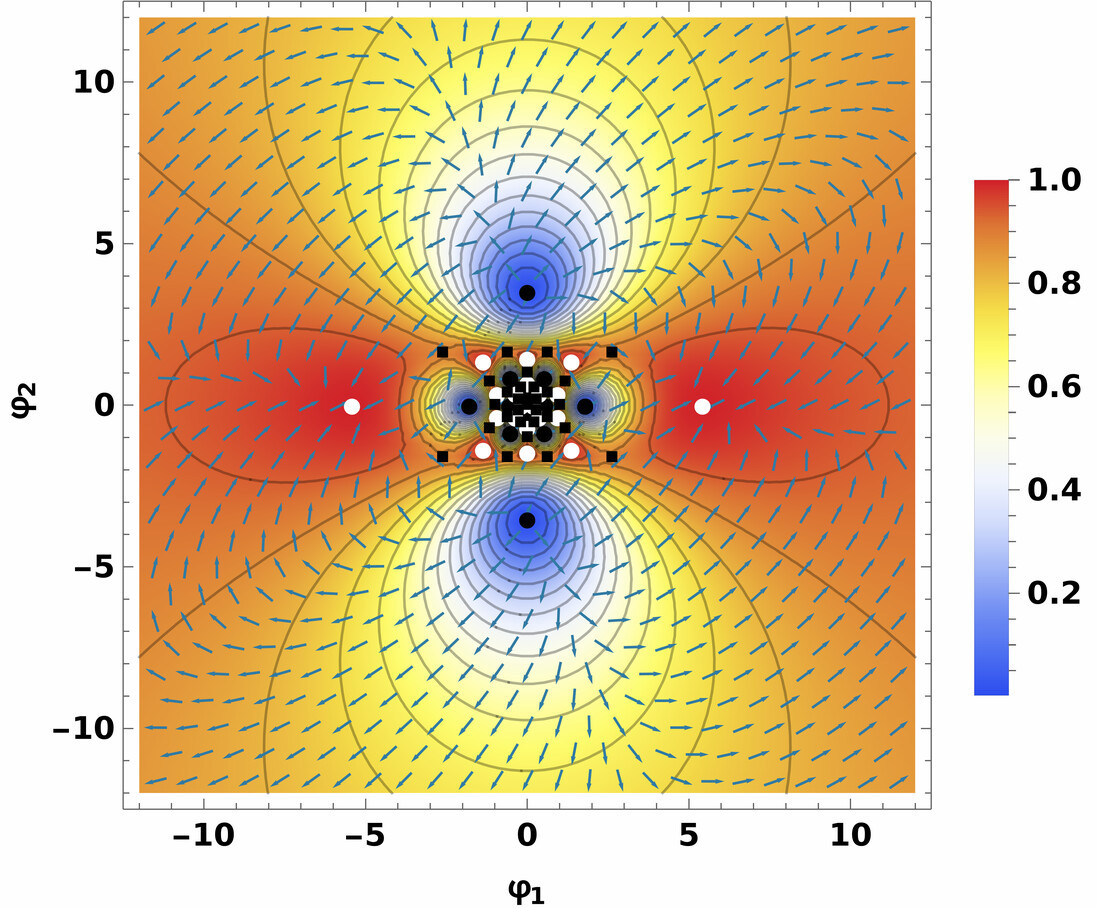}
        \caption{}
        \label{U_2d_icosa_antes_longe}
    \end{subfigure}
    \hfill
    \begin{subfigure}{0.45\textwidth}
        \centering
        \includegraphics[width=\textwidth]{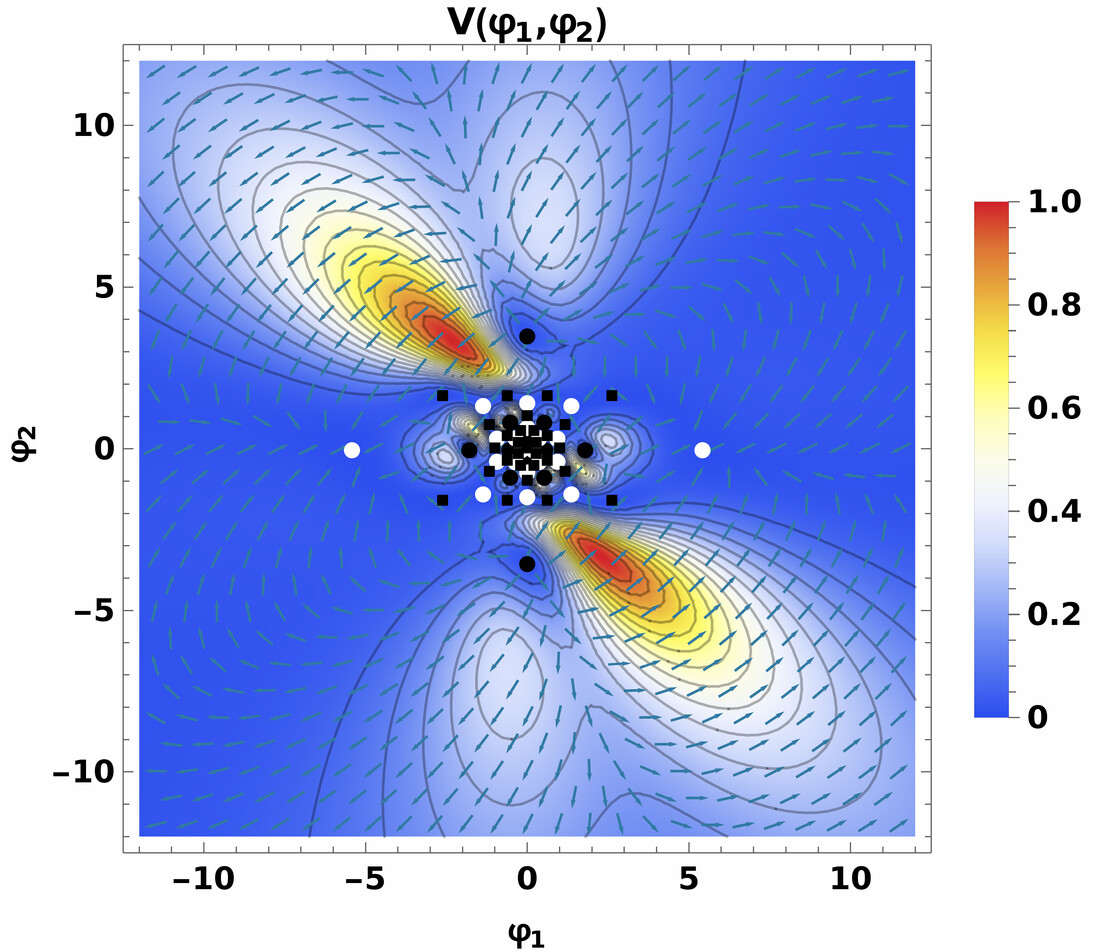}
        \caption{}
        \label{V_2d_icosa_antes_longe}
    \end{subfigure}
    \par\vspace{0.15cm}
    \begin{subfigure}{0.45\textwidth}
        \centering
        \includegraphics[width=\textwidth]{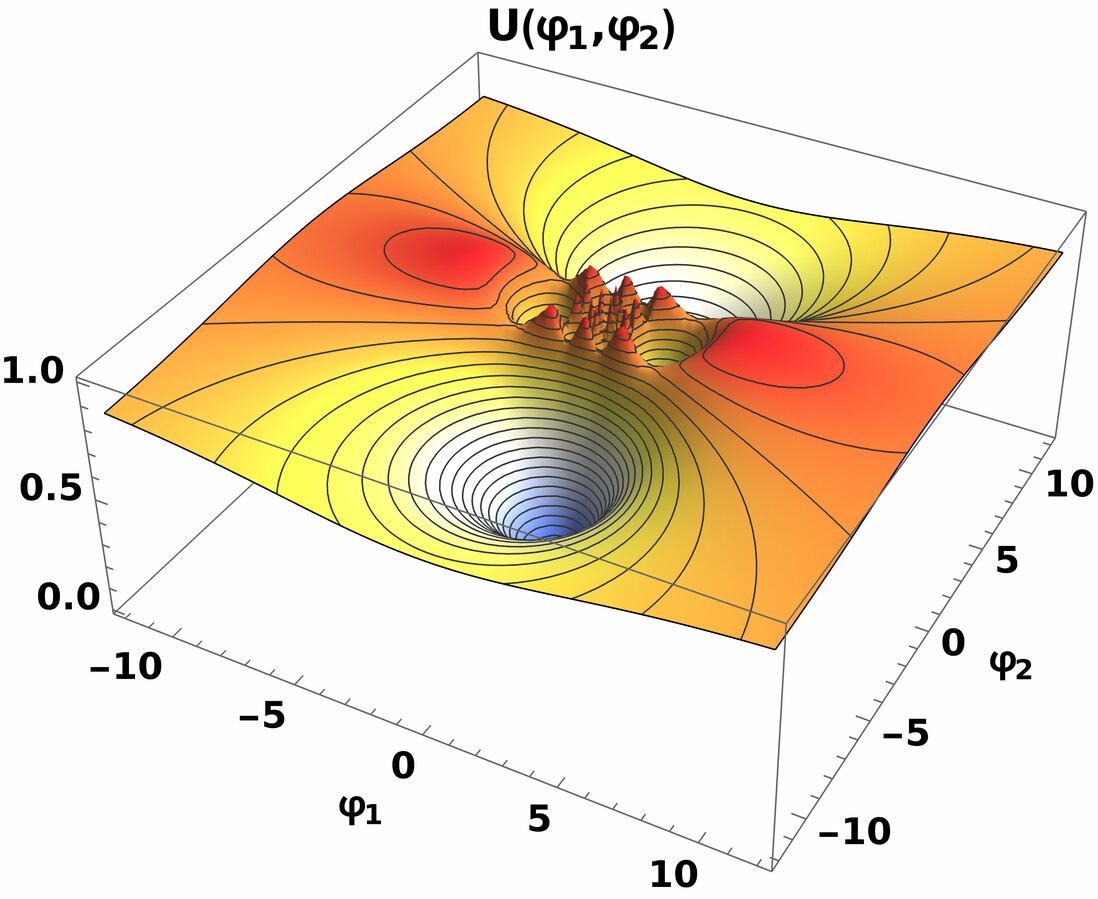}
        \caption{}
    \end{subfigure}
    \hfill
    \begin{subfigure}{0.45\textwidth}
        \centering
        \includegraphics[width=\textwidth]{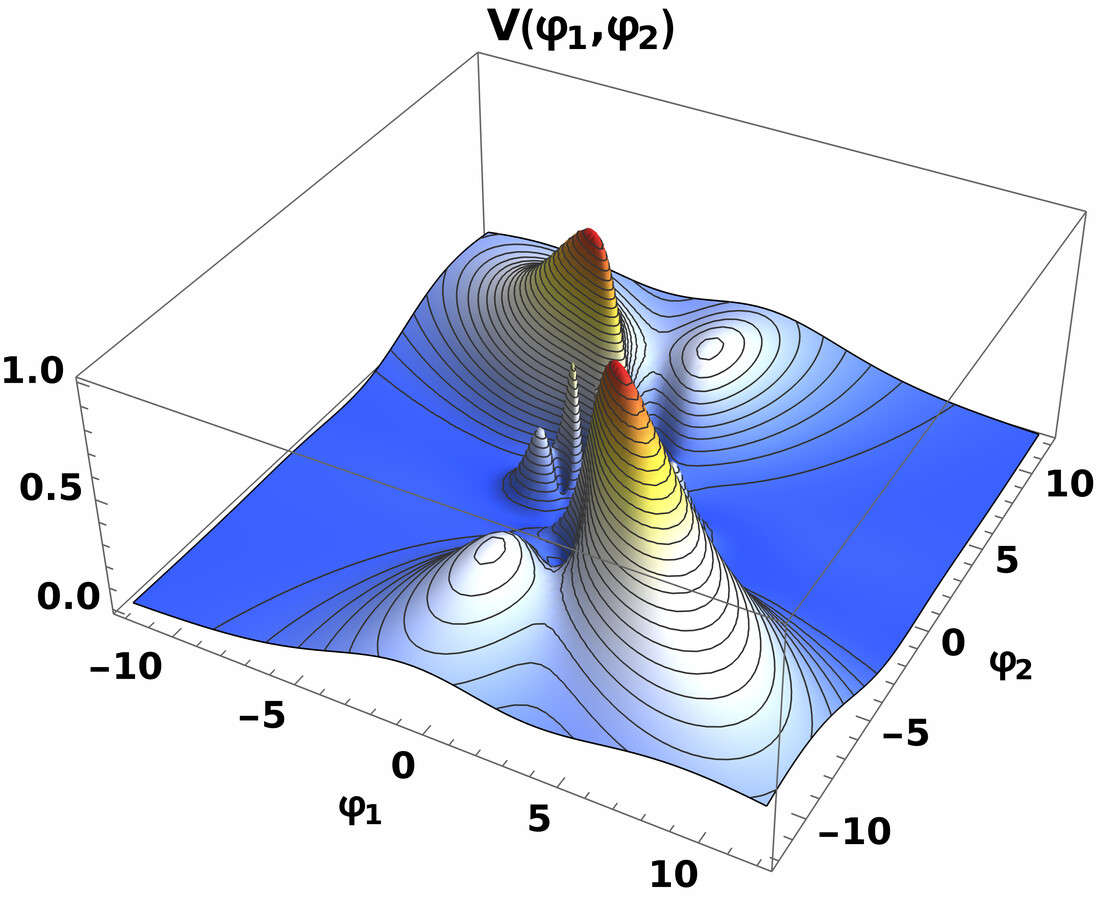}
        \caption{}
    \end{subfigure}
    \caption{Wider view of the unrotated icosahedral model:
    (a,c) pre-potential and (b,d) potential.}
    \label{fig:U_V_icosa_antes_longe}
    
\end{figure}

Figures~\ref{fig:U_V_icosa_antes_perto} and
\ref{fig:U_V_icosa_antes_longe} resolve the same functions near the
origin and over a wider range of field values. The twelve vertex minima
are clustered near the center of the chart, while the site at
stereographic infinity is visible only as the asymptotic behaviour of
the wider view.

For the model analyzed below we apply the rotation \eqref{eq:rot_mat},
so that the vertex set becomes
\begin{equation}
    V'_I=R V_I.
    \label{eq:vertices_icosa_sphere}
\end{equation}
We do not write the rotated vertices explicitly, since their components
are irrational. This rotation is not an element of the icosahedral
group $I$; it moves the regular lattice and places all of its sites,
including the edge-midpoint site previously at the north pole, at finite
stereographic coordinates. By \eqref{eq:rotated_prepotential} the
critical points are carried to $\vb*{\phi}'_*=R\vb*{\phi}_*$ with their
types preserved, and the symmetry group of the rotated pre-potential is
the conjugate $G_R=R I R^{-1}$.

Applying \eqref{eq:U_plato}--\eqref{eq:constant_C} to $V'_I$ gives
\begin{equation}
    U_I(\vb*{\phi})=C_I\prod_{a=1}^{12}\qty(1-v_a\cdot\vb*{\phi}),
    \qquad v_a\in V'_I,\qquad C_I\approx30.616,
    \label{eq:U_icosa}
\end{equation}
whose planar form follows from \eqref{eq:change_coordinates}. We omit
the expanded expression, which is lengthy and offers no additional
insight.

\begin{figure}[!htbp]
    \centering
    \begin{subfigure}{0.45\textwidth}
        \centering
        \includegraphics[width=\textwidth]{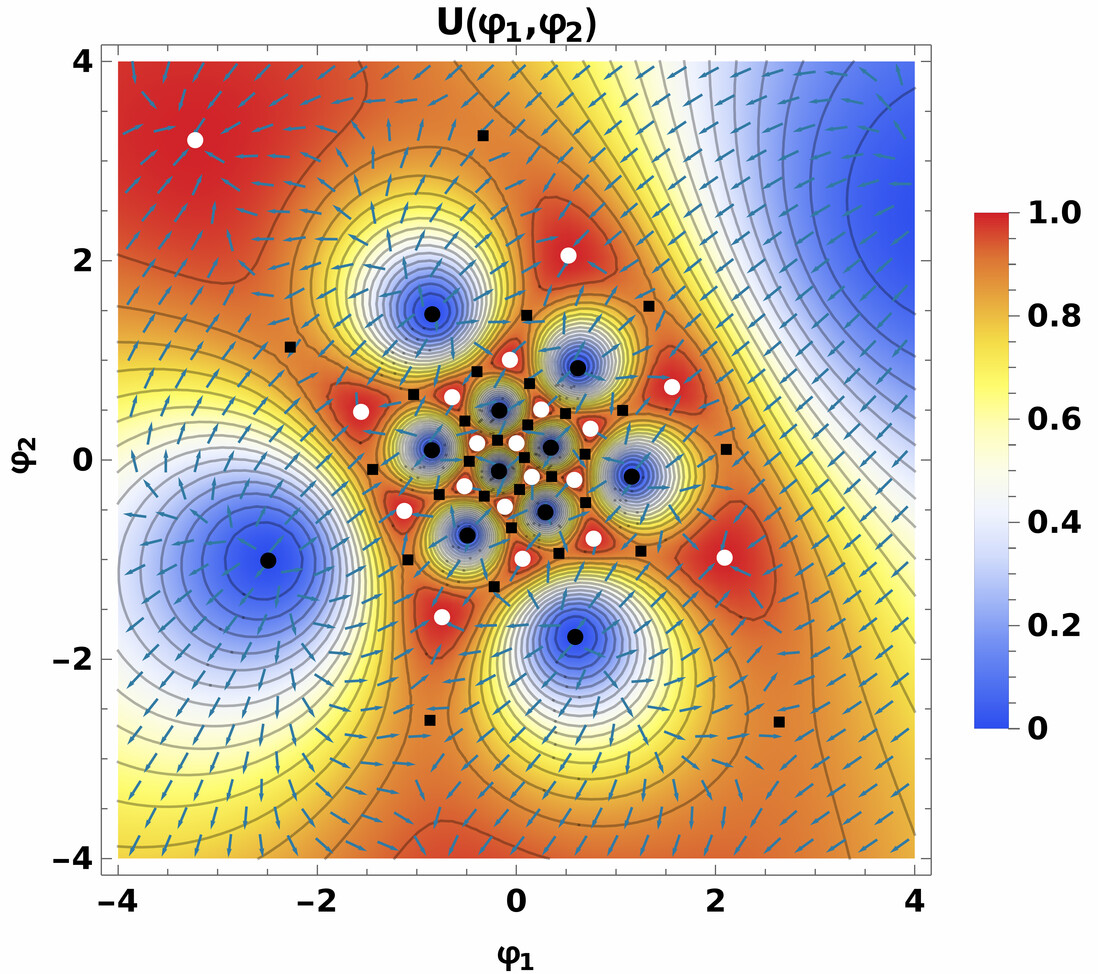}
        \caption{}
        \label{U_2d_icosa}
    \end{subfigure}
    \hfill
    \begin{subfigure}{0.45\textwidth}
        \centering
        \includegraphics[width=\textwidth]{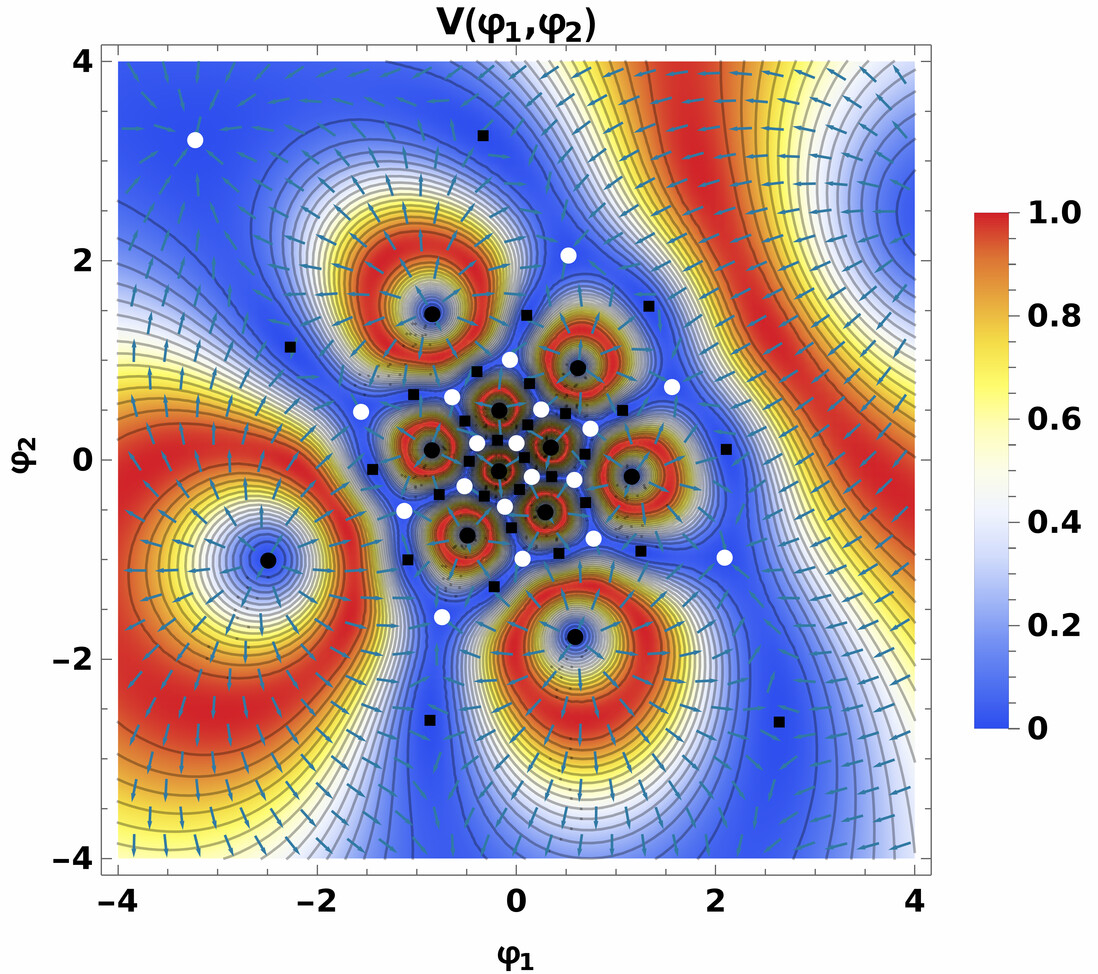}
        \caption{}
        \label{V_2d_icosa}
    \end{subfigure}
    \par\vspace{0.15cm}
    \begin{subfigure}{0.45\textwidth}
        \centering
        \includegraphics[width=\textwidth]{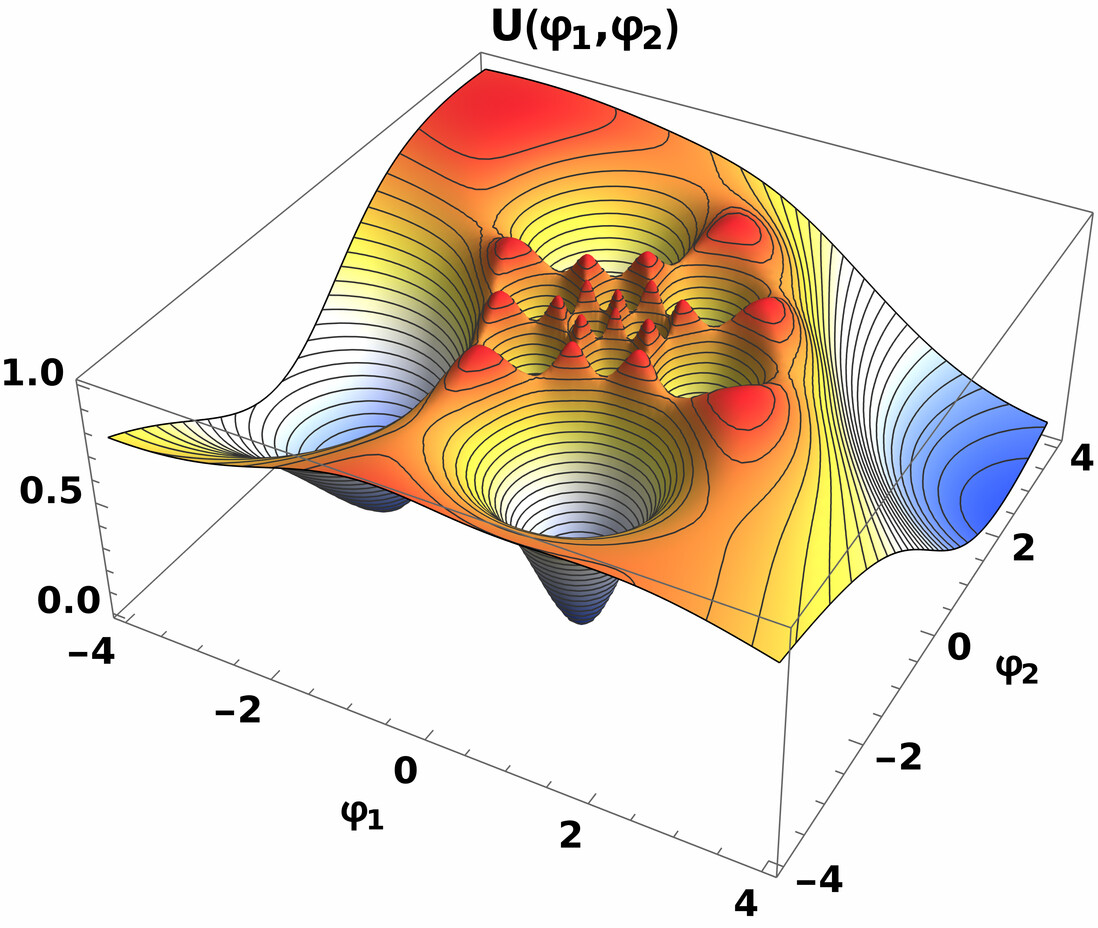}
        \caption{}
        \label{U_3d_icosa}
    \end{subfigure}
    \hfill
    \begin{subfigure}{0.45\textwidth}
        \centering
        \includegraphics[width=\textwidth]{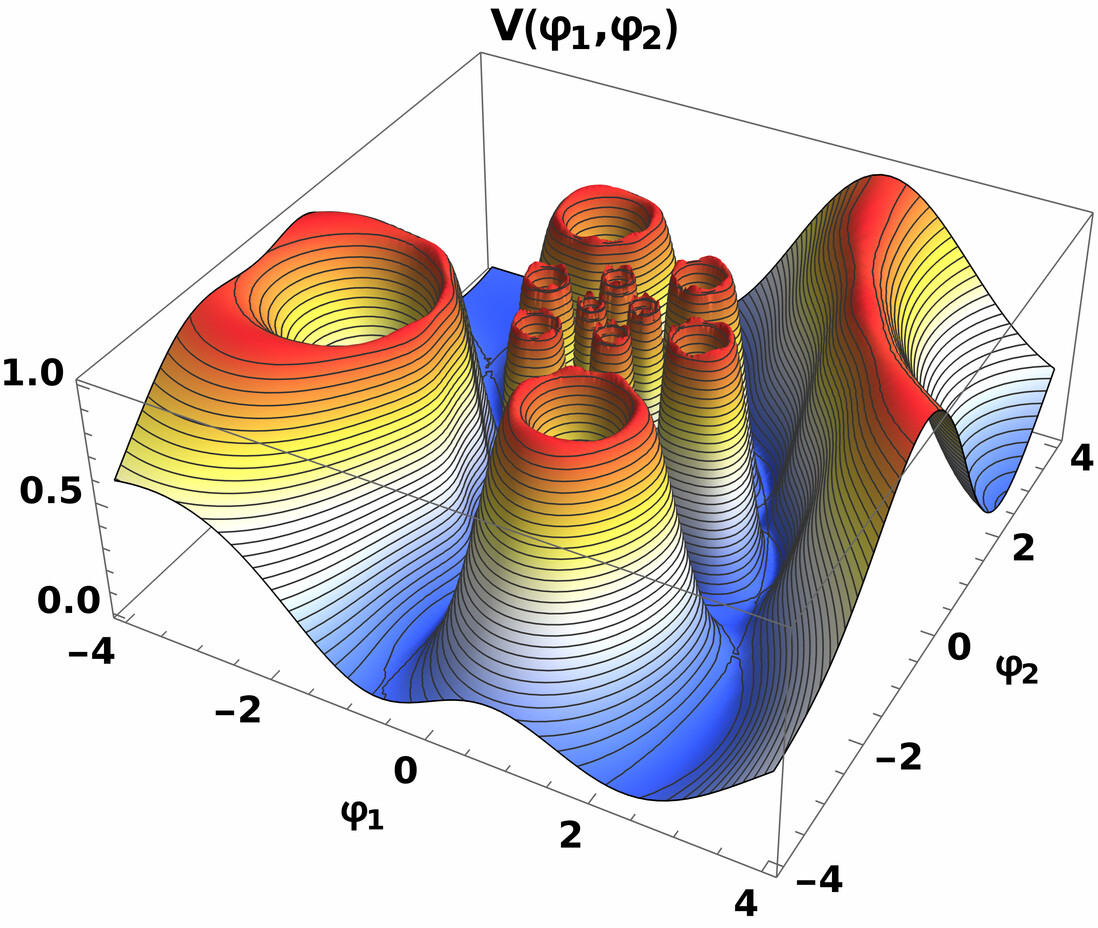}
        \caption{}
        \label{V_3d_icosa}
    \end{subfigure}
    \caption{Icosahedral pre-potential $U_I$ in (a,c) and potential $V$
    in (b,d), after the rotation \eqref{eq:rot_mat}. The upper panels
    show contours, vacuum markers, and the metric gradient of $U_I$;
    the lower panels show the corresponding surfaces.}
    \label{fig:U_V_icosa}
    
\end{figure}

The icosahedral regular lattice contains 62 sites: twelve minima at
$V'_I$, twenty maxima at the vertices of the correspondingly rotated
dodecahedron, and thirty saddle points in the rotated edge-midpoint
directions. Their values are
\begin{equation}
    U_{min}=0,\qquad U_{saddle}\approx1.2247,\qquad U_{max}\approx1.3762.
    \label{eq:valores_U_icosa}
\end{equation}
Since the north pole is a saddle site in the reference orientation, the
normalization gives $U_{saddle}=1$ there, and the value quoted in
\eqref{eq:valores_U_icosa} is the rescaling factor $\lambda$ of
\eqref{eq:rotated_prepotential} produced by re-imposing
\eqref{eq:constant_C} after the rotation.
Figure~\ref{fig:U_V_icosa} shows the rotated model. Its planar
representations differ from those of Figure~\ref{fig:U_V_icosa_antes},
although the critical-point classification is preserved.

\begin{figure}[!htbp]
    \centering
    \begin{subfigure}{0.45\textwidth}
        \centering
        \includegraphics[width=\textwidth]{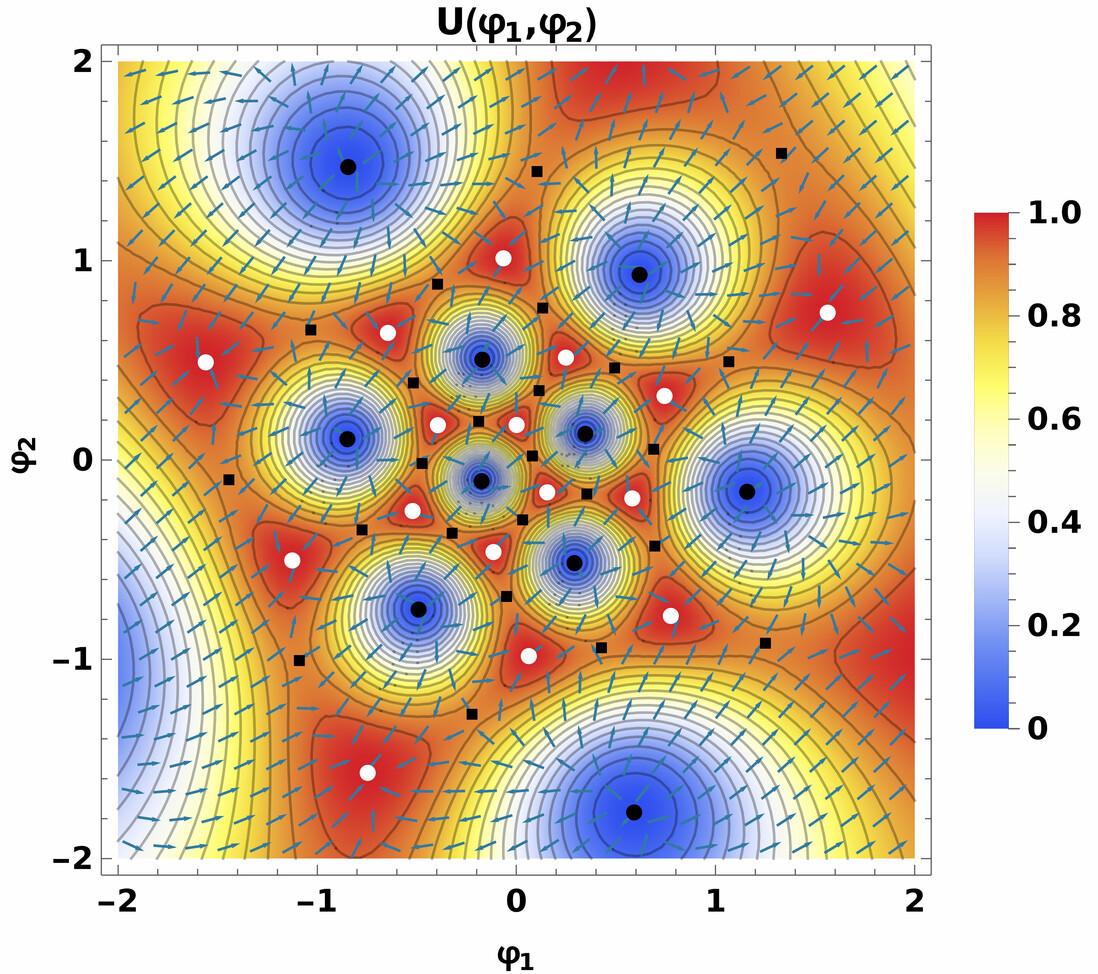}
        \caption{}
        \label{U_2d_icosa_perto}
    \end{subfigure}
    \hfill
    \begin{subfigure}{0.45\textwidth}
        \centering
        \includegraphics[width=\textwidth]{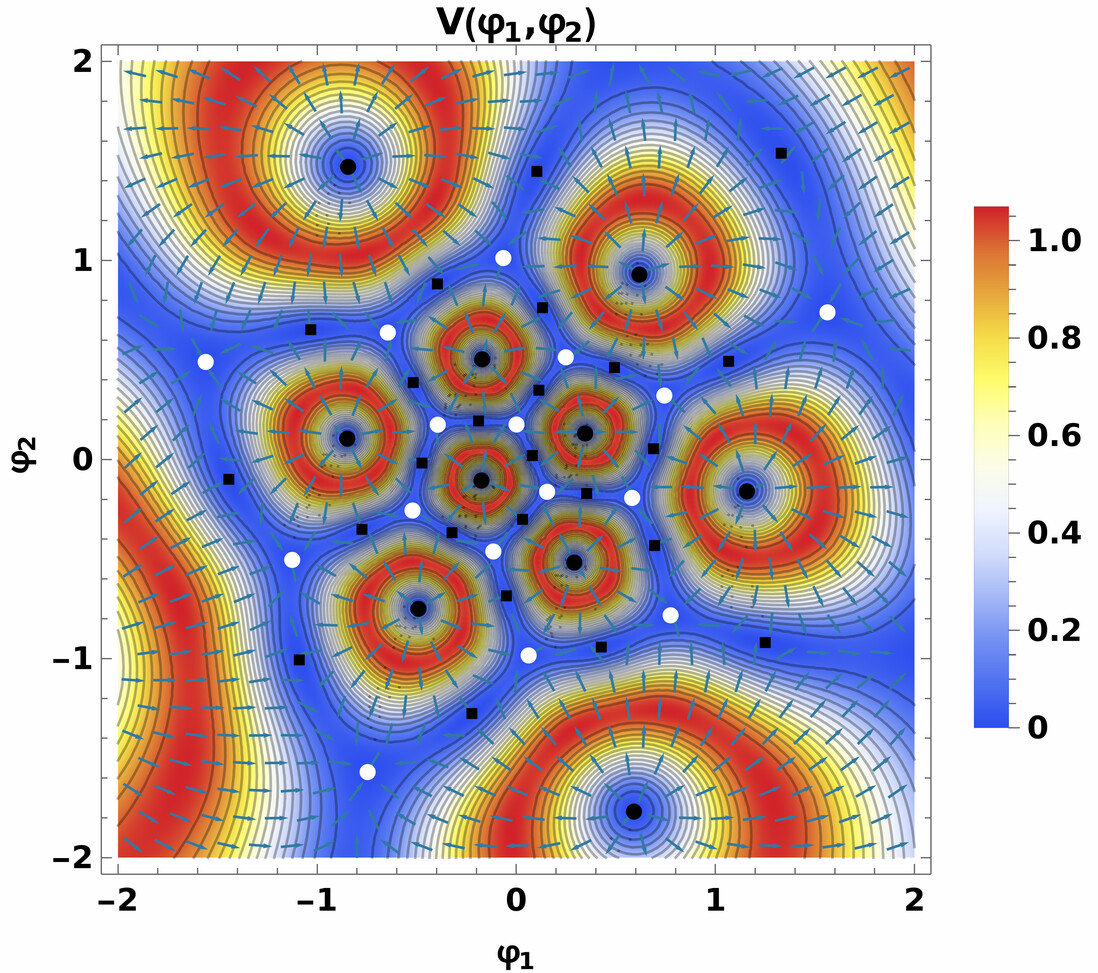}
        \caption{}
        \label{V_2d_icosa_perto}
    \end{subfigure}
    \par\vspace{0.15cm}
    \begin{subfigure}{0.45\textwidth}
        \centering
        \includegraphics[width=\textwidth]{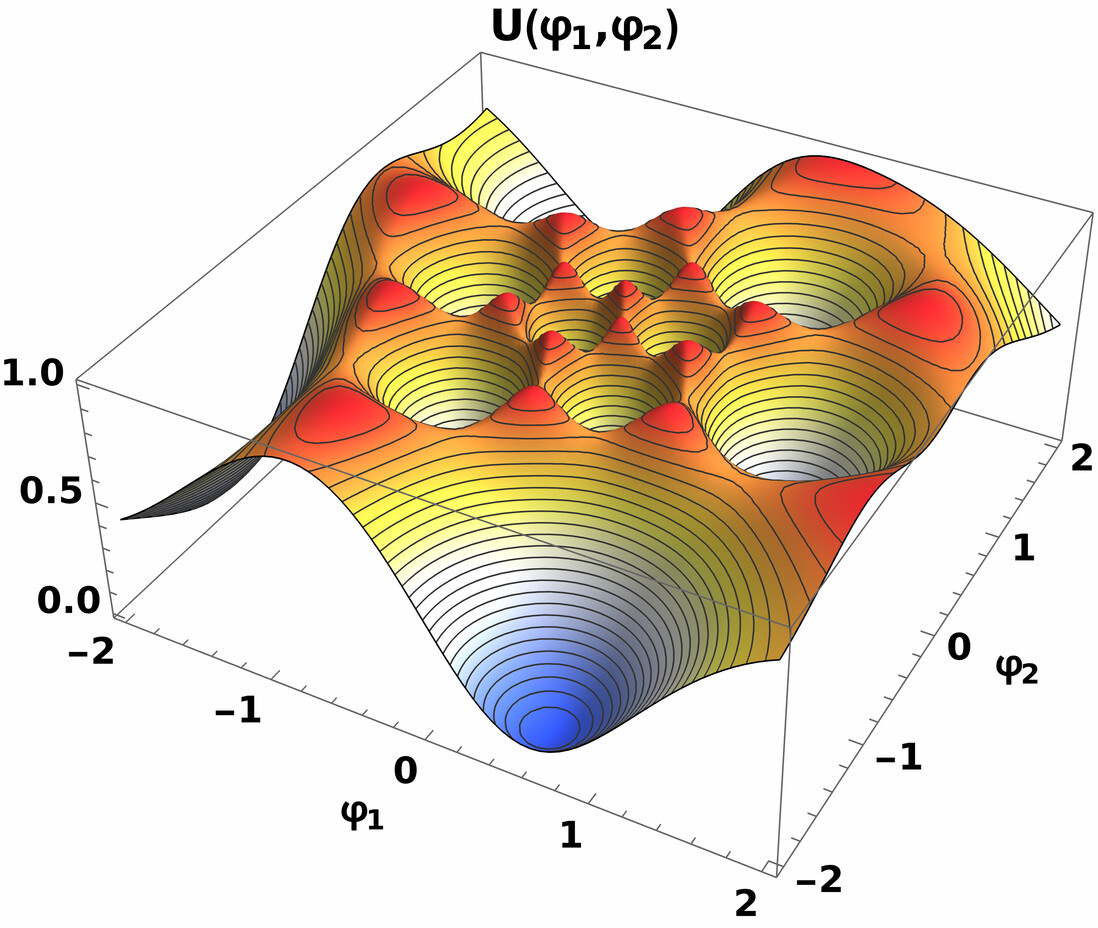}
        \caption{}
        \label{U_3d_icosa_perto}
    \end{subfigure}
    \hfill
    \begin{subfigure}{0.45\textwidth}
        \centering
        \includegraphics[width=\textwidth]{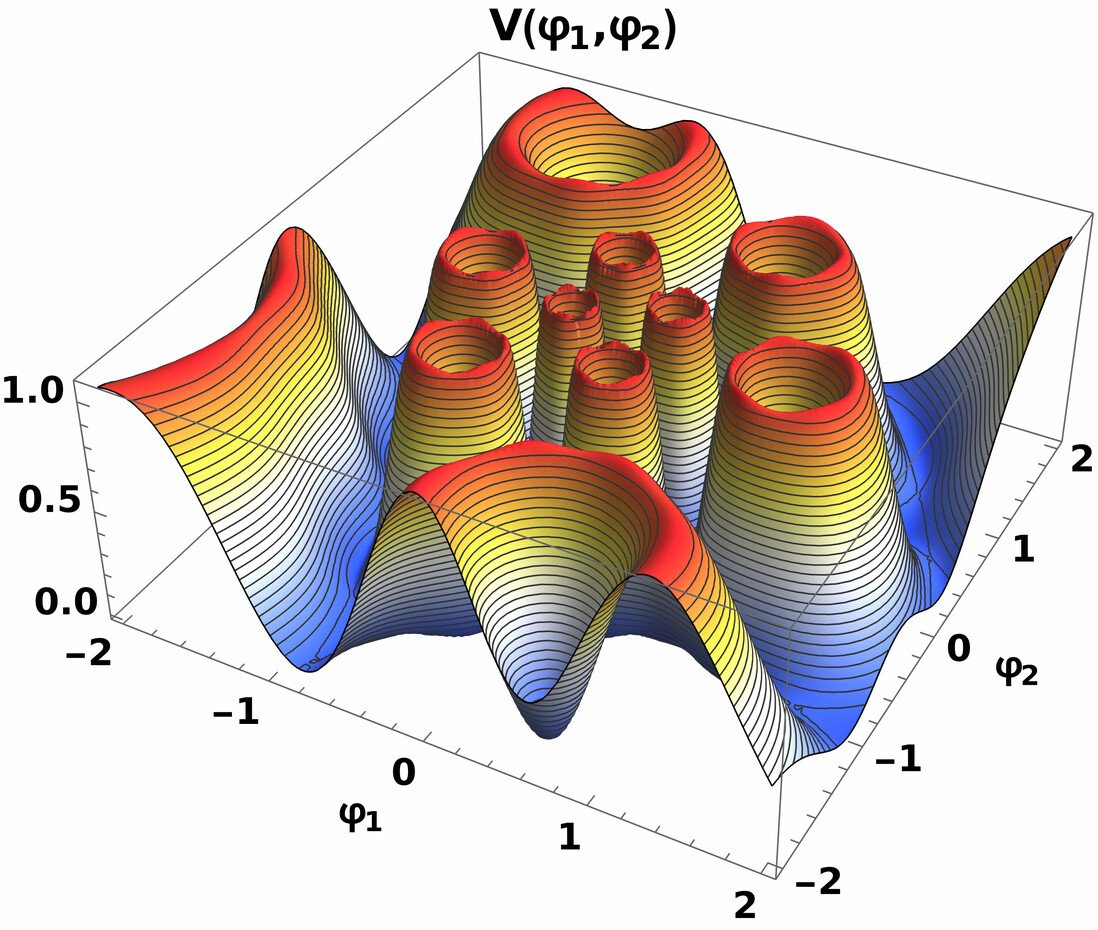}
        \caption{}
        \label{V_3d_icosa_perto}
    \end{subfigure}
    \caption{Central region of the rotated icosahedral model:
    (a,c) pre-potential $U_I$ and (b,d) potential $V$. Markers and
    arrows follow the convention of Figure~\ref{fig:U_V_icosa}.}
    \label{fig:U_V_icosa_perto}
    
\end{figure}

\begin{figure}[!htbp]
    \centering
    \begin{subfigure}{0.45\textwidth}
        \centering
        \includegraphics[width=\textwidth]{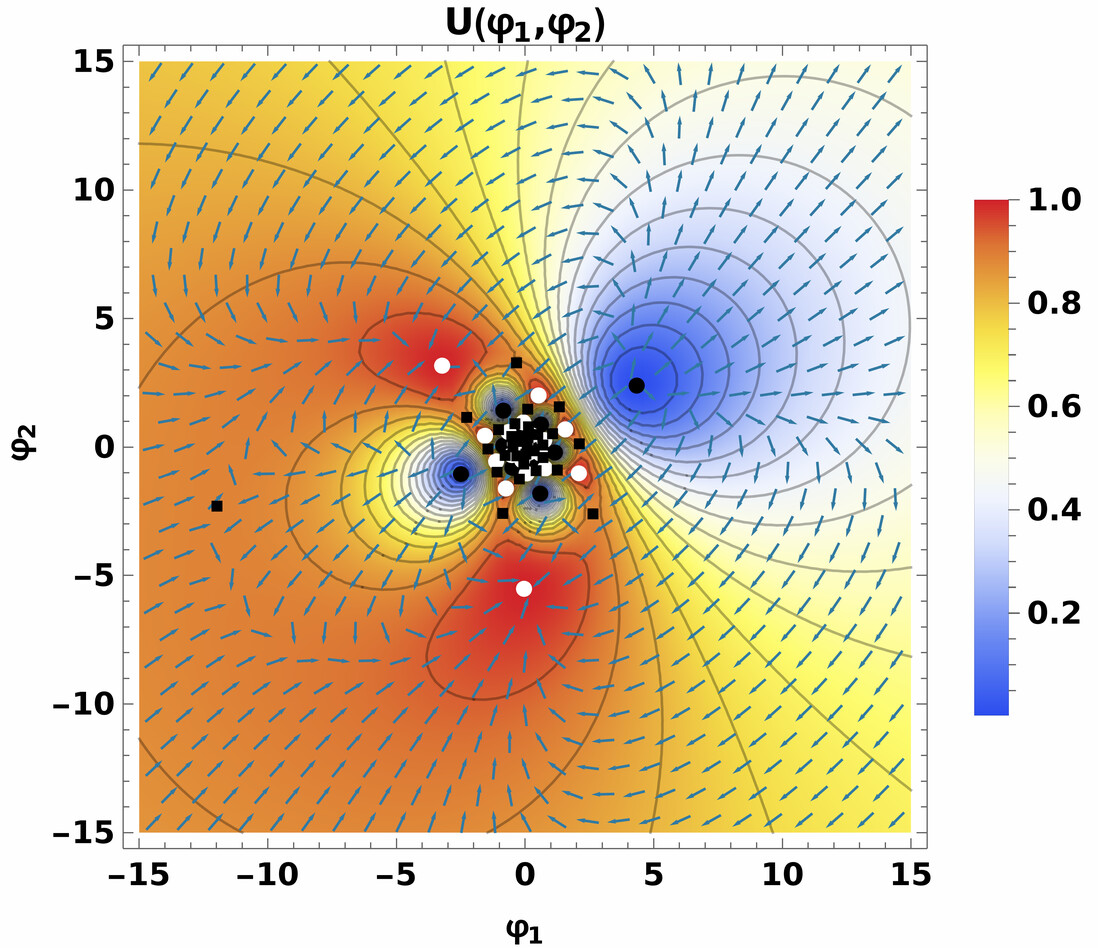}
        \caption{}
        \label{U_2d_icosa_longe}
    \end{subfigure}
    \hfill
    \begin{subfigure}{0.45\textwidth}
        \centering
        \includegraphics[width=\textwidth]{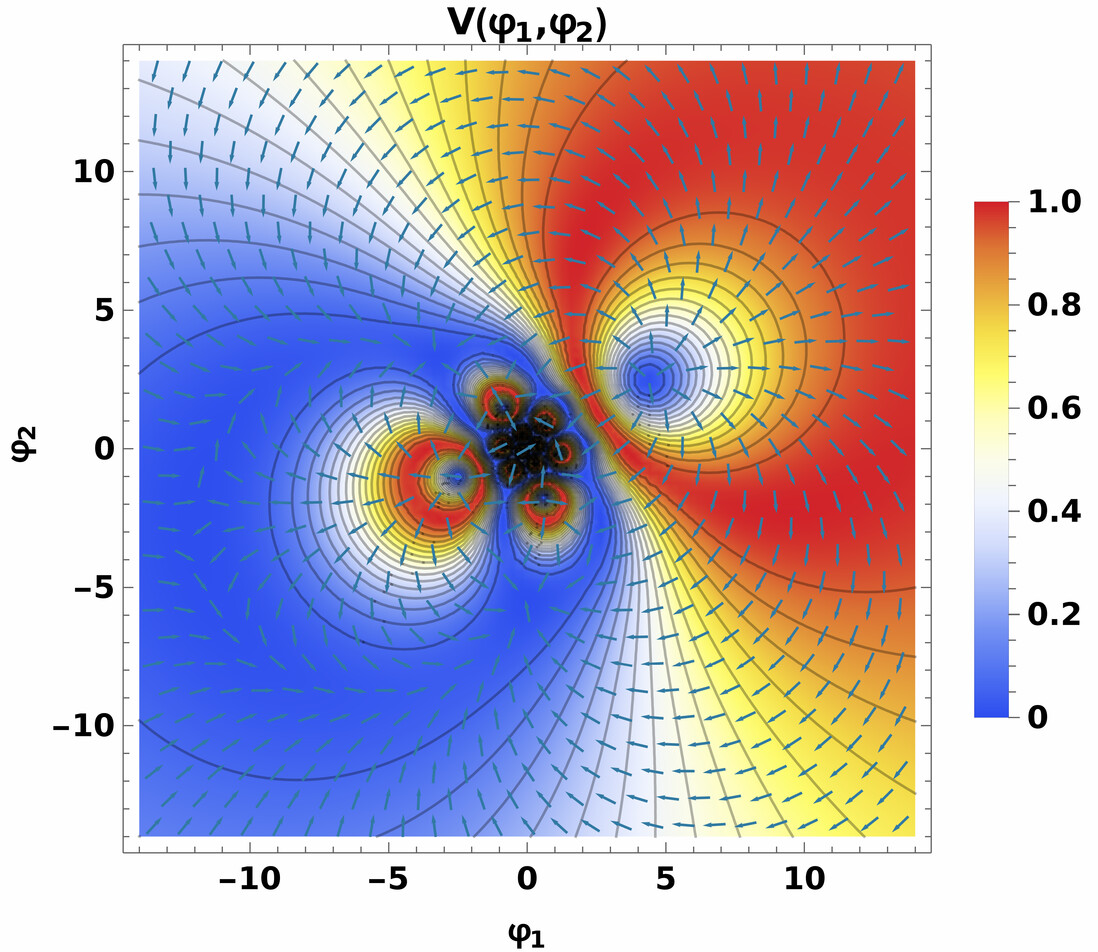}
        \caption{}
        \label{V_2d_icosa_longe}
    \end{subfigure}
    \par\vspace{0.15cm}
    \begin{subfigure}{0.45\textwidth}
        \centering
        \includegraphics[width=\textwidth]{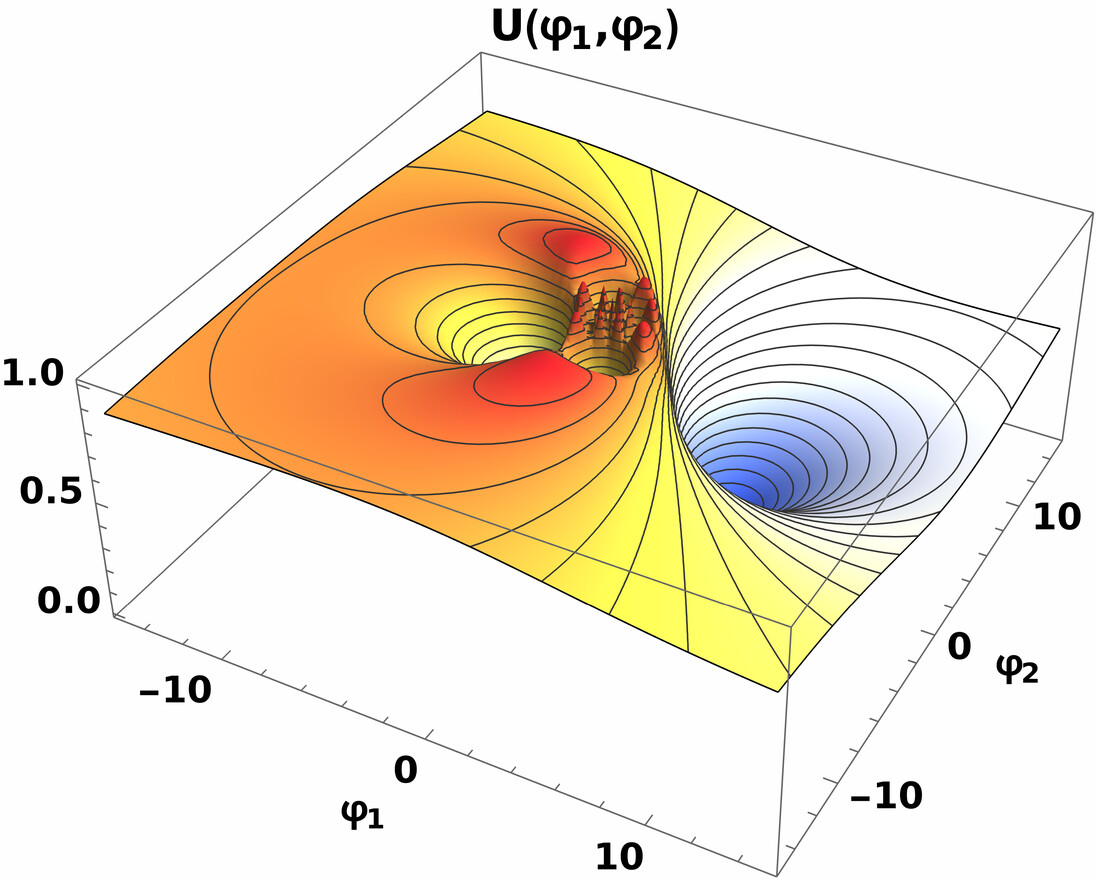}
        \caption{}
        \label{U_3d_icosa_longe}
    \end{subfigure}
    \hfill
    \begin{subfigure}{0.45\textwidth}
        \centering
        \includegraphics[width=\textwidth]{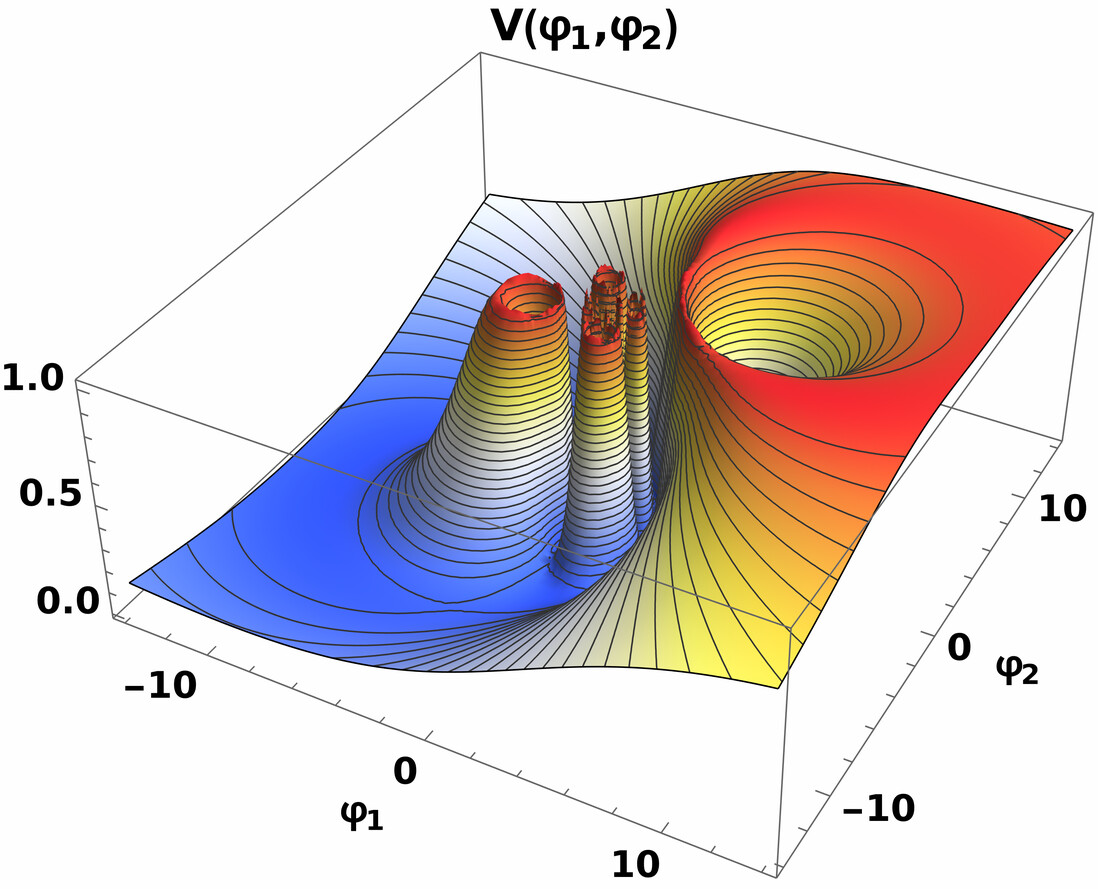}
        \caption{}
        \label{V_3d_icosa_longe}
    \end{subfigure}
    \caption{Wider view of the rotated icosahedral model:
    (a,c) pre-potential $U_I$ and (b,d) potential $V$.}
    \label{fig:U_V_icosa_longe}
    
\end{figure}

Figure~\ref{fig:U_V_icosa_perto} resolves the closely spaced vacua and
the potential barriers near the origin, and
Figure~\ref{fig:U_V_icosa_longe} shows the same functions over a wider
range of field values. With 62 vacua distributed over the sphere, these
two scales are needed to separate the sites projected near the center
from those projected far from it.

\begin{figure}[!htbp]
    \centering
    \begin{subfigure}{0.27\textwidth}
        \centering
        \includegraphics[width=\textwidth]{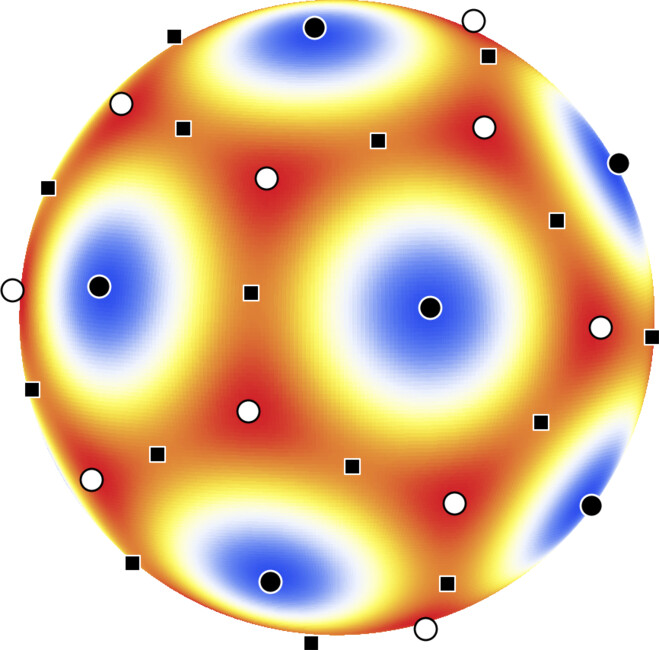}
        \caption{}
        \label{U_icosahedron_sphere}
    \end{subfigure}
    \hspace{0.03\textwidth}
    \begin{subfigure}{0.27\textwidth}
        \centering
        \includegraphics[width=\textwidth]{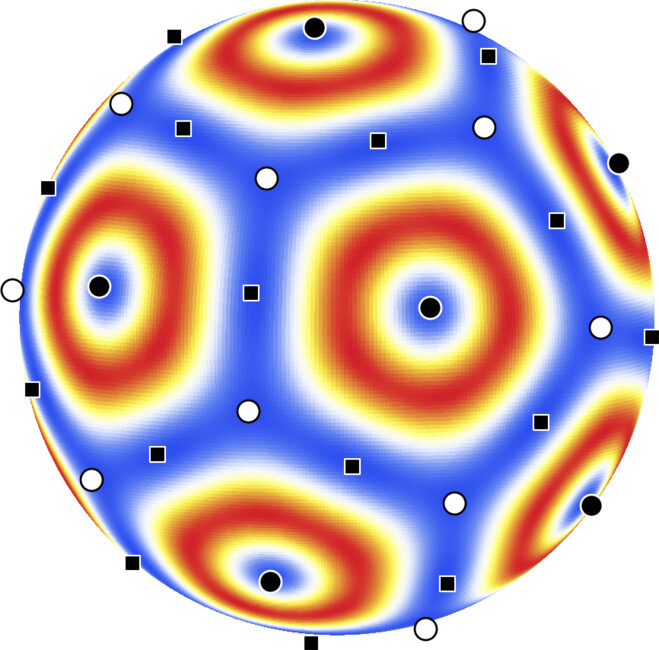}
        \caption{}
        \label{V_icosahedron_sphere}
    \end{subfigure}
    \caption{Icosahedral (a) pre-potential $U_I$ and (b) potential $V$
    represented by color on $\mathbb{S}^2$. Black filled circles,
    white circles, and black squares mark minima, maxima, and saddle
    points of $U_I$, respectively; all are vacua of $V$.}
    \label{fig:U_V_icosahedron_sphere}
    
\end{figure}

In Figure~\ref{fig:U_V_icosahedron_sphere} both functions are displayed
on the unit sphere. This representation is free of the distortion of the
stereographic projection, so the icosahedral symmetry and the regularity
of the lattice are directly visible. All three classes of critical
points of $U_I$ appear as degenerate vacua of $V$.
\FloatBarrier
\subsection{Dodecahedron}
\begin{figure}[!htbp]
    \centering
    \begin{subfigure}{0.45\textwidth}
        \centering
        \includegraphics[width=\textwidth]{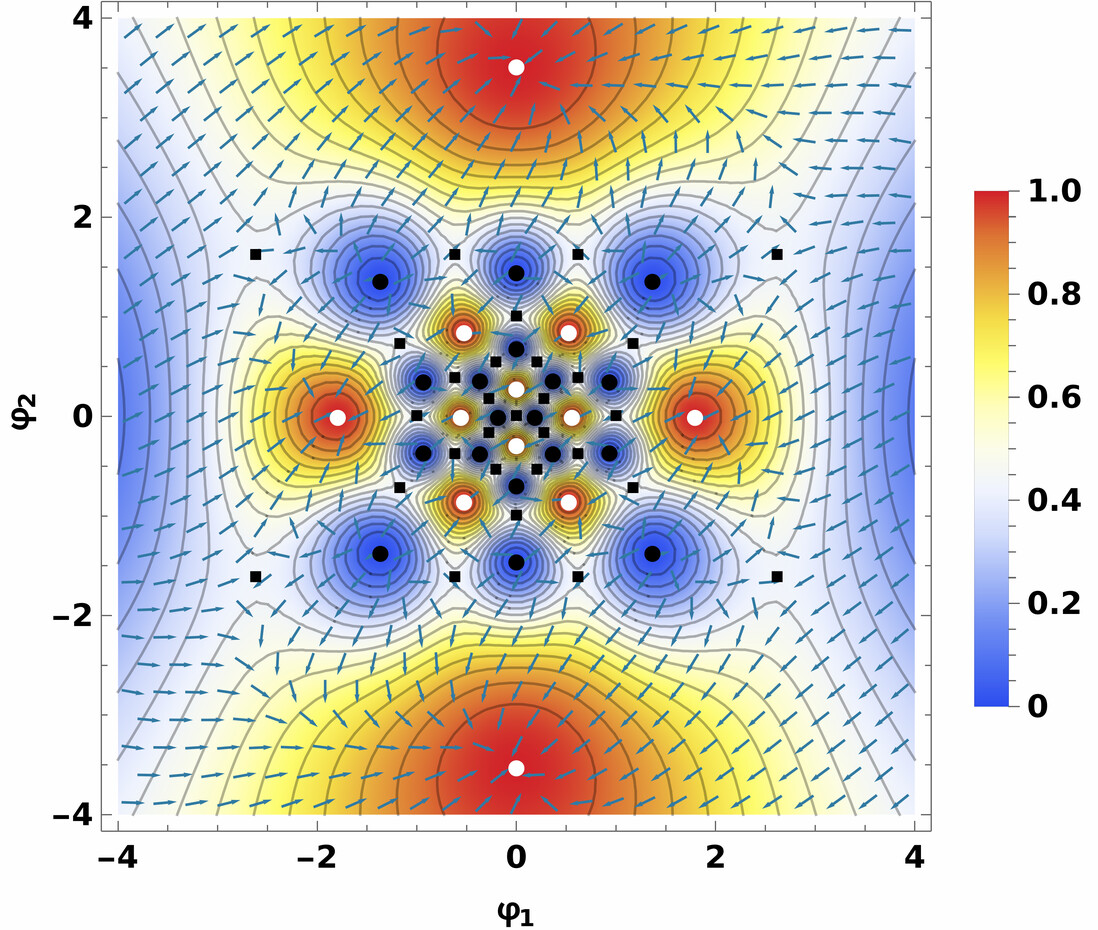}
        \caption{}
        \label{U_2d_dode_antes}
    \end{subfigure}
    \hfill
    \begin{subfigure}{0.45\textwidth}
        \centering
        \includegraphics[width=\textwidth]{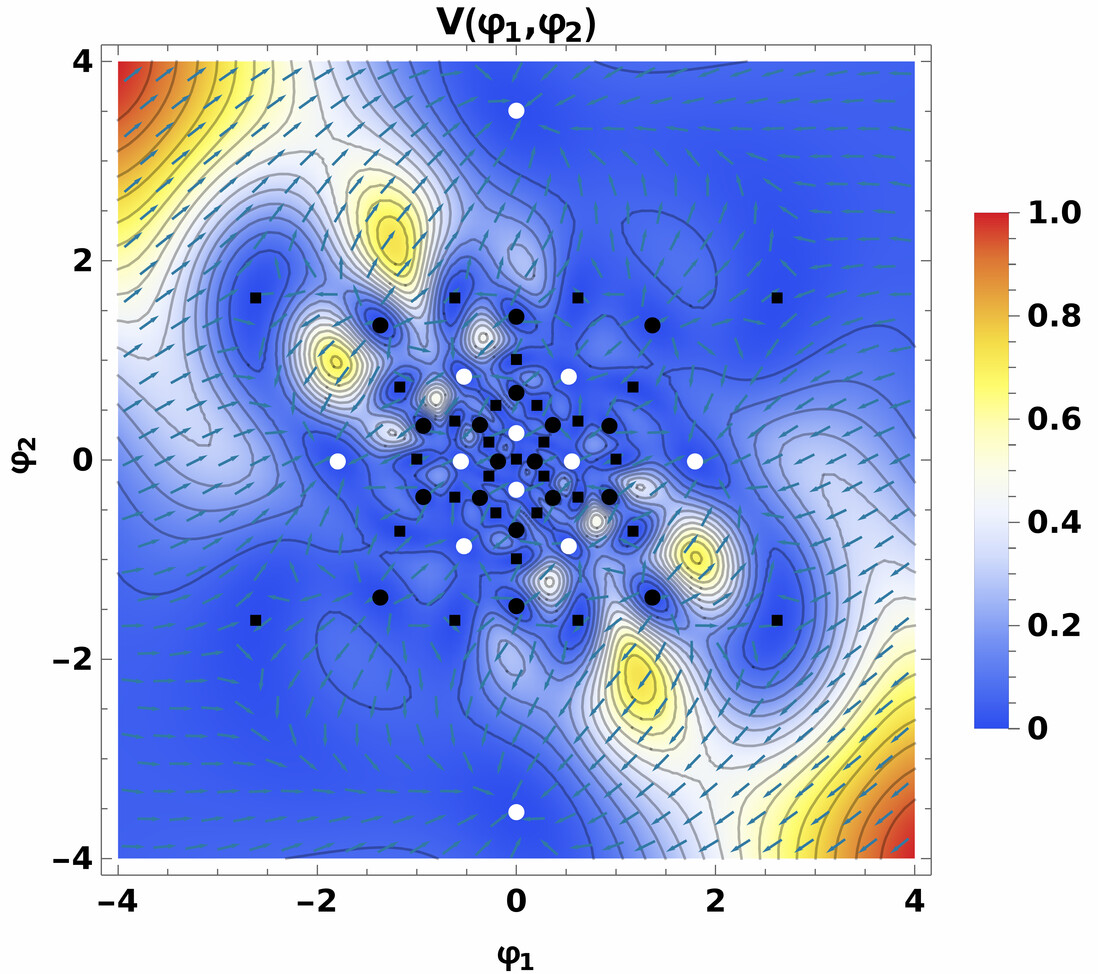}
        \caption{}
        \label{V_2d_dode_antes}
    \end{subfigure}
    \par\vspace{0.15cm}
    \begin{subfigure}{0.45\textwidth}
        \centering
        \includegraphics[width=\textwidth]{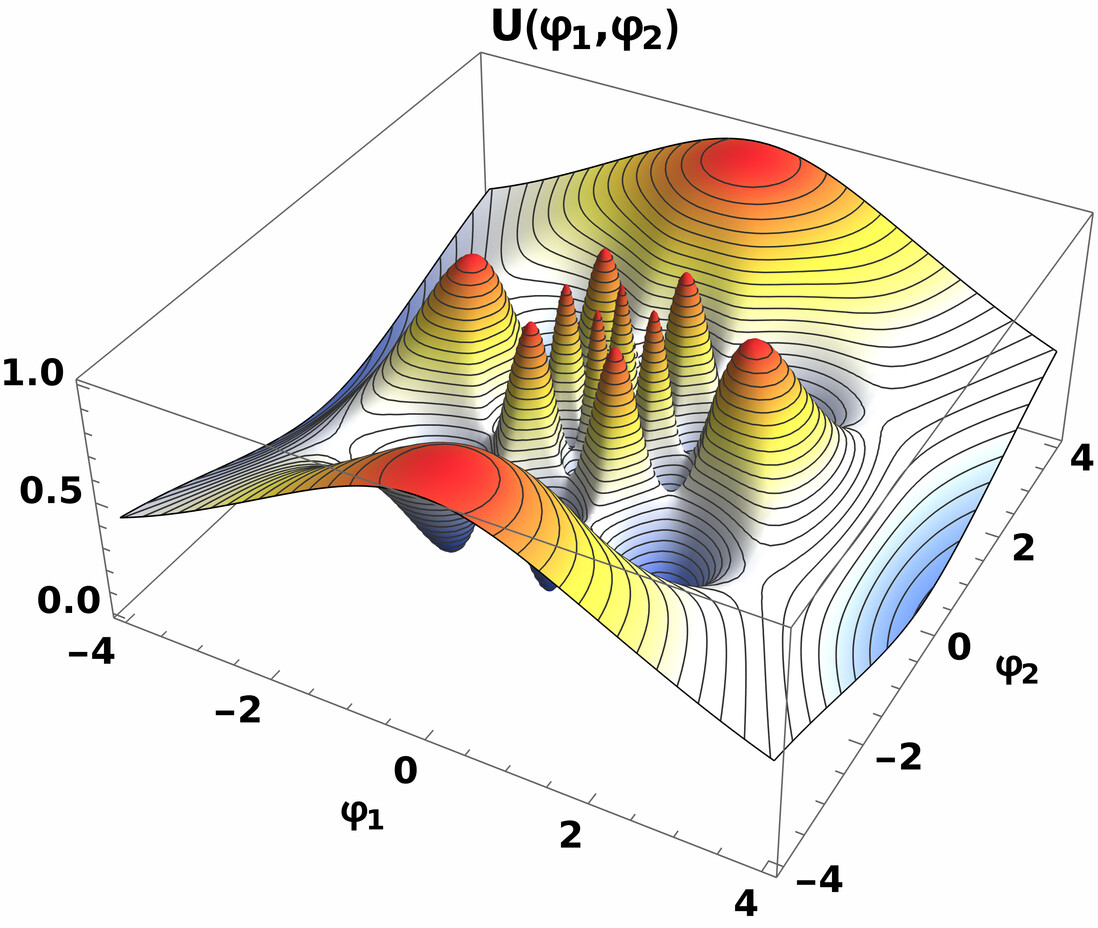}
        \caption{}
    \end{subfigure}
    \hfill
    \begin{subfigure}{0.45\textwidth}
        \centering
        \includegraphics[width=\textwidth]{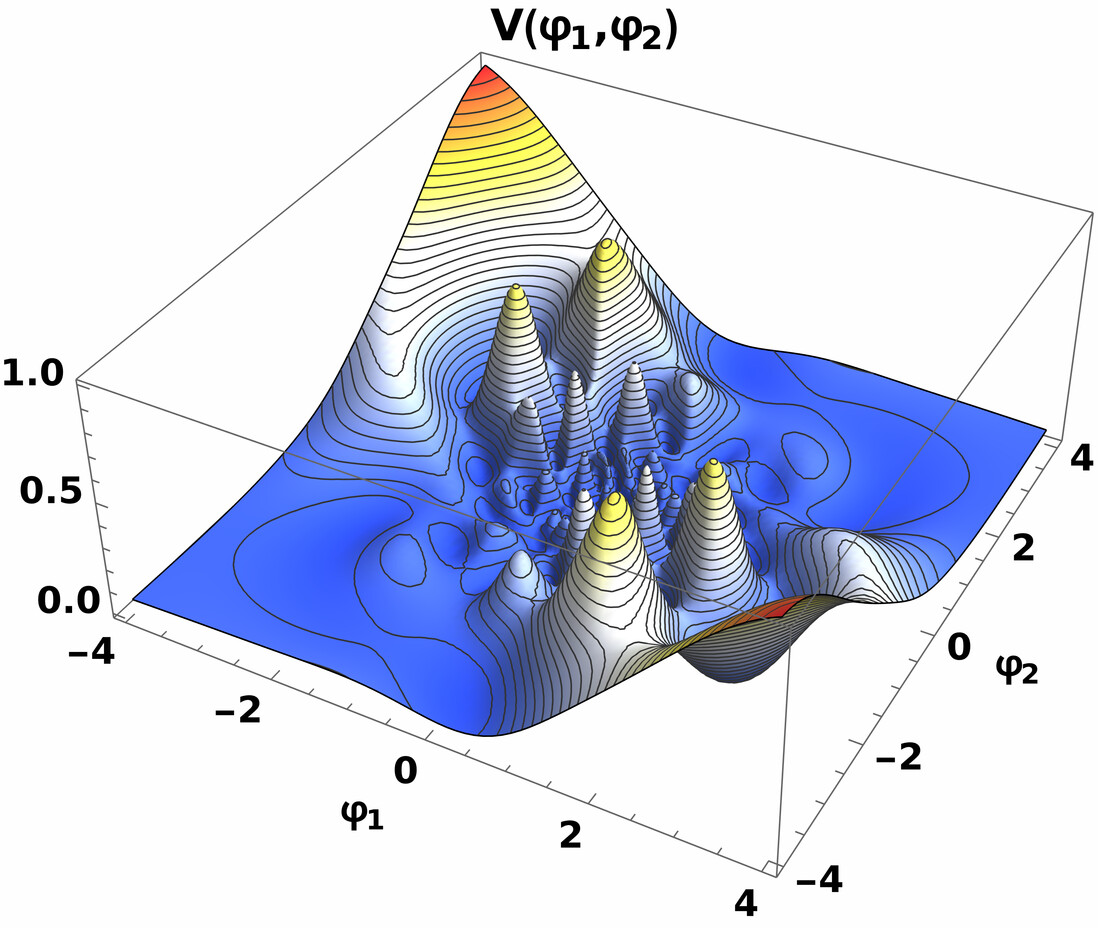}
        \caption{}
    \end{subfigure}
    \caption{Reference plots in the unrotated dodecahedral orientation:
    (a,c) pre-potential and (b,d) potential. The upper panels show
    contours, critical-point markers, and metric-gradient arrows;
    the lower panels show the corresponding surfaces.}
    \label{fig:U_V_dode_antes}
    
\end{figure}

\begin{figure}[!htbp]
    \centering
    \begin{subfigure}{0.45\textwidth}
        \centering
        \includegraphics[width=\textwidth]{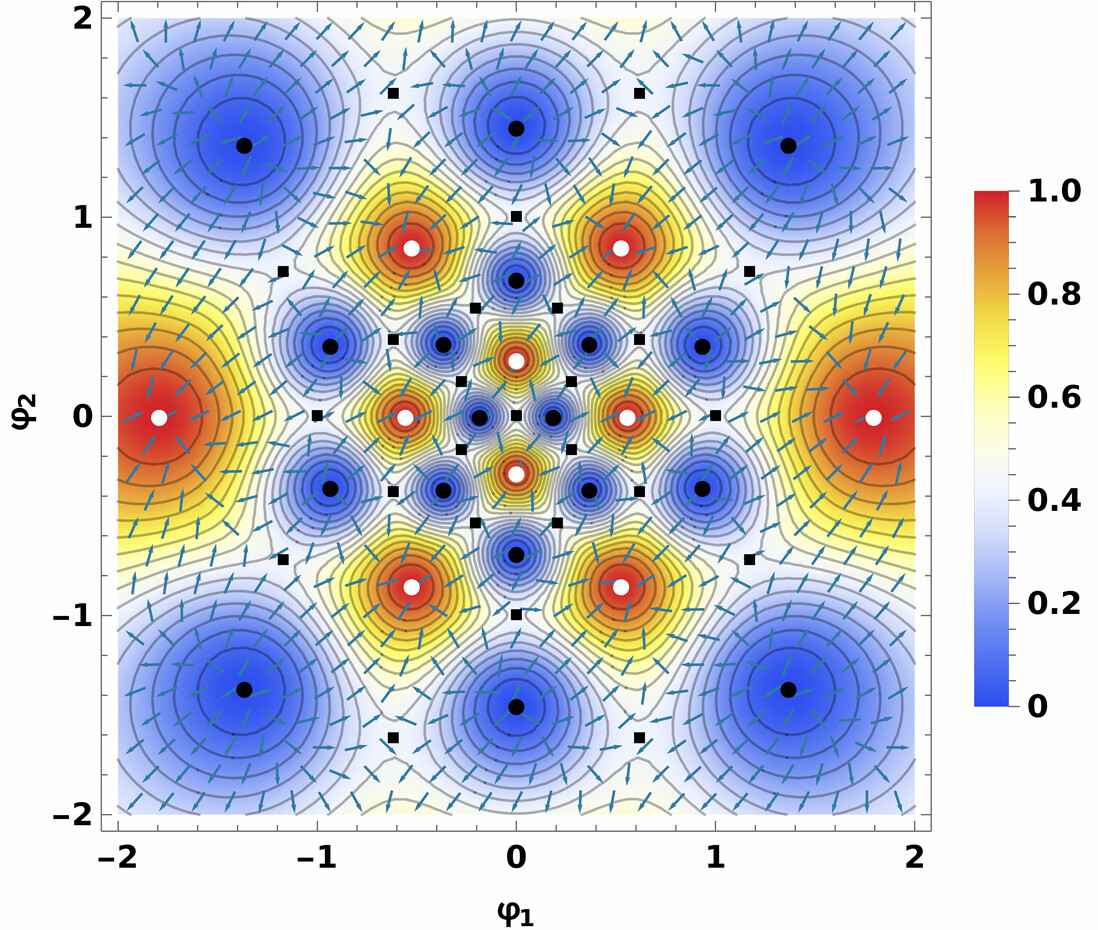}
        \caption{}
        \label{U_2d_dode_antes_perto}
    \end{subfigure}
    \hfill
    \begin{subfigure}{0.45\textwidth}
        \centering
        \includegraphics[width=\textwidth]{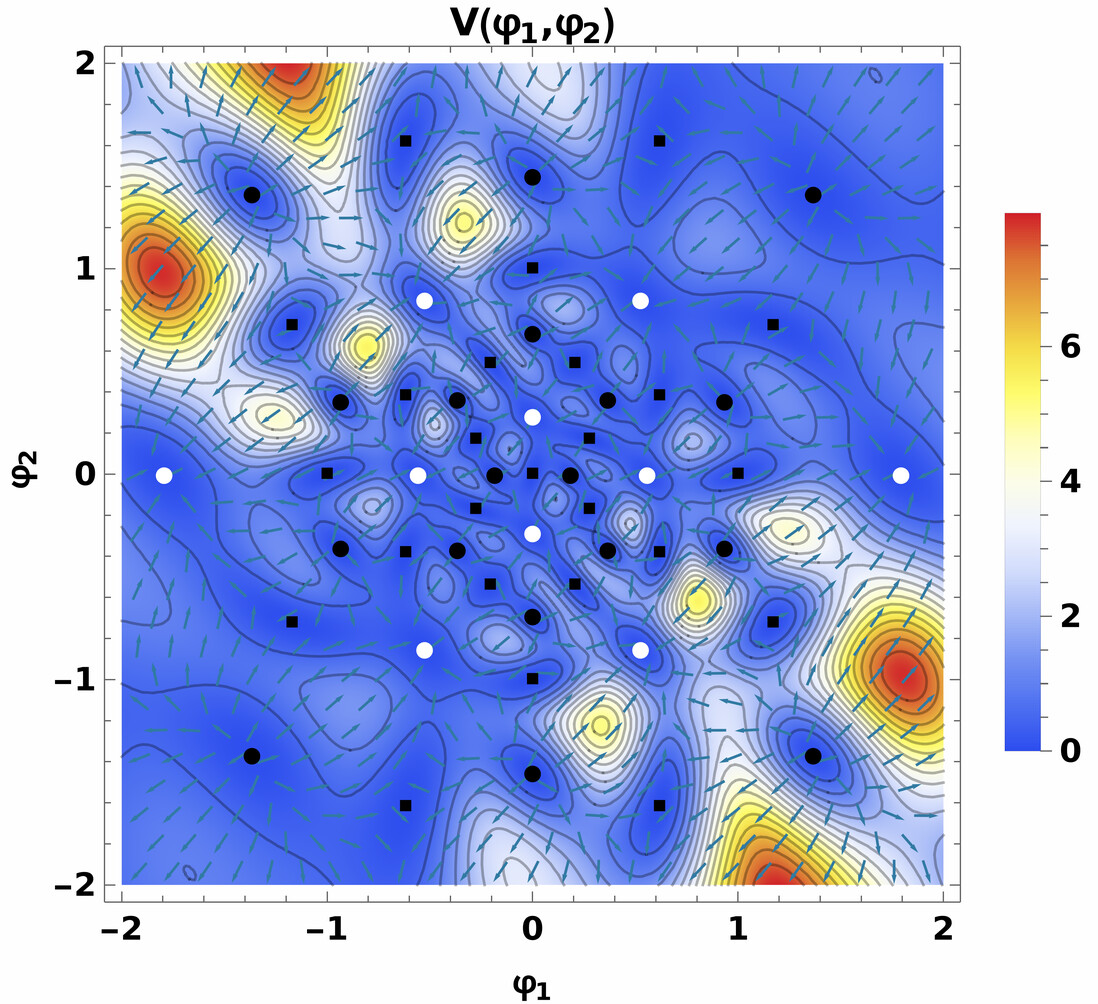}
        \caption{}
        \label{V_2d_dode_antes_perto}
    \end{subfigure}
    \par\vspace{0.15cm}
    \begin{subfigure}{0.45\textwidth}
        \centering
        \includegraphics[width=\textwidth]{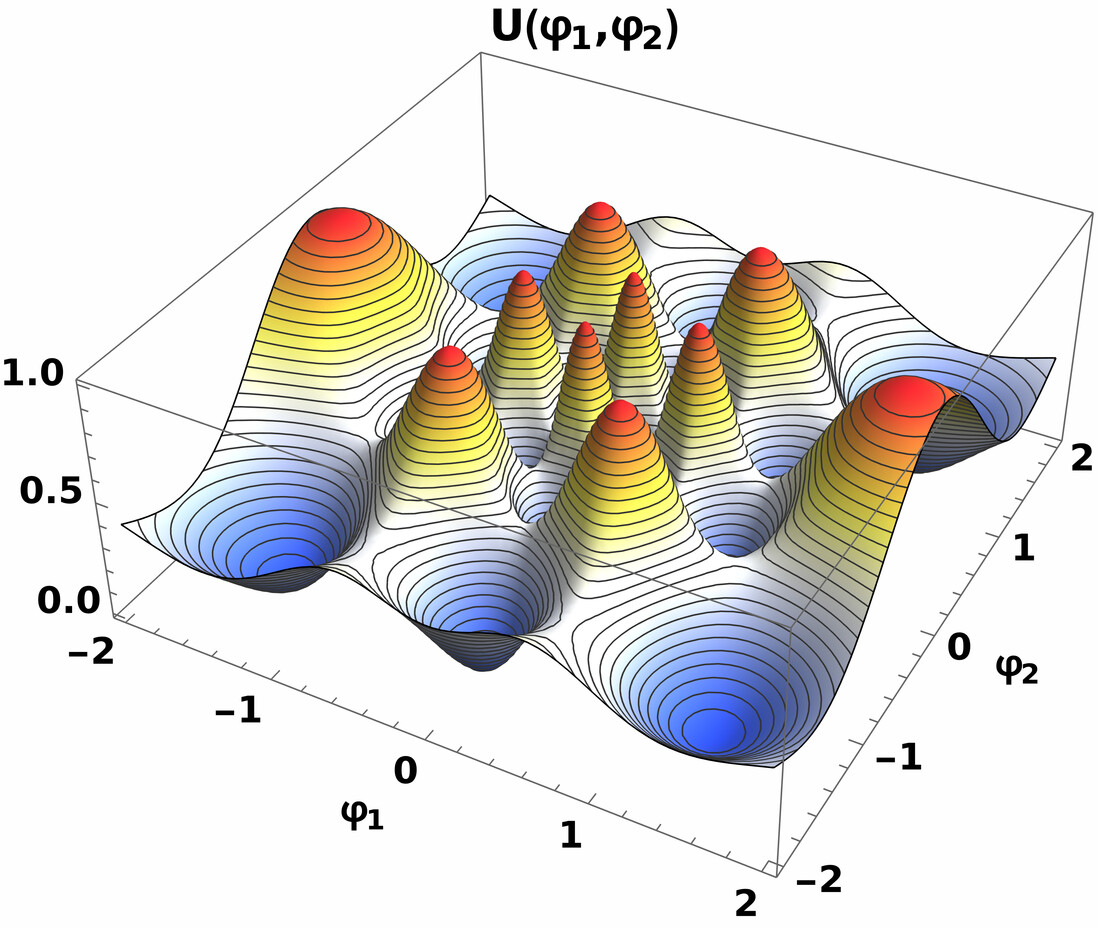}
        \caption{}
        \label{U_3d_dode_antes_perto}
    \end{subfigure}
    \hfill
    \begin{subfigure}{0.45\textwidth}
        \centering
        \includegraphics[width=\textwidth]{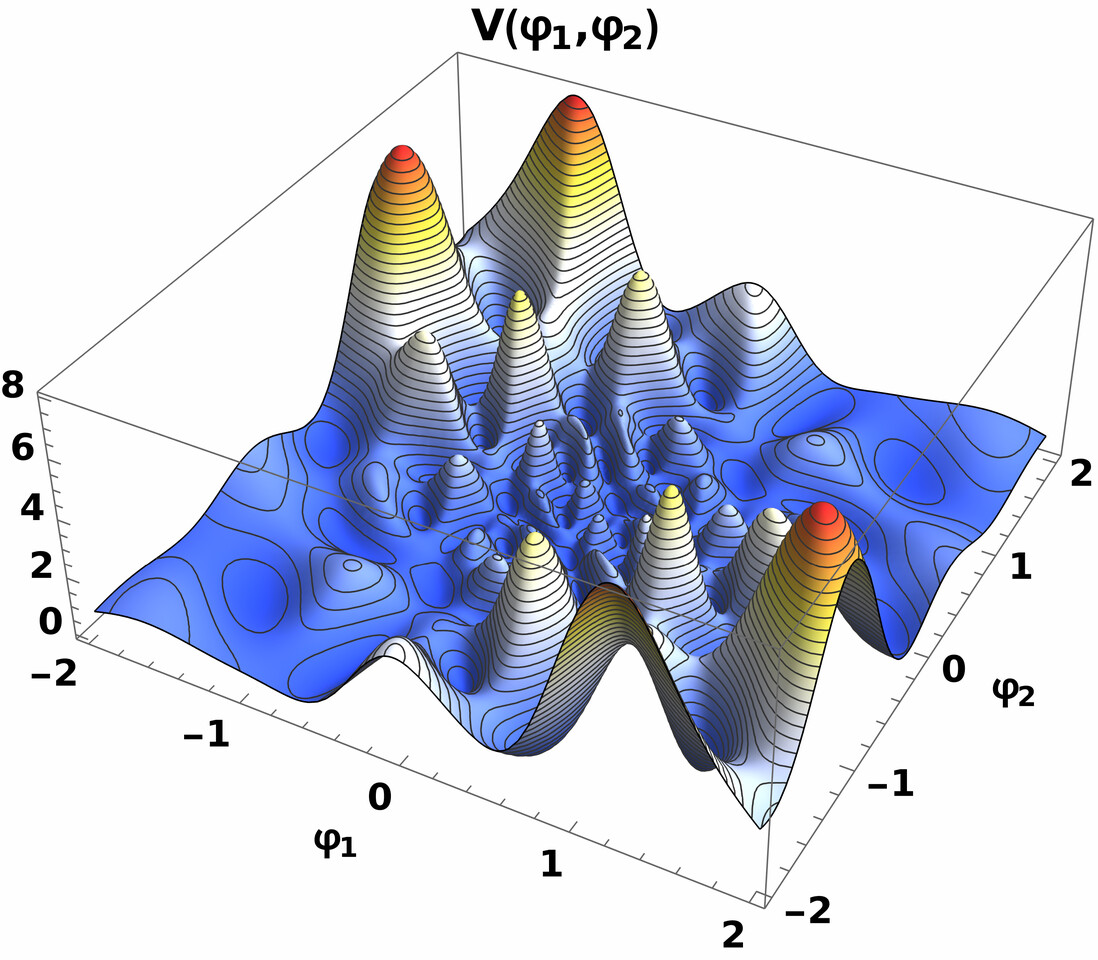}
        \caption{}
        \label{V_3d_dode_antes_perto}
    \end{subfigure}
    \caption{Central region of the unrotated dodecahedral model:
    (a,c) pre-potential and (b,d) potential.}
    \label{fig:U_V_dode_antes_perto}
    
\end{figure}

\begin{figure}[!htbp]
    \centering
    \begin{subfigure}{0.45\textwidth}
        \centering
        \includegraphics[width=\textwidth]{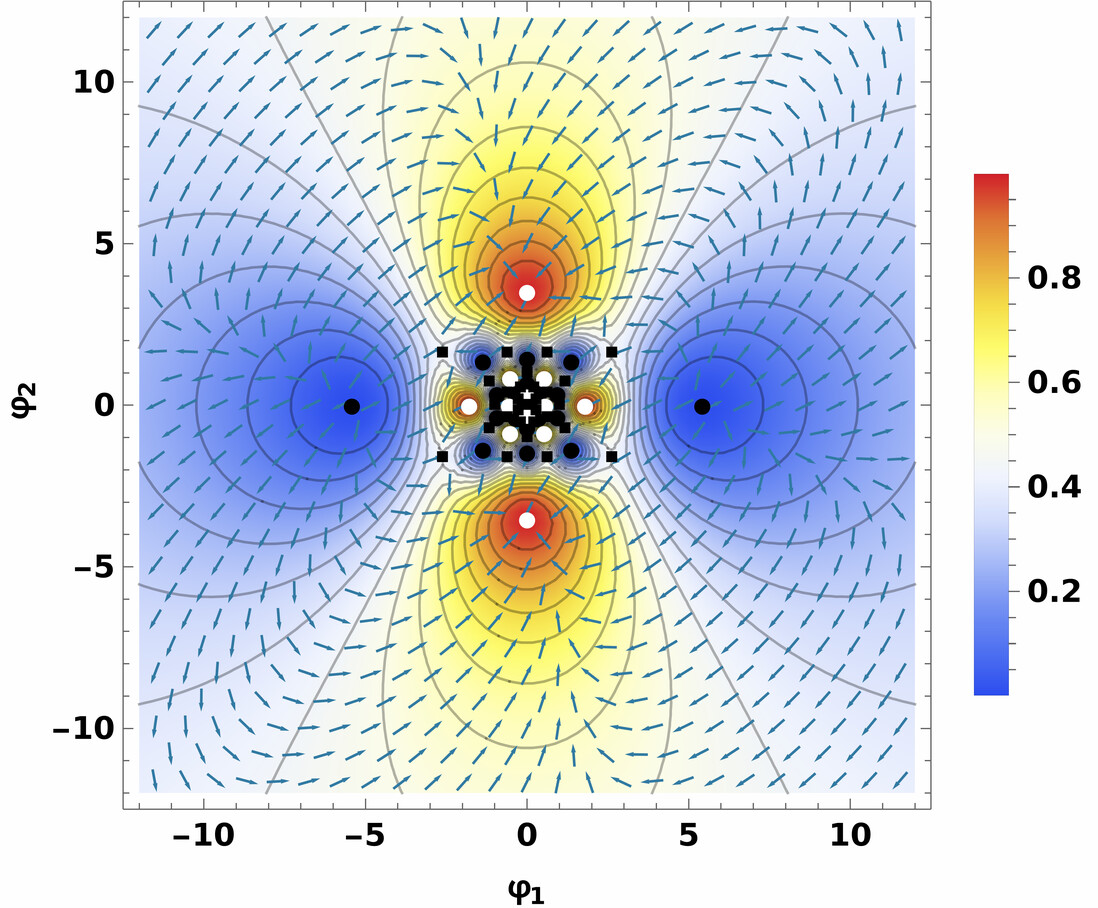}
        \caption{}
        \label{U_2d_dode_antes_longe}
    \end{subfigure}
    \hfill
    \begin{subfigure}{0.45\textwidth}
        \centering
        \includegraphics[width=\textwidth]{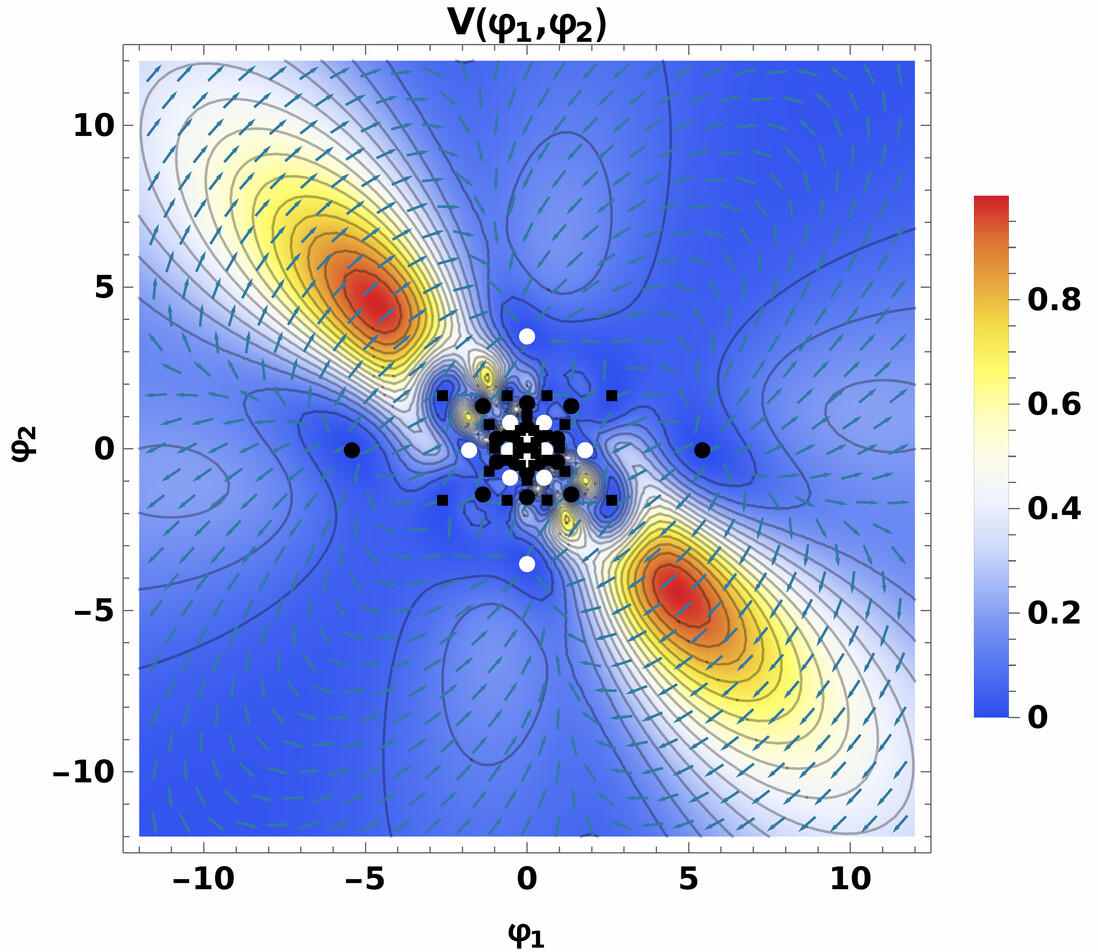}
        \caption{}
        \label{V_2d_dode_antes_longe}
    \end{subfigure}
    \par\vspace{0.15cm}
    \begin{subfigure}{0.45\textwidth}
        \centering
        \includegraphics[width=\textwidth]{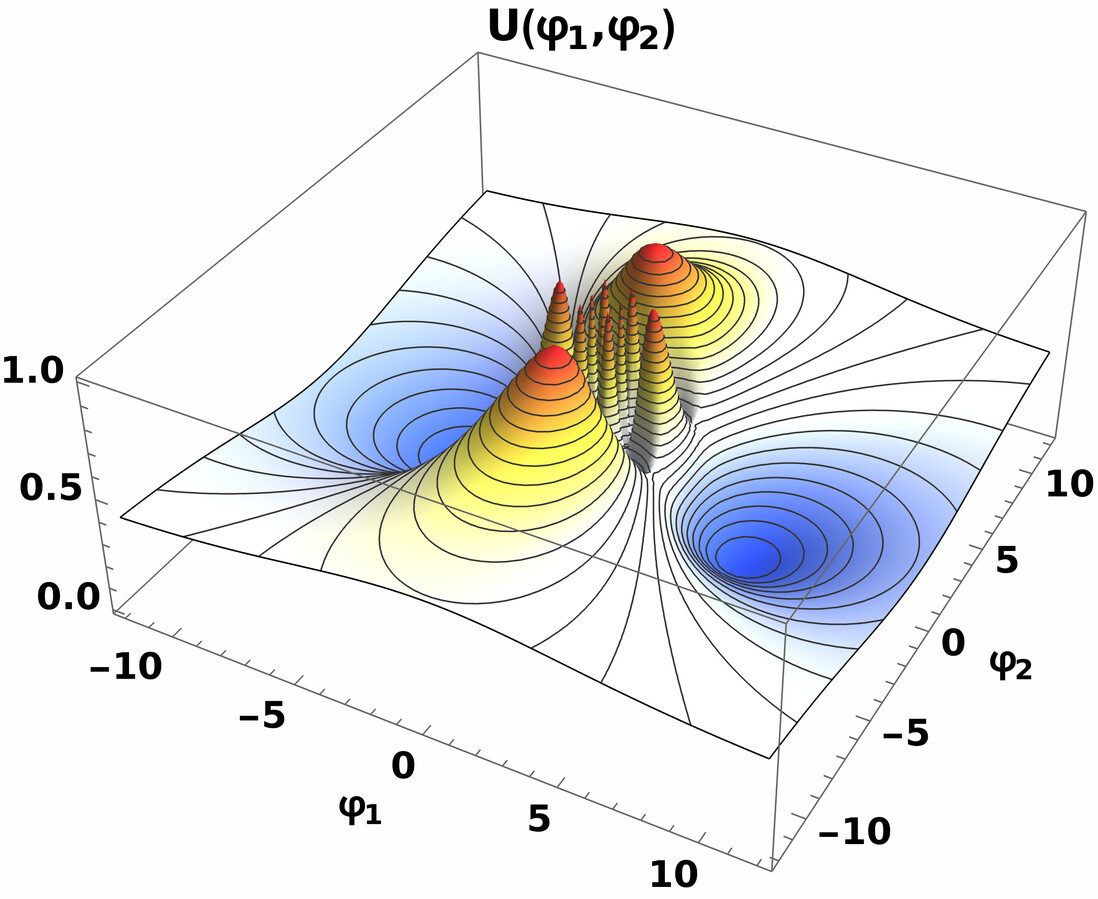}
        \caption{}
        \label{U_3d_dode_antes_longe}
    \end{subfigure}
    \hfill
    \begin{subfigure}{0.45\textwidth}
        \centering
        \includegraphics[width=\textwidth]{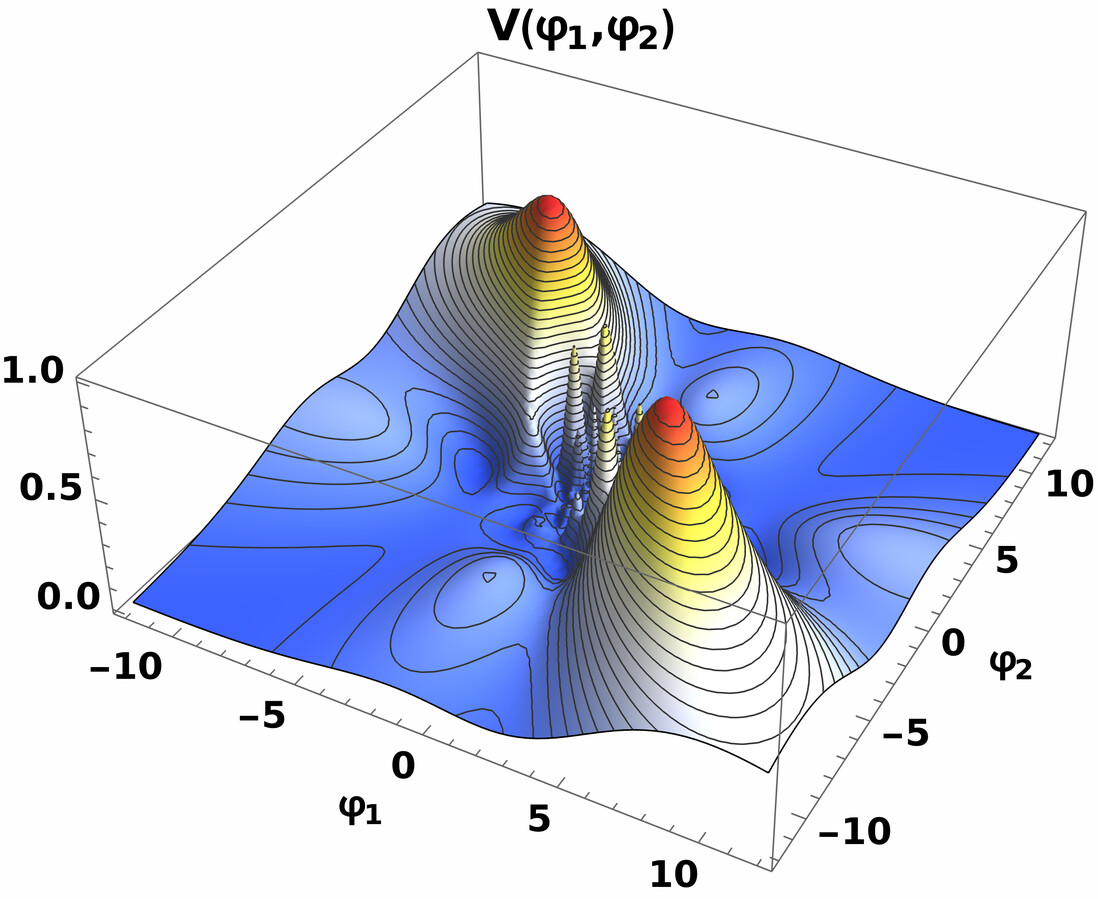}
        \caption{}
        \label{V_3d_dode_antes_longe}
    \end{subfigure}
    \caption{Wider view of the unrotated dodecahedral model:
    (a,c) pre-potential and (b,d) potential.}
    \label{fig:U_V_dode_antes_longe}
    
\end{figure}

The twenty vertices in the reference orientation are
\begin{equation}
    V_D=\dfrac{1}{\sqrt{3}}
    \qty{(\pm1,\pm1,\pm1),
    \qty(0,\pm\tau,\pm\tau^{-1}),
    \qty(\pm\tau^{-1},0,\pm\tau),
    \qty(\pm\tau,\pm\tau^{-1},0)},
\end{equation}
again with all signs chosen independently. The first eight coincide with
the cube vertex set of the previous subsection, reflecting the cube
inscribed in the dodecahedron. Being dual to the icosahedron, the
dodecahedron shares its symmetry axes, so in this orientation the north
pole is again an edge-midpoint direction and a saddle point of the
pre-potential. Figure~\ref{fig:U_V_dode_antes} shows the reference
plots, and Figures~\ref{fig:U_V_dode_antes_perto} and
\ref{fig:U_V_dode_antes_longe} the corresponding close and wider views.

To preserve the common orientation of the dual pair, we apply the same
rotation \eqref{eq:rot_mat}:
\begin{equation}
    V'_D=R V_D.
    \label{eq:vertices_dode_sphere}
\end{equation}
The resulting pre-potential is
\begin{equation}
    U_D(\vb*{\phi})=C_D\prod_{a=1}^{20}\qty(1-v_a\cdot\vb*{\phi}),
    \qquad v_a\in V'_D,\qquad C_D\approx341.078,
    \label{eq:U_dode}
\end{equation}
whose planar form again follows from \eqref{eq:change_coordinates}.
It is invariant under the same conjugate rotation group as $U_I$.
The two pre-potentials share the 62 regular-lattice sites, with minima
and maxima interchanged and the thirty saddle sites in common, as
summarized in Table~\ref{tab:icosa_dode_lattice}.
\begin{table}[htbp]
    \centering
    \begin{tabular}{lccc}
        \hline
        Sites in the common orientation & Number & $U_I$ & $U_D$\\
        \hline
        Icosahedron vertices $V'_I$ & 12 & Minima & Maxima\\
        Dodecahedron vertices $V'_D$ & 20 & Maxima & Minima\\
        Rotated edge-midpoint directions & 30 & Saddles & Saddles\\
        \hline
    \end{tabular}
    \caption{Classification of the shared regular-lattice sites of the
    icosahedron--dodecahedron pair. All 62 sites are vacua of both
    associated potentials.}
    \label{tab:icosa_dode_lattice}
\end{table}

\begin{figure}[!htbp]
    \centering
    \begin{subfigure}{0.45\textwidth}
        \centering
        \includegraphics[width=\textwidth]{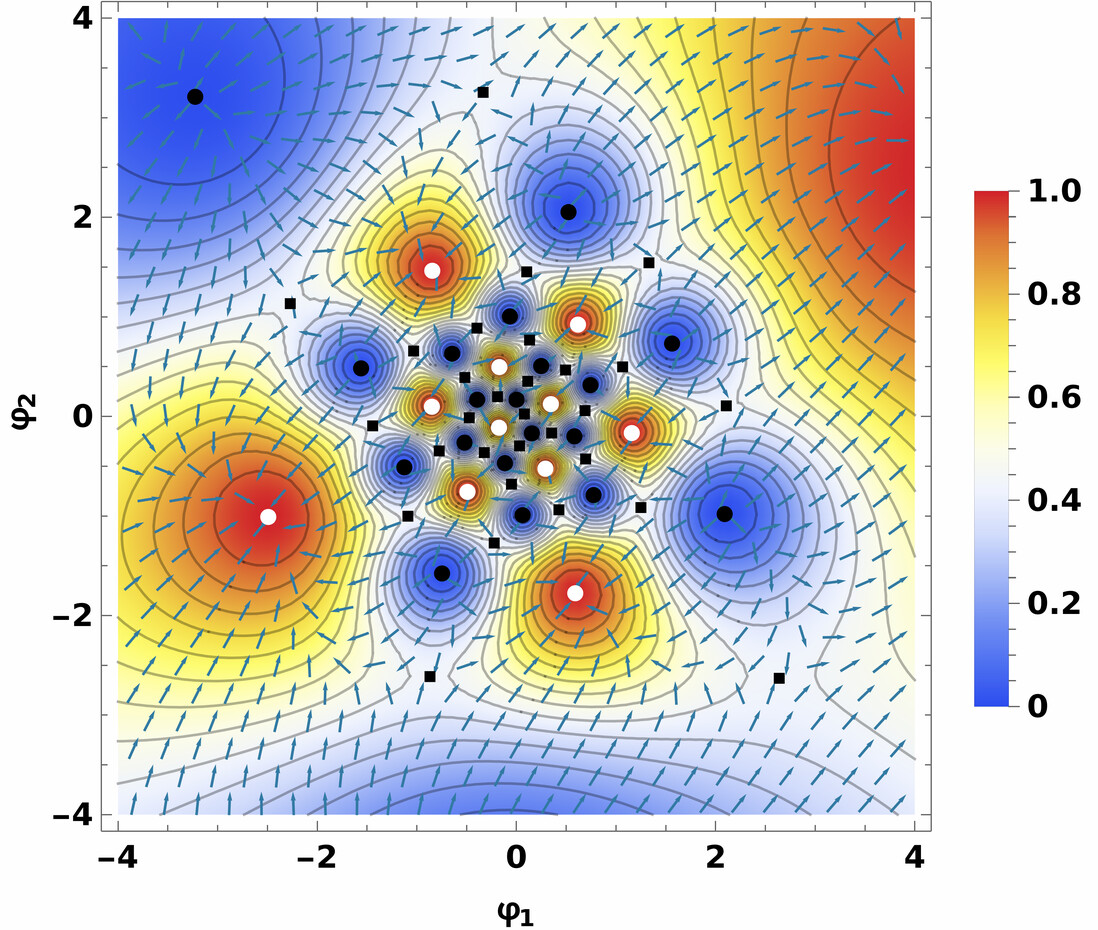}
        \caption{}
        \label{U_2d_dode}
    \end{subfigure}
    \hfill
    \begin{subfigure}{0.45\textwidth}
        \centering
        \includegraphics[width=\textwidth]{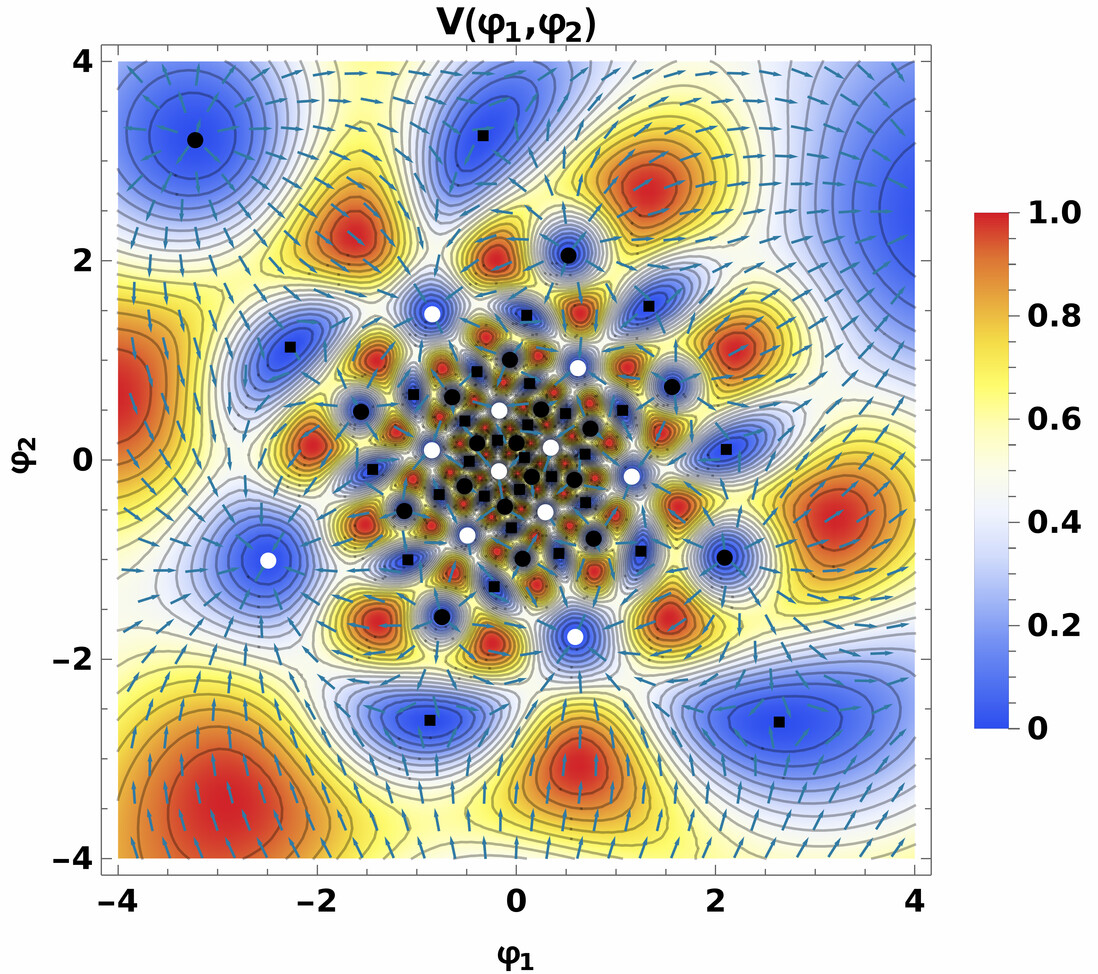}
        \caption{}
        \label{V_2d_dode}
    \end{subfigure}
    \par\vspace{0.15cm}
    \begin{subfigure}{0.45\textwidth}
        \centering
        \includegraphics[width=\textwidth]{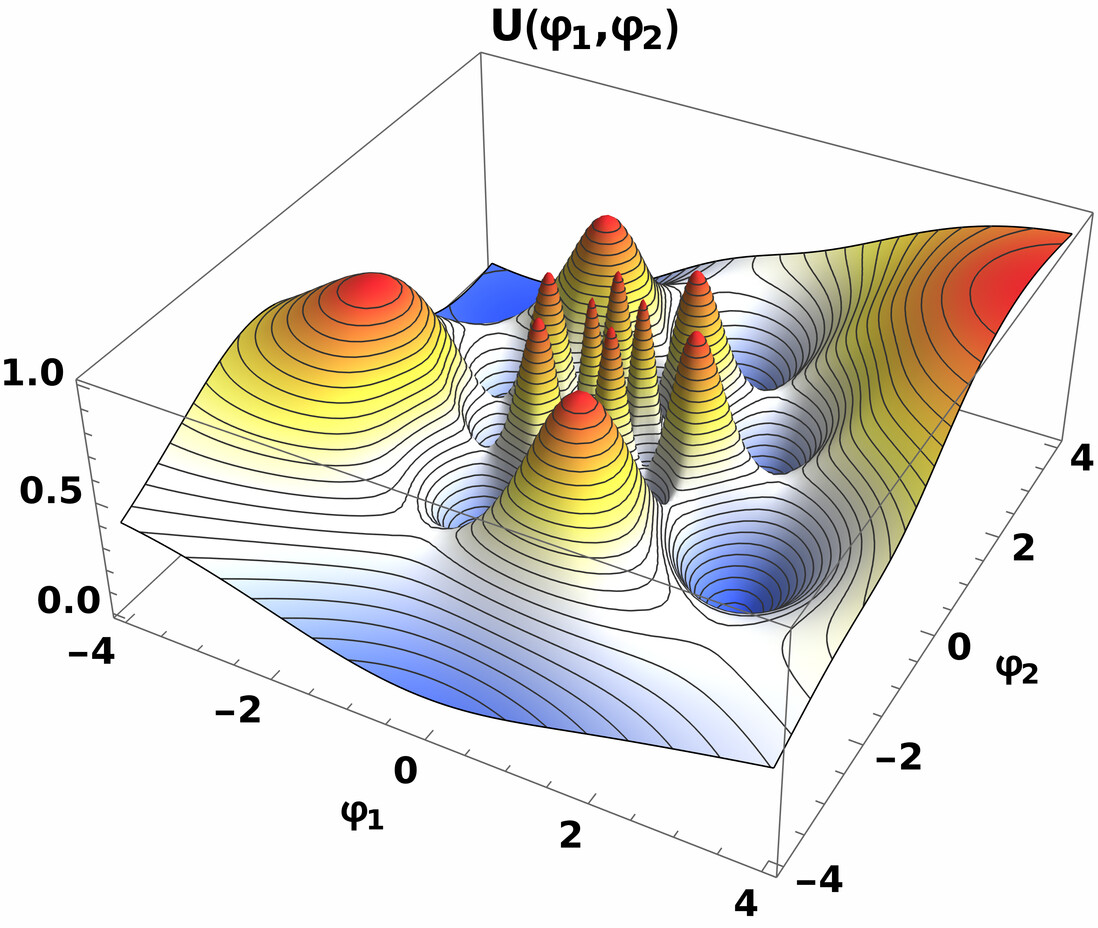}
        \caption{}
        \label{U_3d_dode}
    \end{subfigure}
    \hfill
    \begin{subfigure}{0.45\textwidth}
        \centering
        \includegraphics[width=\textwidth]{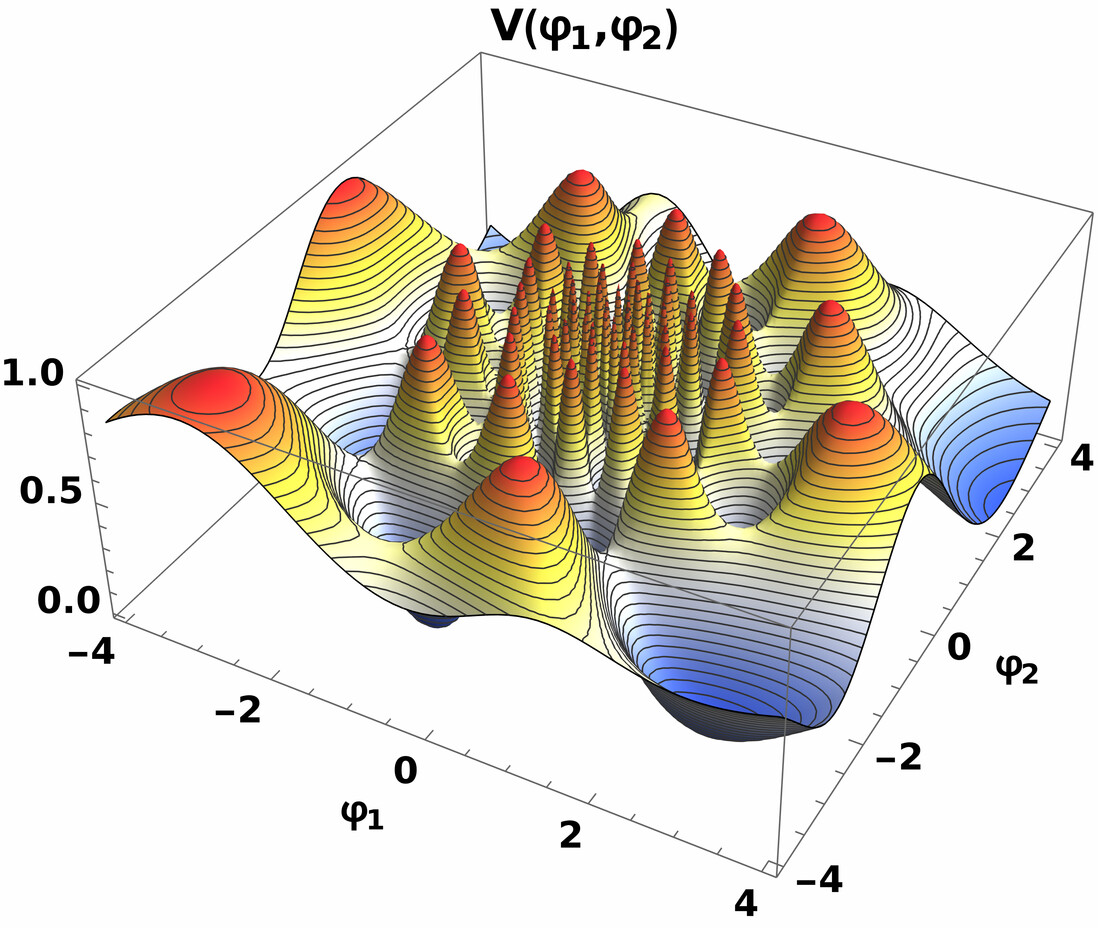}
        \caption{}
        \label{V_3d_dode}
    \end{subfigure}
    \caption{Dodecahedral pre-potential $U_D$ in (a,c) and potential $V$
    in (b,d), after the common rotation \eqref{eq:rot_mat}. The upper
    panels show contours, vacuum markers, and the metric gradient of
    $U_D$; the lower panels show the corresponding surfaces.}
    \label{fig:U_V_dode}
    
\end{figure}

For the dodecahedral model, the critical values are
\begin{equation}
    U_{min}=0,\qquad U_{saddle}\approx0.8318,\qquad U_{max}\approx1.9382.
    \label{eq:valores_U_dode}
\end{equation}
Compared with \eqref{eq:valores_U_icosa}, these values show that the
common vacuum positions support different pre-potential differences and
hence different self-dual energies: the three available values of
$\abs{\Delta U}$ are $1.2247$, $0.1515$ and $1.3762$ for $U_I$, against
$0.8318$, $1.1064$ and $1.9382$ for $U_D$.
Figure~\ref{fig:U_V_dode} shows
the rotated model, and comparison with Figure~\ref{fig:U_V_icosa}
reveals the interchange of minima and maxima together with the resulting
differences in the metric-gradient fields.

\begin{figure}[!htbp]
    \centering
    \begin{subfigure}{0.45\textwidth}
        \centering
        \includegraphics[width=\textwidth]{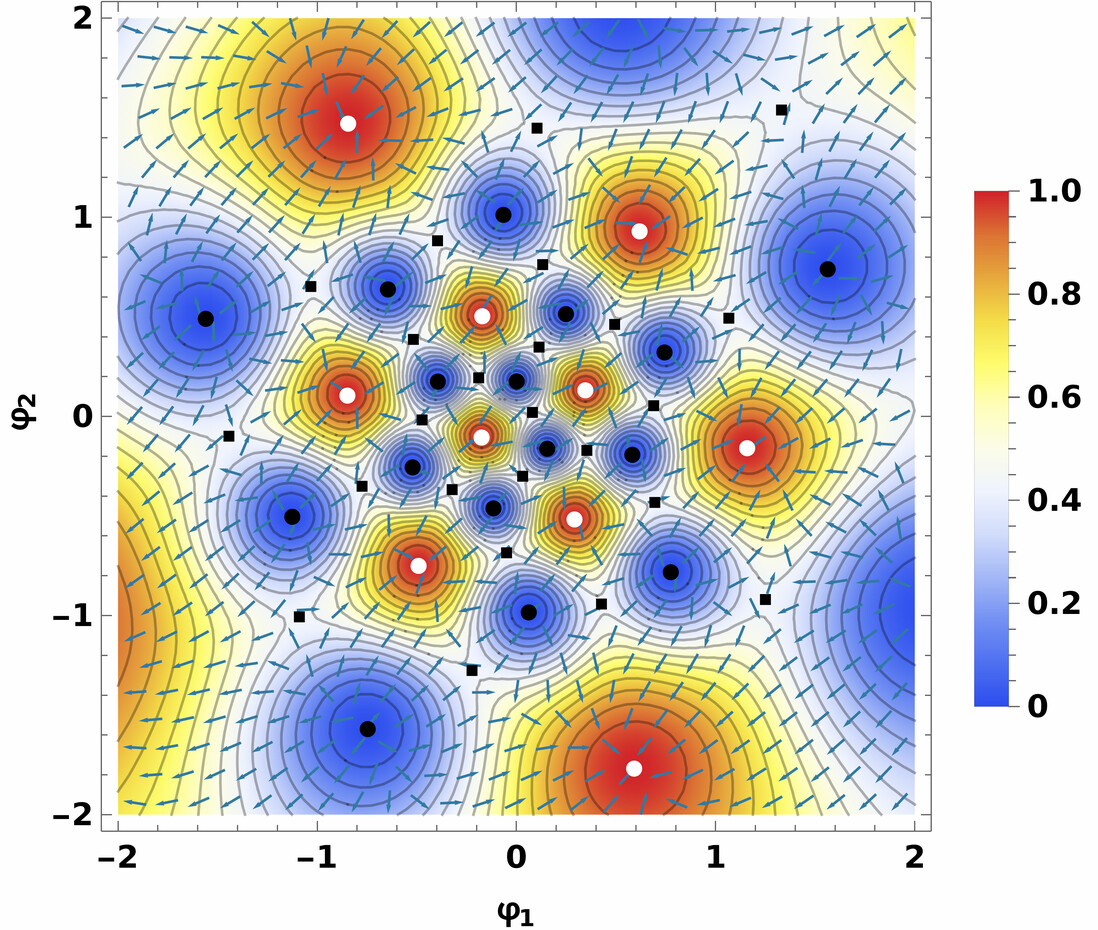}
        \caption{}
        \label{U_2d_dode_perto}
    \end{subfigure}
    \hfill
    \begin{subfigure}{0.45\textwidth}
        \centering
        \includegraphics[width=\textwidth]{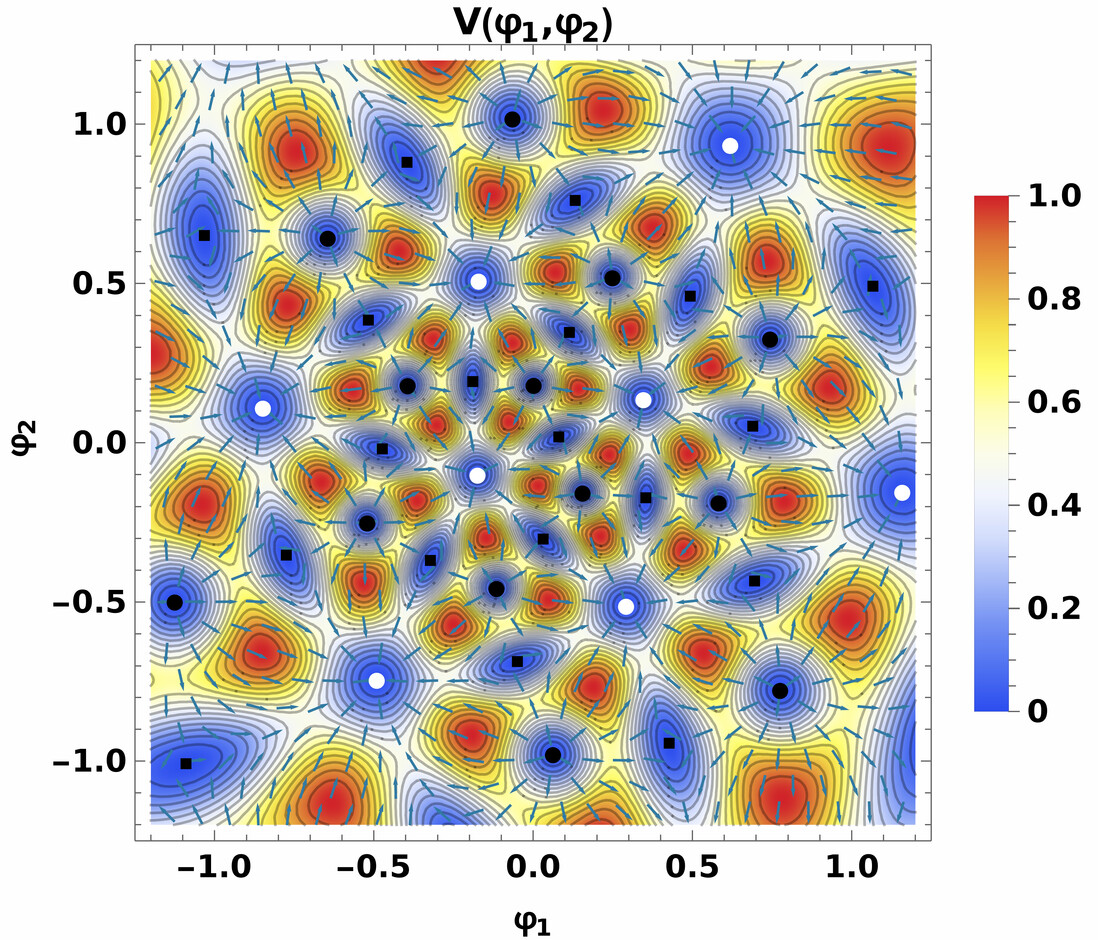}
        \caption{}
        \label{V_2d_dode_perto}
    \end{subfigure}
    \par\vspace{0.15cm}
    \begin{subfigure}{0.45\textwidth}
        \centering
        \includegraphics[width=\textwidth]{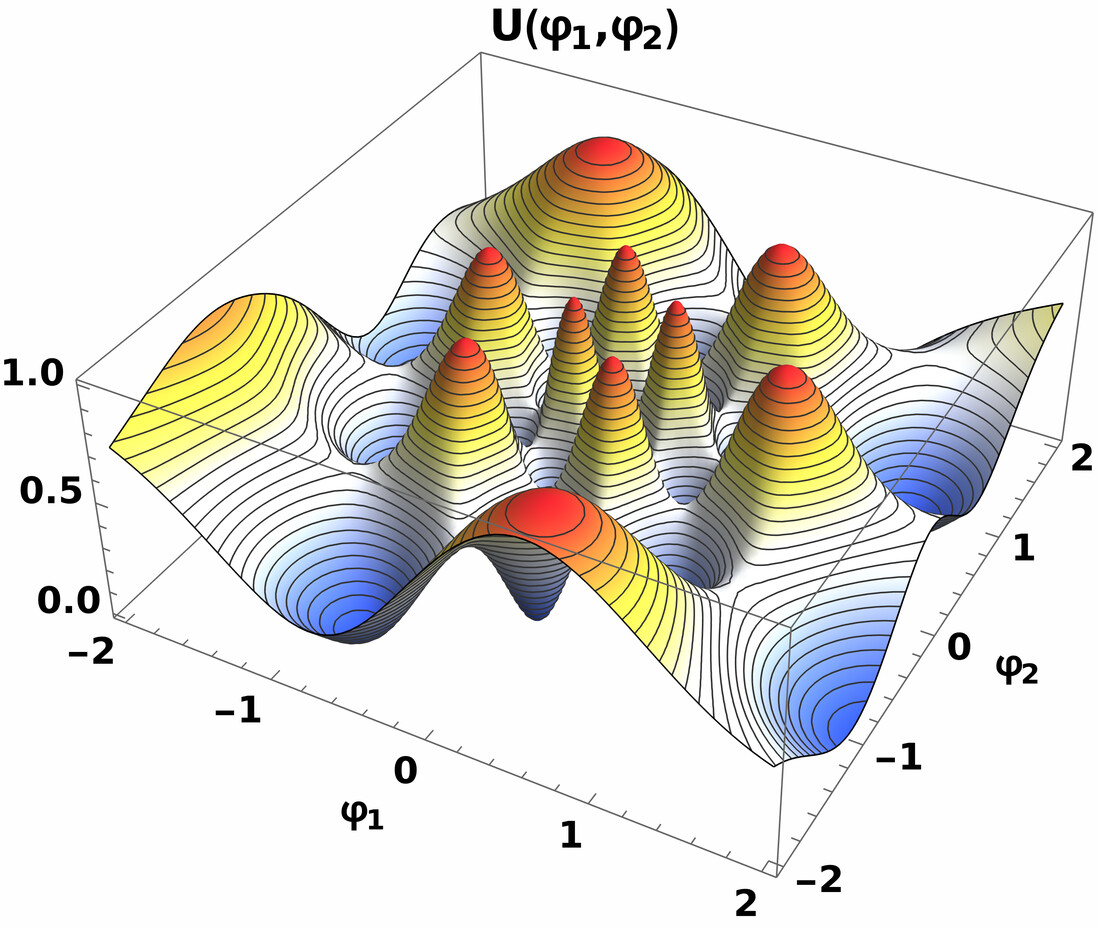}
        \caption{}
        \label{U_3d_dode_perto}
    \end{subfigure}
    \hfill
    \begin{subfigure}{0.45\textwidth}
        \centering
        \includegraphics[width=\textwidth]{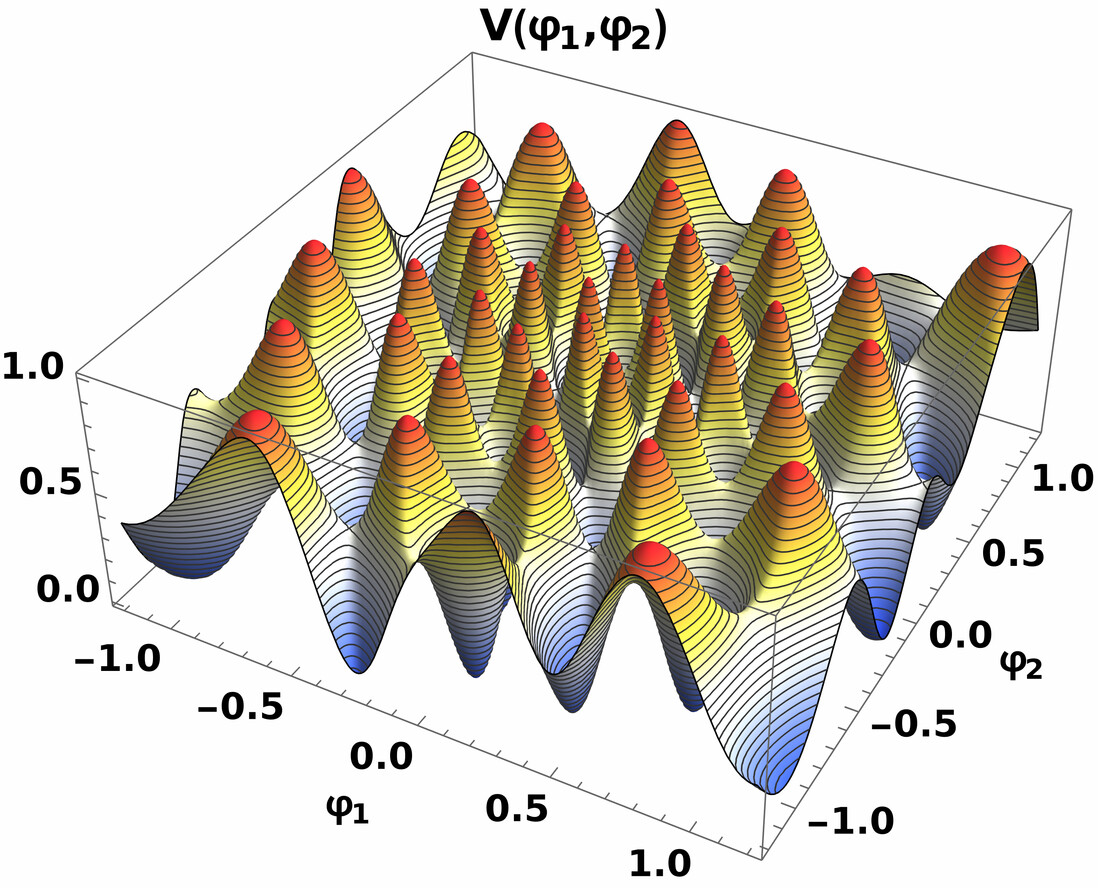}
        \caption{}
        \label{V_3d_dode_perto}
    \end{subfigure}
    \caption{Central region of the rotated dodecahedral model:
    (a,c) pre-potential $U_D$ and (b,d) potential $V$. Markers and
    arrows follow the convention of Figure~\ref{fig:U_V_dode}.}
    \label{fig:U_V_dode_perto}
    
\end{figure}

\begin{figure}[!htbp]
    \centering
    \begin{subfigure}{0.45\textwidth}
        \centering
        \includegraphics[width=\textwidth]{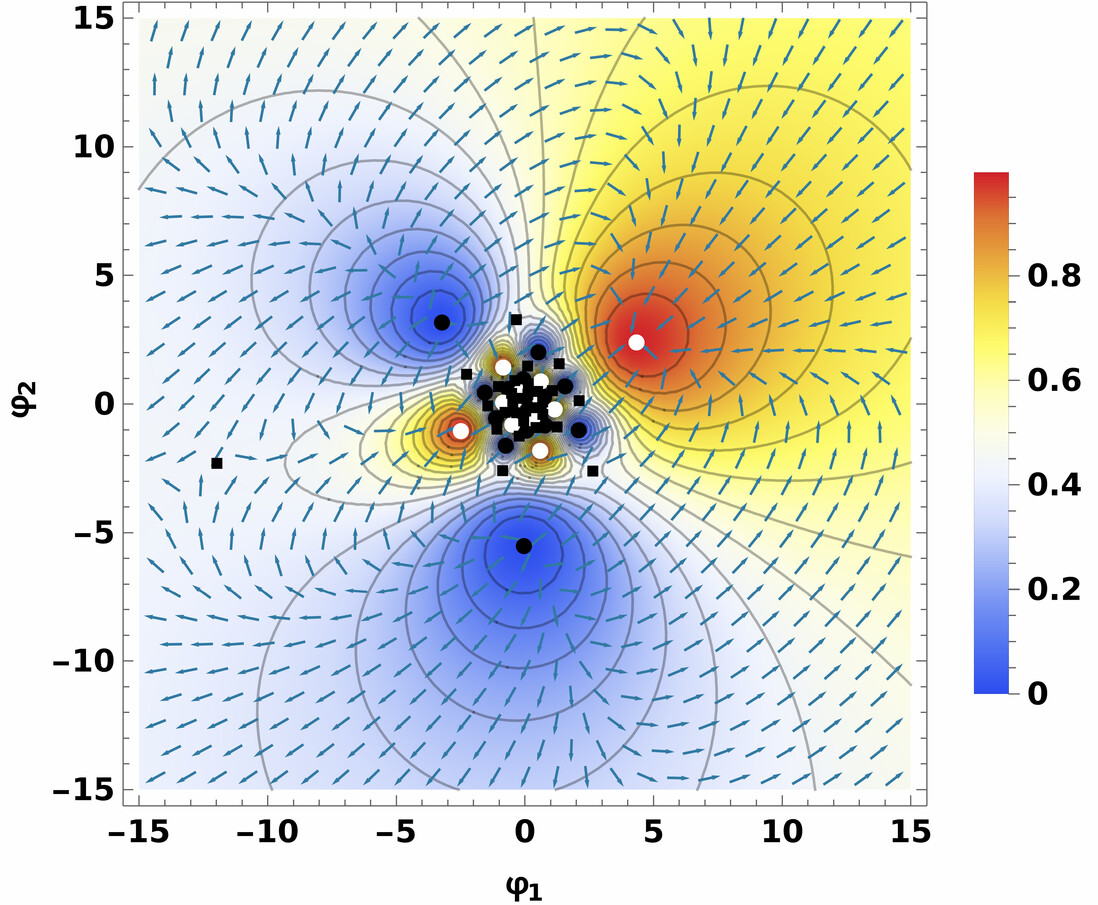}
        \caption{}
        \label{U_2d_dode_longe}
    \end{subfigure}
    \hfill
    \begin{subfigure}{0.45\textwidth}
        \centering
        \includegraphics[width=\textwidth]{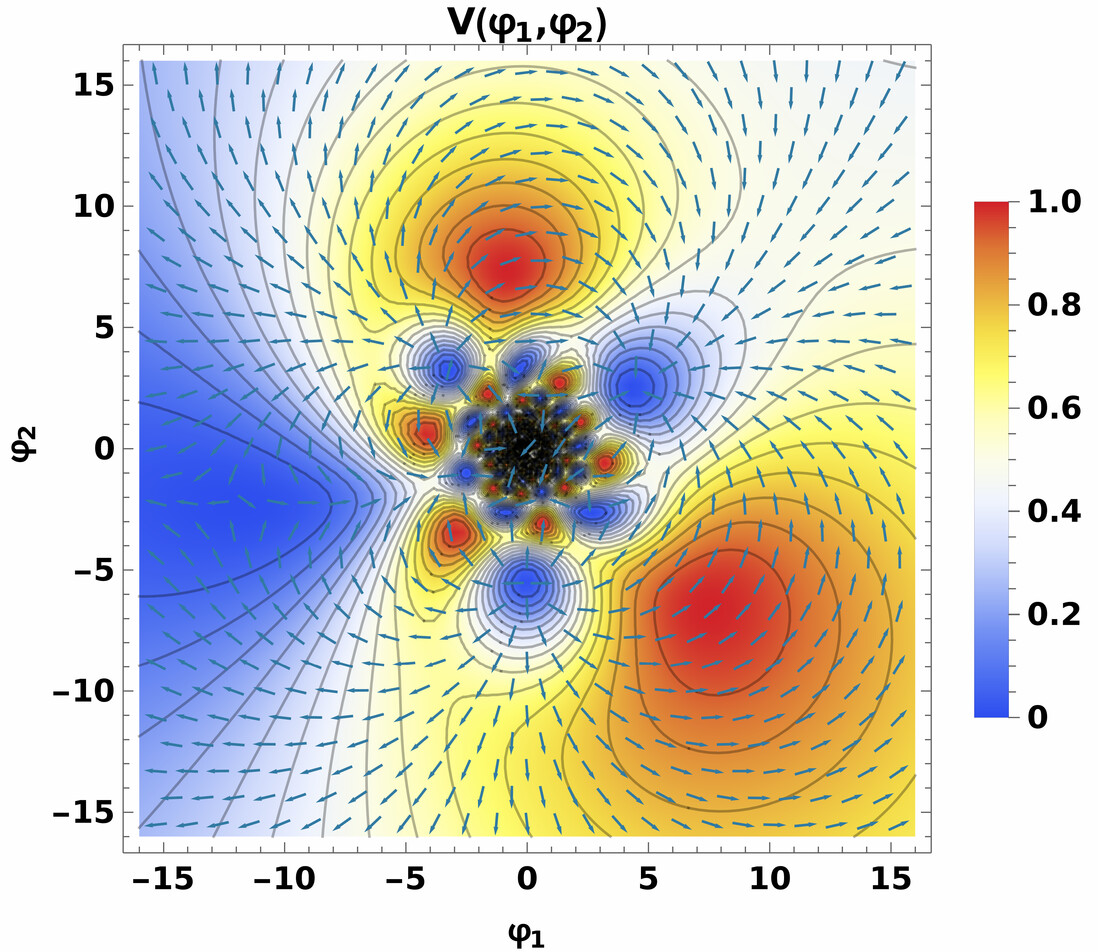}
        \caption{}
        \label{V_2d_dode_longe}
    \end{subfigure}
    \par\vspace{0.15cm}
    \begin{subfigure}{0.45\textwidth}
        \centering
        \includegraphics[width=\textwidth]{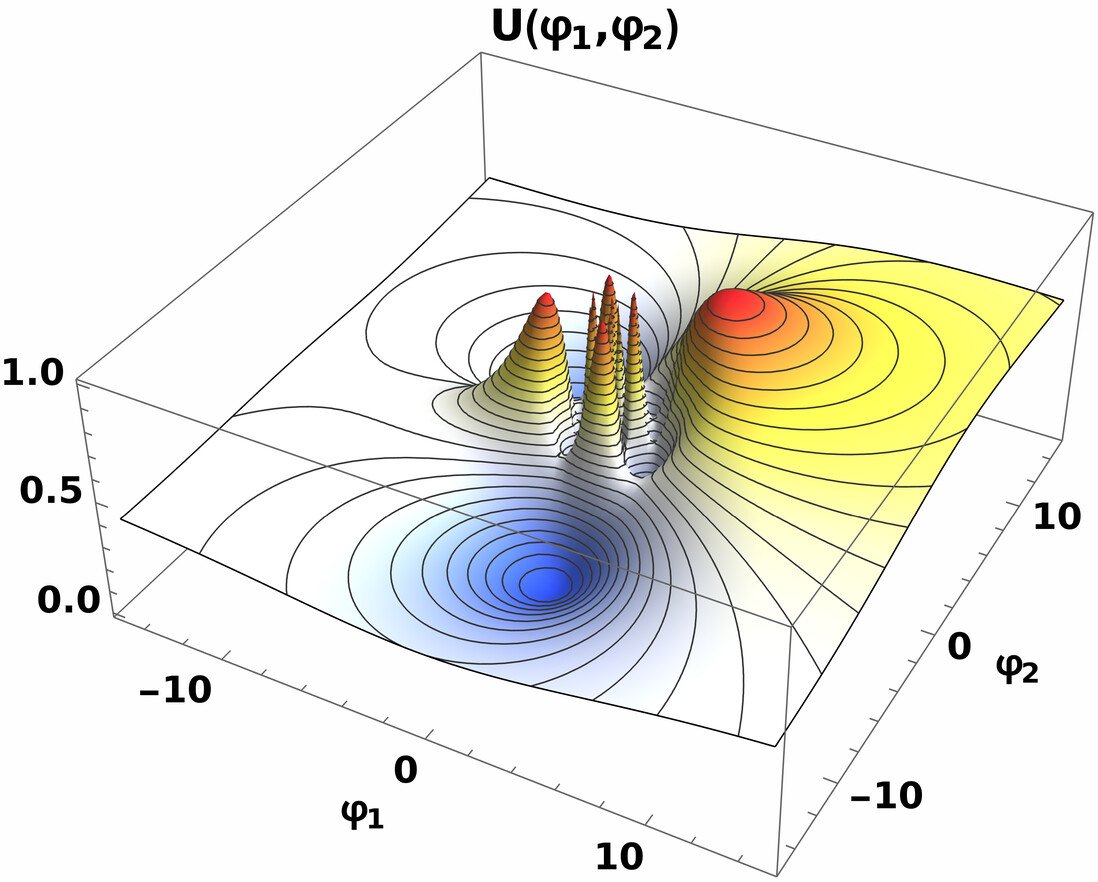}
        \caption{}
        \label{U_3d_dode_longe}
    \end{subfigure}
    \hfill
    \begin{subfigure}{0.45\textwidth}
        \centering
        \includegraphics[width=\textwidth]{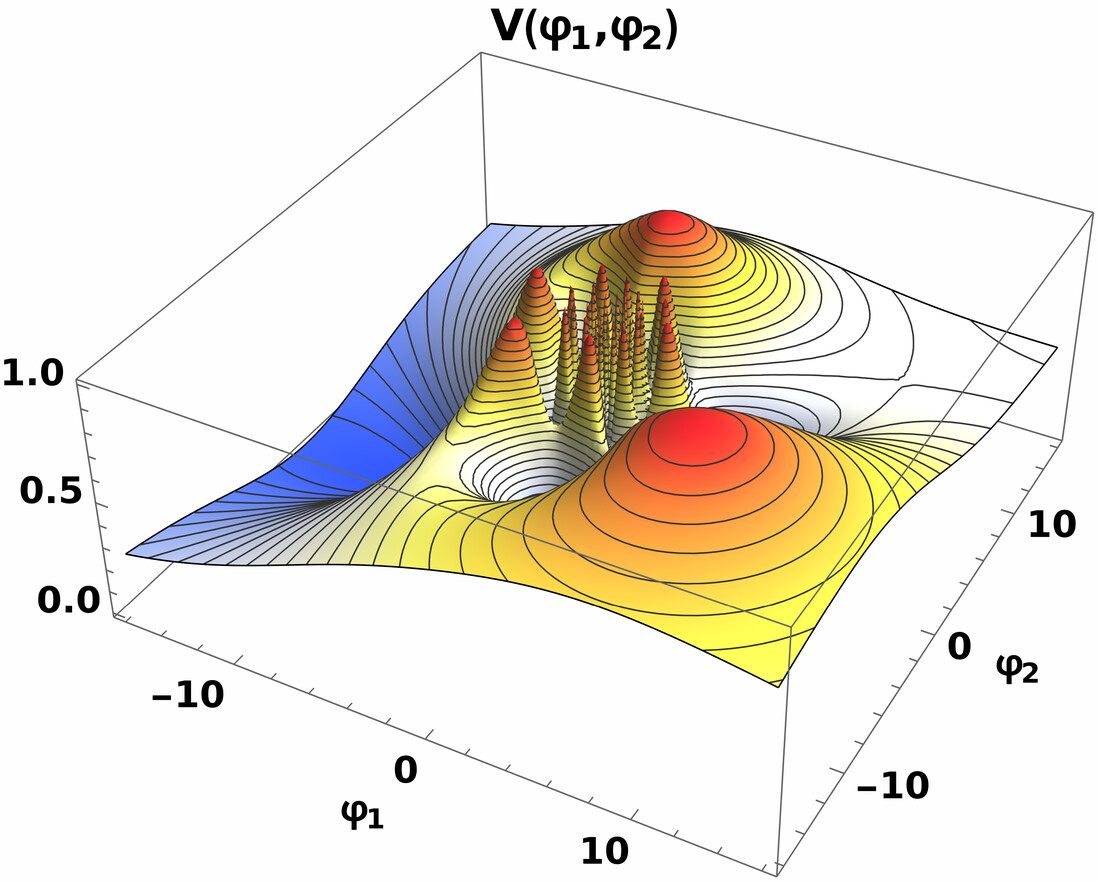}
        \caption{}
        \label{V_3d_dode_longe}
    \end{subfigure}
    \caption{Wider view of the rotated dodecahedral model:
    (a,c) pre-potential $U_D$ and (b,d) potential $V$.}
    \label{fig:U_V_dode_longe}
    
\end{figure}

Figures~\ref{fig:U_V_dode_perto} and \ref{fig:U_V_dode_longe} resolve
the clustered vacua near the origin and the same functions over a wider
field range. These views separate the barriers and metric-gradient
patterns that distinguish the dodecahedral model from the icosahedral
one despite their common vacuum positions.

\begin{figure}[!htbp]
    \centering
    \begin{subfigure}{0.27\textwidth}
        \centering
        \includegraphics[width=\textwidth]{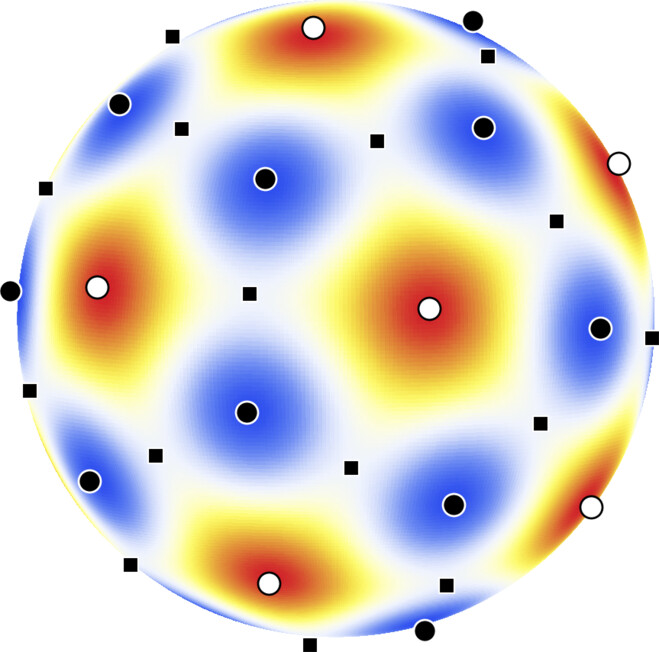}
        \caption{}
        \label{U_dodecahedron_sphere}
    \end{subfigure}
    \hspace{0.03\textwidth}
    \begin{subfigure}{0.27\textwidth}
        \centering
        \includegraphics[width=\textwidth]{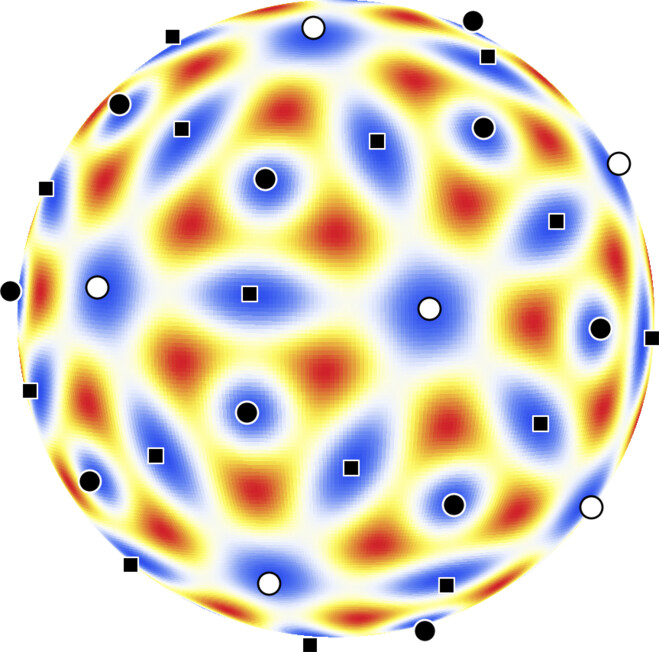}
        \caption{}
        \label{V_dodecahedron_sphere}
    \end{subfigure}
    \caption{Dodecahedral (a) pre-potential $U_D$ and (b) potential $V$
    represented by color on $\mathbb{S}^2$. Black filled circles,
    white circles, and black squares mark minima, maxima, and saddle
    points of $U_D$, respectively; all are vacua of $V$.}
    \label{fig:U_V_dodecahedron_sphere}
    
\end{figure}

The spherical plots in Figure~\ref{fig:U_V_dodecahedron_sphere} make the
dual relation particularly clear when compared with
Figure~\ref{fig:U_V_icosahedron_sphere}: maxima of one pre-potential
coincide with minima of the other, and the two lattices coincide as
sets. For the associated potentials, the common lattice instead consists
entirely of degenerate minima. Their maxima need not coincide, and the
different spatial patterns are visible in panels (b) of the two
spherical figures.
\FloatBarrier

\section{Numerical solutions}
\label{sec:numerical}

In this section, we present numerical solutions of the self-duality
equations \eqref{eq:bps1d} for the five Platonic models and describe
the numerical scheme used to obtain them.

\subsection{Numerical scheme}
We integrate the positive-sign branch of the self-duality equations
\eqref{eq:bps1d},
\begin{equation}
    \frac{\mathrm{d}\vp_a}{\mathrm{d}x}
    =\eta^{-1}_{ab}(\vp)\frac{\partial U}{\partial\vp_b}
    \equiv F_a(\vp),\qquad a=1,2.
    \label{edo_estatica}
\end{equation}
The initial configuration is specified at $x_0=0$. An exact critical
point of $U$ yields a constant solution, so we use small coordinate
displacements of order $10^{-3}$ from sites of the regular lattice.
The displacement vector selects the trajectory passing through the
initial configuration and can affect the asymptotic vacua reached.
Different displacement vectors near the same site can produce different
paths, as illustrated below. Each caption therefore specifies the
reference vacuum site $\vp^{(vac)}$ and the dis\-pla\-ce\-ment used.
We discretize the interval $[-L,L]$
using a uniform grid $x_i=ih$, with $L=50$, mesh spacing $h=10^{-4}$, and
$i=-N,\ldots,N$, where $N=L/h=5\times10^5$.

We use the classical fourth-order Runge--Kutta method \cite{NumRecipes}. Writing
$\mathbf{F}=(F_1,F_2)$ and $\boldsymbol{\vp}_i=(\vp_1(x_i),\vp_2(x_i))$, we
define the increment for a signed step $\Delta x$ by
\begin{equation*}
    \mathbf{K}(\boldsymbol{\vp}_i,\Delta x)
    =\frac{1}{6}\left(\mathbf{k}_1+2\mathbf{k}_2+2\mathbf{k}_3+\mathbf{k}_4\right),
\end{equation*}
where
\begin{equation*}
    \begin{aligned}
        \mathbf{k}_1&=\Delta x\,\mathbf{F}(\boldsymbol{\vp}_i),\\
        \mathbf{k}_2&=\Delta x\,\mathbf{F}\left(
        \boldsymbol{\vp}_i+\frac{\mathbf{k}_1}{2}\right),\\
        \mathbf{k}_3&=\Delta x\,\mathbf{F}\left(
        \boldsymbol{\vp}_i+\frac{\mathbf{k}_2}{2}\right),\\
        \mathbf{k}_4&=\Delta
        x\,\mathbf{F}\left(\boldsymbol{\vp}_i+\mathbf{k}_3\right).
    \end{aligned}
\end{equation*}
Starting from the same configuration at $x=0$, the forward and backward updates
are
\begin{align}
    \boldsymbol{\vp}_{i+1}&=\boldsymbol{\vp}_i+\mathbf{K}(\boldsymbol{\vp}_i,h),
    \label{rk4prafrente}\\
    \boldsymbol{\vp}_{i-1}&=\boldsymbol{\vp}_i+\mathbf{K}(\boldsymbol{\vp}_i,-h).
    \label{rk4pratras}
\end{align}
Both branches solve the same sign choice in \eqref{edo_estatica}; the negative
step changes the integration direction. We perform the fixed number $N$ of steps
in each direction, ending at $x=\pm L$. The fields are already in their
asymptotic regime well before these boundaries. The two branches are combined
into a profile on $[-L,L]$, which approximates the full spatial line.

We evaluate the energy from the static functional \eqref{eq:func_E}, using
finite differences for the field derivatives. Let $D_a(i)$ denote the derivative
approximation,
\begin{equation}
    \begin{aligned}
        D_a(i)&=\frac{\vp_a(i+1)-\vp_a(i-1)}{2h},
        &&-N<i<N,\\
        D_a(-N)&=\frac{-3\vp_a(-N)+4\vp_a(-N+1)-\vp_a(-N+2)}{2h},\\
        D_a(N)&=\frac{3\vp_a(N)-4\vp_a(N-1)+\vp_a(N-2)}{2h}.
    \end{aligned}
    \label{eq:finite_differences}
\end{equation}
These are second-order approximations: for a smooth profile,
$D_a(i)=\mathrm{d}\vp_a/\mathrm{d}x|_{x_i}+O(h^2)$. The numerical energy density
is
\begin{equation}
    \mathcal{E}_i=\frac{1}{2}\eta_{ab}(\boldsymbol{\vp}_i)D_a(i)D_b(i)
    +V(\boldsymbol{\vp}_i),
    \label{num_E_density}
\end{equation}
and the total energy on the truncated interval is estimated with the trapezoidal
rule,
\begin{equation}
    E_{\mathrm{num}}=\frac{h}{2}\sum_{i=-N}^{N-1}
    \left(\mathcal{E}_i+\mathcal{E}_{i+1}\right).
    \label{total_E_num}
\end{equation}
Both boundary points are included with half weight. We compare this estimate
with $\abs{Q}$, computed from the asymptotic vacua through \eqref{eq:Q_1+1}. For
the solutions presented below the absolute difference
$\abs{E_{\mathrm{num}}-\abs{Q}}$ lies between $10^{-8}$ and $10^{-6}$, consistent
with the self-dual energy relation \eqref{eq:igualdade_carga_energia}.

The two contributions to \eqref{num_E_density} coincide on a self-dual solution:
substituting \eqref{eq:bps1d} in the gradient term gives
$\frac{1}{2}\eta_{ab}\dv{\vp_a}{x}\dv{\vp_b}{x}=V(\vp)$ by \eqref{eq:potential},
so the energy density is twice the potential along the solution,
\begin{equation}
    \mathcal{E}(x)=2V(\vp(x)),
    \label{eq:dens_2V}
\end{equation}
and is bounded by $\max_x\mathcal{E}\leq 2V_{max}$, where $V_{max}$ is the
largest value of $V$, quoted for each model in Section~\ref{sec:maxV}. Since
$\mathcal{E}_i$ in \eqref{num_E_density} is built from finite differences of the
numerical solution, its agreement with $2V(\vp_i)$ is a test of the solution
itself, independent of the self-duality equations used to generate it.

Since $U$ is constant on each symmetry orbit of the regular lattice, the
absolute pre-potential difference is fixed by the orbits containing the two
endpoints. Table~\ref{tab:charges} lists these differences for the five models.
Whenever a self-dual connection exists, its energy is the corresponding value.
\begin{table}[htbp]
    \centering
    \begin{tabular}{lccc}
        \hline
        Model & minimum--saddle & saddle--maximum & minimum--maximum\\
        \hline
        Tetrahedron  & $1$      & $1/3$    & $4/3$   \\
        Octahedron   & $1.306$  & $0.242$  & $1.548$ \\
        Cube         & $1.138$  & $0.885$  & $2.023$ \\
        Icosahedron  & $1.2247$ & $0.1515$ & $1.3762$\\
        Dodecahedron & $0.8318$ & $1.1064$ & $1.9382$\\
        \hline
    \end{tabular}
    \caption{Absolute pre-potential differences between the three critical-point
    orbits, obtained from \eqref{eq:valores_U_tetra}, \eqref{eq:valores_U_octa},
    \eqref{eq:valores_U_cube}, \eqref{eq:valores_U_icosa}, and
    \eqref{eq:valores_U_dode}. These are the charge magnitudes and energies of
    self-dual connections between the corresponding orbits, when such
    connections exist.}
    \label{tab:charges}
\end{table}

The examples below show that solutions with the same charge can follow
different paths in field space and have different energy-density
profiles. For the positive-sign branch of \eqref{edo_estatica},
\begin{equation*}
    \int_{x_1}^{x_2}\mathcal{E}(x)\,\mathrm{d}x
    =U(\boldsymbol{\vp}(x_2))-U(\boldsymbol{\vp}(x_1)).
\end{equation*}
Thus, pre-potential differences determine integrated energies, whereas
peak heights and widths depend on the potential sampled along the
trajectory through \eqref{eq:dens_2V}. If a profile has two transitions
separated by a region close to a saddle, their energy contributions can
be approximated by the minimum--saddle and saddle--maximum differences
in Table~\ref{tab:charges}. A nonconstant solution cannot reach an exact
critical point at finite $x$ and then leave it, by uniqueness of the
initial-value problem. Connections with saddle-point asymptotics were
not obtained by the numerical procedure for the octahedral, cubic,
icosahedral, and dodecahedral models; the examples below are restricted
to the connections found numerically.

\subsection{Maxima of the potential}
\label{sec:maxV}
\begin{figure}[htb]
    \centering
    \begin{subfigure}{0.45\textwidth}
        \centering
        \includegraphics[width=\textwidth]{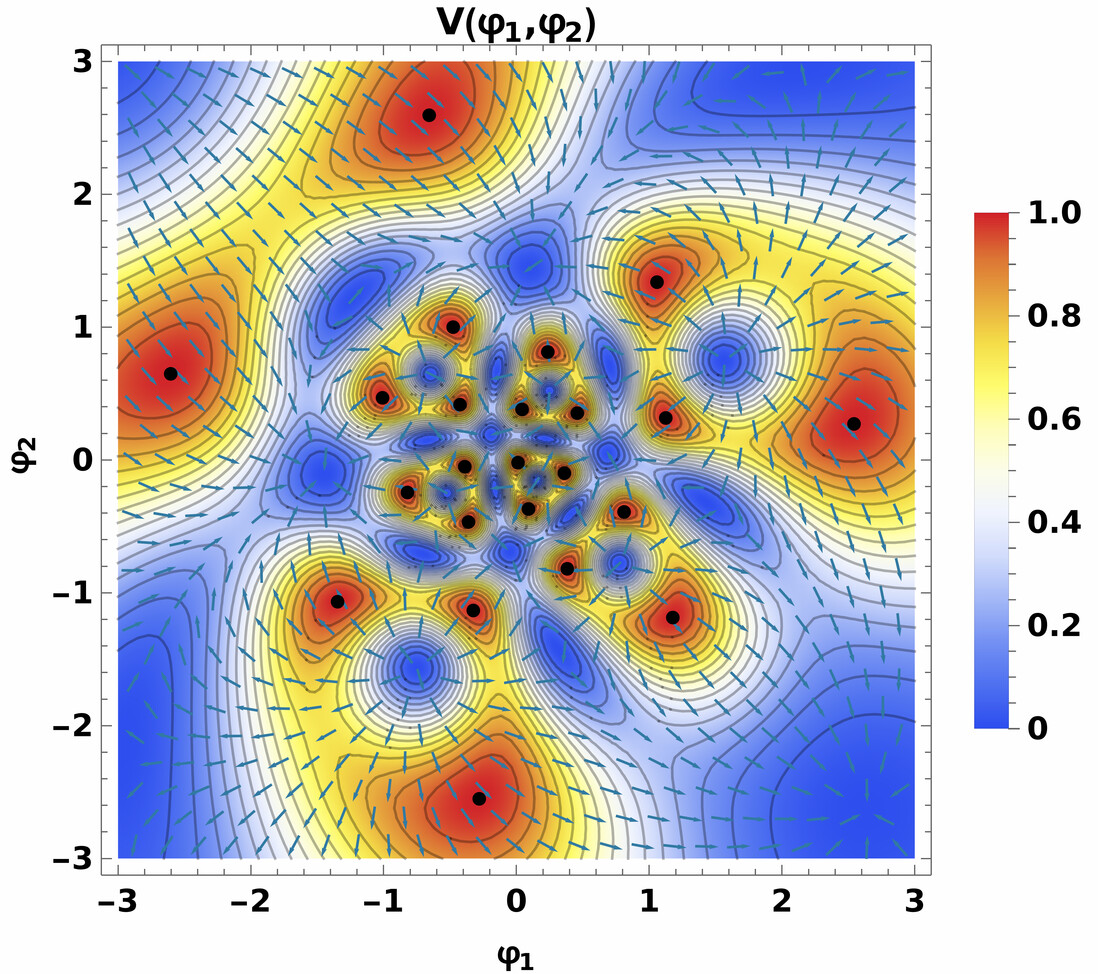}
        \caption{}
        \label{V2d_cube_maxPot}
    \end{subfigure}
    \hfill
    \begin{subfigure}{0.45\textwidth}
        \centering
        \includegraphics[width=\textwidth]{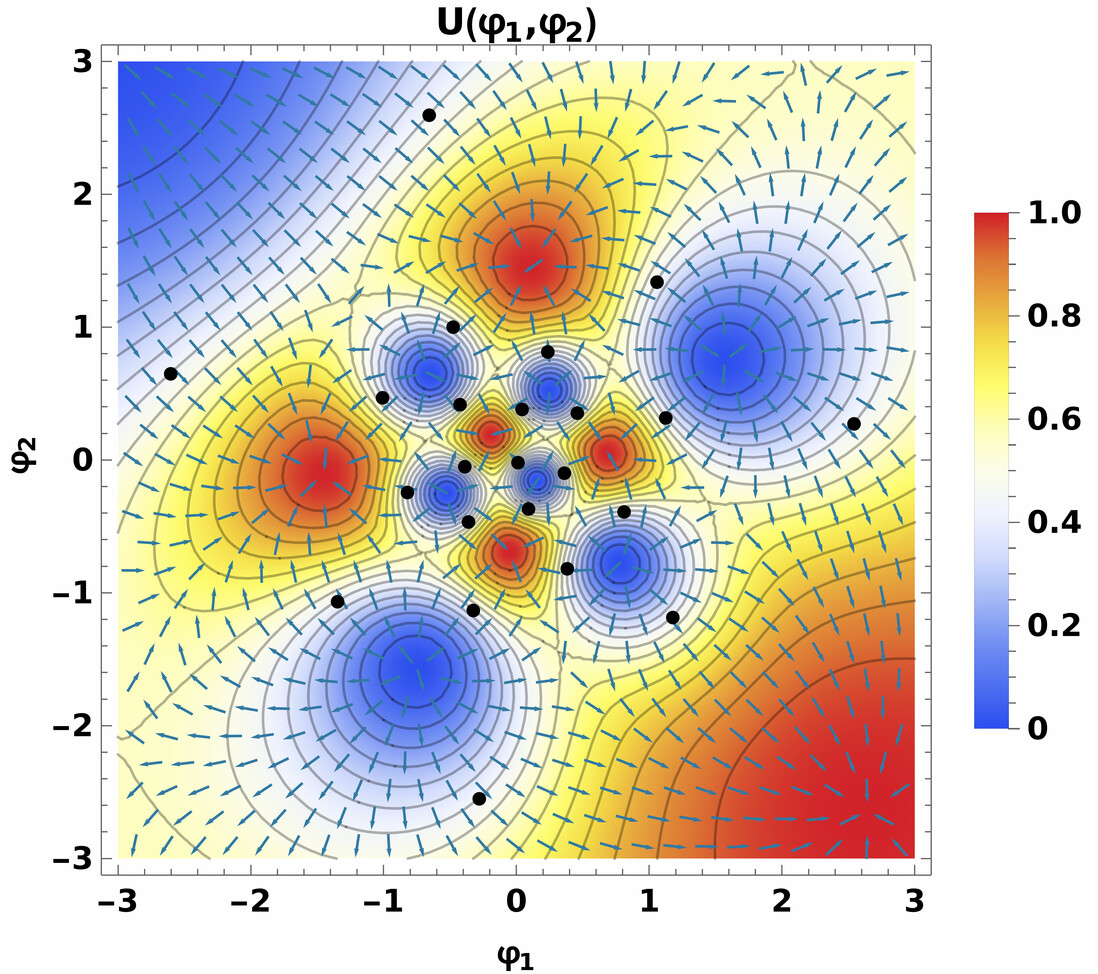}
        \caption{}
        \label{U2d_cube_maxPot}
    \end{subfigure}
    \par\vspace{0.15cm}
    \begin{subfigure}{0.45\textwidth}
        \centering
        \includegraphics[width=\textwidth]{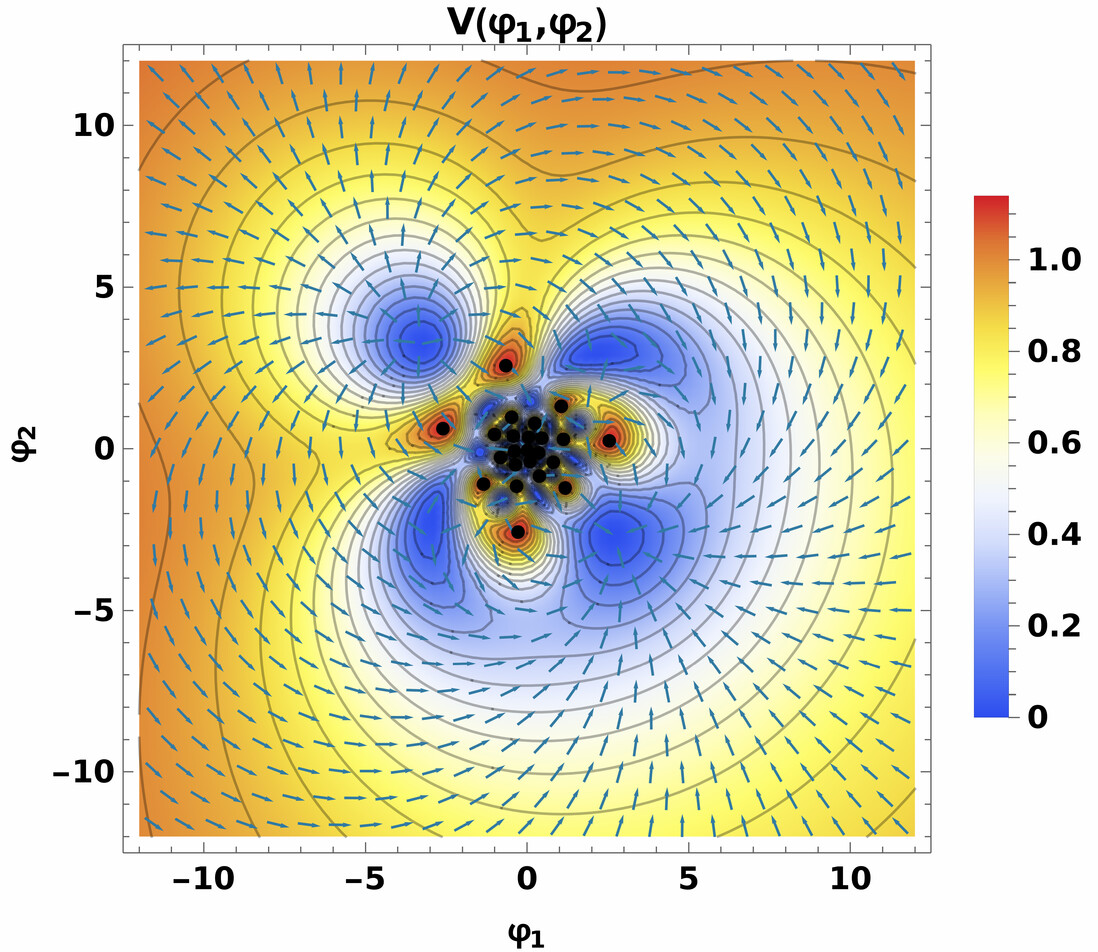}
        \caption{}
    \end{subfigure}
    \hfill
    \begin{subfigure}{0.45\textwidth}
        \centering
        \includegraphics[width=\textwidth]{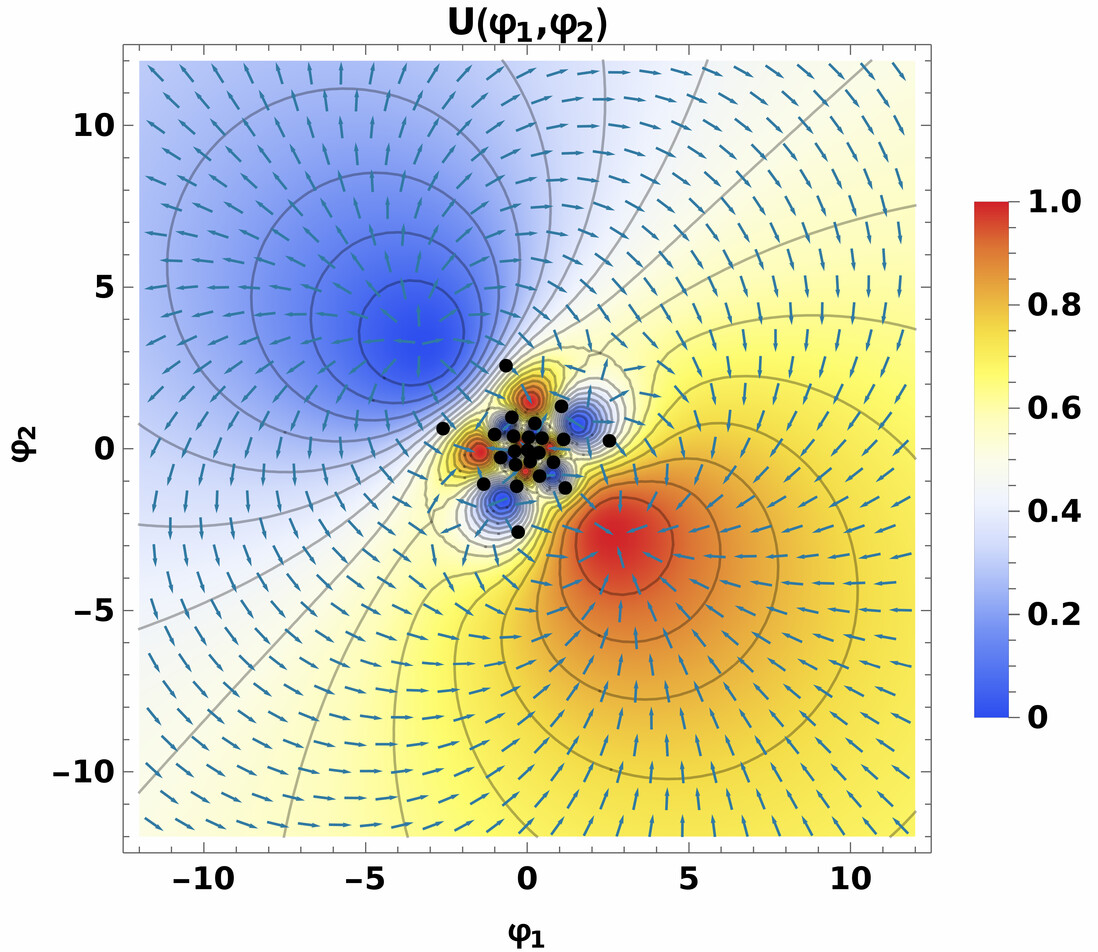}
        \caption{}
    \end{subfigure}
    \caption{Cubic (a,c) potential $V$ and (b,d) pre-potential $U_C$ as
    functions of $(\vp_1,\vp_2)$, at two scales. Black circles mark the
    maxima of $V$, all of which have the same value.}
    \label{fig:maxPot_cube}
   
\end{figure}

\begin{figure}[htb]
    \centering
    \begin{subfigure}{0.49\textwidth}
        \centering
        \includegraphics[width=\textwidth]{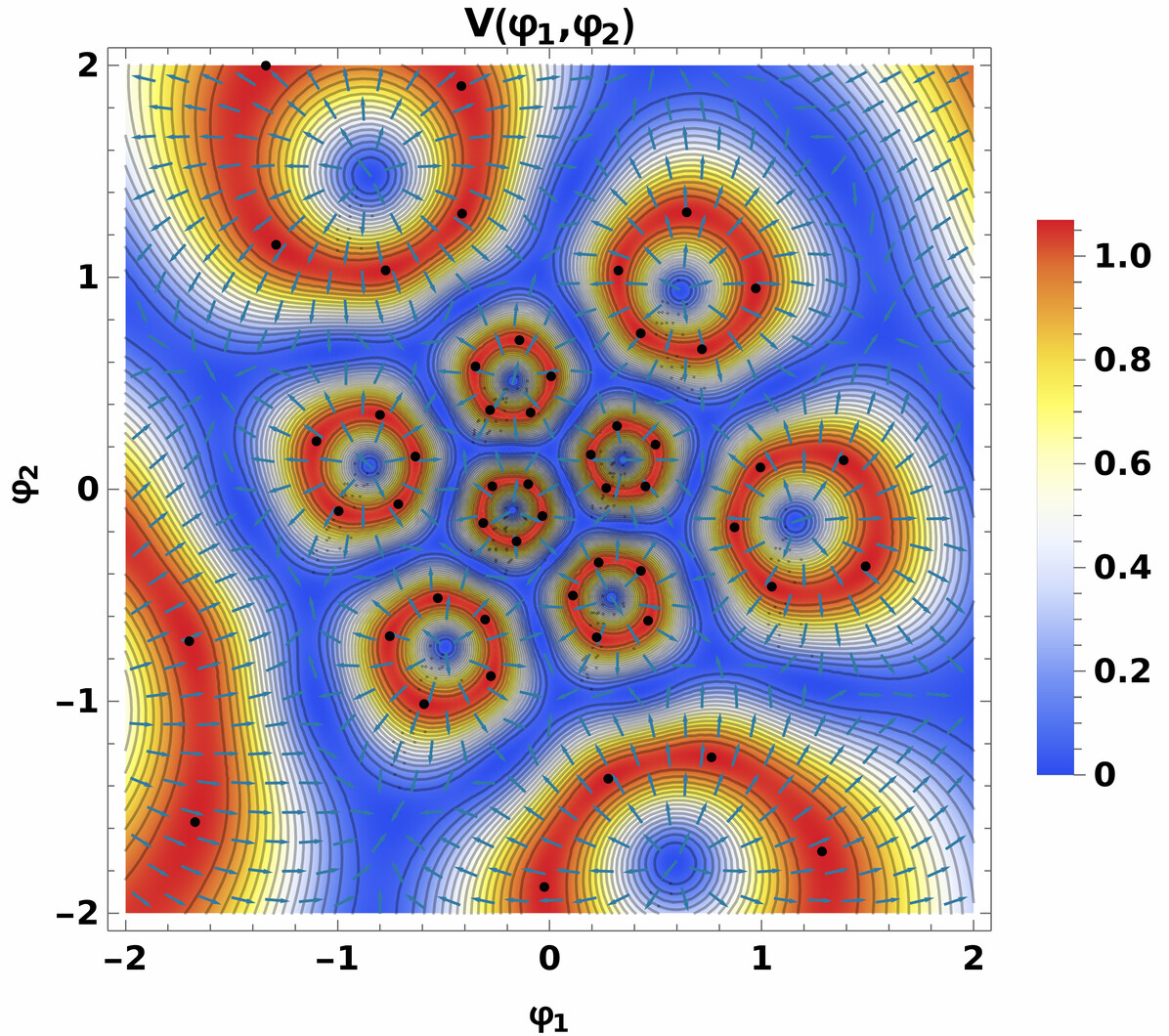}
        \caption{}
        \label{V2d_icosa_maxPot}
    \end{subfigure}
    \hfill
    \begin{subfigure}{0.49\textwidth}
        \centering
        \includegraphics[width=\textwidth]{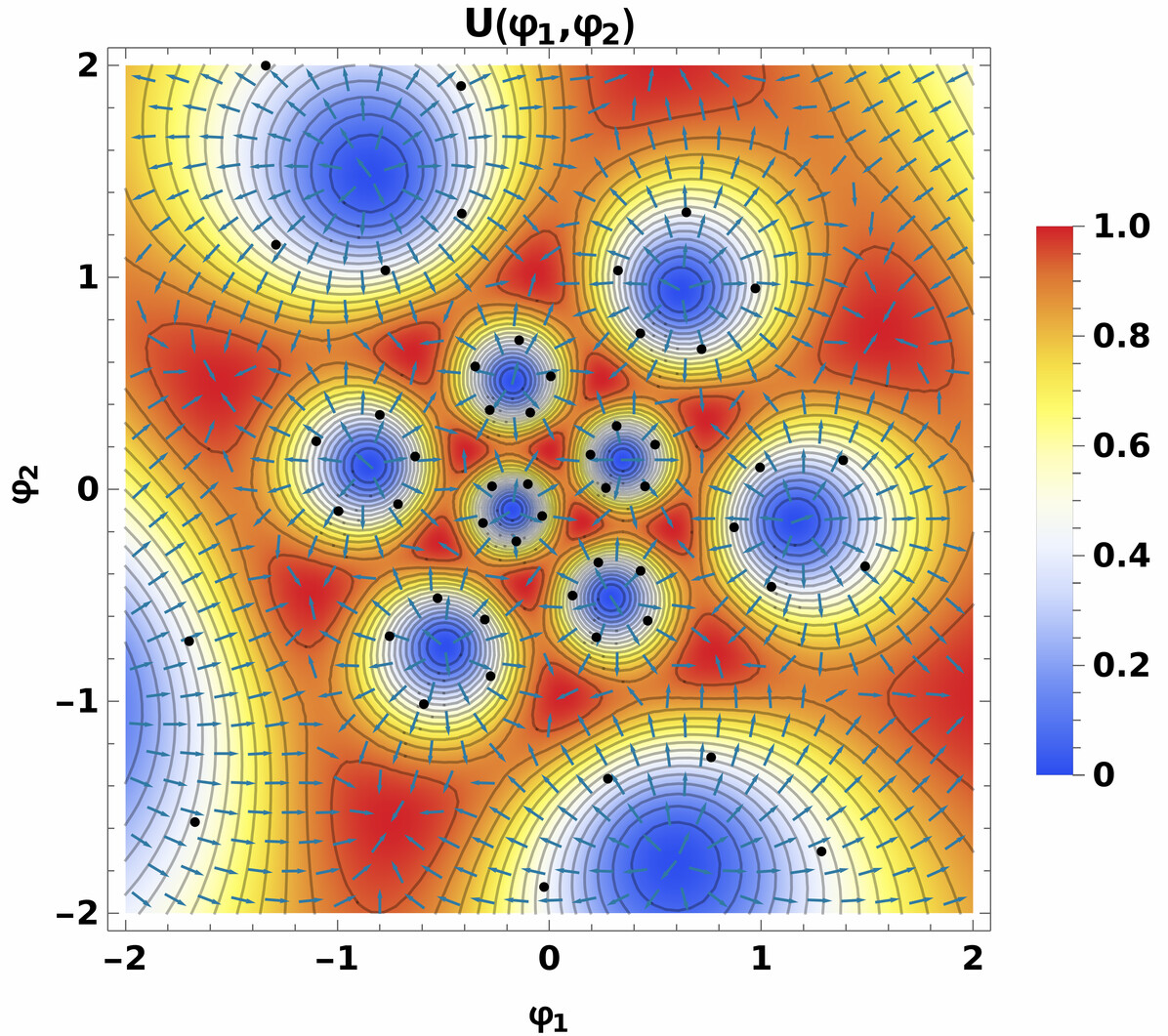}
        \caption{}
        \label{U2d_icosa_maxPot}
    \end{subfigure}
    \par\vspace{0.15cm}
    \begin{subfigure}{0.49\textwidth}
        \centering
        \includegraphics[width=\textwidth]{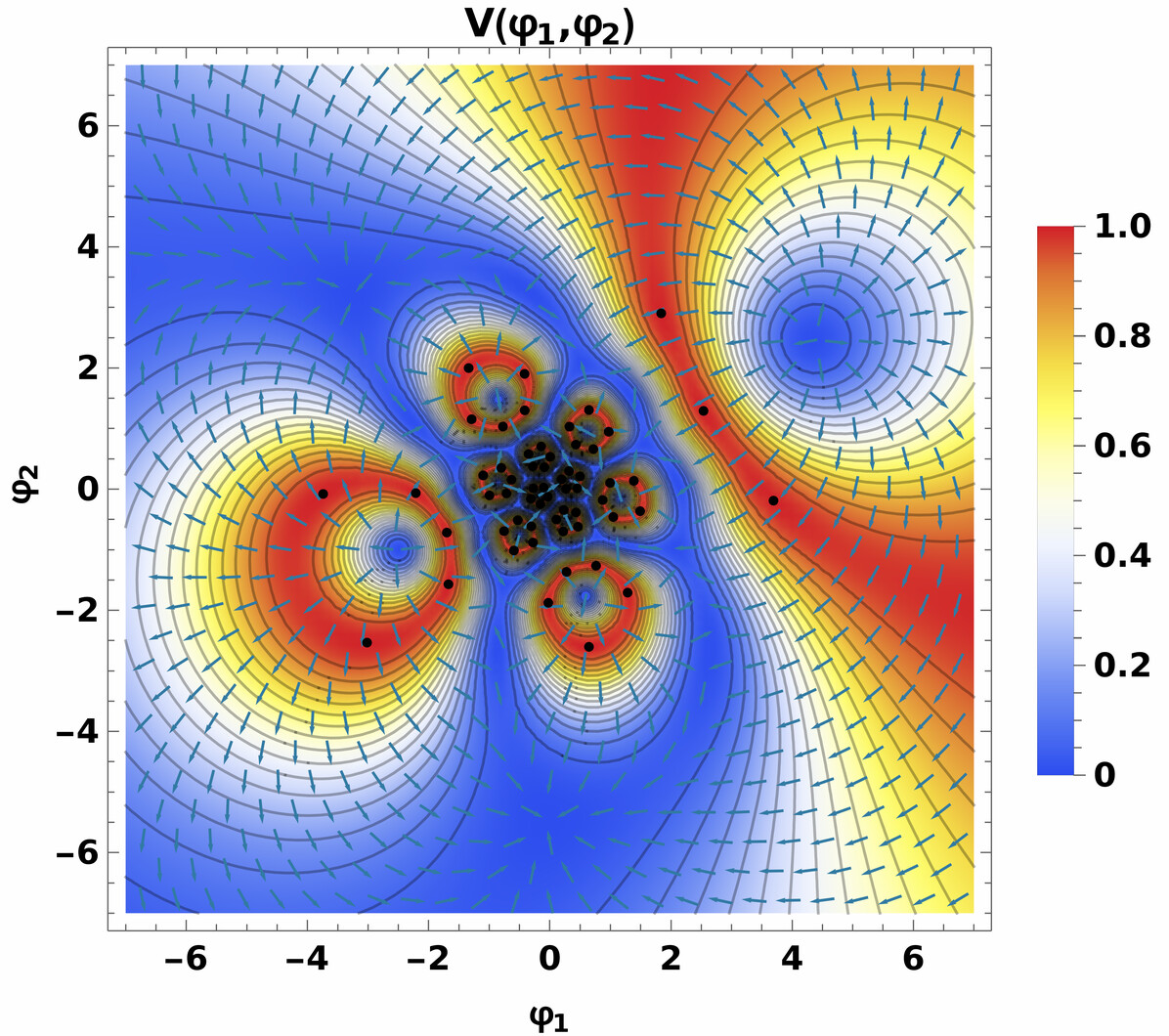}
        \caption{}
    \end{subfigure}
    \hfill
    \begin{subfigure}{0.49\textwidth}
        \centering
        \includegraphics[width=\textwidth]{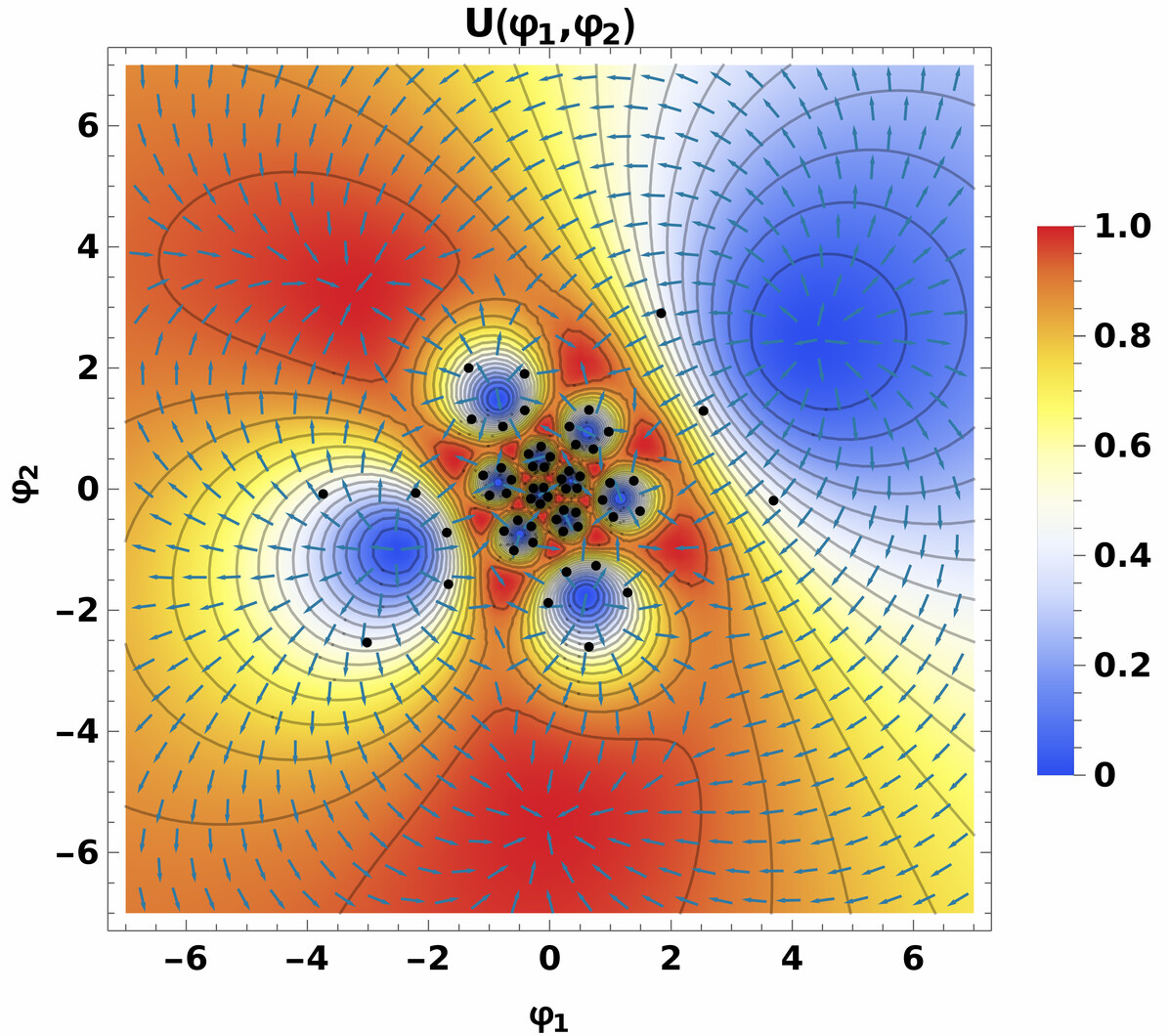}
        \caption{}
    \end{subfigure}
    \caption{Icosahedral (a,c) potential $V$ and (b,d) pre-potential
    $U_I$ as functions of $(\vp_1,\vp_2)$, at two scales. Black circles
    mark the maxima of $V$, all of which have the same value.}
    \label{fig:maxPot_icosa}
    
\end{figure}

\begin{figure}[htb]
    \centering
    \begin{subfigure}{0.49\textwidth}
        \centering
        \includegraphics[width=\textwidth]{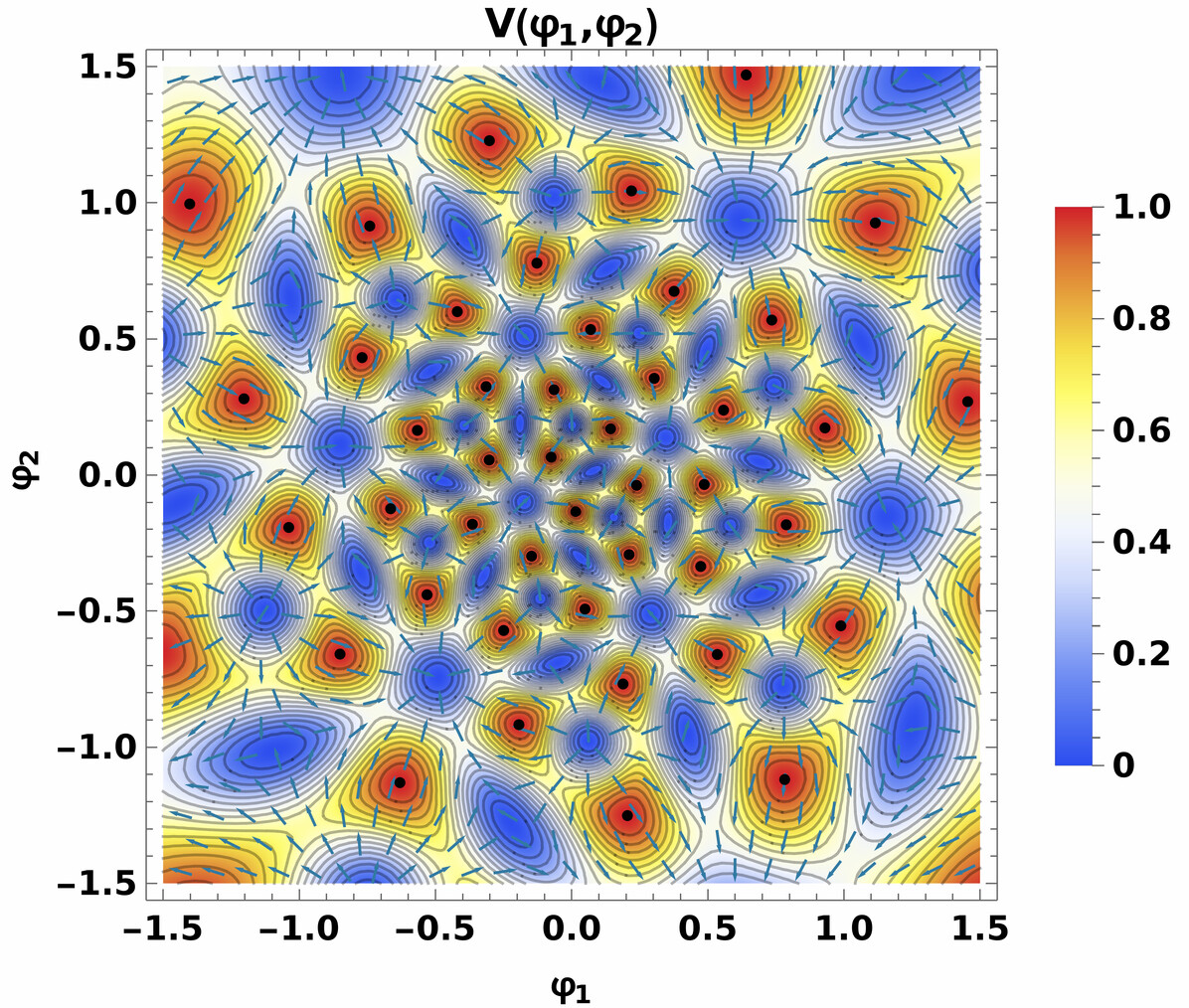}
        \caption{}
        \label{V2d_dode_maxPot}
    \end{subfigure}
    \hfill
    \begin{subfigure}{0.49\textwidth}
        \centering
        \includegraphics[width=\textwidth]{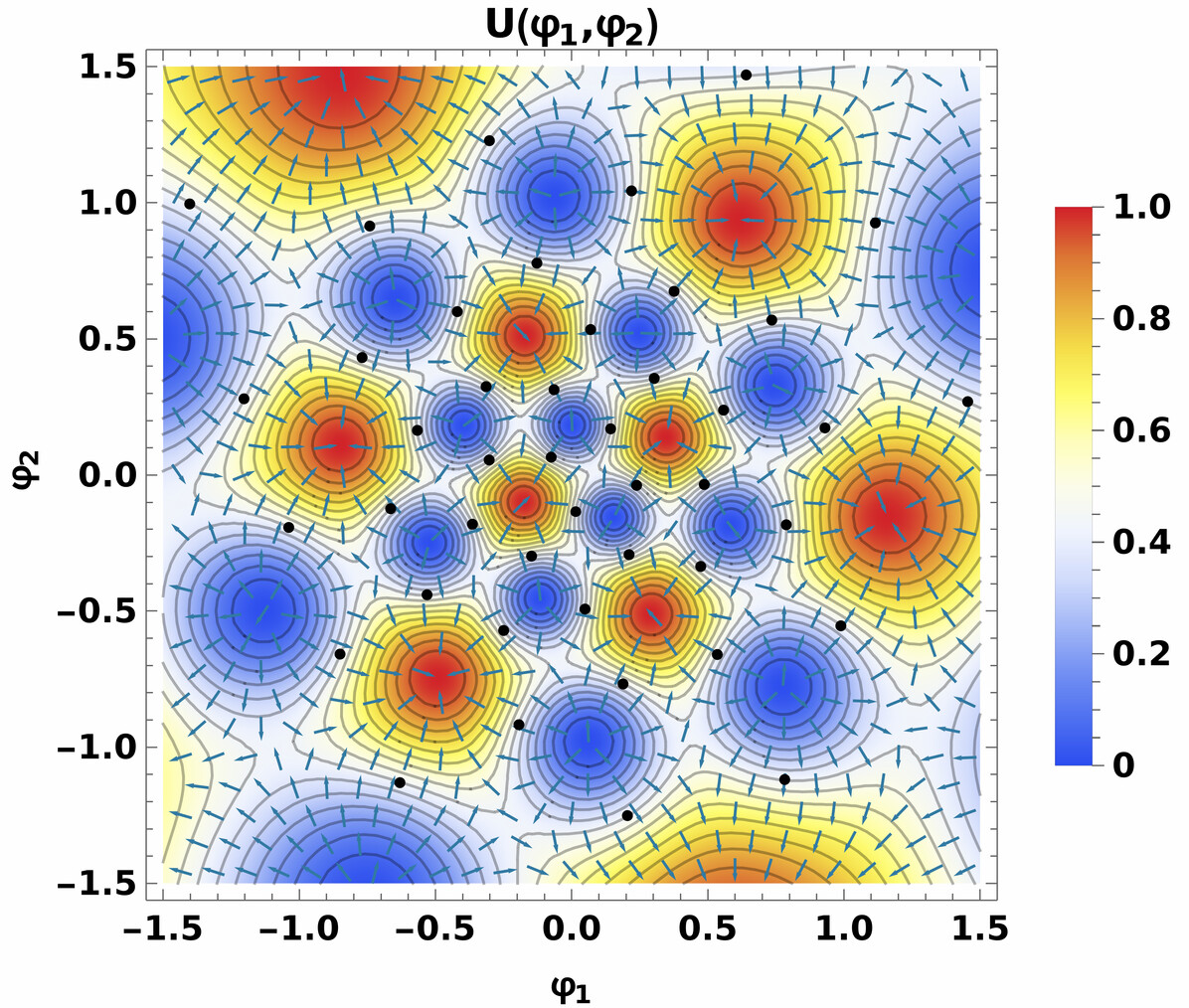}
        \caption{}
        \label{U2d_dode_maxPot}
    \end{subfigure}
    \par\vspace{0.15cm}
    \begin{subfigure}{0.49\textwidth}
        \centering
        \includegraphics[width=\textwidth]{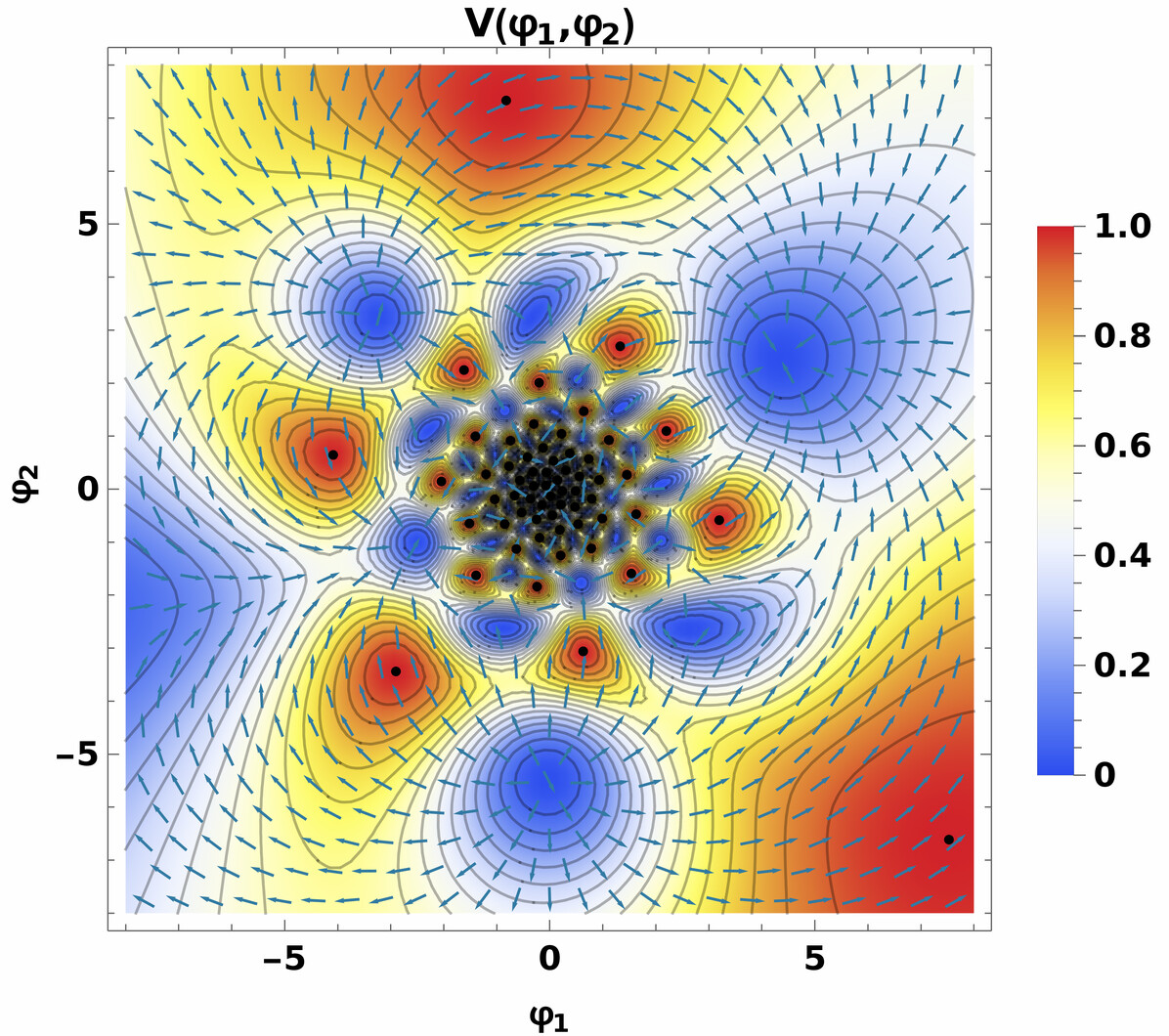}
        \caption{}
    \end{subfigure}
    \hfill
    \begin{subfigure}{0.49\textwidth}
        \centering
        \includegraphics[width=\textwidth]{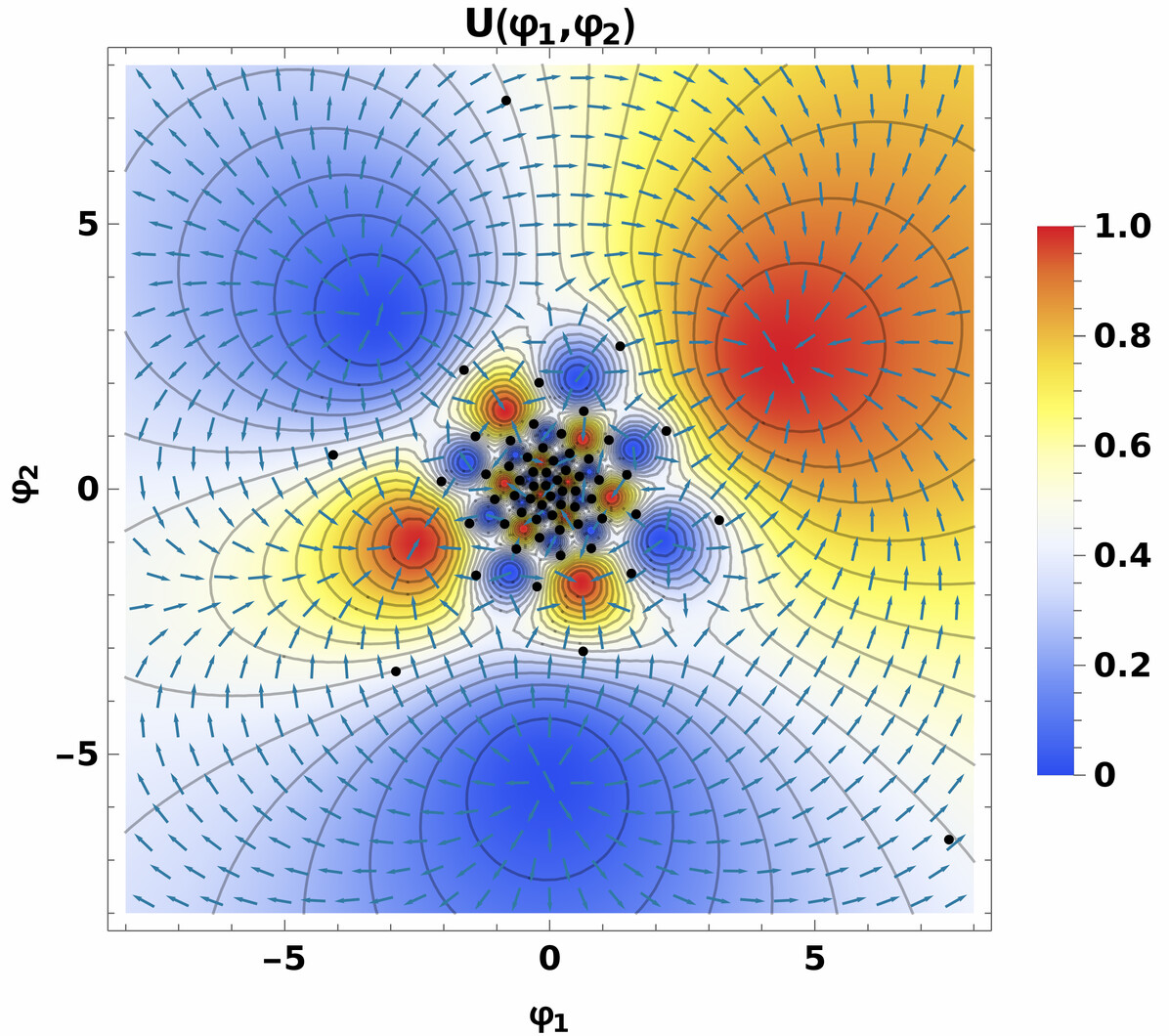}
        \caption{}
    \end{subfigure}
    \caption{Dodecahedral (a,c) potential $V$ and (b,d) pre-potential
    $U_D$ as functions of $(\vp_1,\vp_2)$, at two scales. Black circles
    mark the maxima of $V$, all of which have the same value.}
    \label{fig:maxPot_dode}
    
\end{figure}

Each potential has positive maxima that are distinct from the sites of
the regular lattice. By \eqref{eq:potential}, $V$ vanishes at every critical
point of $U$ and is one half of the squared norm of its gradient with
respect to the target-space metric $\eta_{ab}$. The height of these
maxima therefore measures the variation of $U$ with respect to spherical
distance. Their positions are determined numerically and are marked by
black circles in Figures~\ref{fig:maxPot_tetra}, \ref{fig:maxPot_octa},
\ref{fig:maxPot_cube}, \ref{fig:maxPot_icosa}, and \ref{fig:maxPot_dode}.

The invariance of $U$ under the proper rotation group $G$ of the vertex
set, together with the rotational invariance of the target-space metric,
implies that $V$ is constant on every orbit of $G$. Here the group acts
in the orientation used for each model, including the applied rotation
of the vertices. The maxima identified numerically in each model belong
to a single orbit. Since they lie away from the rotation axes defining
the regular lattice, no nonidentity element of $G$ fixes an individual
maximum. Each such orbit therefore contains $|G|$ distinct points:
twelve for the tetrahedron, twenty-four for the octahedron and the cube,
and sixty for the icosahedron and the dodecahedron, in accordance with
Table~\ref{tab:platonic_groups}. Symmetry accounts for the equality of
the potential values within each of these sets.

Stereographic projection changes the planar appearance of this
structure while preserving the symmetry of the model. Rotations that
fix the projection pole act as ordinary rotations in the plane; the
remaining elements of $G$ generally induce nonlinear transformations
of the planar coordinates. The maxima can therefore appear unevenly
spaced in the planar plots even though they are related by rotations
on the sphere.

A maximum close to the north pole is projected far from the planar
origin. In the octahedral, cubic, and icosahedral models, one member of
the orbit lies outside the window in which the individual maxima are
marked. The same projection effect produces the extended features seen
at large $|\vp|$ in the octahedral and cubic potentials. These features
reflect the coordinate distortion near the projection pole.
Appendix~\ref{app:maxV} lists the tabulated planar coordinates and the
corresponding values of $V$.

\subsection{Tetrahedron}
We solve the self-duality equations for the tetrahedral pre-potential
\eqref{eq:U_tetra} projected onto the $(\vp_1,\vp_2)$ plane. In these
coordinates the minima of $U$ are
\begin{equation}
    \qty(\vp_1^{(vac)},\vp_2^{(vac)}) = \dfrac{1}{2}
    \qty{\qty(\pm(\sqrt{3}+1),\pm(\sqrt{3}+1)),
    \qty(\pm(\sqrt{3}-1),\mp(\sqrt{3}-1))},
    \label{eq:minima_tetra}
\end{equation}
the maxima are
\begin{equation}
    \qty(\vp_1^{(vac)},\vp_2^{(vac)}) = \dfrac{1}{2}
    \qty{\qty(\pm(\sqrt{3}-1),\pm(\sqrt{3}-1)),
    \qty(\pm(\sqrt{3}+1),\mp(\sqrt{3}+1))},
    \label{eq:maxima_tetra}
\end{equation}
and the saddle points are
\begin{equation}
    \qty(\vp_1^{(vac)},\vp_2^{(vac)}) = \qty{(\pm 1,0),(0,\pm 1),(0,0)},
    \label{eq:saddle_tetra}
\end{equation}
the sixth saddle being the one at the north pole, which
\eqref{eq:change_coordinates} sends to infinity. The values of $U$ on
the three orbits are given in \eqref{eq:valores_U_tetra}. For the
remaining four models we omit the corresponding planar expressions,
which are obtained in the same way.
\begin{figure}[htb]
    \centering
    \begin{subfigure}{0.52\textwidth}
        \centering
        \includegraphics[height=6cm]{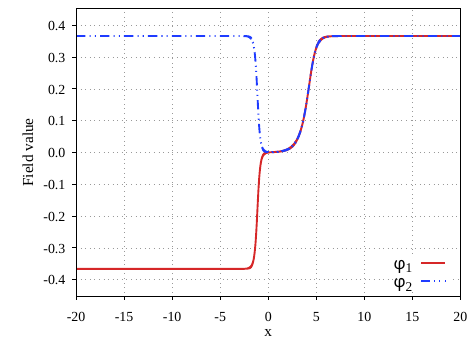}
        \caption{}
        \label{(0,0)_tetra_campos}
    \end{subfigure}
    \hfill
    \begin{subfigure}{0.47\textwidth}
        \centering
        \includegraphics[height=6cm]{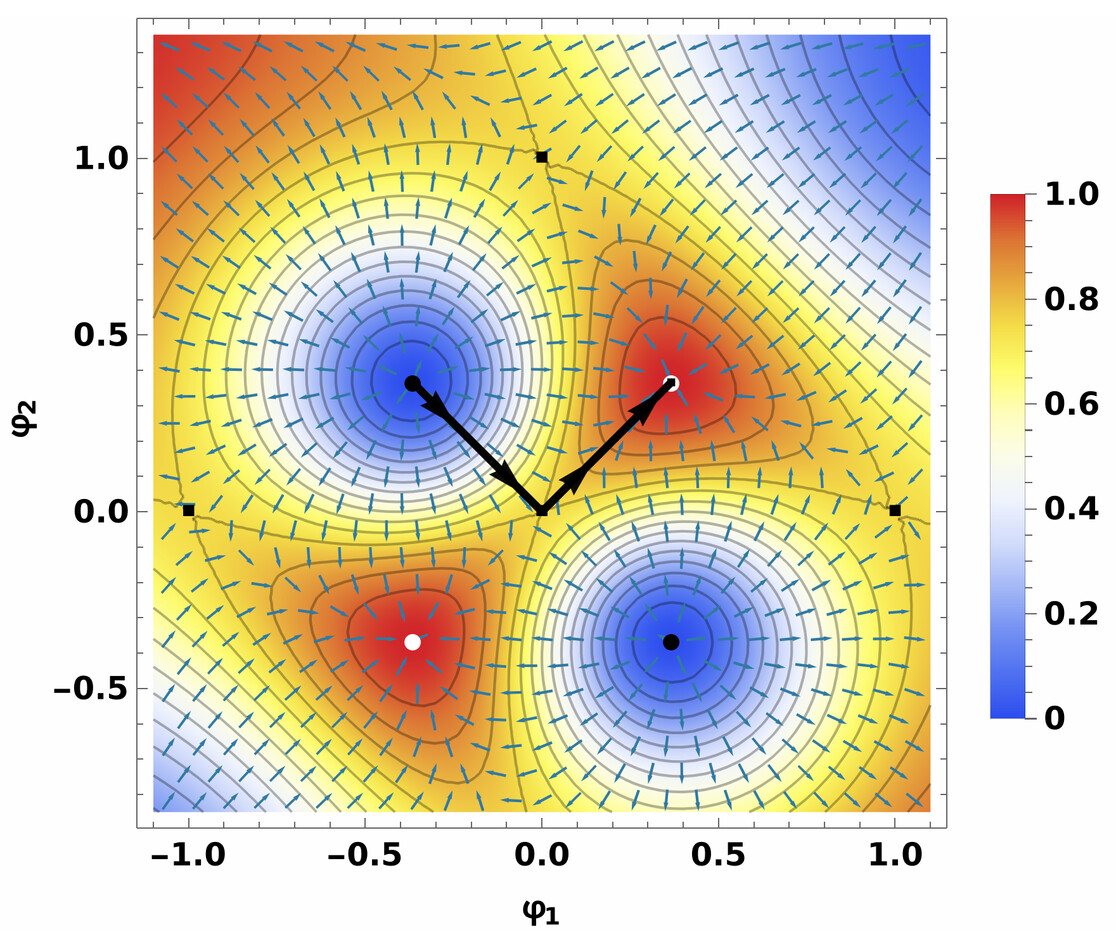}
        \caption{}
        \label{(0,0)_tetra_path}
    \end{subfigure}
    \caption{Solution with $Q=4/3$ from the saddle
    $\vp^{(vac)}=(0,0)$ displaced by $(0,10^{-3})$: (a) the fields
    $(\vp_1,\vp_2)$ as functions of $x$ and (b) the path they follow on the
    pre-potential $U_T$.}
    \label{fig:field_path_tetra_(0,0)}
   
\end{figure}

\begin{figure}[htb]
    \centering
    \begin{subfigure}{0.52\textwidth}
        \centering
        \includegraphics[height=6cm]{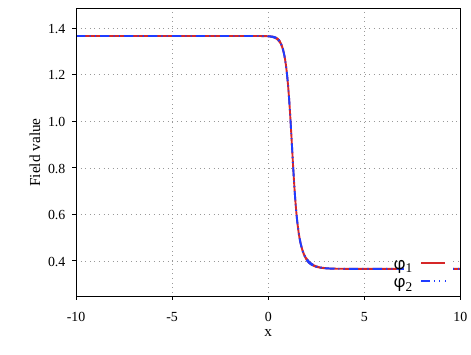}
        \caption{}
        \label{(s3+1,s3+1)_tetra_campos}
    \end{subfigure}
    \hfill
    \begin{subfigure}{0.47\textwidth}
        \centering
        \includegraphics[height=6cm]{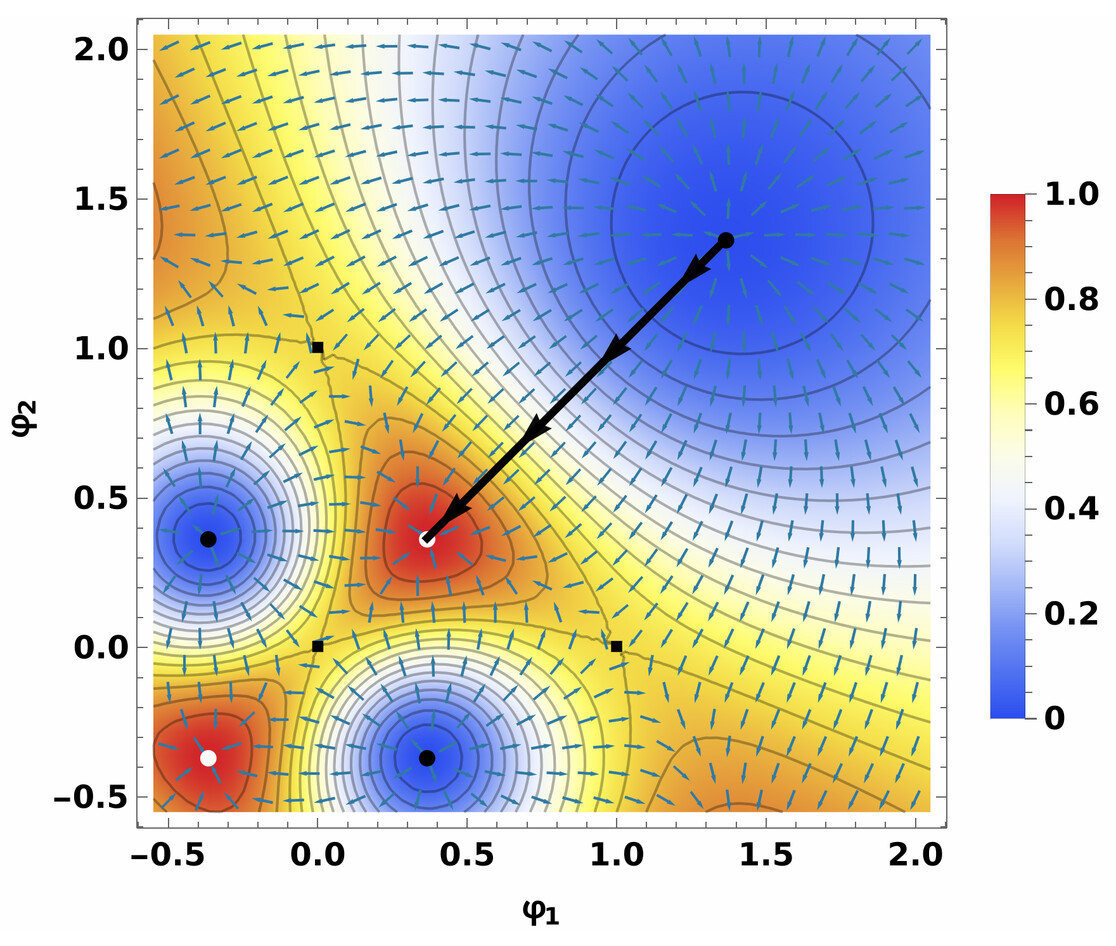}
        \caption{}
        \label{(s3+1,s3+1)_tetra_path}
    \end{subfigure}
    \caption{Solution with $Q=4/3$ from the minimum
    $\vp^{(vac)}=\qty((\sqrt{3}+1)/2,(\sqrt{3}+1)/2)$ displaced by
    $(-10^{-3},-10^{-3})$: (a) the fields as functions of $x$ and (b) their
    path on $U_T$.}
    \label{fig:(s3+1,s3+1)_field_path_tetra}
    
\end{figure}

Two solutions with the same charge $Q=4/3$ are shown in
Figures~\ref{fig:field_path_tetra_(0,0)}
and~\ref{fig:(s3+1,s3+1)_field_path_tetra}, both interpolating between a
minimum and a maximum of $U_T$. The first connects
\begin{equation}
    \qty(\vp_1,\vp_2) :
    \frac{1}{2}\qty(-(\sqrt{3}-1),(\sqrt{3}-1))
    \qquad\longrightarrow\qquad
    \frac{1}{2}\qty((\sqrt{3}-1),(\sqrt{3}-1)),
\end{equation}
passing close to the saddle point at the origin. The second starts
near the minimum
$\qty((\sqrt{3}+1)/2,(\sqrt{3}+1)/2)$ and reaches the same maximum
without approaching an intermediate saddle. The paths in
Figures~\ref{(0,0)_tetra_path} and
\ref{(s3+1,s3+1)_tetra_path} follow increasing values of $U_T$,
as required by the positive-sign self-duality equations. Their initial
minima belong to the same symmetry orbit, so the two solutions have
the same charge by \eqref{eq:Q_1+1} despite their different endpoints
at $x\rightarrow-\infty$.
\begin{figure}[htb]
    \centering
    \begin{subfigure}{0.49\textwidth}
        \centering
        \includegraphics[width=\textwidth]{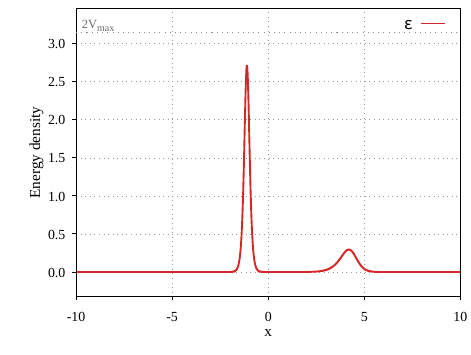}
        \caption{}
        \label{(0,0)_tetra_energia}
    \end{subfigure}
    \hfill
    \begin{subfigure}{0.49\textwidth}
        \centering
        \includegraphics[width=\textwidth]{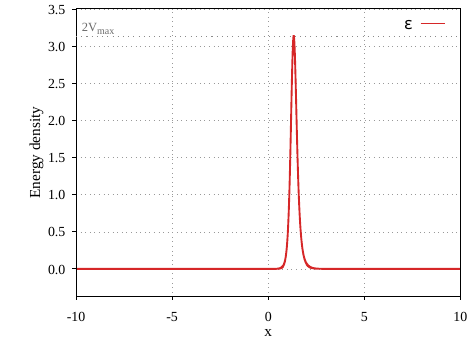}
        \caption{}
        \label{(s3+1,s3+1)_tetra_energia}
    \end{subfigure}
    \caption{Energy density of the two tetrahedral solutions with
    $Q=4/3$: (a) from $\vp^{(vac)}=(0,0)$ displaced by $(0,10^{-3})$ and
    (b) from $\vp^{(vac)}=\qty((\sqrt{3}+1)/2,(\sqrt{3}+1)/2)$ displaced by
    $(-10^{-3},-10^{-3})$.}
    \label{fig:tetra_energia}
   
\end{figure}

The two profiles in Figure~\ref{fig:tetra_energia} differ accordingly.
The first has two peaks separated by a region in which the fields
remain close to the saddle. The critical-value differences $1$ and
$1/3$ describe the limiting energy contributions of the two transitions
as this intermediate configuration approaches the saddle. The second
has a single localized peak. Both have total energy $E=4/3$.
The second solution runs along the symmetry line $\vp_1=\vp_2$ and
passes through a maximum of $V$, so it saturates the density bound,
$\max_x\mathcal{E}=2V_{max}\approx3.13857$.
More generally, applying a symmetry of the model to an existing BPS
solution produces another solution with the same charge and energy;
distinct profiles can also have equal energy when their endpoints
belong to the same pair of critical-point orbits.
\begin{figure}[htb]
    \centering
    \begin{subfigure}{0.49\textwidth}
        \centering
        \includegraphics[width=\textwidth]{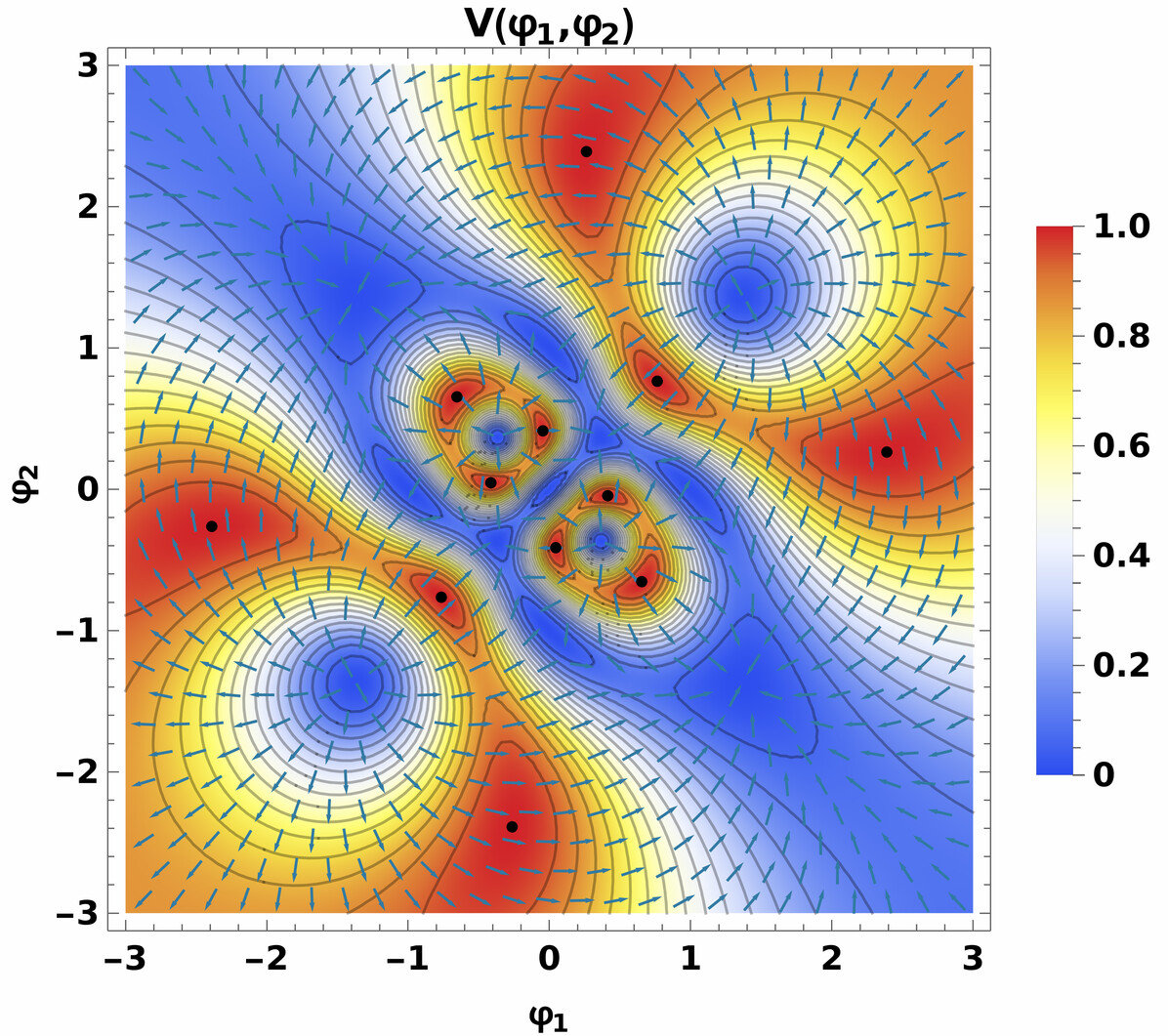}
        \caption{}
        \label{V2d_tetra_maxPot}
    \end{subfigure}
    \hfill
    \begin{subfigure}{0.49\textwidth}
        \centering
        \includegraphics[width=\textwidth]{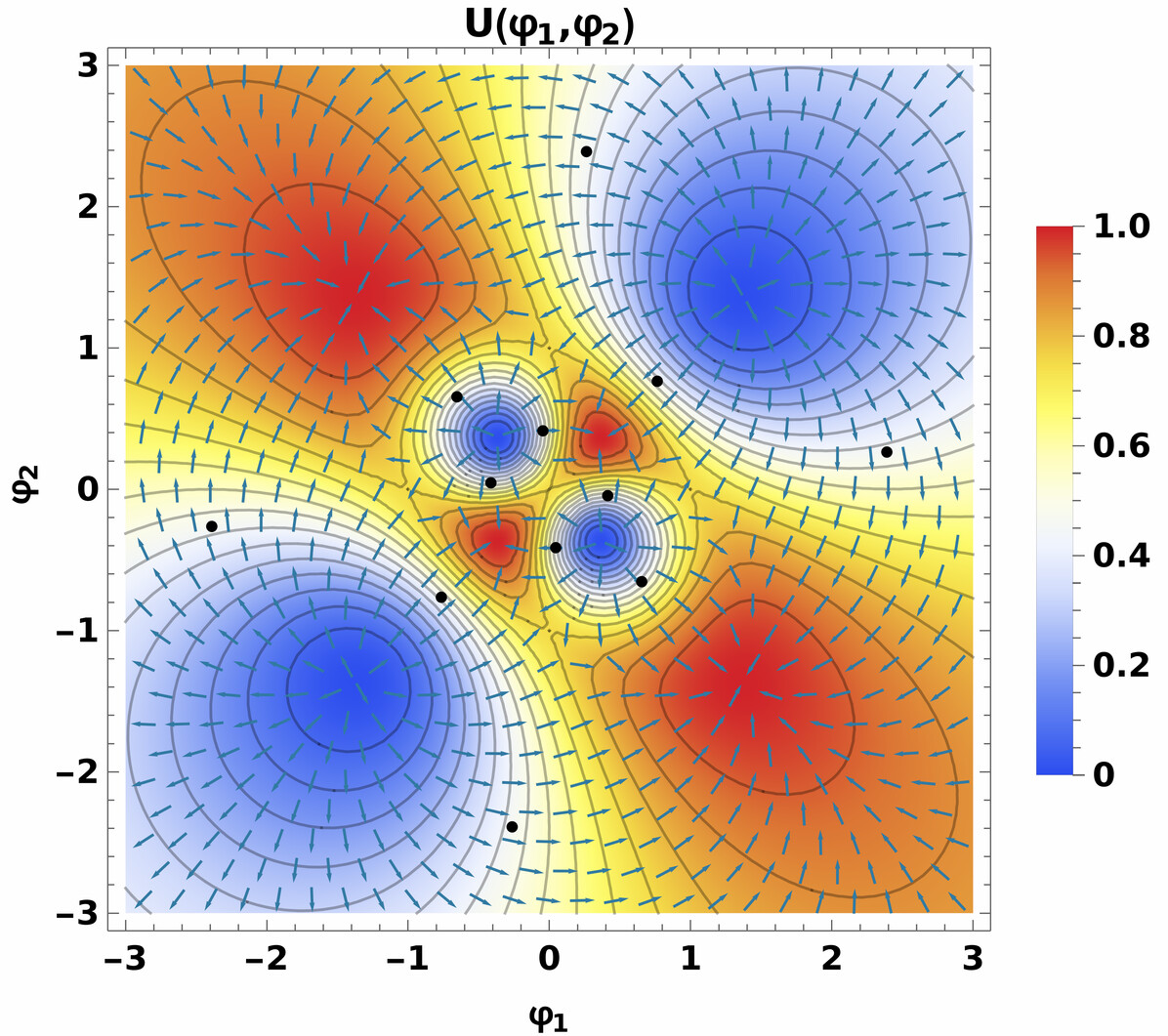}
        \caption{}
        \label{U2d_tetra_maxPot}
    \end{subfigure}
    \caption{Tetrahedral (a) potential $V$ and (b) pre-potential $U_T$
    as functions of $(\vp_1,\vp_2)$. Black circles mark the maxima of
    $V$, all of which have the same value.}
    \label{fig:maxPot_tetra}
   
\end{figure}

Figure~\ref{fig:maxPot_tetra} shows the maxima of $V$ for this model.
There are twelve of them, the order of the tetrahedral rotation group,
all with $V=1.56928$, and all lie within the plotted region. They are
invariant under inversion through the origin and form rings around the
minima of $U_T$.
\FloatBarrier
\subsection{Octahedron}
For the octahedral model we solve the self-duality equations for
\eqref{eq:U_octa} projected onto the plane.
\begin{figure}[htb]
    \centering
    \begin{subfigure}{0.49\textwidth}
        \centering
        \includegraphics[width=\textwidth]{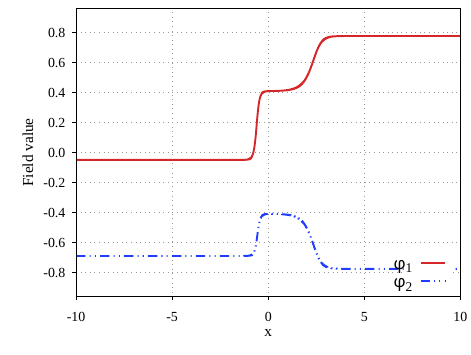}
        \caption{}
        \label{(sq6,sq6)_octa_campos}
    \end{subfigure}
    \hfill
    \begin{subfigure}{0.49\textwidth}
        \centering
        \includegraphics[width=\textwidth]{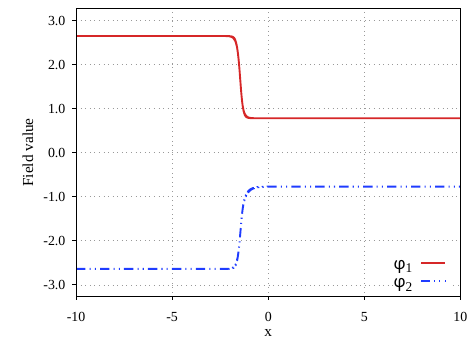}
        \caption{}
        \label{(4-sq6)_octa_campos}
    \end{subfigure}
    \caption{Fields $(\vp_1,\vp_2)$ as functions of $x$ with $Q=1.548$,
    both displaced by $(0,-10^{-3})$, from (a) the saddle
    $\vp^{(vac)}=\qty(\dfrac{1}{\sqrt{6}},\dfrac{-1}{\sqrt{6}})$
    and (b) the maximum
    $\vp^{(vac)}=\qty(\dfrac{4-\sqrt{6}}{2},\dfrac{-4+\sqrt{6}}{2})$.}
    \label{fig:field_campos_octa1}
   
\end{figure}
\begin{figure}[htb]
    \centering
    \begin{subfigure}{0.5\textwidth}
        \centering
        \includegraphics[width=\textwidth]{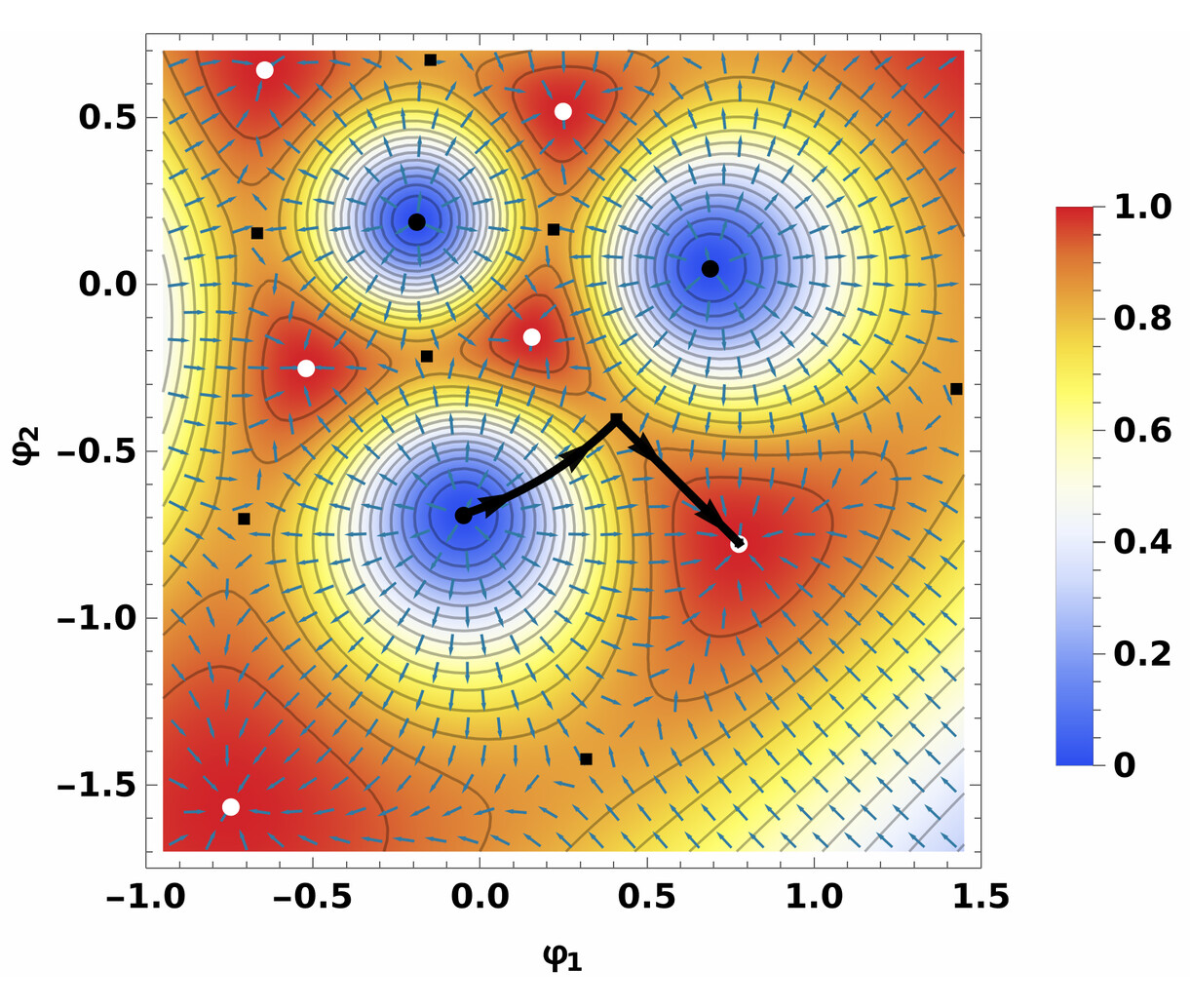}
        \caption{}
        \label{(sq6,sq6)_octa_path}
    \end{subfigure}
    \hfill
    \begin{subfigure}{0.49\textwidth}
        \centering
        \includegraphics[width=\textwidth]{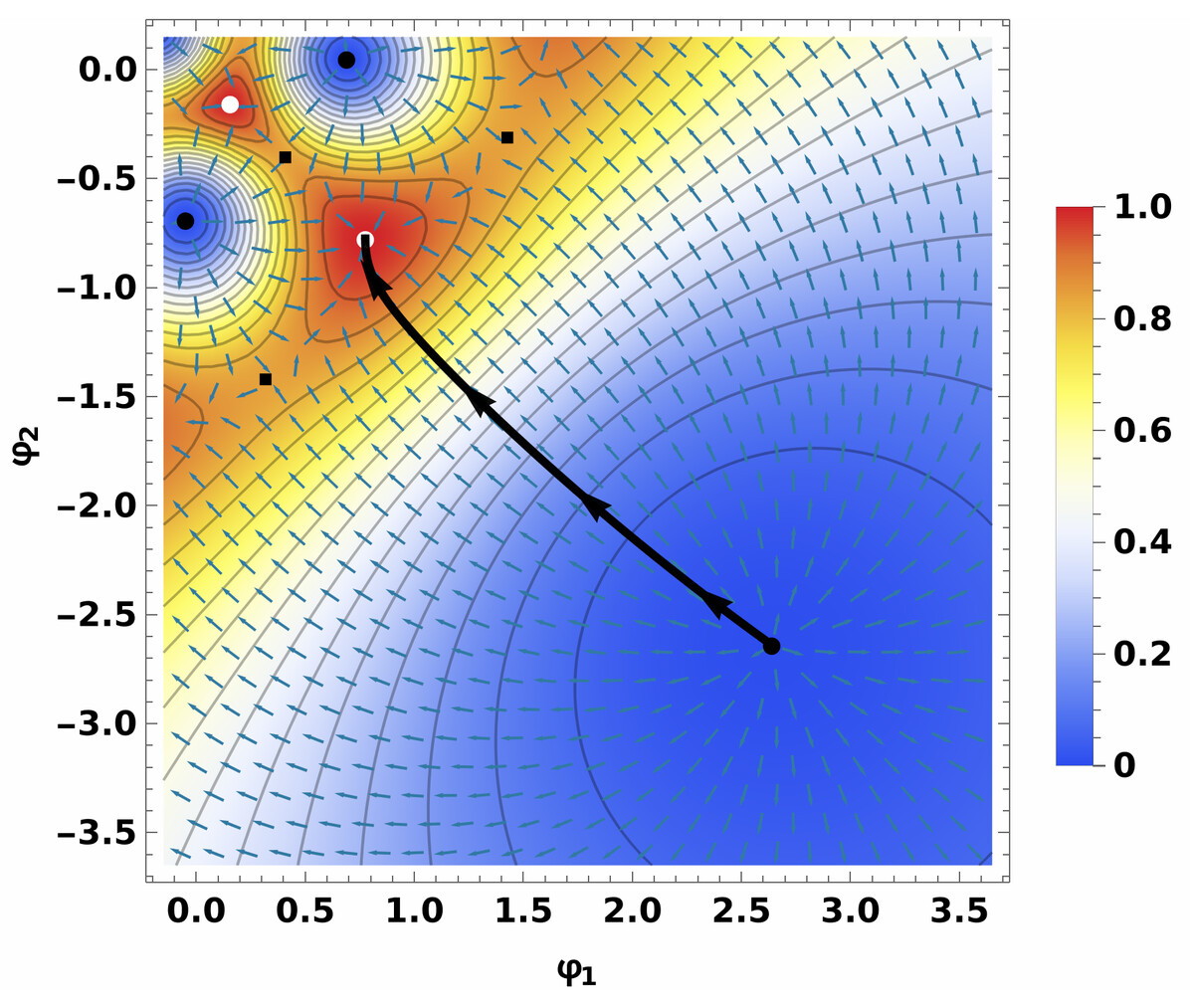}
        \caption{}
        \label{(4-sq6)_octa_path}
    \end{subfigure}
    \caption{Paths of the two octahedral solutions on $U_O$, for the
    same starting points as
    Figure~\ref{fig:field_campos_octa1}.}
    \label{fig:field_path_octa1}
   
\end{figure}
\begin{figure}[htb]
    \centering
    \begin{subfigure}{0.49\textwidth}
        \centering
        \includegraphics[width=\textwidth]{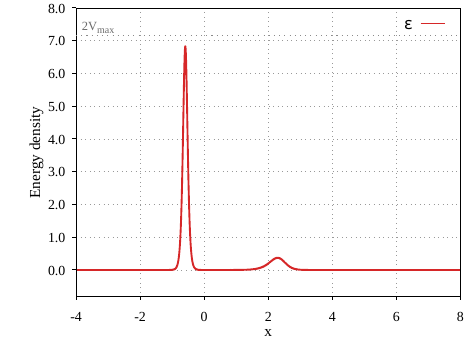}
        \caption{}
        \label{(sq6,sq6)_octa_energia}
    \end{subfigure}
    \hfill
    \begin{subfigure}{0.49\textwidth}
        \centering
        \includegraphics[width=\textwidth]{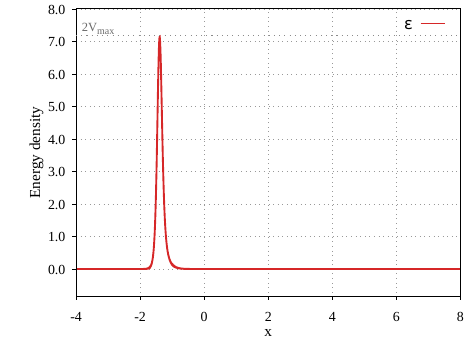}
        \caption{}
        \label{(4-sq6)_octa_energia}
    \end{subfigure}
    \caption{Energy density of the two octahedral solutions with
    $Q=1.548$, for the same starting points as
    Figure~\ref{fig:field_campos_octa1}.}
    \label{fig:field_energia_octa1}
   
\end{figure}

Figure~\ref{fig:field_campos_octa1} shows two configurations with the
same charge $Q\approx1.548$. The first connects
\begin{equation}
    \qty(\vp_1,\vp_2) :
    \qty(-\frac{2-\sqrt{3}}{4+\sqrt{2}},
    -\frac{2+\sqrt{3}}{4+\sqrt{2}})
    \qquad\longrightarrow\qquad
    \qty(\frac{4-\sqrt{6}}{2},\frac{-4+\sqrt{6}}{2}),
\end{equation}
passing close to the saddle at
$\qty(1/\sqrt{6},-1/\sqrt{6})$. The differences
$U_{saddle}-U_{min}\approx1.306$ and
$U_{max}-U_{saddle}\approx0.242$ provide the limiting energy
contributions when the profile separates into two transitions near
this saddle. The second reaches the same maximum from a different
minimum of the same orbit,
\begin{equation}
    \qty(\vp_1,\vp_2) :
    \qty(\frac{2+\sqrt{3}}{\sqrt{2}},
    -\frac{2+\sqrt{3}}{\sqrt{2}})
    \qquad\longrightarrow\qquad
    \qty(\frac{4-\sqrt{6}}{2},\frac{-4+\sqrt{6}}{2}),
    \label{path_octa2}
\end{equation}
without approaching an intermediate saddle. Their energy densities in
Figure~\ref{fig:field_energia_octa1} show two peaks and a single peak,
respectively, while both have total energy $E\approx1.548$.
The paths in Figure~\ref{fig:field_path_octa1} illustrate how different
trajectories can carry the same charge because $U$ has the same value
at minima belonging to the same orbit.
\begin{figure}[htb]
    \centering
    \begin{subfigure}{0.45\textwidth}
        \centering
        \includegraphics[width=\textwidth]{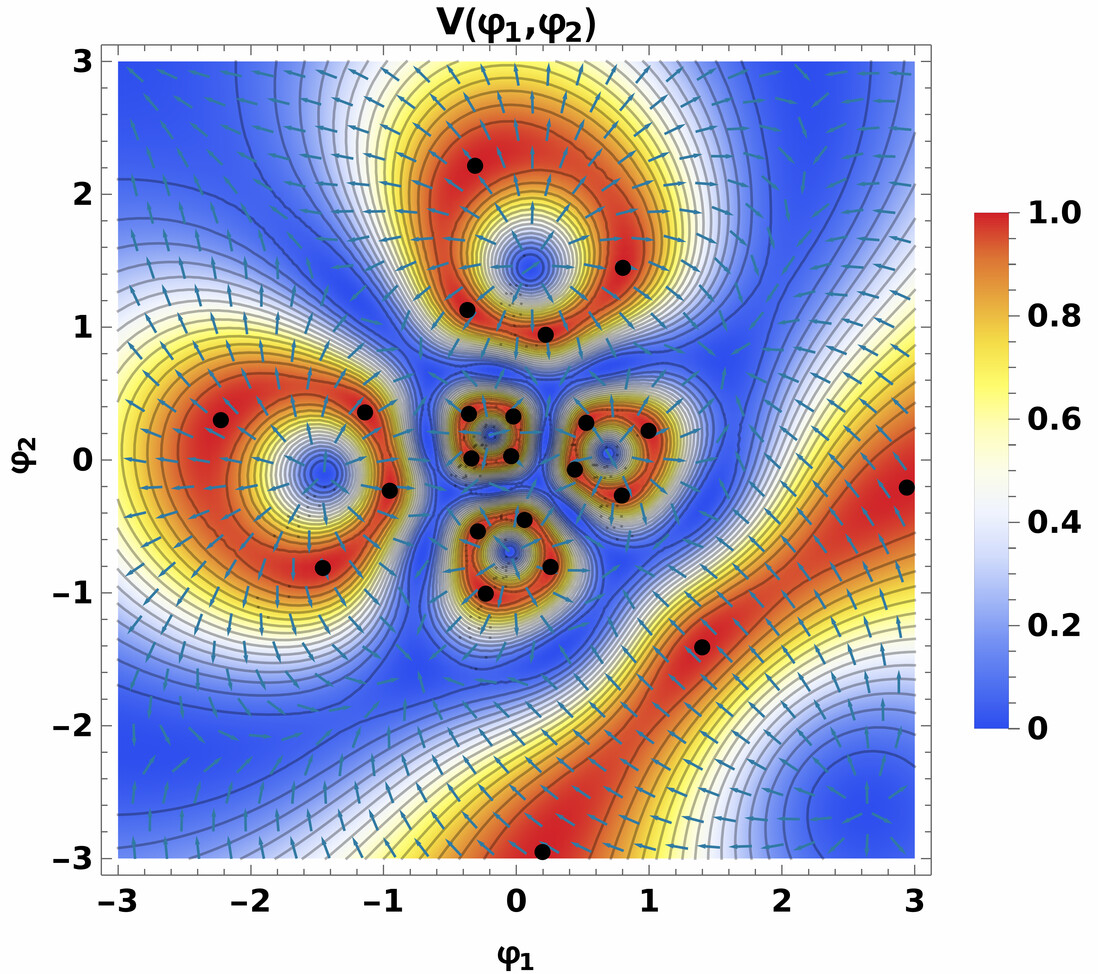}
        \caption{}
        \label{V2d_octa_maxPot}
    \end{subfigure}
    \hfill
    \begin{subfigure}{0.45\textwidth}
        \centering
        \includegraphics[width=\textwidth]{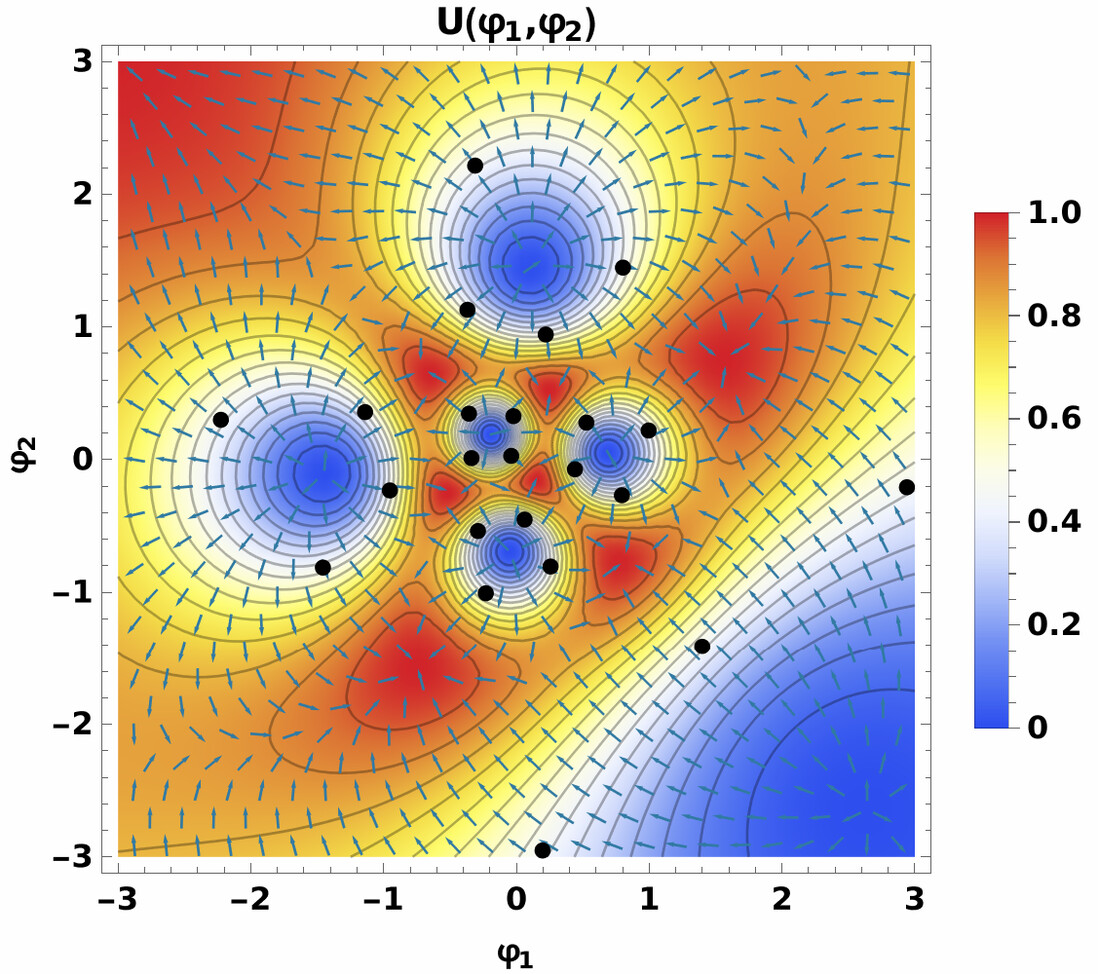}
        \caption{}
        \label{U2d_octa_maxPot}
    \end{subfigure}
    \par\vspace{0.15cm}
    \begin{subfigure}{0.45\textwidth}
        \centering
        \includegraphics[width=\textwidth]{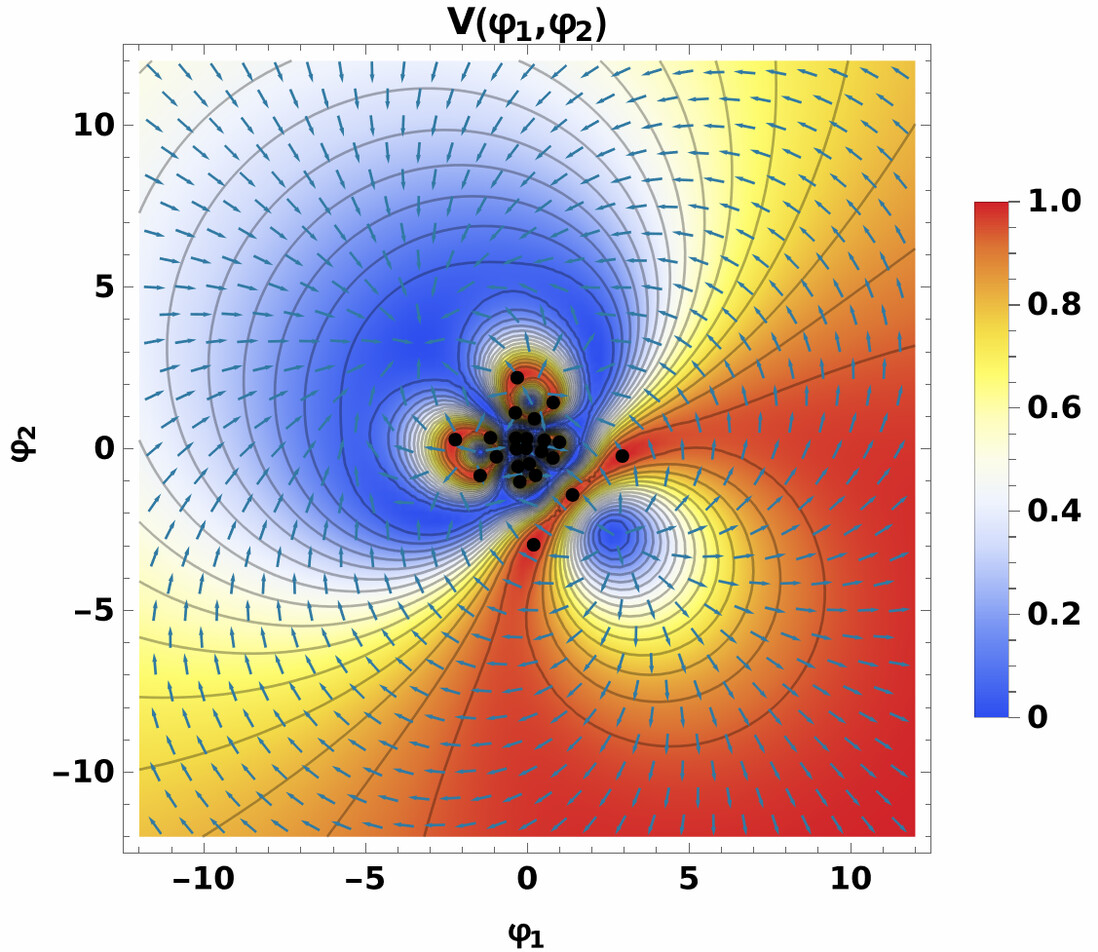}
        \caption{}
        \label{V2d_octa_maxPot_longe}
    \end{subfigure}
    \hfill
    \begin{subfigure}{0.45\textwidth}
        \centering
        \includegraphics[width=\textwidth]{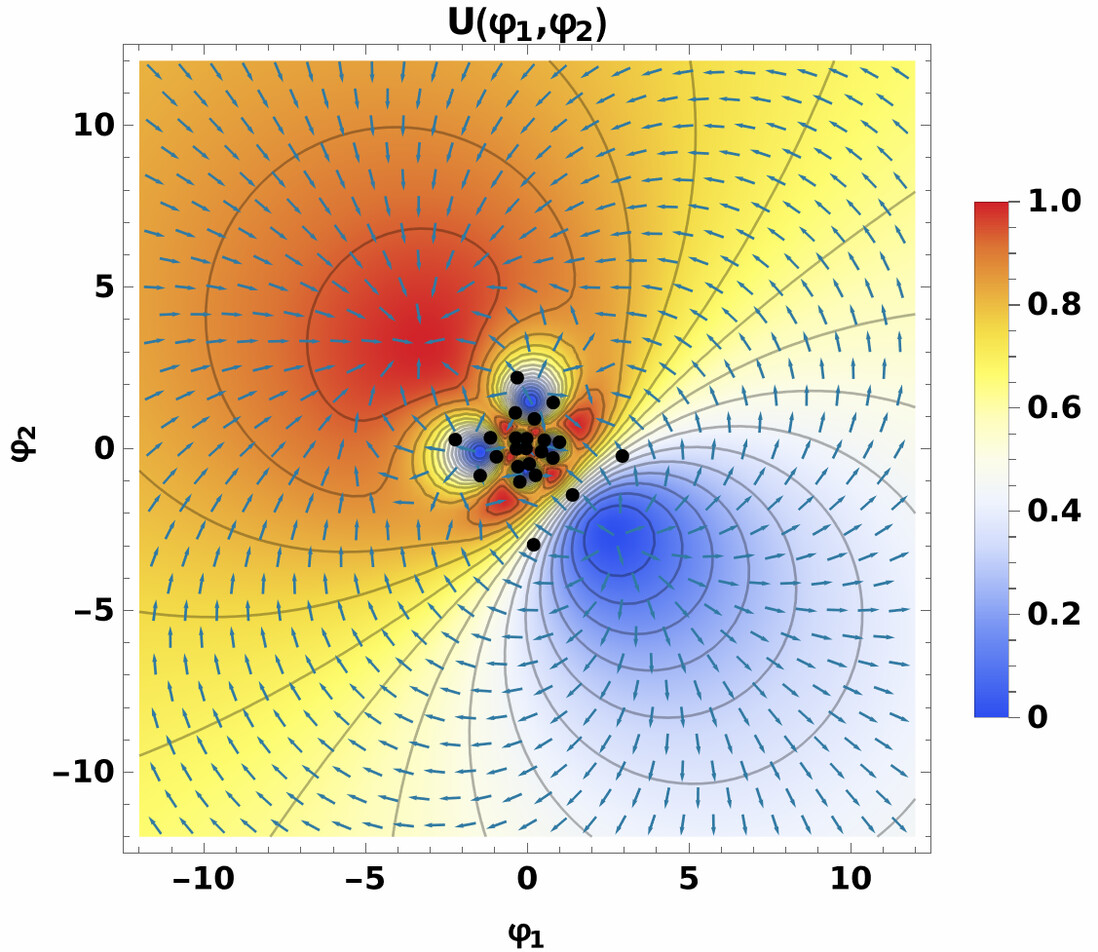}
        \caption{}
    \end{subfigure}
    \caption{Octahedral (a,c) potential $V$ and (b,d) pre-potential
    $U_O$ as functions of $(\vp_1,\vp_2)$, at two scales. Black circles
    mark the maxima of $V$, all of which have the same value.}
    \label{fig:maxPot_octa}
    
\end{figure}

The octahedral potential has twenty-four maxima, all with
$V\approx3.58693$. One lies beyond the window in which individual
maxima are marked in Figure~\ref{fig:maxPot_octa}, so twenty-three
are listed in Appendix~\ref{app:maxV}. The extended features in
Figures~\ref{V2d_octa_maxPot_longe} and \ref{V_3d_octa_longe}
reflect the stereographic distortion near the projection pole.
The planar pattern is invariant under reflection about the
antidiagonal $\vp_2=-\vp_1$,
$(\vp_1,\vp_2)\mapsto(-\vp_2,-\vp_1)$, consistently with the symmetry
identified in the construction of $U_O$.
\FloatBarrier
\subsection{Cube}
For the cubic model we solve the self-duality equations for
\eqref{eq:U_cube} projected onto the plane. Its regular lattice
coincides with that of the octahedron, with minima and maxima
interchanged, as summarized in Table~\ref{tab:octa_cube_lattice}.
\begin{figure}[htb]
    \centering
    \begin{subfigure}{0.49\textwidth}
        \centering
        \includegraphics[width=\textwidth]{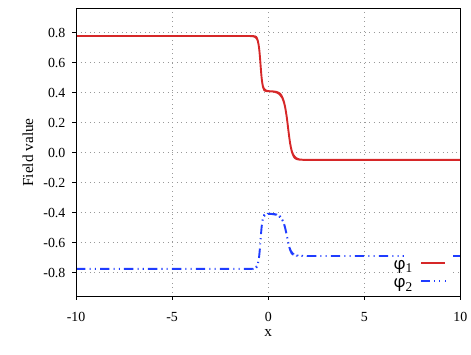}
        \caption{}
        \label{(sq6,sq6)_cube_campos}
    \end{subfigure}
    \hfill
    \begin{subfigure}{0.49\textwidth}
        \centering
        \includegraphics[width=\textwidth]{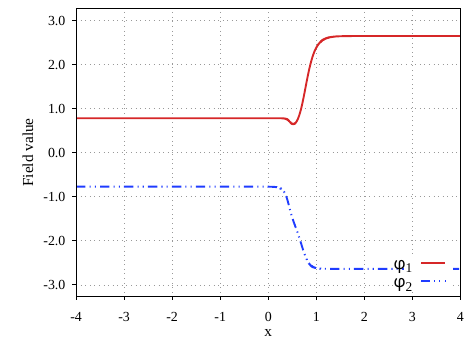}
        \caption{}
        \label{(4-sq6)_cube_campos}
    \end{subfigure}
    \caption{Fields $(\vp_1,\vp_2)$ as functions of $x$ with $Q=2.023$,
    both displaced by $(0,-10^{-3})$, from (a) the saddle
    $\vp^{(vac)}=\qty(\dfrac{1}{\sqrt{6}},\dfrac{-1}{\sqrt{6}})$
    and (b) the minimum
    $\vp^{(vac)}=\qty(\dfrac{4-\sqrt{6}}{2},\dfrac{-4+\sqrt{6}}{2})$.}
    \label{fig:field_campos_cube1}
   
\end{figure}
\begin{figure}[htb]
    \centering
    \begin{subfigure}{0.5\textwidth}
        \centering
        \includegraphics[width=\textwidth]{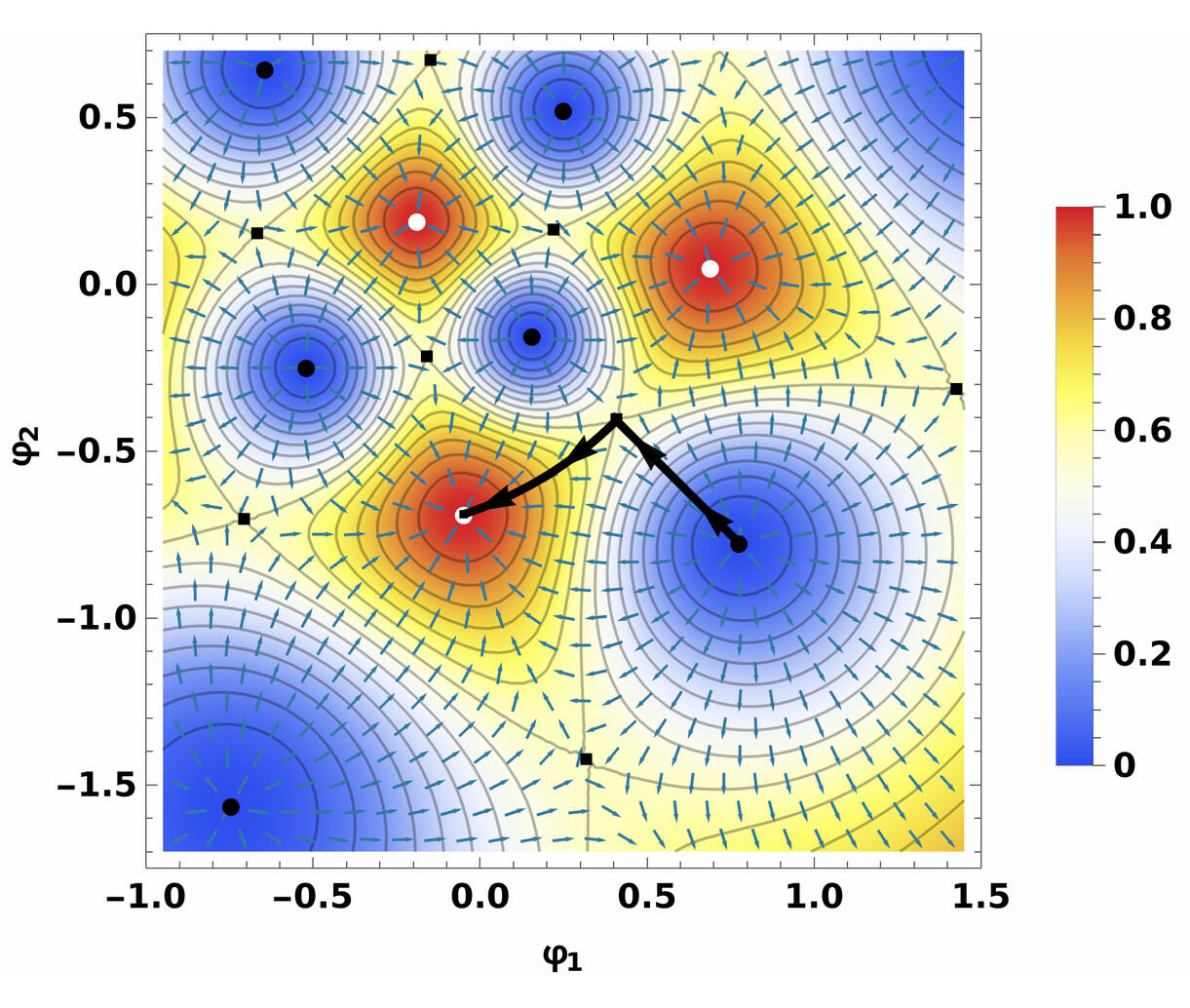}
        \caption{}
        \label{(sq6,sq6)_cube_path}
    \end{subfigure}
    \hfill
    \begin{subfigure}{0.485\textwidth}
        \centering
        \includegraphics[width=\textwidth]{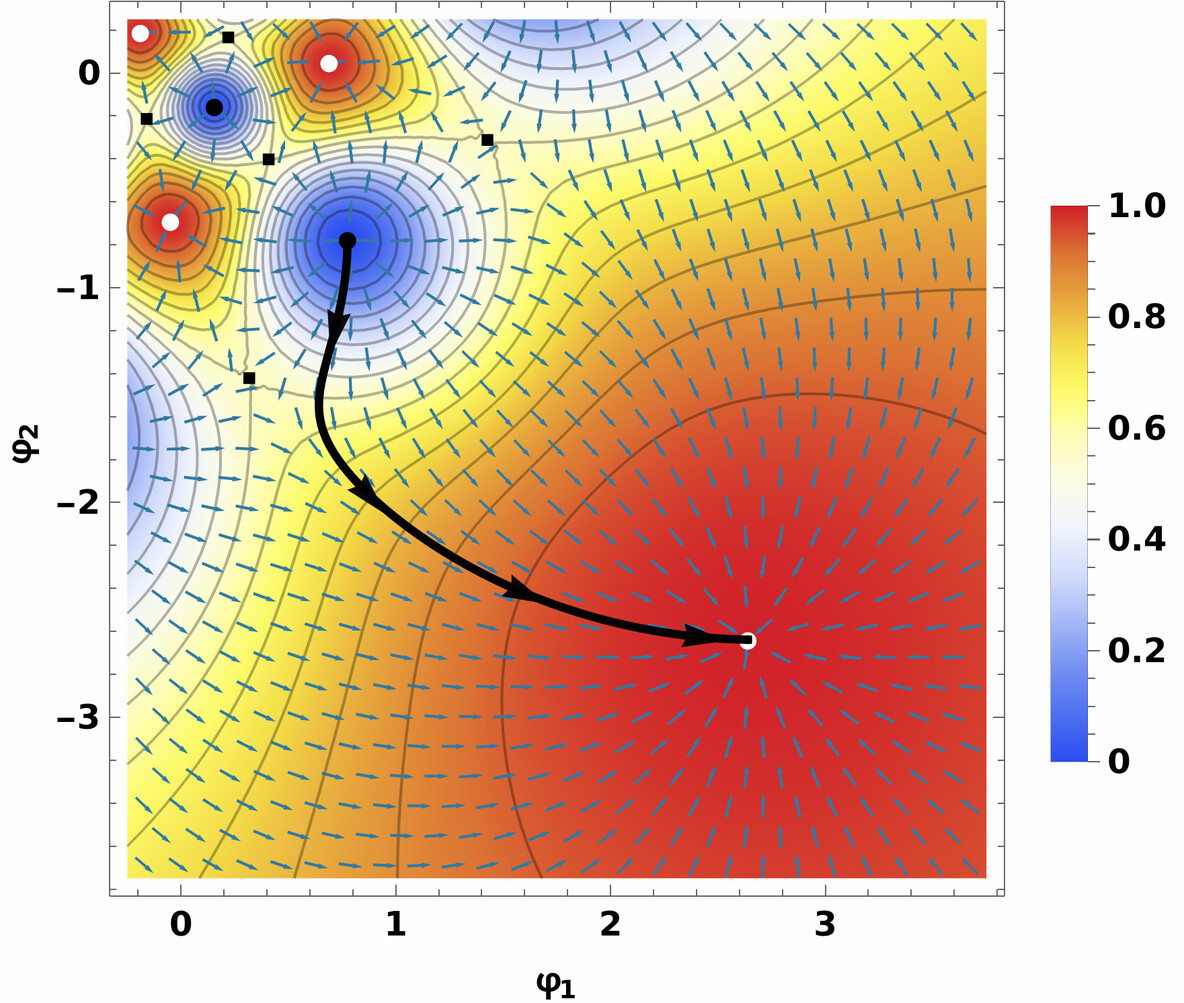}
        \caption{}
        \label{(4-sq6)_cube_path}
    \end{subfigure}
    \caption{Paths of the two cubic solutions on $U_C$, for the same
    starting points as Figure~\ref{fig:field_campos_cube1}.}
    \label{fig:field_path_cube1}
   
\end{figure}
\begin{figure}[htb]
    \centering
    \begin{subfigure}{0.49\textwidth}
        \centering
        \includegraphics[width=\textwidth]{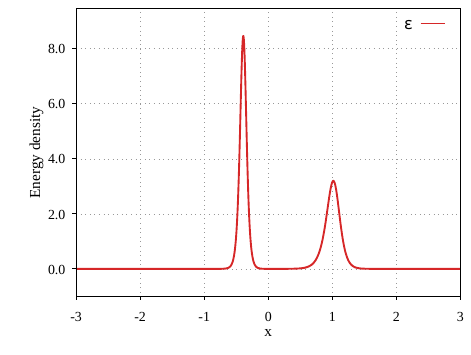}
        \caption{}
        \label{(sq6,sq6)_cube_energia}
    \end{subfigure}
    \hfill
    \begin{subfigure}{0.49\textwidth}
        \centering
        \includegraphics[width=\textwidth]{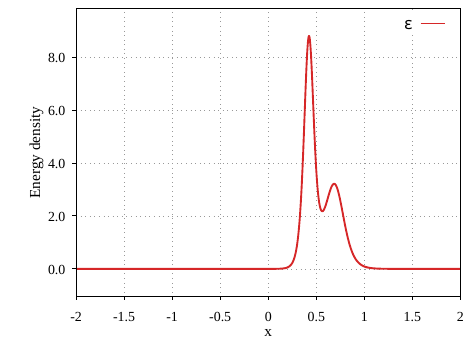}
        \caption{}
        \label{(4-sq6)_cube_energia}
    \end{subfigure}
    \caption{Energy density of the two cubic solutions with $Q=2.023$,
    for the same starting points as
    Figure~\ref{fig:field_campos_cube1}.}
    \label{fig:field_energia_cube1}
   
\end{figure}

Figure~\ref{fig:field_campos_cube1} shows two configurations with
$Q\approx2.023$. The first connects
\begin{equation}
    \qty(\vp_1,\vp_2) :
    \qty(\frac{4-\sqrt{6}}{2},\frac{-4+\sqrt{6}}{2})
    \qquad\longrightarrow\qquad
    \qty(-\frac{2-\sqrt{3}}{4+\sqrt{2}},
    -\frac{2+\sqrt{3}}{4+\sqrt{2}}),
    \label{path_cube1}
\end{equation}
passing close to the saddle at $\qty(1/\sqrt{6},-1/\sqrt{6})$.
The differences $U_{saddle}-U_{min}\approx1.138$ and
$U_{max}-U_{saddle}\approx0.885$ characterize the limiting integrated
energies of transitions separated near this saddle. The unequal peak
heights in Figure~\ref{(sq6,sq6)_cube_energia} instead follow from
the values of $V$ sampled along the trajectory through
\eqref{eq:dens_2V}. The second solution leaves the same minimum and
reaches a different maximum of the same orbit,
\begin{equation}
    \qty(\vp_1,\vp_2) :
    \qty(\frac{4-\sqrt{6}}{2},\frac{-4+\sqrt{6}}{2})
    \qquad\longrightarrow\qquad
    \qty(\frac{2+\sqrt{3}}{\sqrt{2}},
    -\frac{2+\sqrt{3}}{\sqrt{2}}),
    \label{path_cube2}
\end{equation}
without passing close to a saddle. Its endpoints are those of the
octahedral connection \eqref{path_octa2}, with their roles exchanged.
The smaller secondary peak in Figure~\ref{(4-sq6)_cube_energia}
corresponds to a second local maximum of $V$ along the path in
Figure~\ref{(4-sq6)_cube_path}, as follows from \eqref{eq:dens_2V}.
\begin{figure}[htb]
    \centering
    \begin{subfigure}{0.49\textwidth}
        \centering
        \includegraphics[width=\textwidth]{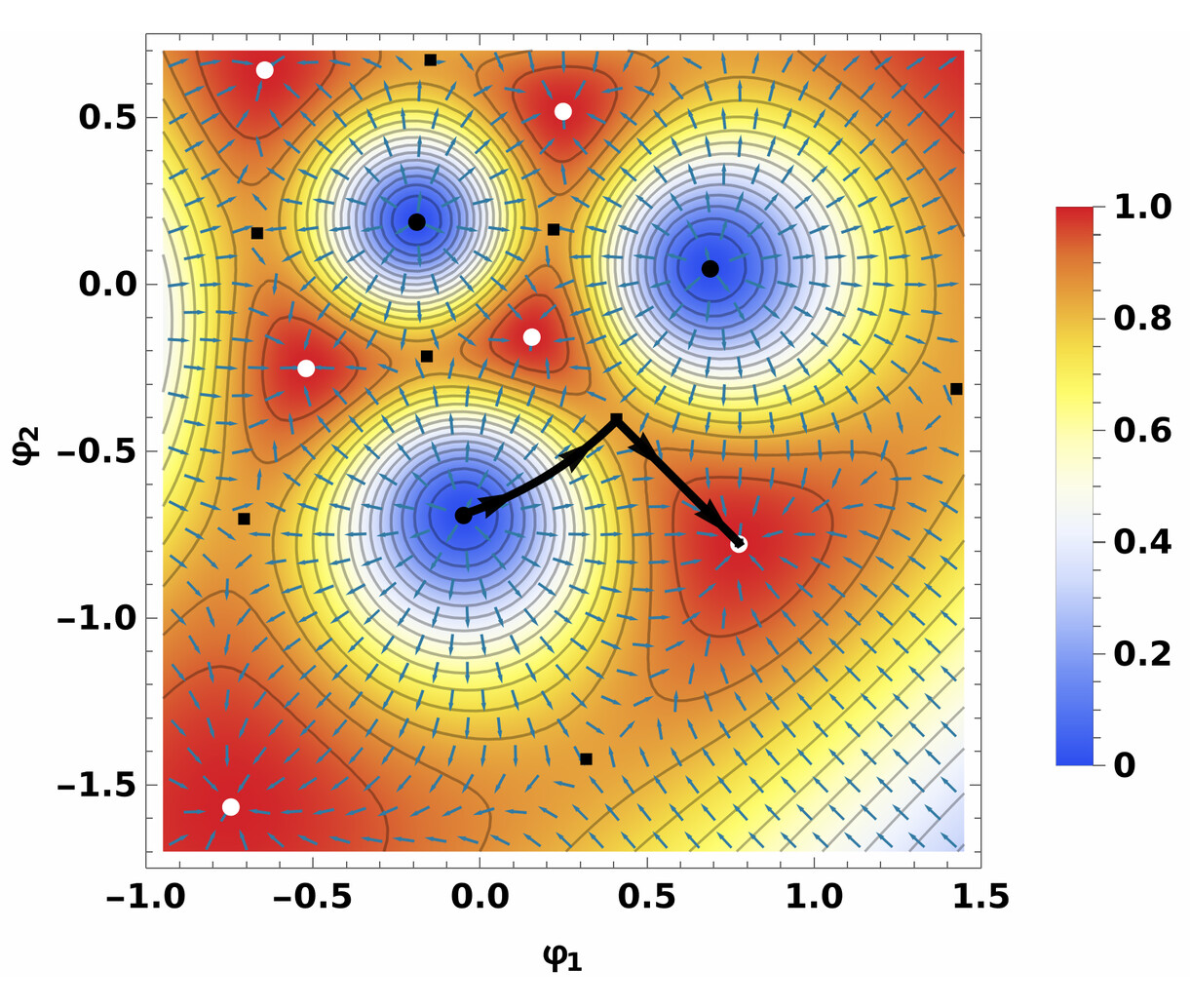}
        \caption{}
        \label{sq6_octa_path}
    \end{subfigure}
    \hfill
    \begin{subfigure}{0.49\textwidth}
        \centering
        \includegraphics[width=\textwidth]{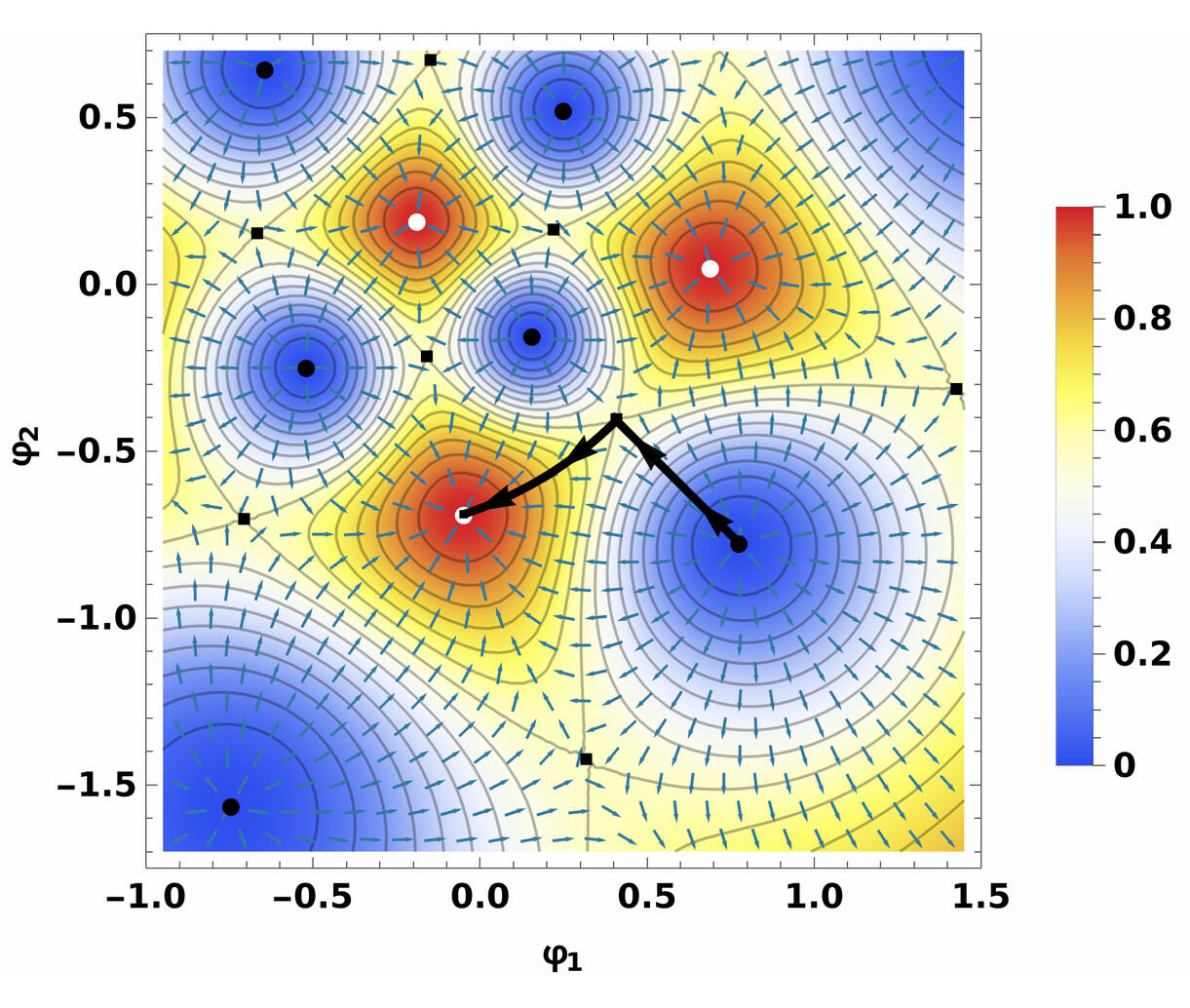}
        \caption{}
        \label{sq6_cube_path}
    \end{subfigure}
    \caption{Paths taken on the pre-potential of the (a) octahedral and
    (b) cubic models, using the same initial configuration obtained from
    the saddle $\vp^{(vac)}=\qty(1/\sqrt{6},-1/\sqrt{6})$
    displaced by $(0,-10^{-3})$.}
    \label{fig:comparison_octa_cube}
    
\end{figure}

Figure~\ref{fig:comparison_octa_cube} compares two examples whose
endpoints exchange their minimum and maximum roles in the dual
models. Their trajectories differ because the two pre-potentials
generate different metric-gradient fields. Their charges and energies
also differ, as shown in Table~\ref{tab:charges}. These examples
illustrate a correspondence between the displayed connections; the
shared vacuum set alone does not establish such a correspondence for
all solutions. In particular, the comparison involves two different
models, whereas reversing the BPS sign within a fixed model reverses
the charge while preserving its modulus and the energy.

The cubic potential also has twenty-four maxima, all with
$V\approx5.81692$, one of them projected outside the window in which
individual maxima are marked. Its planar pattern has the same
reflection symmetry as the octahedral model,
$(\vp_1,\vp_2)\mapsto(-\vp_2,-\vp_1)$, about the antidiagonal
$\vp_2=-\vp_1$.
\FloatBarrier

\subsection{Icosahedron}
For the icosahedral model we solve the self-duality equations for
\eqref{eq:U_icosa} projected onto the plane.
\begin{figure}[htb]
    \centering
    \begin{subfigure}{0.49\textwidth}
        \centering
        \includegraphics[width=\textwidth]{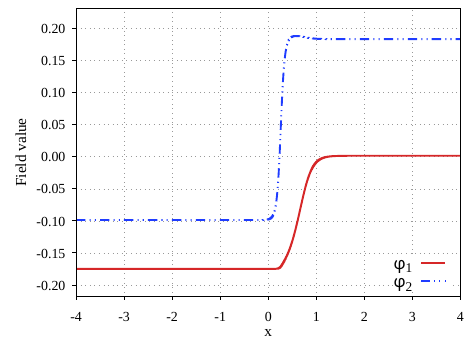}
        \caption{}
        \label{minima_4(1)_icosa_campos}
    \end{subfigure}
    \hfill
    \begin{subfigure}{0.49\textwidth}
        \centering
        \includegraphics[width=\textwidth]{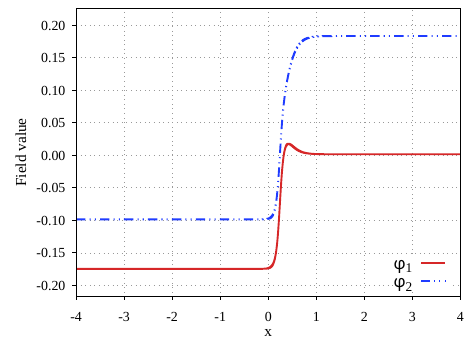}
        \caption{}
        \label{minima_4(2)_icosa_campos}
    \end{subfigure}
    \caption{Fields $(\vp_1,\vp_2)$ as functions of $x$ with $Q=1.376$,
    from the minimum $\vp^{(vac)}=\qty(-0.17475,-0.0987035)$ displaced by
    (a) $(0,10^{-3})$ and (b) $(10^{-3},10^{-3})$.}
    \label{fig:field_campos_icosa1}
    
\end{figure}
\begin{figure}[htb]
    \centering
    \begin{subfigure}{0.49\textwidth}
        \centering
        \includegraphics[width=\textwidth]{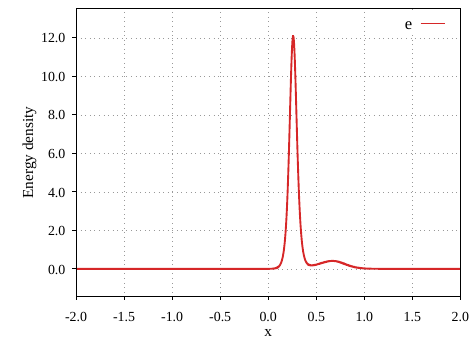}
        \caption{}
        \label{minima_4(1)_icosa_energia}
    \end{subfigure}
    \hfill
    \begin{subfigure}{0.49\textwidth}
        \centering
        \includegraphics[width=\textwidth]{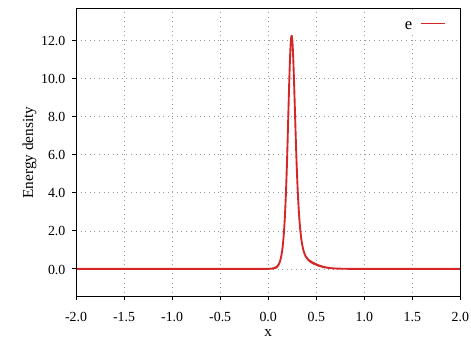}
        \caption{}
        \label{minima_4(2)_icosa_energia}
    \end{subfigure}
    \caption{Energy density of the solutions of
    Figure~\ref{fig:field_campos_icosa1}.}
    \label{fig:field_energia_icosa1}
    
\end{figure}
\begin{figure}[htb]
    \centering
    \begin{subfigure}{0.49\textwidth}
        \centering
        \includegraphics[width=\textwidth]{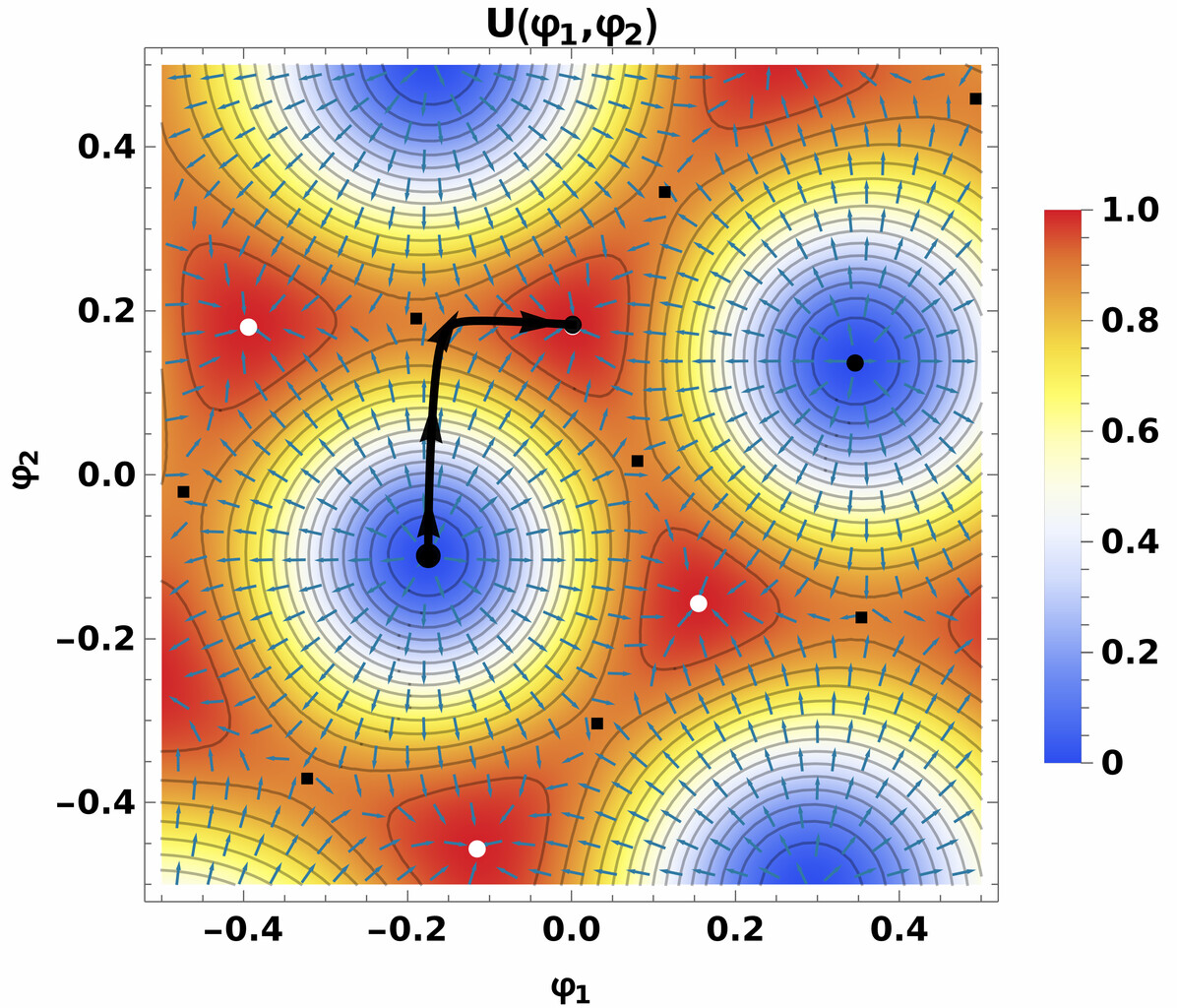}
        \caption{}
        \label{minima_4(1)_icosa_path}
    \end{subfigure}
    \hfill
    \begin{subfigure}{0.49\textwidth}
        \centering
        \includegraphics[width=\textwidth]{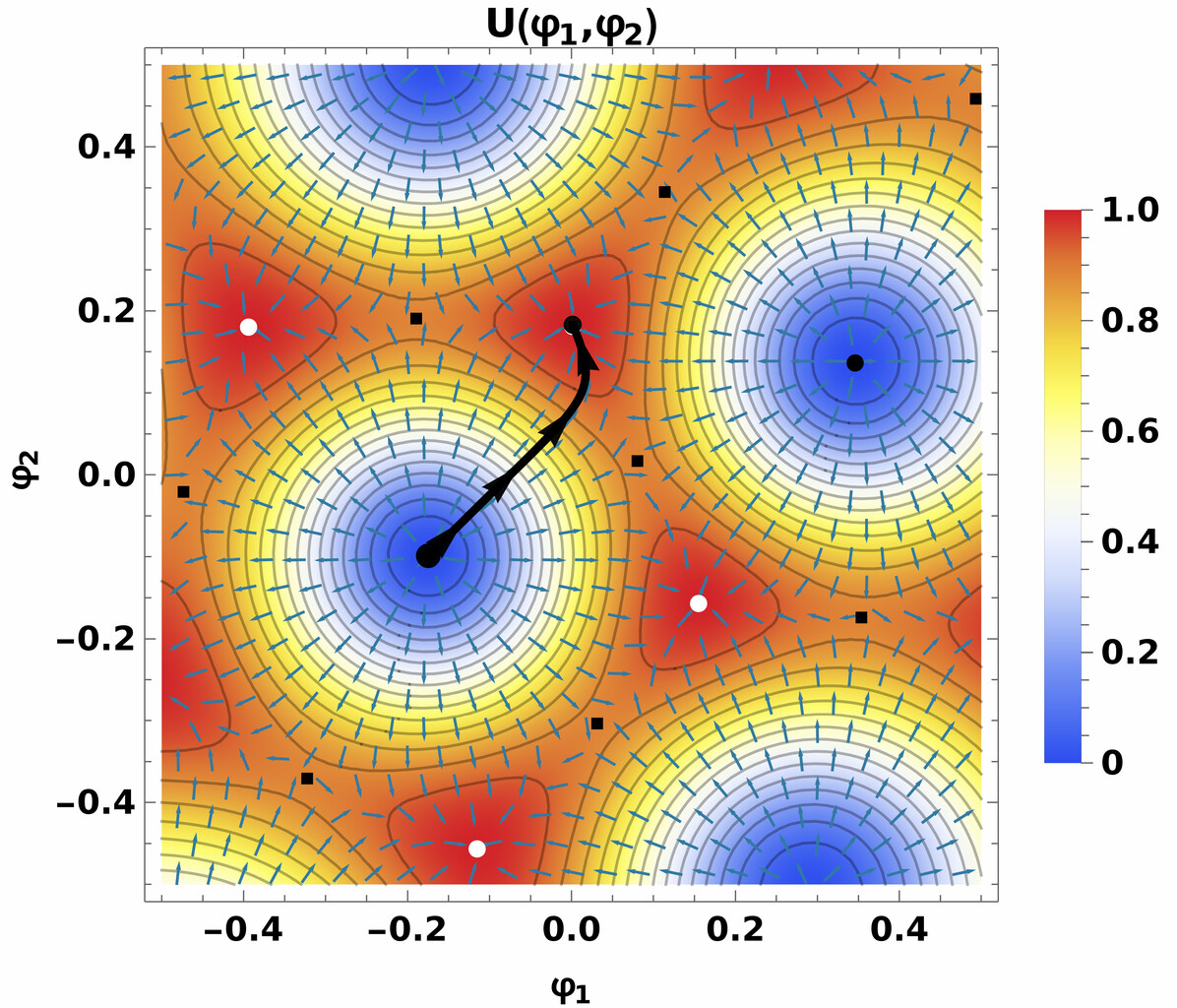}
        \caption{}
        \label{minima_4(2)_icosa_path}
    \end{subfigure}
    \caption{Paths of the solutions of
    Figure~\ref{fig:field_campos_icosa1} on $U_I$.}
    \label{fig:field_path_icosa1}
    
\end{figure}

The two solutions in Figure~\ref{fig:field_campos_icosa1} use different
displacements from the same reference point and connect the same pair
of vacua along different paths. In
Figure~\ref{fig:field_energia_icosa1}, the first profile has a small
secondary bump associated with the passage near a saddle shown in
Figure~\ref{fig:field_path_icosa1}; the second has a single peak.
The critical-value differences $1.2247$ and $0.1515$ describe the
limiting energy contributions of the minimum--saddle and
saddle--maximum transitions. The heights of the observed peaks are
determined by $V$ along each trajectory, according to
\eqref{eq:dens_2V}.

\begin{figure}[htb]
    \centering
    \begin{subfigure}{0.49\textwidth}
        \centering
        \includegraphics[width=\textwidth]{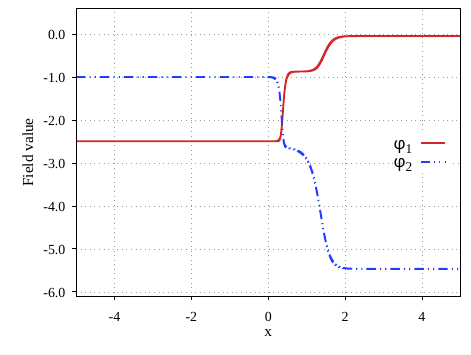}
        \caption{}
        \label{minima_11(5)_icosa_campos}
    \end{subfigure}
    \hfill
    \begin{subfigure}{0.49\textwidth}
        \centering
        \includegraphics[width=\textwidth]{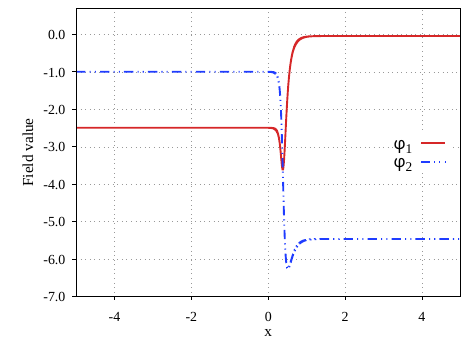}
        \caption{}
        \label{minima_11(6)_icosa_campos}
    \end{subfigure}
    \caption{Fields $(\vp_1,\vp_2)$ as functions of $x$ with $Q=1.376$,
    from the minimum $\vp^{(vac)}=\qty(-2.49144,-0.995506)$ displaced by
    (a) $(0,-10^{-3})$ and (b) $(-10^{-3},-10^{-3})$.}
    \label{fig:field_campos_icosa2}
    
\end{figure}
\begin{figure}[htb]
    \centering
    \begin{subfigure}{0.49\textwidth}
        \centering
        \includegraphics[width=\textwidth]{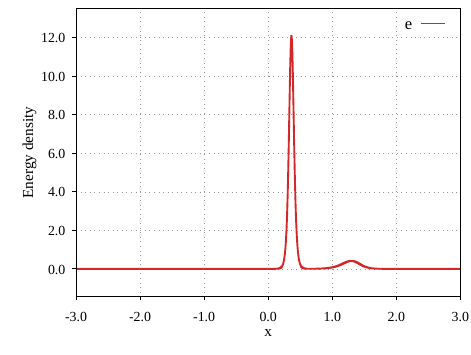}
        \caption{}
        \label{minima_11(5)_icosa_energia}
    \end{subfigure}
    \hfill
    \begin{subfigure}{0.49\textwidth}
        \centering
        \includegraphics[width=\textwidth]{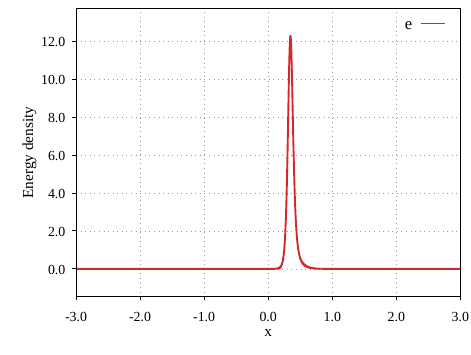}
        \caption{}
        \label{minima_11(6)_icosa_energia}
    \end{subfigure}
    \caption{Energy density of the solutions of
    Figure~\ref{fig:field_campos_icosa2}.}
    \label{fig:field_energia_icosa2}
   
\end{figure}
\begin{figure}[htb]
    \centering
    \begin{subfigure}{0.49\textwidth}
        \centering
        \includegraphics[width=\textwidth]{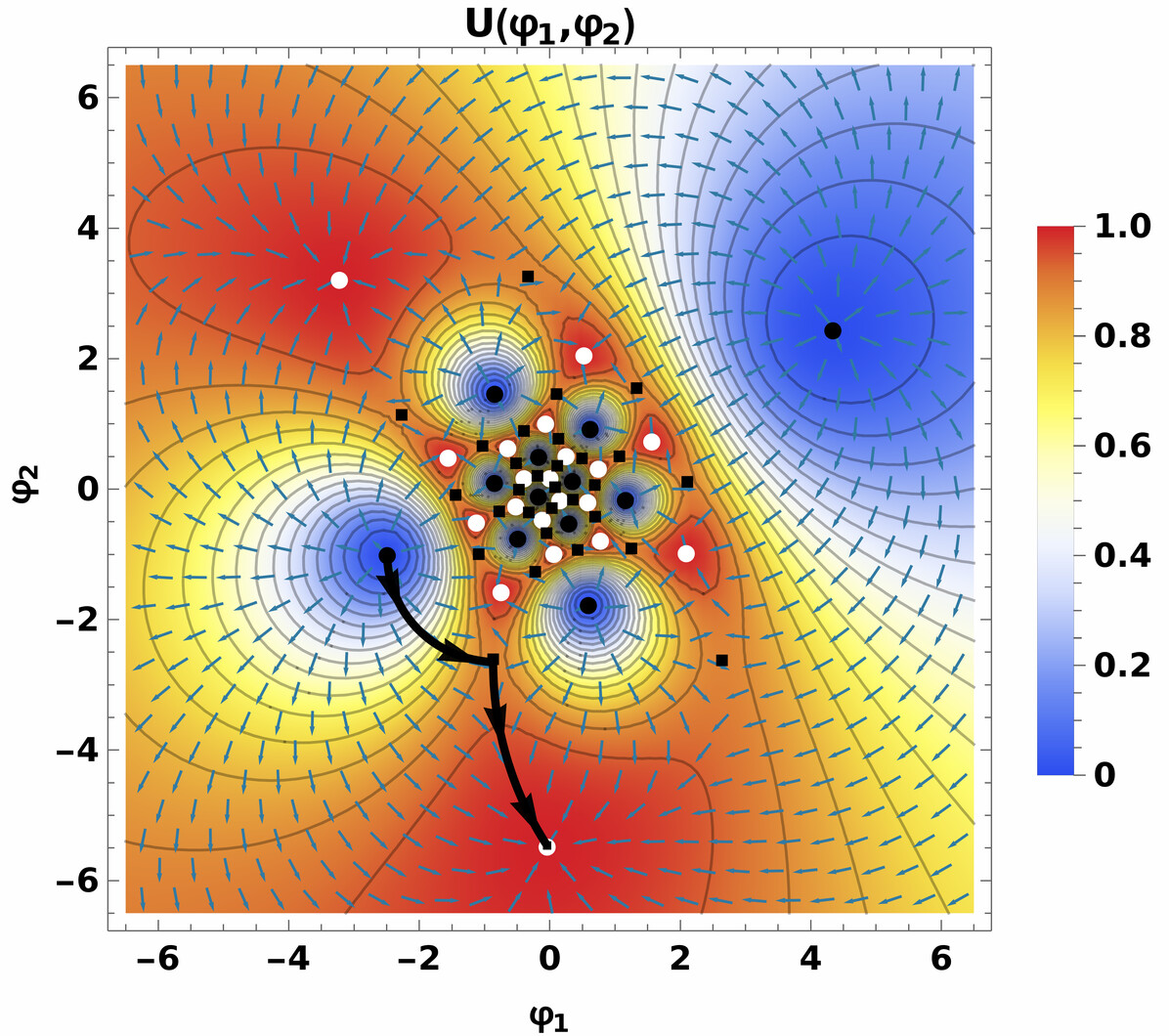}
        \caption{}
        \label{minima_11(5)_icosa_path}
    \end{subfigure}
    \hfill
    \begin{subfigure}{0.49\textwidth}
        \centering
        \includegraphics[width=\textwidth]{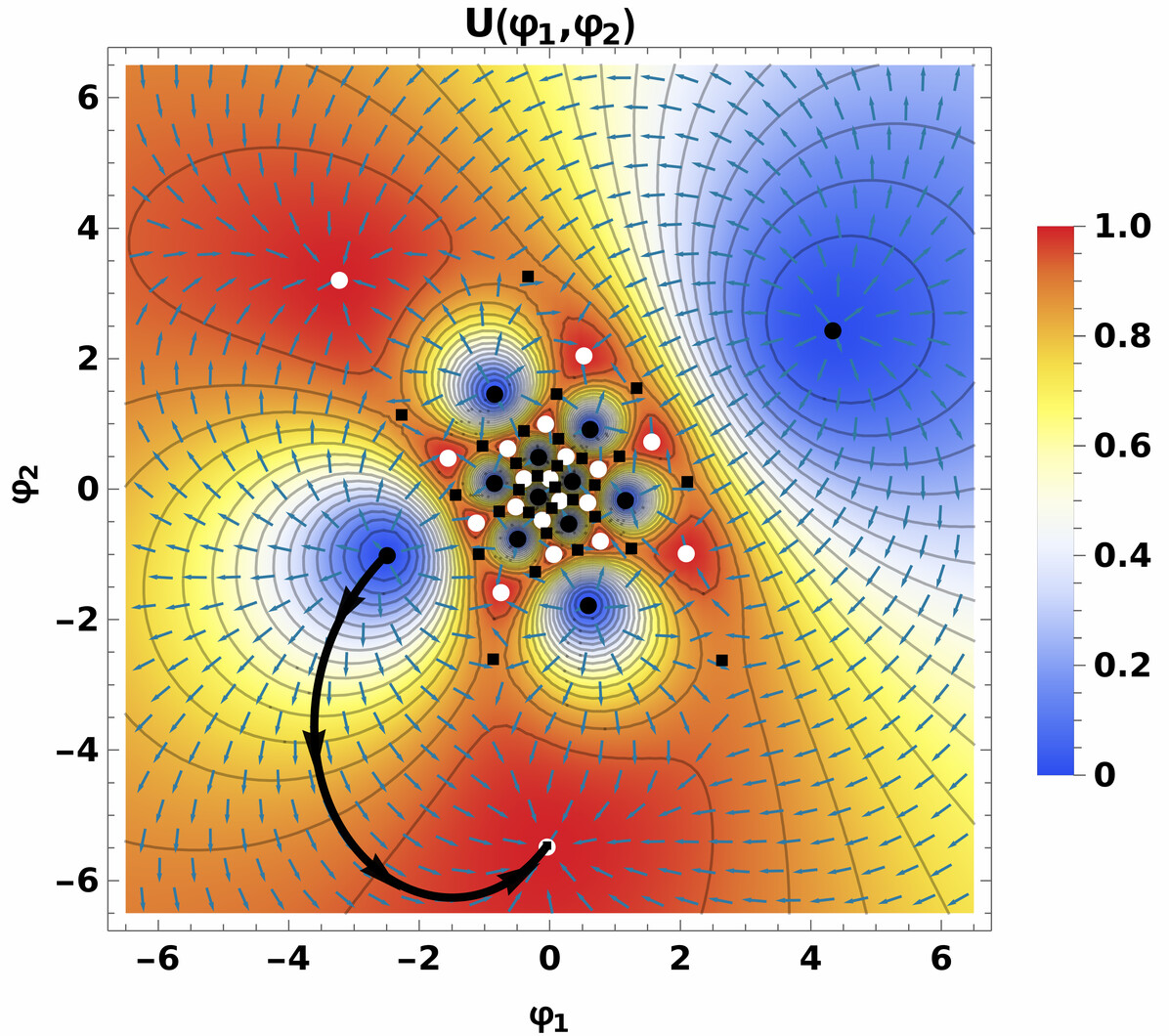}
        \caption{}
        \label{minima_11(6)_icosa_path}
    \end{subfigure}
    \caption{Paths of the solutions of
    Figure~\ref{fig:field_campos_icosa2} on $U_I$.}
    \label{fig:field_path_icosa2}
   
\end{figure}

Figure~\ref{fig:field_campos_icosa2} shows another pair with the same
charge in a different region of the target space. The equality of the
energies follows from the values of $U_I$ at the endpoint orbits.
One trajectory passes close to a saddle, while the other reaches the
maximum without a comparable intermediate passage, as shown in
Figure~\ref{fig:field_path_icosa2}. Their energy densities in
Figure~\ref{fig:field_energia_icosa2} exhibit a secondary bump and a
single peak, respectively. These profiles are determined by the
potential along the trajectories through \eqref{eq:dens_2V}.

The icosahedral potential has sixty maxima, all with
$V\approx6.15632$, forming the orbit described in
Section~\ref{sec:maxV}. One lies outside the plotted region, so
fifty-nine are listed in Appendix~\ref{app:maxV}.
The dotted line in Figure~\ref{fig:diagonal_maxima_icosa} serves as a
visual guide to the approximate alignment apparent in this plotting
window. It does not define an exact planar reflection symmetry. The
exact symmetry is inherited from the rotated icosahedral group and is
generally represented by nonlinear transformations of the
stereographic coordinates.

\begin{figure}[htb]
    \centering
    \includegraphics[width=0.6\textwidth]{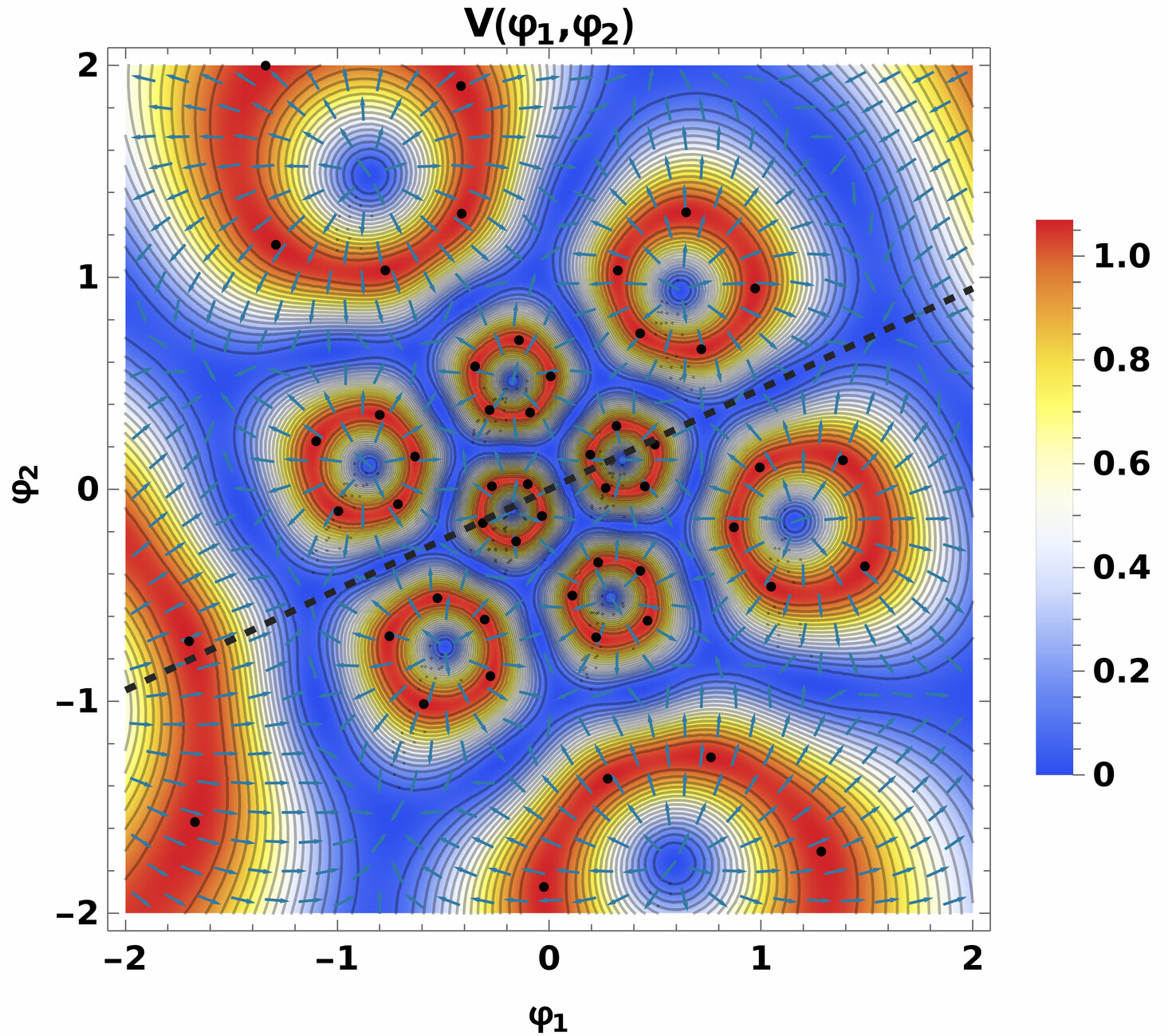}
    \caption{Icosahedral potential with its maxima marked. The dotted line
is a visual guide to the approximate alignment of the plotted pattern.}
    \label{fig:diagonal_maxima_icosa}
\end{figure}
\FloatBarrier

\subsection{Dodecahedron}
For the dodecahedral model we solve the self-duality equations for
\eqref{eq:U_dode} projected onto the plane, using the same starting
points as for the icosahedron. Since the two models share their regular
lattice with minima and maxima interchanged, a point close to a minimum
of $U_I$ is close to a maximum of $U_D$.
\begin{figure}[htb]
    \centering
    \begin{subfigure}{0.49\textwidth}
        \centering
        \includegraphics[width=\textwidth]{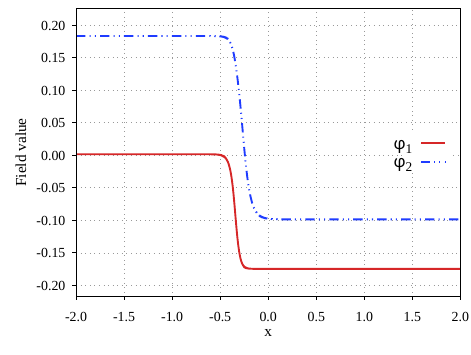}
        \caption{}
        \label{minima_4(1)_dode_campos}
    \end{subfigure}
    \hfill
    \begin{subfigure}{0.49\textwidth}
        \centering
        \includegraphics[width=\textwidth]{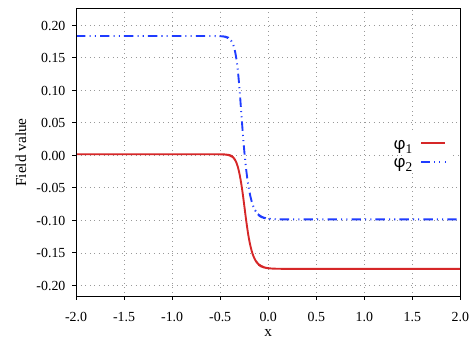}
        \caption{}
        \label{minima_4(2)_dode_campos}
    \end{subfigure}
    \caption{Fields $(\vp_1,\vp_2)$ as functions of $x$ with $Q=1.938$,
    from the maximum $\vp^{(vac)}=\qty(-0.17475,-0.0987035)$ displaced by
    (a) $(0,10^{-3})$ and (b) $(10^{-3},10^{-3})$.}
    \label{fig:field_campos_dode1}
   
\end{figure}
\begin{figure}[htb]
    \centering
    \begin{subfigure}{0.49\textwidth}
        \centering
        \includegraphics[width=\textwidth]{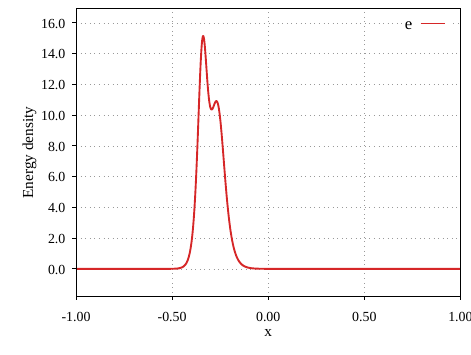}
        \caption{}
        \label{minima_4(1)_dode_energia}
    \end{subfigure}
    \hfill
    \begin{subfigure}{0.49\textwidth}
        \centering
        \includegraphics[width=\textwidth]{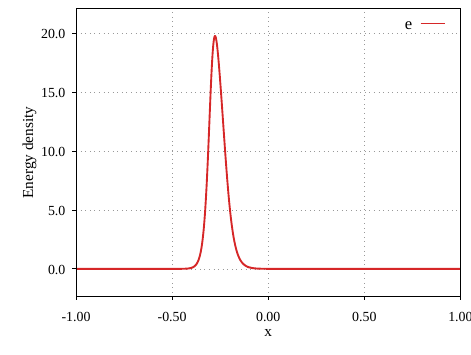}
        \caption{}
        \label{minima_4(2)_dode_energia}
    \end{subfigure}
    \caption{Energy density of the solutions of
    Figure~\ref{fig:field_campos_dode1}.}
    \label{fig:field_energia_dode1}
   
\end{figure}
\begin{figure}[htb]
    \centering
    \begin{subfigure}{0.49\textwidth}
        \centering
        \includegraphics[width=\textwidth]{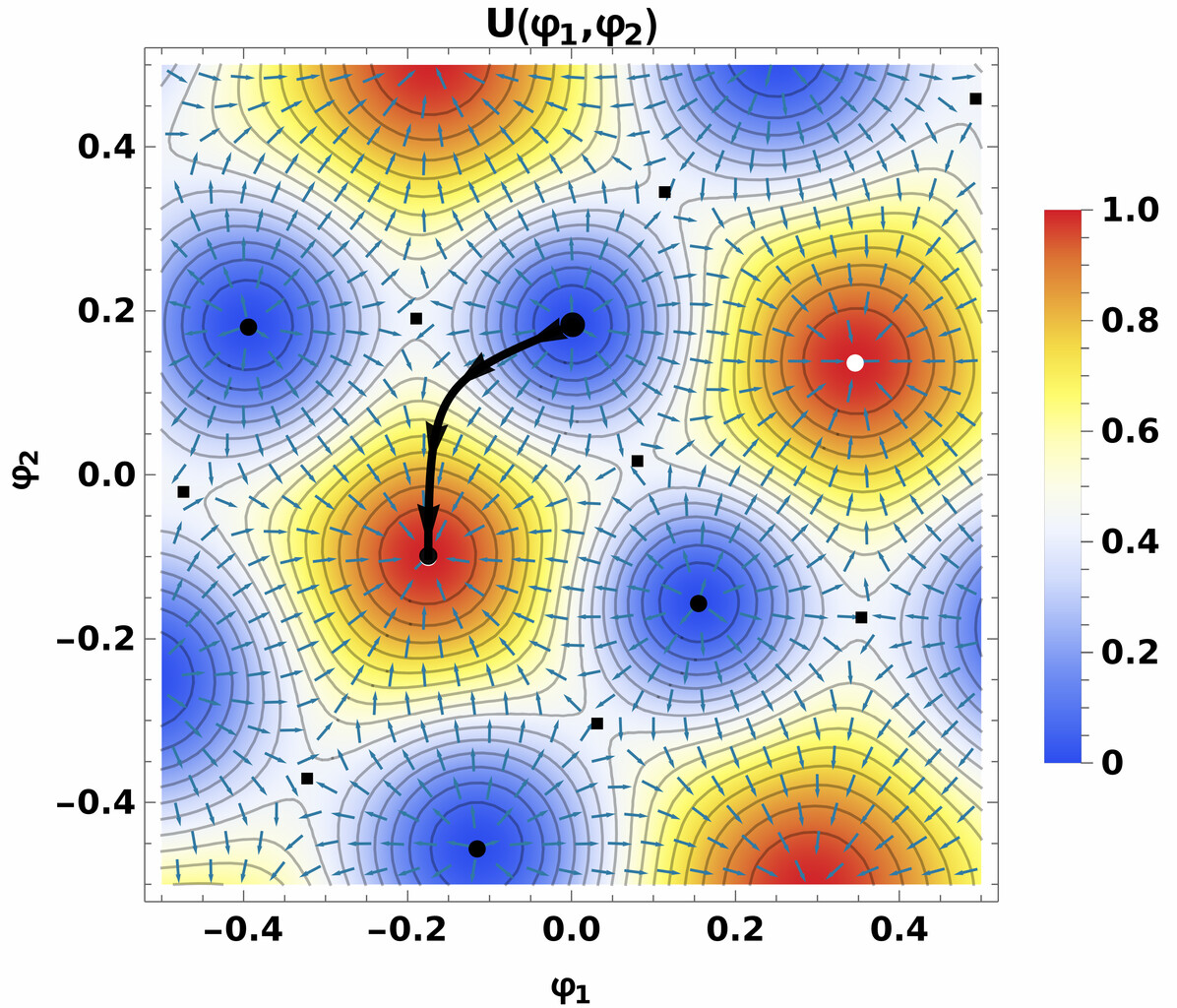}
        \caption{}
        \label{minima_4(1)_dode_path}
    \end{subfigure}
    \hfill
    \begin{subfigure}{0.49\textwidth}
        \centering
        \includegraphics[width=\textwidth]{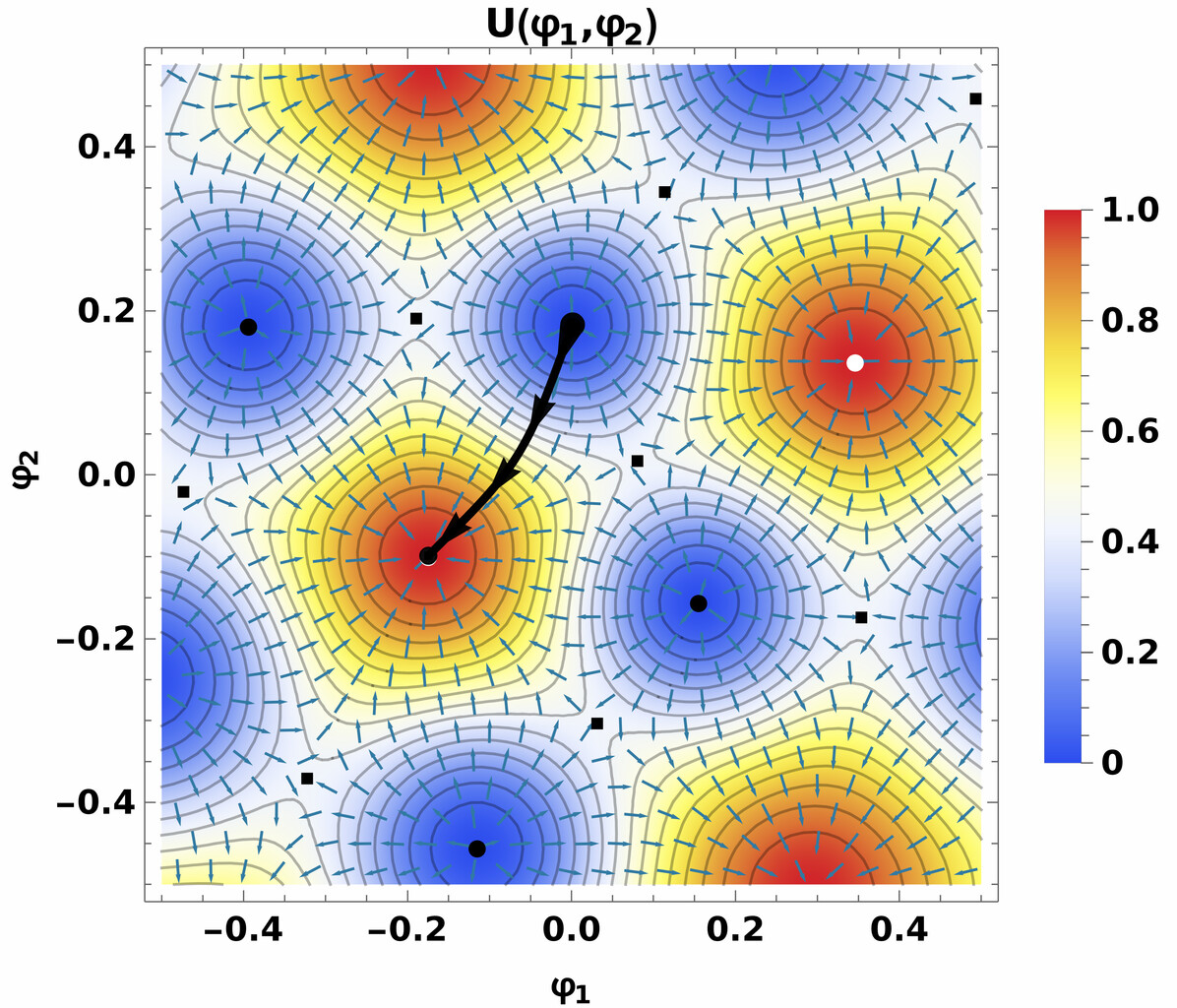}
        \caption{}
        \label{minima_4(2)_dode_path}
    \end{subfigure}
    \caption{Paths of the solutions of
    Figure~\ref{fig:field_campos_dode1} on $U_D$.}
    \label{fig:field_path_dode1}
    
\end{figure}

The pair in Figure~\ref{fig:field_campos_dode1} connects the same
endpoints along different paths. The first energy density in
Figure~\ref{fig:field_energia_dode1} has two overlapping peaks, and
its trajectory passes close to a saddle in
Figure~\ref{fig:field_path_dode1}. The second has a single peak.
The critical-value differences $0.8318$ and $1.1064$ characterize the
limiting integrated energies of two transitions separated near the
saddle. The comparable heights observed in the first profile depend
on the potential sampled along that particular trajectory, through
\eqref{eq:dens_2V}.
\begin{figure}[htb]
    \centering
    \begin{subfigure}{0.49\textwidth}
        \centering
        \includegraphics[width=\textwidth]{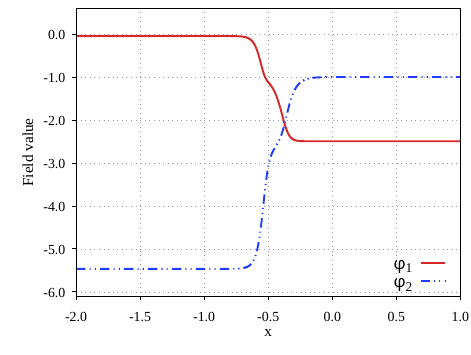}
        \caption{}
        \label{minima_11(5)_dode_campos}
    \end{subfigure}
    \hfill
    \begin{subfigure}{0.49\textwidth}
        \centering
        \includegraphics[width=\textwidth]{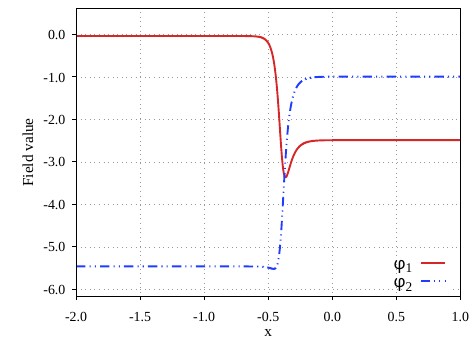}
        \caption{}
        \label{minima_11(6)_dode_campos}
    \end{subfigure}
    \caption{Fields $(\vp_1,\vp_2)$ as functions of $x$ with $Q=1.938$,
    from the maximum $\vp^{(vac)}=\qty(-2.49144,-0.995506)$ displaced by
    (a) $(0,-10^{-3})$ and (b) $(-10^{-3},-10^{-3})$.}
    \label{fig:field_campos_dode2}
   
\end{figure}
\begin{figure}[htb]
    \centering
    \begin{subfigure}{0.49\textwidth}
        \centering
        \includegraphics[width=\textwidth]{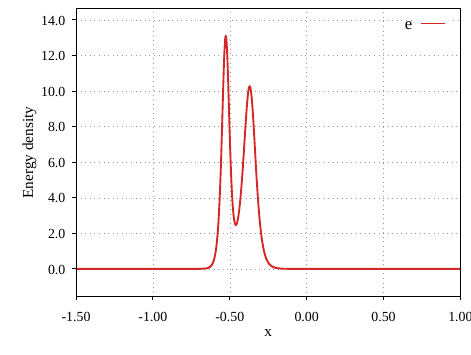}
        \caption{}
        \label{minima_11(5)_dode_energia}
    \end{subfigure}
    \hfill
    \begin{subfigure}{0.49\textwidth}
        \centering
        \includegraphics[width=\textwidth]{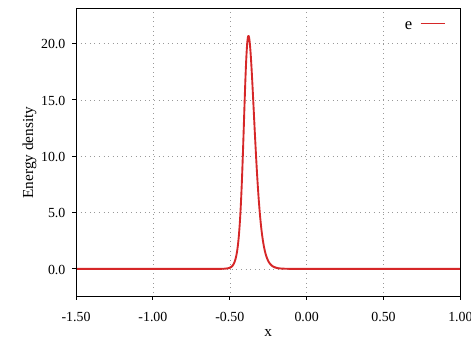}
        \caption{}
        \label{minima_11(6)_dode_energia}
    \end{subfigure}
    \caption{Energy density of the solutions of
    Figure~\ref{fig:field_campos_dode2}.}
    \label{fig:field_energia_dode2}
    
\end{figure}
\begin{figure}[htb]
    \centering
    \begin{subfigure}{0.49\textwidth}
        \centering
        \includegraphics[width=\textwidth]{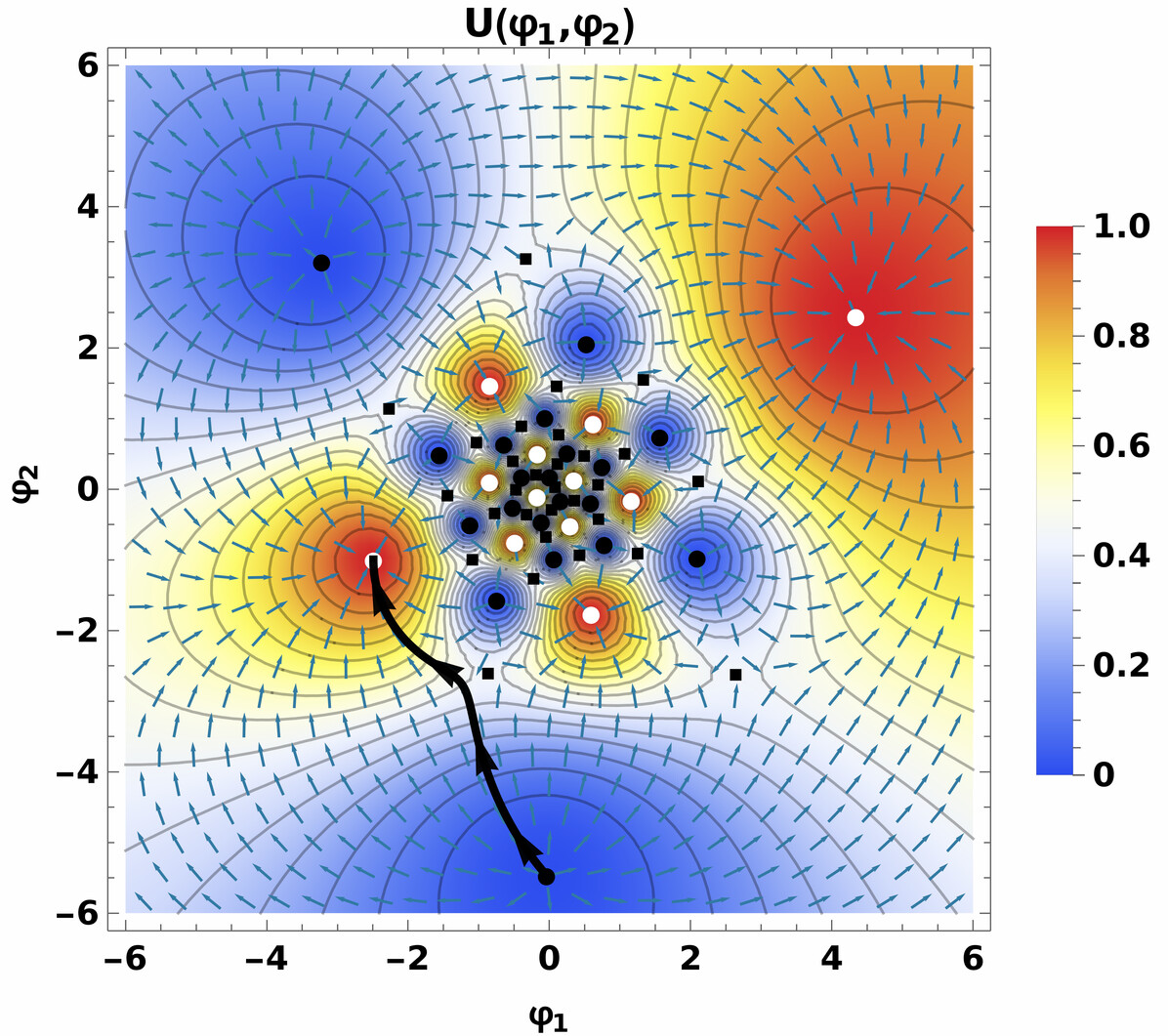}
        \caption{}
        \label{minima_11(5)_dode_path}
    \end{subfigure}
    \hfill
    \begin{subfigure}{0.49\textwidth}
        \centering
        \includegraphics[width=\textwidth]{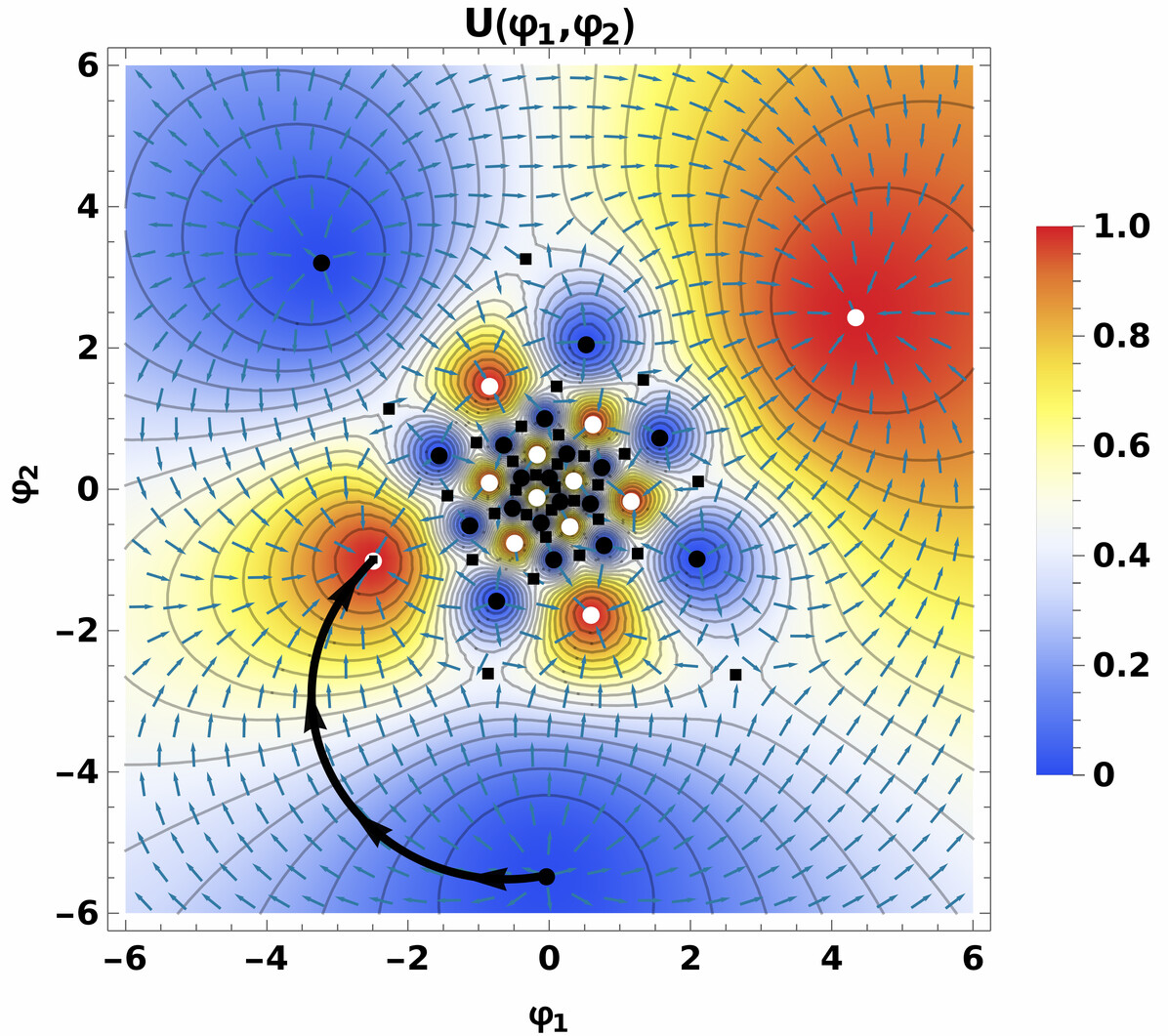}
        \caption{}
        \label{minima_11(6)_dode_path}
    \end{subfigure}
    \caption{Paths of the solutions of
    Figure~\ref{fig:field_campos_dode2} on $U_D$.}
    \label{fig:field_path_dode2}
    
\end{figure}

The second pair, in Figure~\ref{fig:field_campos_dode2}, uses initial
configurations close to a maximum of $U_D$. The paths in
Figure~\ref{fig:field_path_dode2} show a passage near a saddle in one
case and a direct connection in the other, with the corresponding
two-peak and single-peak profiles in
Figure~\ref{fig:field_energia_dode2}.
\begin{figure}[htb]
    \centering
    \begin{subfigure}{0.49\textwidth}
        \centering
        \includegraphics[width=\textwidth]{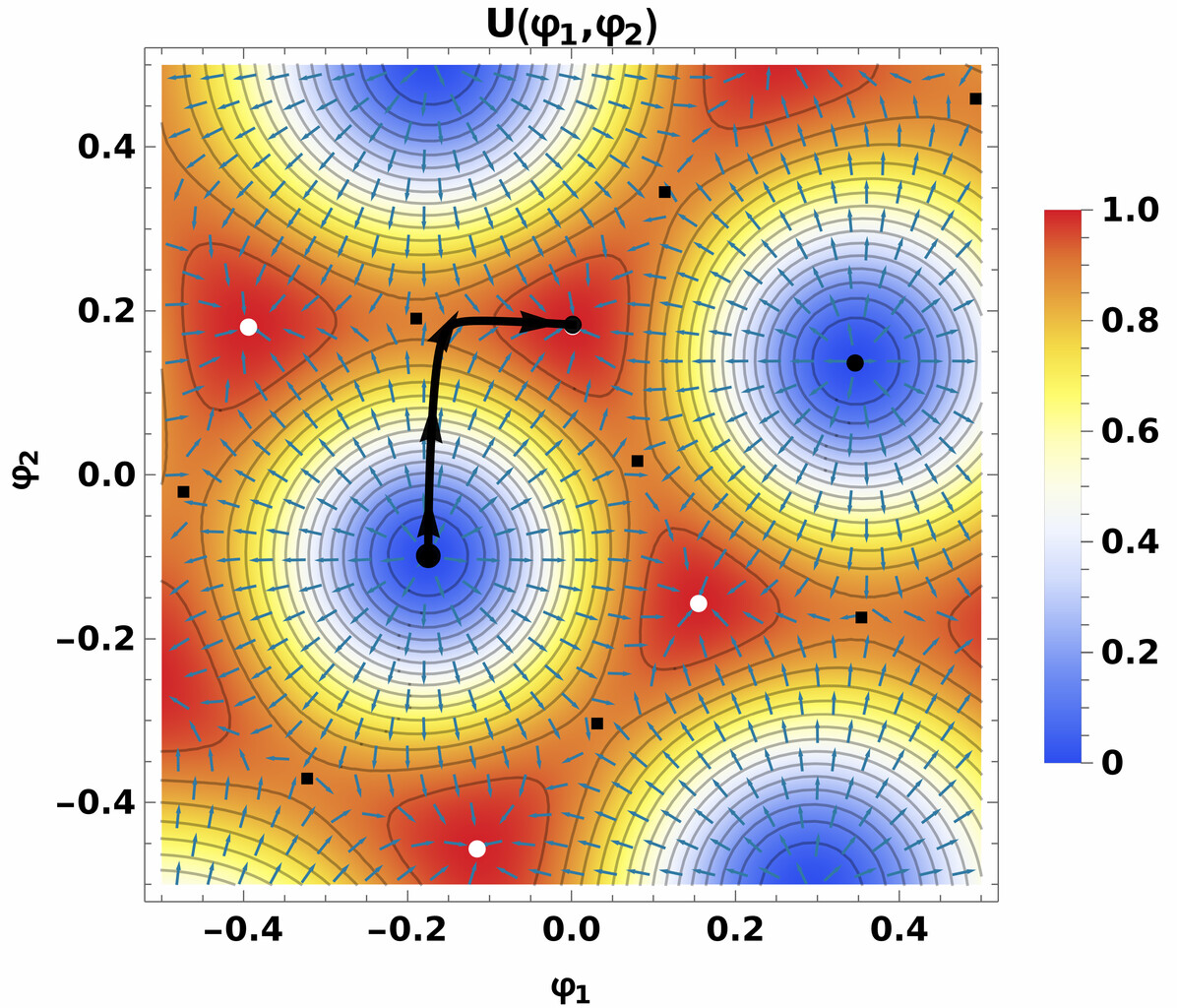}
        \caption{}
    \end{subfigure}
    \hfill
    \begin{subfigure}{0.49\textwidth}
        \centering
        \includegraphics[width=\textwidth]{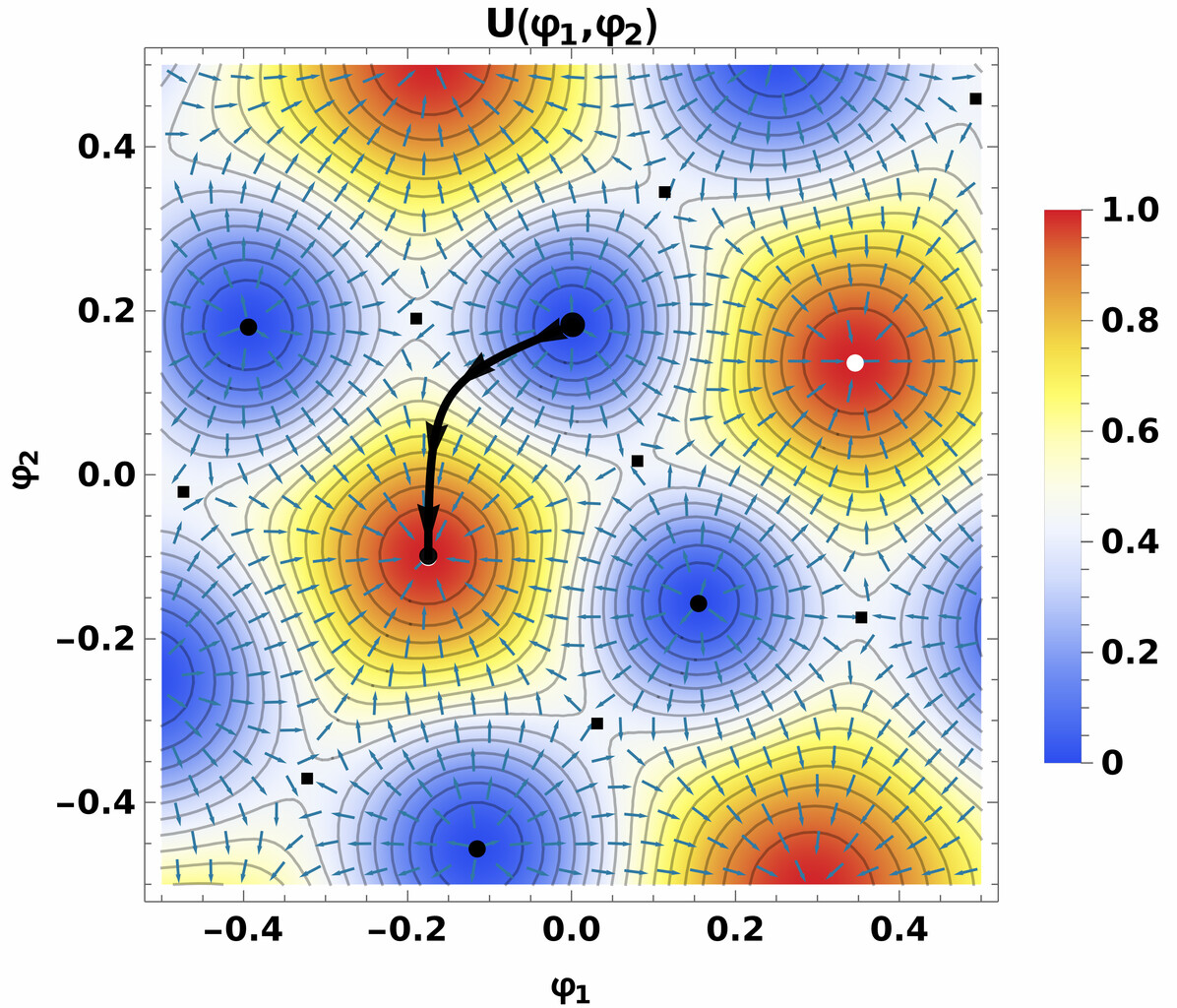}
        \caption{}
    \end{subfigure}
    \par\vspace{0.15cm}
    \begin{subfigure}{0.49\textwidth}
        \centering
        \includegraphics[width=\textwidth]{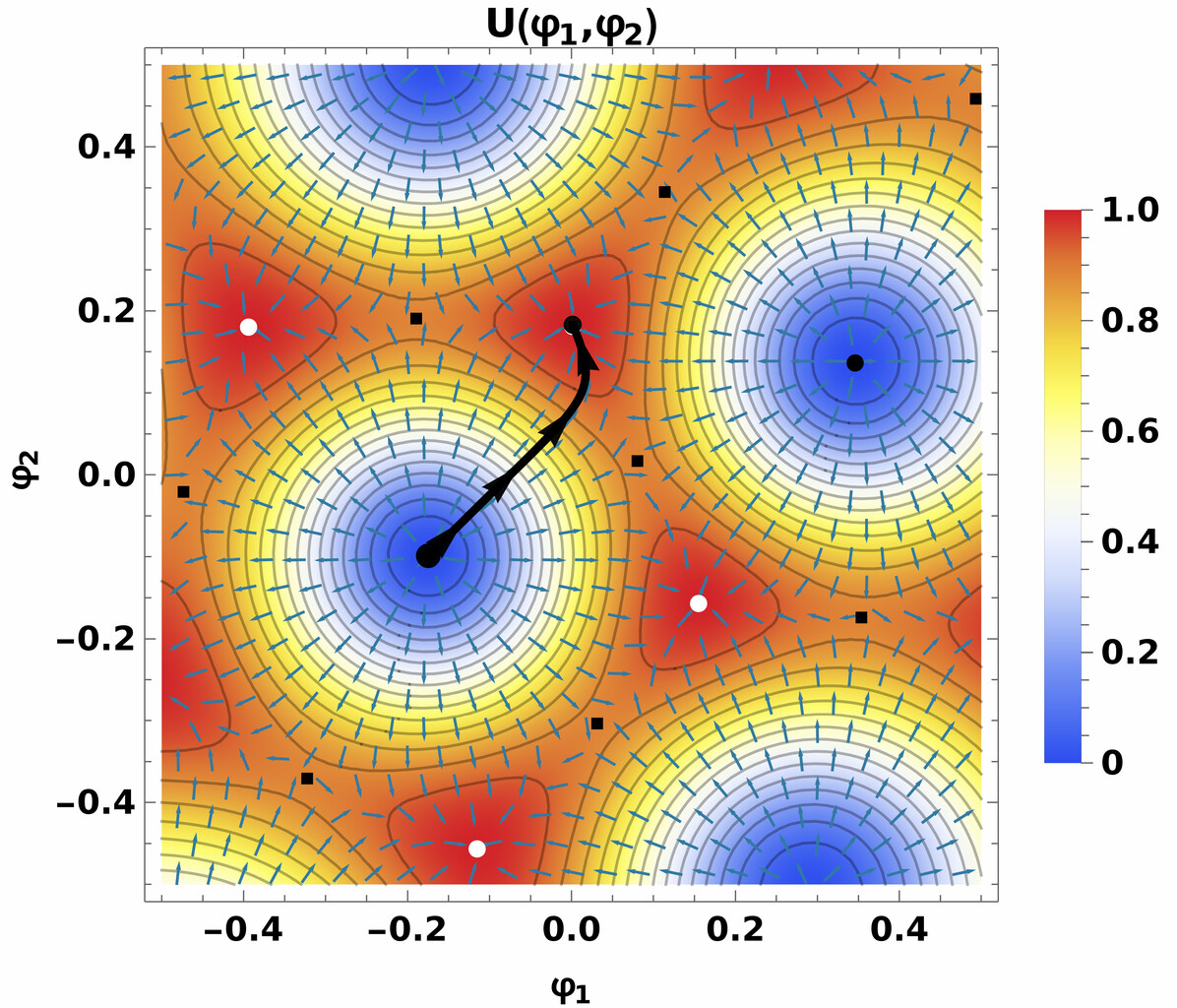}
        \caption{}
    \end{subfigure}
    \hfill
    \begin{subfigure}{0.49\textwidth}
        \centering
        \includegraphics[width=\textwidth]{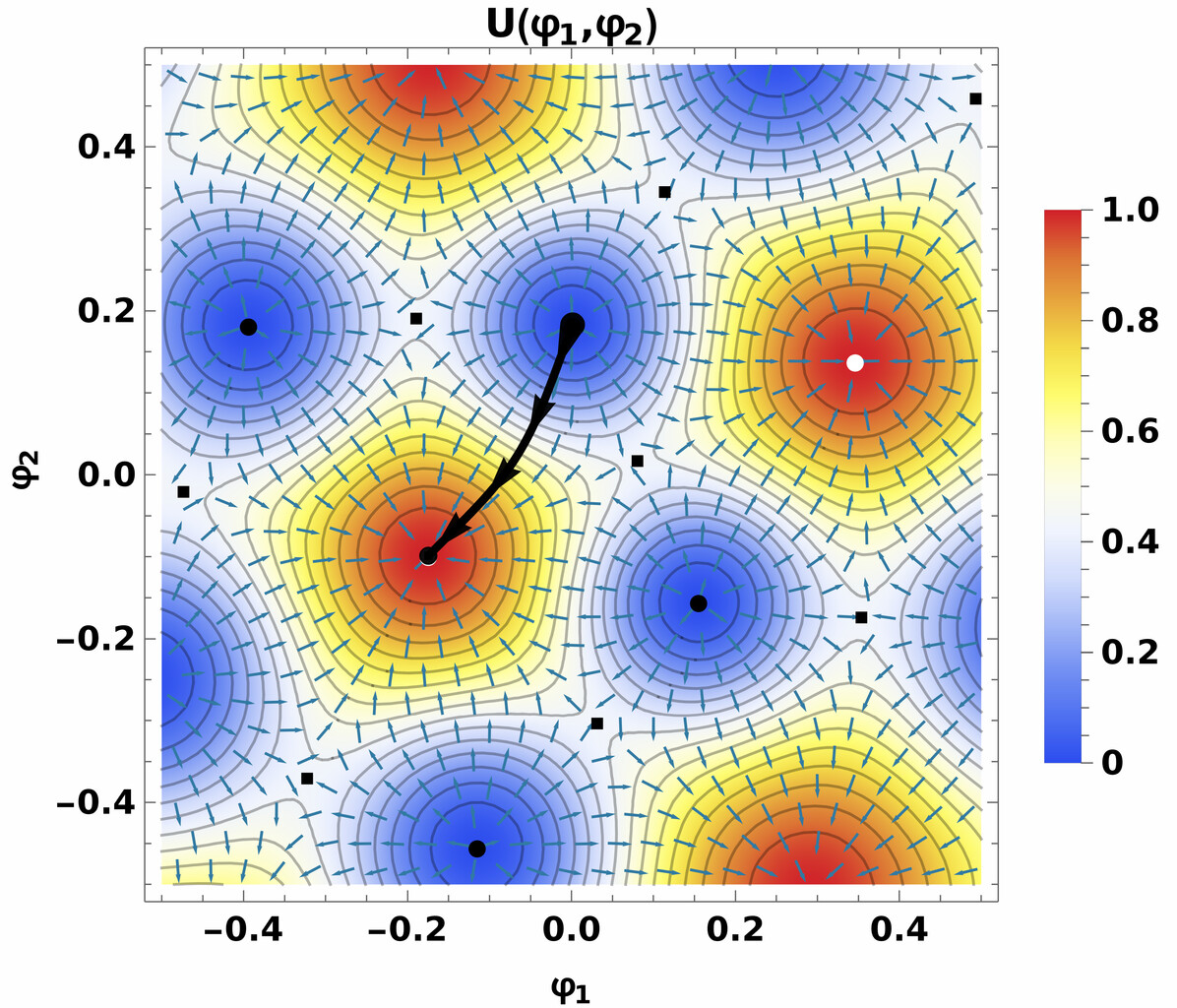}
        \caption{}
    \end{subfigure}
    \caption{Paths of the (a,c) icosahedral and (b,d) dodecahedral
    solutions from the site $\vp^{(vac)}=\qty(-0.17475,-0.0987035)$,
    displaced by $(0,10^{-3})$ in (a,b) and by $(10^{-3},10^{-3})$ in (c,d).}
    \label{fig:comparison_path_icosa_dode1}
   
\end{figure}
\begin{figure}[htb]
    \centering
    \begin{subfigure}{0.49\textwidth}
        \centering
        \includegraphics[width=\textwidth]{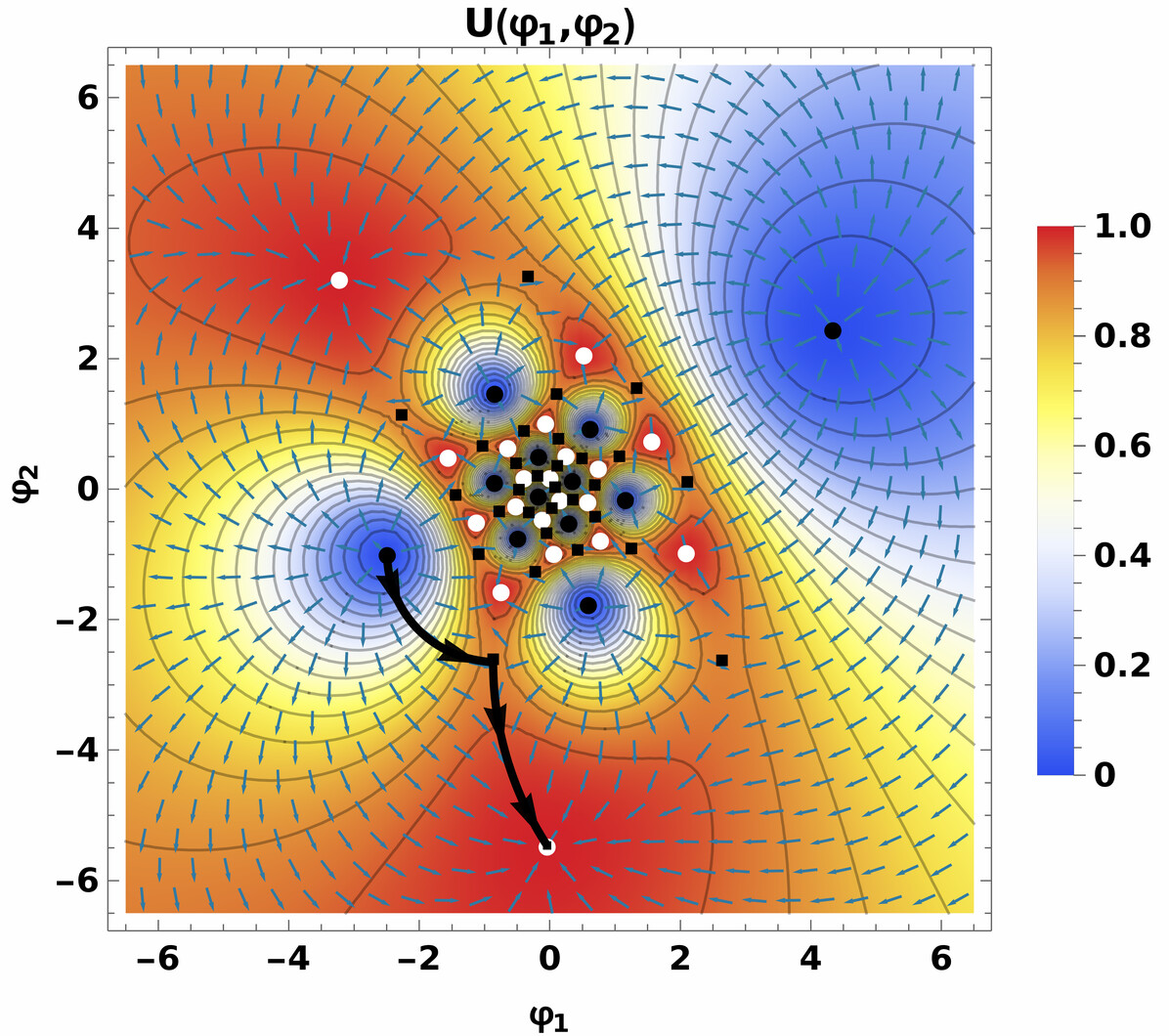}
        \caption{}
    \end{subfigure}
    \hfill
    \begin{subfigure}{0.49\textwidth}
        \centering
        \includegraphics[width=\textwidth]{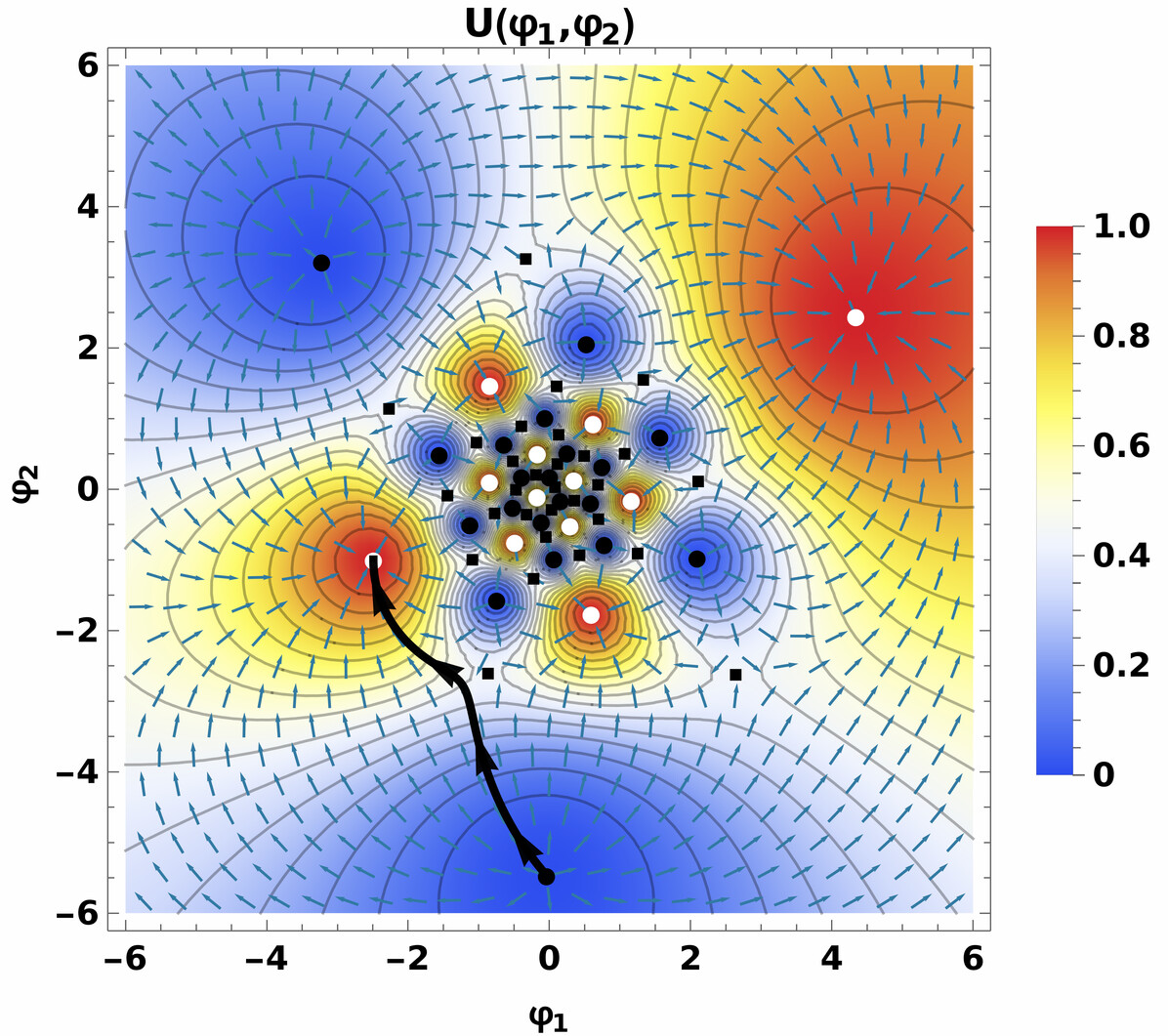}
        \caption{}
    \end{subfigure}
    \par\vspace{0.15cm}
    \begin{subfigure}{0.49\textwidth}
        \centering
        \includegraphics[width=\textwidth]{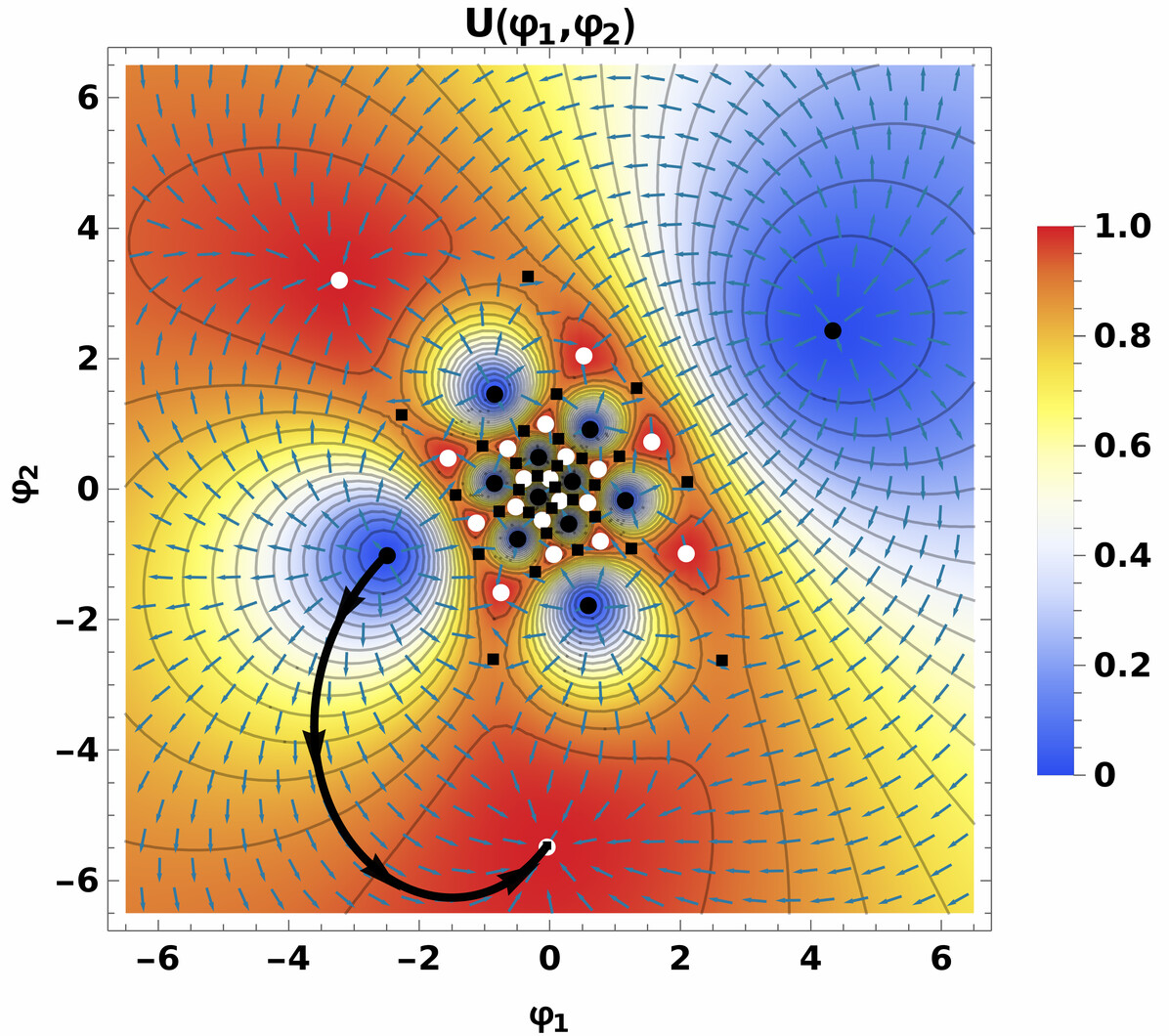}
        \caption{}
    \end{subfigure}
    \hfill
    \begin{subfigure}{0.49\textwidth}
        \centering
        \includegraphics[width=\textwidth]{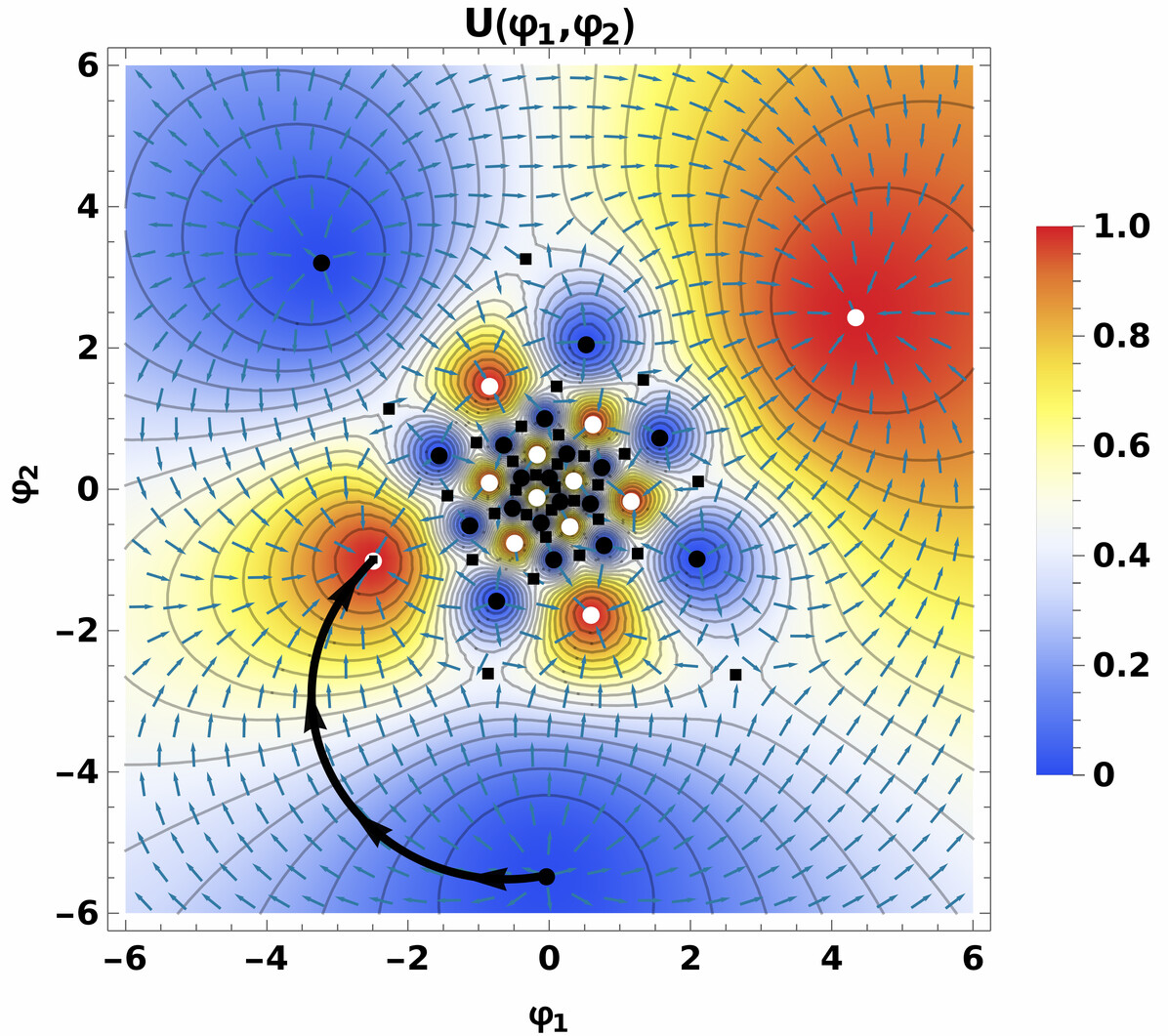}
        \caption{}
    \end{subfigure}
    \caption{Paths of the (a,c) icosahedral and (b,d) dodecahedral
    solutions from the site $\vp^{(vac)}=\qty(-2.49144,-0.995506)$,
    displaced by $(0,-10^{-3})$ in (a,b) and by $(-10^{-3},-10^{-3})$ in (c,d).}
    \label{fig:comparison_path_icosa_dode2}
    
\end{figure}

Figures~\ref{fig:comparison_path_icosa_dode1} and
\ref{fig:comparison_path_icosa_dode2} compare selected connections
in the two dual models. In these examples the endpoints exchange
their minimum and maximum roles, while the trajectories and
energy-density profiles differ. The total charges also differ because
the two pre-potentials have different critical values, as listed in
Table~\ref{tab:charges}. The common regular lattice fixes the vacuum
positions, but the existence and shape of a connection depend on the
metric-gradient field of each model. A correspondence between all
their self-dual connections would require a separate argument.

The dodecahedral potential has sixty maxima, all with $V\approx10.62609$, and
here none of them is projected outside the plotted region, so the whole
orbit appears in Appendix~\ref{app:maxV}. Their planar arrangement is
the least regular of the five models, and the rings seen around the
minima of the other pre-potentials are deformed into sharp peaks
surrounding them, as the three-dimensional plot in
Figure~\ref{V_3d_dode} shows.
\FloatBarrier

\section{Conclusion}
\label{sec:conclusion}

We have constructed a family of $(1+1)$-dimensional $\IC\IP^1$ sigma
models with potentials derived from Platonic geometry through the
generalized self-duality framework of \cite{lafBPS,laf_1+1}. The unit
vertices of each solid define a smooth pre-potential on
$\mathbb{S}^2$, and the round target-space metric determines the
potential through \eqref{eq:potential}. This provides a geometric
prescription for constructing models with prescribed vacuum sites and
self-dual sectors.

The rotational symmetry of each pre-potential makes every site of the
corresponding regular lattice a critical point. For the five individual
models, the vertices are minima, while the numerical analysis
identifies the dual vertices as maxima and the edge-midpoint sites as
saddles. The equality of the pre-potential values within each symmetry
orbit follows directly from group invariance. Consequently, existing
self-dual connections between the same pair of critical-point orbits
have equal energies. Dual solids share their vacuum positions, while
their distinct pre-potentials produce different charges and
metric-gradient fields.

Stereographic projection expresses these models in two real fields
while preserving the Platonic symmetry through its induced action
on the planar coordinates. Rotating the vertex set rotates the critical
points and conjugates the symmetry group. At fixed normalization this
preserves the critical values and self-dual energies. If the
normalization is imposed again after the rotation, the pre-potential
and the energies are rescaled by the factor in
\eqref{eq:rotated_prepotential}.

The numerical solutions exhibit distinct trajectories and
energy-density profiles with the same total energy, in agreement with
$E=\abs{Q}$. Passages close to saddle points can produce profiles with
two peaks. Their integrated energies are controlled by pre-potential
differences, while their local heights and widths follow from the
potential along the trajectory through \eqref{eq:dens_2V}.
The comparisons between dual models demonstrate these differences for
the connections found numerically.

The positive maxima of $V$ identified numerically form a single orbit
in each model. Their trivial stabilizers give twelve points for the
tetrahedral model, twenty-four for the octahedral and cubic models, and
sixty for the icosahedral and dodecahedral models. Symmetry explains
their equal values, although an explicit geometric characterization
of their positions remains to be developed.

Further work can explore the mixed pre-potentials in
\eqref{eq:U_possibilidades}, including changes in critical-point types
and the possible appearance of additional vacua. Determining the full
set of self-dual connections and studying time evolution and soliton
scattering would extend the present static analysis and provide ways
to investigate the possible connection of these models with integrability.

\appendix

\section{Coordinates of the maxima of the potential}
\label{app:maxV}
The tables below list the stereographic coordinates of the maxima of $V$
for the five models, obtained numerically. For each model the potential
takes the same value at every one of these points, quoted in the caption,
and the number of points is the order of the rotation group of the solid,
except where one of them is projected outside the region covered by the
plots of Section~\ref{sec:maxV}, in which case it is absent from the
table. The vertex orientations are those used for the numerical solutions.
The rotation \eqref{eq:rot_mat} is applied to all solids except
the tetrahedron.

\begin{table}[htbp]
    \centering
    \footnotesize
    \setlength{\tabcolsep}{3pt}
    \begin{tabular}{r r r r r r}
        \hline
        \# & $\vp_1$ & $\vp_2$ & \# & $\vp_1$ & $\vp_2$\\
        \hline
        1 & $-2.38951$ & $-0.26318$ & 7 & $-0.76432$ & $-0.76432$\\
        2 & $0.76432$ & $0.76432$ & 8 & $-0.41348$ & $0.04554$\\
        3 & $0.65418$ & $-0.65418$ & 9 & $0.26318$ & $2.38951$\\
        4 & $-0.26318$ & $-2.38951$ & 10 & $-0.65418$ & $0.65418$\\
        5 & $0.04554$ & $-0.41348$ & 11 & $-0.04554$ & $0.41348$\\
        6 & $0.41348$ & $-0.04554$ & 12 & $2.38951$ & $0.26318$\\
        \hline
    \end{tabular}
    \caption{Maxima of $V$ for the tetrahedron model, at which $V=1.56928$. The tetrahedral rotation group has order 12.}
    \label{tab:maxV_tetra}
\end{table}

\begin{table}[htbp]
    \centering
    \footnotesize
    \setlength{\tabcolsep}{3pt}
    \begin{tabular}{r r r r r r r r r}
        \hline
        \# & $\vp_1$ & $\vp_2$ & \# & $\vp_1$ & $\vp_2$ & \# & $\vp_1$ & $\vp_2$\\
        \hline
        1 & $-0.95343$ & $-0.22077$ & 9 & $-0.23051$ & $-0.99547$ & 17 & $-0.33858$ & $0.02266$\\
        2 & $0.25714$ & $-0.79409$ & 10 & $0.80218$ & $1.45734$ & 18 & $0.79409$ & $-0.25714$\\
        3 & $-0.28987$ & $-0.52662$ & 11 & $-1.13978$ & $0.36908$ & 19 & $2.94034$ & $-0.19678$\\
        4 & $0.19678$ & $-2.94034$ & 12 & $-0.31094$ & $2.22569$ & 20 & $-1.45734$ & $-0.80218$\\
        5 & $-0.03942$ & $0.03942$ & 13 & $-0.02266$ & $0.33858$ & 21 & $0.99547$ & $0.23051$\\
        6 & $0.44070$ & $-0.06157$ & 14 & $0.22077$ & $0.95343$ & 22 & $1.39927$ & $-1.39927$\\
        7 & $-2.22569$ & $0.31094$ & 15 & $0.52662$ & $0.28987$ & 23 & $-0.35733$ & $0.35733$\\
        8 & $-0.36908$ & $1.13978$ & 16 & $0.06157$ & $-0.44070$ &  &  & \\
        \hline
    \end{tabular}
    \caption{Maxima of $V$ for the octahedron model, at which $V=3.58693$. The octahedral rotation group has order 24. The remaining point of the orbit is projected outside the plotted region and is not listed.}
    \label{tab:maxV_octa}
\end{table}

\begin{table}[htbp]
    \centering
    \footnotesize
    \setlength{\tabcolsep}{3pt}
    \begin{tabular}{r r r r r r r r r}
        \hline
        \# & $\vp_1$ & $\vp_2$ & \# & $\vp_1$ & $\vp_2$ & \# & $\vp_1$ & $\vp_2$\\
        \hline
        1 & $-0.42444$ & $0.42444$ & 9 & $-0.27971$ & $-2.54255$ & 17 & $2.54255$ & $0.27971$\\
        2 & $0.04275$ & $0.38860$ & 10 & $0.45883$ & $0.36080$ & 18 & $-1.34671$ & $-1.05898$\\
        3 & $-0.32436$ & $-1.12489$ & 11 & $-0.82074$ & $-0.23666$ & 19 & $-1.00803$ & $0.47608$\\
        4 & $0.23666$ & $0.82074$ & 12 & $0.81111$ & $-0.38308$ & 20 & $-0.47608$ & $1.00803$\\
        5 & $0.36117$ & $-0.09108$ & 13 & $-0.65650$ & $2.60320$ & 21 & $1.05898$ & $1.34671$\\
        6 & $1.12489$ & $0.32436$ & 14 & $0.09108$ & $-0.36117$ & 22 & $0.38308$ & $-0.81111$\\
        7 & $0.01202$ & $-0.01202$ & 15 & $1.17803$ & $-1.17803$ & 23 & $-2.60320$ & $0.65650$\\
        8 & $-0.36080$ & $-0.45883$ & 16 & $-0.38860$ & $-0.04275$ &  &  & \\
        \hline
    \end{tabular}
    \caption{Maxima of $V$ for the cube model, at which $V=5.81692$. The octahedral rotation group has order 24. The remaining point of the orbit is projected outside the plotted region and is not listed.}
    \label{tab:maxV_cube}
\end{table}

\begin{table}[htbp]
    \centering
    \footnotesize
    \setlength{\tabcolsep}{3pt}
    \begin{tabular}{r r r r r r r r r r r r}
        \hline
        \# & $\vp_1$ & $\vp_2$ & \# & $\vp_1$ & $\vp_2$ & \# & $\vp_1$ & $\vp_2$ & \# & $\vp_1$ & $\vp_2$\\
        \hline
        1 & $-3.01594$ & $-2.53373$ & 16 & $0.32440$ & $1.03204$ & 31 & $-1.69982$ & $-0.71700$ & 46 & $-0.30432$ & $-0.61473$\\
        2 & $1.97421$ & $7.38647$ & 17 & $-1.33839$ & $1.99866$ & 32 & $-0.15577$ & $-0.24585$ & 47 & $0.27655$ & $-1.36586$\\
        3 & $-0.10109$ & $0.02412$ & 18 & $0.23132$ & $-0.34543$ & 33 & $0.99424$ & $0.10296$ & 48 & $0.46451$ & $-0.62023$\\
        4 & $-1.09982$ & $0.22651$ & 19 & $0.43076$ & $-0.38510$ & 34 & $-0.59211$ & $-1.01369$ & 49 & $0.64680$ & $-2.60342$\\
        5 & $0.97240$ & $0.94754$ & 20 & $-0.63321$ & $0.15444$ & 35 & $-0.99512$ & $-0.10305$ & 50 & $-0.27011$ & $0.01392$\\
        6 & $-0.52750$ & $-0.51402$ & 21 & $-0.75417$ & $-0.69342$ & 36 & $0.10972$ & $-0.50160$ & 51 & $0.64680$ & $1.30654$\\
        7 & $0.42964$ & $0.73554$ & 22 & $-0.27718$ & $-0.88183$ & 37 & $0.76373$ & $-1.26487$ & 52 & $0.71853$ & $0.66065$\\
        8 & $-0.41618$ & $1.90260$ & 23 & $2.53716$ & $1.29164$ & 38 & $-3.74033$ & $-0.07979$ & 53 & $1.83897$ & $2.90240$\\
        9 & $0.19438$ & $0.16330$ & 24 & $0.45175$ & $0.01341$ & 39 & $-0.71399$ & $-0.07033$ & 54 & $-0.14240$ & $0.70331$\\
        10 & $1.04835$ & $-0.45980$ & 25 & $-1.29025$ & $1.15350$ & 40 & $-0.03377$ & $-0.12636$ & 55 & $1.49057$ & $-0.36355$\\
        11 & $-0.28105$ & $0.37385$ & 26 & $-0.79999$ & $0.35087$ & 41 & $0.87224$ & $-0.17964$ & 56 & $-0.02492$ & $-1.87650$\\
        12 & $-1.67224$ & $-1.57030$ & 27 & $0.22182$ & $-0.69851$ & 42 & $-0.31302$ & $-0.15935$ & 57 & $1.28478$ & $-1.70901$\\
        13 & $-0.77359$ & $1.03294$ & 28 & $0.26723$ & $0.00570$ & 43 & $-0.08988$ & $0.36178$ & 58 & $3.69243$ & $-0.19029$\\
        14 & $-0.41298$ & $1.30047$ & 29 & $1.38713$ & $0.13663$ & 44 & $0.49943$ & $0.21067$ & 59 & $-2.21167$ & $-0.06567$\\
        15 & $0.31778$ & $0.29841$ & 30 & $0.00707$ & $0.53281$ & 45 & $-0.34982$ & $0.57937$ &  &  & \\
        \hline
    \end{tabular}
    \caption{Maxima of $V$ for the icosahedron model, at which $V=6.15632$. The icosahedral rotation group has order 60. The remaining point of the orbit is projected outside the plotted region and is not listed.}
    \label{tab:maxV_icosa}
\end{table}

\begin{table}[htbp]
    \centering
    \footnotesize
    \setlength{\tabcolsep}{3pt}
    \begin{tabular}{r r r r r r r r r r r r}
        \hline
        \# & $\vp_1$ & $\vp_2$ & \# & $\vp_1$ & $\vp_2$ & \# & $\vp_1$ & $\vp_2$ & \# & $\vp_1$ & $\vp_2$\\
        \hline
        1 & $0.64118$ & $1.46917$ & 16 & $0.20505$ & $-1.25140$ & 31 & $0.78259$ & $-1.11897$ & 46 & $0.06977$ & $0.53441$\\
        2 & $7.52936$ & $-6.61254$ & 17 & $0.23883$ & $-0.03771$ & 32 & $-0.56642$ & $0.16424$ & 47 & $-0.41972$ & $0.60013$\\
        3 & $-4.08525$ & $0.64511$ & 18 & $-0.63049$ & $-1.13043$ & 33 & $-0.24018$ & $-1.83985$ & 48 & $-0.19353$ & $-0.91751$\\
        4 & $-2.90228$ & $-3.43937$ & 19 & $-0.20042$ & $2.00827$ & 34 & $-0.12752$ & $0.77821$ & 49 & $0.48699$ & $-0.03452$\\
        5 & $0.04920$ & $-0.49303$ & 20 & $1.53810$ & $-1.59142$ & 35 & $3.19640$ & $-0.58587$ & 50 & $0.22010$ & $1.04348$\\
        6 & $0.18881$ & $-0.76809$ & 21 & $-1.40244$ & $0.99555$ & 36 & $-0.24953$ & $-0.57176$ & 51 & $-1.38991$ & $-1.62701$\\
        7 & $0.78840$ & $-0.18359$ & 22 & $-0.30268$ & $0.05548$ & 37 & $0.01511$ & $-0.13469$ & 52 & $-0.07498$ & $0.06585$\\
        8 & $1.11542$ & $0.92575$ & 23 & $0.30354$ & $0.35532$ & 38 & $1.62856$ & $-0.47222$ & 53 & $0.62983$ & $-3.06073$\\
        9 & $-0.36416$ & $-0.18155$ & 24 & $0.37633$ & $0.67472$ & 39 & $-2.04315$ & $0.14482$ & 54 & $-0.74127$ & $0.91441$\\
        10 & $-1.20315$ & $0.28017$ & 25 & $0.73518$ & $0.56919$ & 40 & $0.93048$ & $0.17318$ & 55 & $0.55793$ & $0.23859$\\
        11 & $-0.31400$ & $0.32489$ & 26 & $1.45467$ & $0.26972$ & 41 & $-1.03873$ & $-0.19332$ & 56 & $0.98874$ & $-0.55455$\\
        12 & $-0.06450$ & $0.31345$ & 27 & $0.21107$ & $-0.29312$ & 42 & $-0.85044$ & $-0.65843$ & 57 & $1.32723$ & $2.69534$\\
        13 & $2.19942$ & $1.09651$ & 28 & $-1.61780$ & $2.24664$ & 43 & $0.53497$ & $-0.65992$ & 58 & $-1.51524$ & $-0.64798$\\
        14 & $0.14330$ & $0.16982$ & 29 & $-0.53085$ & $-0.44059$ & 44 & $-0.82228$ & $7.33208$ & 59 & $-0.30180$ & $1.22774$\\
        15 & $-0.66460$ & $-0.12323$ & 30 & $0.47413$ & $-0.33657$ & 45 & $-0.76936$ & $0.43151$ & 60 & $-0.14704$ & $-0.29861$\\
        \hline
    \end{tabular}
    \caption{Maxima of $V$ for the dodecahedron model, at which $V=10.62609$. The icosahedral rotation group has order 60.}
    \label{tab:maxV_dode}
\end{table}

\FloatBarrier

\end{document}